\documentclass[12pt,a4paper]{article}
\usepackage{amsmath,amsfonts,amssymb,tikz}
\usetikzlibrary{calc,cd}
\usepackage[utf8]{inputenc}
\numberwithin{equation}{section}

\usepackage{upgreek}
\usepackage{color}
\usepackage{array}
\usepackage[maxnames=10, sorting=none]{biblatex}
\usepackage[pdfencoding=unicode]{hyperref}

\newcommand{\ri}{\mathrm{i}}
\newcommand{\st}{\mathrm{t}}
\newcommand{\sm}{\mathrm{m}}
\newcommand{\se}{\mathrm{e}}
\newcommand{\VI}{\scalebox{.6}{VI}}
\newcommand{\V}{\scalebox{.6}{V}}
\newcommand{\sV}{\scalebox{.4}{V}}
\newcommand{\IV}{\scalebox{.6}{IV}}
\newcommand{\sIV}{\scalebox{.4}{IV}}
\newcommand{\III}{\scalebox{.6}{III}}
\newcommand{\sIII}{\scalebox{.4}{III}}
\newcommand{\II}{\scalebox{.6}{II}}
\newcommand{\sII}{\scalebox{.4}{II}}
\newcommand{\I}{\scalebox{.6}{I}}
\newcommand{\sI}{\scalebox{.4}{I}}

\newcommand{\NS}{\scalebox{.5}{NS}}

\newcommand{\D}{\scriptscriptstyle{D}}

\newcolumntype{M}[1]{>{\centering\arraybackslash}m{#1}}

\title{Refined Painlev\'e/gauge theory correspondence\\and quantum tau functions
}
\author{Giulio Bonelli\thanks{bonelli@sissa.it},\qquad Anton Shchechkin\thanks{shch145@gmail.com},\qquad Alessandro Tanzini\thanks{tanzini@sissa.it}\\{\small
International School for Advanced Studies (SISSA), via Bonomea 265, 34136 Trieste, Italy}\\ {\small National Institute for Nuclear Physics (INFN), Section of Trieste, 
via Valerio 2, 34127 Trieste, Italy}\\ {\small Institute for Geometry and Physics (IGAP), via Beirut 2, 34151 Trieste, Italy}}

\date{}

\begin{document}

\maketitle

\begin{abstract}

In this paper we study strong coupling asymptotic expansions
of $\mathcal{N}=2$ four-dimensional $SU(2)$ gauge theory partition functions in general $\Omega$-background.
This is done by refining the Painlev\'e/gauge theory correspondence
in terms of quantum Painlev\'e equations, obtained from $\mathbb{C}^2/\mathbb{Z}_2$ blowup relations. We present a general ansatz and a systematic analysis of the expansions of the gauge theory partition functions by solving the above equations around the strong coupling singularities, including Argyres-Douglas points. We check our results via refined holomorphic anomaly equations and compare with irregular Virasoro conformal blocks.  

\end{abstract}

\begin{quote}
\tableofcontents
\end{quote}

\section{Introduction}
\label{sec:intro}

Since the seminal papers \cite{GIL12, GIL13}, the connection between 
supersymmetric gauge theories and Painlev\'e differential equations has been established with increasing detail \cite{BS18,BGT16,BLMST16,GMS20,N20}.

The main statement is that the Zak transform of a gauge theory partition function 
in self-dual $\Omega$-background
solves some Painlev\'e equation in the Hirota bilinear tau form, the specific equation being determined by the specific gauge theory under consideration.

The correspondence can be looked at from different viewpoints. In the regime in the parameter space where the gauge theory multi-instanton counting algorithm \cite{N02} gives explicit expansions for 
the partition functions --- namely in the small time regime --- it offers a novel way to explicitly formulate 
closed convergent expansions for the Painlev\'e tau functions.
On the other hand, the fact that the partition function of a given theory obeys the corresponding Painlev\'e equation can be used to calculate the partition function itself in strongly coupled regimes, which are typically inaccessible with perturbative methods, unless
some strong\,/\,weak coupling duality is available.

This latter viewpoint is the one leading to learn about
strongly coupled fixed points of the gauge theory, such as Argyres-Douglas theories \cite{AD95}.
This was the main subject of \cite{BLMST16}.
It looks important to extend these results to general $\Omega$-background in order to have access to other regimes of the gauge theory, such as refined topological strings, gauge theories on compact manifolds 
\cite{P07, ICMP, HH12, BBRT14, BBRT15, BBRT16, BFMRST20},
and the study of the Nekrasov-Shatashvili limit \cite{NS09}.
In this paper we propose that this can be achieved 
by formulating the correspondence in terms of a proper quantum deformation of Painlev\'e equations.
The outcome is a plethora of novel well defined presentations of the gauge theory partition functions in general $\Omega$-background at strong coupling. 

One of the proofs \cite{BS14} of the Painlev\'e/gauge correspondence is based on blowup equations for supersymmetric gauge theories on (the minimal resolution of) $\mathbb{C}^2/\mathbb{Z}_2$.
 These were studied in \cite{BMT1107,BMT1106,BBFLT11}
 in the context of a super Liouville extension of AGT correspondence by 
 generalizing the blowup equations on $\mathbb{C}^2$
 of Nakajima and Yoshioka \cite{NY03L}.
 The $\mathbb{C}^2/\mathbb{Z}_2$
 blowup relations in general $\Omega$-background provide a set of deformed bilinear relations which we propose as a quantization of the 
 corresponding Painlev\'e equations in the Hirota bilinear tau form.
 
Starting from a $\mathbb{C}^2/\mathbb{Z}_2$ blowup equation for the $SU(2)$ $N_f=4$ gauge theory, which is a quantum deformation of Painlev\'e VI, one can follow the renormalization group flow by integrating out heavy degrees of freedom and access nonconformal Lagrangian theories or Argyres-Douglas (AD) strongly coupled points in the gauge theory moduli space. We realize the counterpart of this renormalization group flow in terms of a coalescence of quantum Painlev\'e equations modeled on the classical one. This allows us to systematically formulate quantum Painlev\'e equations in bilinear tau forms for the whole set of gauge theories that can be reached via the renormalization group flow.

For the set of fixed points admitting a Lagrangian formulation,
we provide weak coupling\,/\,small time expansions of the quantum Painlev\'e tau functions by a suitable quantum deformation of the Zak transform ansatz (following the lines of \cite{BGM17}), in which the monodromy parameters describing the initial data satisfy the canonical commutation algebra \eqref{as_comm_rel}. 
Due to the noncommutativity of the monodromy space parameters, the solutions we obtain are operator-like Zak transforms, while the corresponding partition function has a nonzero finite or infinite radius of convergence \cite{ABT22,ILTy14}.

The central result of this paper is a set of strong coupling\,/\,large time tau function expansions of quantum Painlev\'e equations both for the Lagrangian theories with $0\leq N_f\leq4$ in the magnetic frame and for the AD points, which in this setting admit only strong coupling\,/\,large time nontrivial expansions. All this follows by a suitable quantum deformation of the ansatz for the self-dual $\Omega$-background case proposed in \cite{BLMST16} and generalizes the results of \cite[Sec. 7, App. D]{GMS20} 
where the strong coupling\,/\,large time expansion of the quantum Painlev\'e III$_3$ ($N_f=0$) tau function was given up to the seventh order. 

Contrary to the weakly coupled cases, these strong coupling expansions are only asymptotic and display a nontrivial structure of canonical rays (see Fig.~\ref{fig3}) related to that of the associated linear system. By deforming the solutions of the self-dual case \cite{BLMST16}, we study two types of strong coupling expansions which are characterized by a linear or a quadratic exponential asymptotic behavior in the expansion variable, see e.g. Sections \ref{ssec:exp_3l} and \eqref{ssec:exp_3s} respectively for the quantum PV case. From the gauge theory viewpoint, these two classes of expansions correspond to the presence of multiple or single light particles respectively in the spectrum of the theory in the magnetic frame.

Let us notice that Nekrasov-like combinatorial formulae are not available at the moment for the strong coupling expansions, the only attempt as far as we know is \cite[Conj. 5.3]{N16} for $N_f=3$ case.
There are also no simple analogs of the holomorphic decoupling limits of the massive hypermultiplets. Limit from $N_f=4$ to the strong coupling of $N_f=3$ was studied in \cite{LNR18} and, further, to the $N_f=2$ symmetric case --- in \cite{BIPT22}. In order to check our formulae we therefore use the refined holomorphic anomaly equations for the topological string amplitudes geometrically engineering the gauge theory \cite{HKK11}. In revisiting these formulae we obtain a novel compact conjectural form \eqref{F2} for the general genus two amplitude of elliptic mirror geometry. 

A complementary approach to these expansions is provided via the AGT correspondence \cite{AGT09} and the description of the Painlev\'e tau functions in terms of overlap of irregular states \cite{GT12,NU19,PP23,IILZ25,HNNT24,PP25,G09,BMT1112,FMP23} and/or irregular vertex operators \cite{N15,N18} of the Virasoro algebra.
We use this to perform further checks of our formulas for the linear/multiple light particles asymptotic expansion cases. Let us remark that a Liouville description of the quadratic/single light particle case is not known in the literature. 

\paragraph{Structure of the paper.}
In Section~\ref{sec:blowup_part_funct} we review the $\mathbb{C}^2/\mathbb{Z}_2$ blowup relations for $N_f{=}4$ SUSY $\mathcal{N}{=}2$ $SU(2)$ partition functions, writing them for arbitrary $\Omega$-background as bilinear relations on quantum tau functions.
In Section~\ref{sec:coalesc} we give a detailed description of the coalescence for the blowup relations corresponding to the renormalization group flows to the AD superconformal $H_{0,1,2}$ points and to the Lagrangian theories with $N_f<4$ (see Fig.~\ref{fig:gauge_coalescence}).
In Section~\ref{sec:quant_exp} we describe the general ansatz we use to form the solutions at the strong coupling regime and
compute the particular ones case by case.
In Section~\ref{sec:holom_anom} we review the holomorphic anomaly setting, give the general genus two amplitude and check the expansions found in the previous section, while in Section~
\ref{sec:comparison} we compare our results with two CFT approaches.
Section~\ref{sec:further} is devoted to remarks and open questions. In Appendix~\ref{sec:notations} we provide some reminders about notations.
In Appendix~\ref{sec:gamma} we review the double Gamma function and its properties, in Appendix~\ref{sec:BS14} we review how 
$\mathbb{C}^2/\mathbb{Z}_2$ blowup relations
are obtained in terms of Virasoro and Super Virasoro representation theory
and in Appendix~\ref{sec:Painleve} we review details of the classical Painlev\'e equations coalescence.

\paragraph{Acknowledgements.}
We would like to thank Mikhail Bershtein, Irina Bobrova,  Pavlo Gavrylenko, Amir Kashani-Poor,
Oleg Lisovyy, Hajime Nagoya, Rubik Poghossian for fruitful and insightful discussions.
We are especially grateful to Amir Kashani-Poor, Oleg Lisovyy, Hajime Nagoya and Rubik Poghossian who provided us with their computations, giving us opportunity to further compare our results with theirs.
We are grateful to Mikhail Bershtein for the reading of the paper and for his suggestions.

A preliminary version of the results contained in this paper was presented at the poster session of StringMath24 by A.S., who thanks the participants for comments and discussions. 

This research is partially supported by the INFN Research Projects GAST and ST$\&$FI, by 
MIUR PRIN Grant 2020KR4KN2 "String Theory as a bridge between Gauge Theories and Quantum Gravity".  
G.B. and A.T. acknowledge funding from the EU project Caligola HORIZON-MSCA-2021-SE-01), Project ID: 101086123 and
COST Action CaLISTA CA21109 supported by COST (European Cooperation in Science and Technology).
A.T. acknowledges GNFM INdAM project. 

\section{\texorpdfstring{$\mathbb{C}^2/\mathbb{Z}_2$}{ℂ²/ℤ₂} blowup relations}\label{sec:blowup_part_funct}
\subsection{\texorpdfstring{$\mathcal{N}=2$ $SU(2)$}{N=2 SU(2)} partition functions}
\label{ssec:part_funct}

\paragraph{Explicit definitions.} The full $\mathcal{N}=2$ four-dimensional SUSY $SU(2)$ partition function $\mathcal{Z}^{[N_f]}$ 
with $0\leq N_f\leq4$ hypermultiplets in the (anti-) fundamental representation of the gauge group is given by the product of three factors
\begin{equation}\label{Zstr}
\mathcal{Z}^{[N_f]}=\mathcal{Z}_{cl}^{[N_f]}\mathcal{Z}^{[N_f]}_{1-loop}\mathcal{Z}^{[N_f]}_{inst}.
\end{equation}
It depends on parameters $\epsilon_1, \epsilon_2$ of the $\Omega$-background (we also denote $\epsilon=\epsilon_1+\epsilon_2$),  complexified gauge coupling $t$,  vacuum expectation value $(a_1, a_2)=(a,-a)$ and mass variables $m_1,\ldots m_{N_f}$, together denoted below as $m_{1,\ldots N_f}$. Notice that in this subsection we mainly follow~\cite{AGT09}. 
 
The above factors of the partition function are given in the weak coupling by formulas \footnote{We use the physical masses $m_f$ which are connected with the ones used in \cite{AGT09} by $\mu_f=\epsilon/2+m_f|_{f=1,2}, \, \,\mu_f=\epsilon/2-m_f|_{f=3,4}$. 
We also use a slightly different form for the classical and 1-loop parts with respect to that in \cite{AGT09}.} 
	\begin{align}\label{Zcl}
	&\mathcal{Z}^{[N_f]}_{cl}(a;\epsilon_1,\epsilon_2|t)=t^{\frac{\epsilon^2/4-a^2}{\epsilon_1\epsilon_2}},\\ \label{Z1loop}
&\mathcal{Z}^{[N_f]}_{1-loop}(a,m_{1,\ldots N_f};\epsilon_1,\epsilon_2)=\prod\limits_{\pm}\frac{\prod\limits_{f=1}^{N_f}\exp\gamma_{\epsilon_1,\epsilon_2}(m_f\pm a-\epsilon/2)}{\exp\gamma_{\epsilon_1,\epsilon_2}(\pm 2a)}, \\ \label{Zinst}
	&\mathcal{Z}^{[N_f]}_{inst}(a,m_{1,\ldots N_f};\epsilon_1,\epsilon_2|t)=\sum_{Y^+,Y^-}\frac{\prod\limits_{f=1}^{N_f} Z_{fund}(a,m_f;\epsilon_1,\epsilon_2|Y^+,Y^-)}{Z_{vec}(a;\epsilon_1,\epsilon_2|Y^+,Y^-)} t^{|Y^+|+|Y^-|}.
 \end{align}
 1-loop factor $\mathcal{Z}^{[N_f]}_{1-loop}$ is expressed in terms of  double gamma function $\gamma_{\epsilon_1,\epsilon_2}(x)$ presented in App.~\ref{sec:gamma}. 
 In instanton counting factor $\mathcal{Z}_{inst}^{[N_f]}$ the sum runs over all pairs of integer partitions $Y^{+}, Y^{-}$ whose number of boxes is denoted by $|Y^{\pm}|$. 
 The factors in each summand of $\mathcal{Z}^{[N_f]}_{inst}$ are given by
 \begin{equation}
   Z_{fund}(a,m;\epsilon_1,\epsilon_2|Y^+,Y^-)=\prod_{\pm}\prod_{(i,j)\in Y^{\pm}} (m\pm a-\epsilon/2+\epsilon_1 i+\epsilon_2 j),
   \end{equation}
   \vspace{-0.7cm}
    \begin{multline}
  Z_{vec}(a;\epsilon_1,\epsilon_2|Y^+,Y^-)=\prod_{\pm}\left(\prod_{(i,j)\in Y^{\pm}}\Big(\epsilon_2 (Y^{\pm}_i{-}j{+}1)-\epsilon_1 (\tilde{Y}^{\pm}_j{-}i)\Big)\prod_{(i,j)\in Y^{\pm}}\Big(\epsilon_1 (\tilde{Y}^{\pm}_j{-}i{+}1)-\epsilon_2 (Y^{\pm}_i{-}j)\Big)\right.  \\ \times  \left.\prod_{(i,j)\in Y^{\pm}}\Big(\pm 2a+\epsilon_2 (Y^{\pm}_i{-}j{+}1)-\epsilon_1 (\tilde{Y}^{\mp}_j{-}i)\Big)\prod_{(i,j)\in Y^{\mp}}\Big(\pm 2a+\epsilon_1 (\tilde{Y}^{\pm}_j{-}i{+}1)-\epsilon_2 (Y^{\mp}_i{-}j)\Big)\right),
\end{multline} 
where $\tilde{Y}$ denotes the transposed of $Y$. 

\paragraph{Symmetries and homogeneity.} These classical, 1-loop and instanton factors are symmetric by construction under permutations of the masses $m_{1,\ldots N_f}$, $\epsilon_1\leftrightarrow\epsilon_2$\footnote{This symmetry is not visible directly from \eqref{Zinst}.}, and $a\mapsto-a$, actually the symmetry is even bigger, see, e.g., discussion related to expression \eqref{symm_form}. Also the partition function is a homogeneous function of $a$, $\epsilon_{1,2}$, $m_{1,\ldots N_f}$ and $t$. More precisely, for any $\Lambda\in\mathbb{C}$ we have
\begin{equation}
\mathcal{Z}^{[N_f]}_{cl}(\Lambda a;\Lambda \epsilon_1,\Lambda \epsilon_2|\Lambda^{4{-}N_f}t)=\Lambda^{(4{-}N_f)\frac{\epsilon^2/4-a^2}{\epsilon_1\epsilon_2}}\mathcal{Z}^{[N_f]}_{cl}(a;\epsilon_1,\epsilon_2|t),
\end{equation}
and, due to \eqref{gamma_homogen},
\begin{equation}\label{Z1loop_scale}
\mathcal{Z}^{[N_f]}_{1-loop}(\Lambda a,\Lambda m_1,\ldots \Lambda m_{N_f};\Lambda\epsilon_1,\Lambda\epsilon_2)=\Lambda^{\frac{1{-}N_f}6+\frac{(4{-}N_f)a^2-w_2^{[N_f]}+\frac{N_f{+}2}{12}\epsilon^2}{\epsilon_1\epsilon_2}}\mathcal{Z}^{[N_f]}_{1-loop}(a,m_1,\ldots m_{N_f};\epsilon_1,\epsilon_2)    
\end{equation}
and
\begin{equation}\label{Zinst_scale}
\mathcal{Z}^{[N_f]}_{inst}(\Lambda a,\Lambda m_1,\ldots \Lambda m_{N_f};\Lambda\epsilon_1,\Lambda\epsilon_2|\Lambda^{4{-}N_f}t)=\mathcal{Z}^{[N_f]}_{inst}(a,m_1,\ldots m_{N_f};\epsilon_1,\epsilon_2|t)    
\end{equation}
due to
\begin{align}
 Z_{fund}(\Lambda a,\Lambda m;\Lambda\epsilon_1,\Lambda\epsilon_2|Y^+,Y^-)&=\Lambda^{|Y^+|+|Y^-|}Z_{fund}(a,m;\epsilon_1,\epsilon_2|Y^+,Y^-),\\
 Z_{vec}(\Lambda a;\Lambda\epsilon_1,\Lambda\epsilon_2|Y^+,Y^-)&=\Lambda^{4(|Y^+|+|Y^-|)}Z_{vec}(a;\epsilon_1,\epsilon_2|Y^+,Y^-).
\end{align}
In the following we keep in mind this property and refer to it as the 
(mass) dimension. So $a$, $\epsilon_{1,2}$, $m_{1,\ldots N_f}$ have mass dimension $1$ and $t$ has mass dimension $4{-}N_f$. 

\subsection{\texorpdfstring{$\mathbb{C}^2/\mathbb{Z}_2$}{ℂ²/ℤ₂} blowup relations for the \texorpdfstring{$N_f=4$}{N\_f=4} partition function}
\label{ssec:blowup}

\paragraph{Formulation.}
The $N_f=4$ SUSY $\mathcal{N}=2$ $SU(2)$ instanton partition function equals the $4$-point Virasoro regular conformal block~\cite{AGT09}
upon a proper identification of the parameters (in App.~\ref{sec:BS14} it is \eqref{AGT_block} under the dictionary \eqref{AGT_dictionary}, \eqref{AGT_dictionary_masses}). 
In~\cite{BS14}, by exploiting the representation theory of $\mathcal{N}=1$ Super Virasoro algebra, some bilinear relations on $4$-point Virasoro conformal blocks were obtained. 
These relations have the form of $\mathbb{C}^2/\mathbb{Z}_2$ blowup relations\footnote{Note that the form of (one side of) $\mathbb{C}^2$ blowup relations differs from \eqref{blowup_gen_form_split} in the arguments of the partition functions inside the bilinear operator: factors $2$ is absent in the shift of $a$ and in the $\Omega$-background, see, e.g., \cite[(5.2, 5.3)]{NY03L}.} already mentioned in Introduction and read 
\begin{equation}\label{blowup_gen_form_split}
\sum_{n\in\mathbb{Z}+\frac{\mathfrak{p}}2}\mathfrak{D}_{\epsilon_1,\epsilon_2}^{[4],k}\Bigl(\mathcal{Z}^{[4]}(a+2n\epsilon_1;m_{1,2,3,4};2\epsilon_1,\epsilon_2{-}\epsilon_1|t), \mathcal{Z}^{[4]}(a+2n\epsilon_2;m_{1,2,3,4};\epsilon_1{-}\epsilon_2,2\epsilon_2|t) \Bigr)=0, \quad \mathfrak{p}=0,1,
\end{equation}
 where $\mathfrak{D}^{[4],k}_{\epsilon_1,\epsilon_2}$ are bilinear differential operators  of order $k=1,3,4$ in $\ln t$. They are given by
\begin{equation}\label{D13}
\mathfrak{D}^{[4],1}_{\epsilon_1,\epsilon_2}=D^1_{\epsilon_1,\epsilon_2[\ln t]}, \qquad
\mathfrak{D}^{[4],3}_{\epsilon_1,\epsilon_2}=D^3_{\epsilon_1,\epsilon_2[\ln t]}-\epsilon\,\frac{1+t}{1-t}D^2_{\epsilon_1,\epsilon_2[\ln t]},
\end{equation}
\begin{multline}\label{D4}
\mathfrak{D}^{[4],4}_{\epsilon_1,\epsilon_2}=D^4_{\epsilon_1,\epsilon_2[\ln t]}+2\,\frac{1+t}{1-t} \left(\epsilon_1\epsilon_2\frac{d}{d\ln t}\right)D^2_{\epsilon_1,\epsilon_2[\ln t]}+\frac{t}{(1-t)^2}\left(\epsilon_1\epsilon_2\frac{d}{d\ln t}\right)^2D^0_{\epsilon_1,\epsilon_2[\ln t]}\\
-\frac{t^2}{1-t}\left(\frac{\epsilon_1\epsilon_2+\epsilon^2}{t^2}+\frac{e_2^{[4]}+e_1^{[4]}\epsilon+6\epsilon^2}t+\frac{\frac14e_1^{[4]}(e_1^{[4]}+2\epsilon)+\epsilon_1\epsilon_2+6\epsilon^2}{1-t}\right)D^2_{\epsilon_1,\epsilon_2[\ln t]}
\\ 
-\frac{t^2}{4(1-t)^2}\left(\frac{e_1^{[4]}(e_1^{[4]}+2\epsilon)}{1-t}+\frac{2(e_2^{[4]}+e_1^{[4]}\epsilon+\epsilon^2)}t\right)\left(\epsilon_1\epsilon_2\frac{d}{d\ln t}\right)D^0_{\epsilon_1,\epsilon_2[\ln t]}+
\\
+\frac{t^2}{4(1-t)^2}\left(\frac{e_4^{[4]}}t+\frac{e_1^{[4]}(e_3^{[4]}+e_2^{[4]}\epsilon+e_1^{[4]}\epsilon^2+\epsilon^3)}{2(1-t)}\right) D^0_{\epsilon_1,\epsilon_2[\ln t]},
\end{multline}
where we used the generalized Hirota differential operators
$D^n_{\epsilon_1,\epsilon_2[x]}$ 
defined by the power expansion
\begin{equation}\label{Hirota}
f(x+\epsilon_1 y)\, g(x+\epsilon_2 y)=\sum_{n=0}^{+\infty} D^n_{\epsilon_1,\epsilon_2[x]} (f(x),g(x))\, \frac{y^n}{n!}.  
\end{equation}
Operator $\mathfrak{D}^{[4],4}_{\epsilon_1,\epsilon_2}$ depends on the masses via elementary symmetric polynomials $e_{1,2,3,4}^{[4]}(m_{1,2,3,4})$
\begin{multline}\label{e1234}
e_1^{[4]}(m_{1,2,3,4})=m_1{+}m_2{+}m_3{+}m_4, \qquad  e_2^{[4]}(m_{1,2,3,4})=m_1m_2{+}m_1m_3{+}m_1m_4{+}m_2m_3{+}m_2m_4{+}m_3m_4, \\
e_3^{[4]}(m_{1,2,3,4})=m_1m_2m_3{+}m_1m_2m_4{+}m_1m_3m_4{+}m_2m_3m_4,\qquad e_4^{[4]}(m_{1,2,3,4})=m_1m_2m_3m_4.
\end{multline}
In~\cite{BS14} relations \eqref{blowup_gen_form_split} in terms of the Virasoro conformal blocks are given by formulas (A.3), (A.5), (4.35) for $k=1,3,4$ respectively. The precise connection of \eqref{blowup_gen_form_split} with \cite{BS14} is described in details in App.~\ref{sec:BS14}.

\paragraph{$\mathbb{C}^2/\mathbb{Z}_2$ blowup relations on the instanton part.}
Note that $\mathbb{C}^2/\mathbb{Z}_2$ blowup relations \eqref{blowup_gen_form_split} can be rewritten as bilinear relations just on instanton part $\mathcal{Z}^{[4]}_{inst}$.
Indeed, by substituting \eqref{Zstr} into \eqref{blowup_gen_form_split} with $\mathcal{Z}^{[4]}_{cl}$ given by \eqref{Zcl} and $\mathcal{Z}^{[4]}_{1-loop}$ given by \eqref{Z1loop}, one obtains relations
\begin{equation}\label{blowup_gen_form_inst}
\sum_{n\in\mathbb{Z}+\frac{\mathfrak{p}}2}\mathfrak{l}_n^{[4]}\, t^{2n^2}\,\mathfrak{D}_{\substack{\epsilon_1,\epsilon_2\\inst}}^{[4],k}\Bigl(\mathcal{Z}_{inst}^{[4]}(a+2n\epsilon_1;m_{1,2,3,4};2\epsilon_1,\epsilon_2{-}\epsilon_1|t), \mathcal{Z}_{inst}^{[4]}(a+2n\epsilon_2;m_{1,2,3,4};\epsilon_1{-}\epsilon_2,2\epsilon_2|t) \Bigr)=0, 
\end{equation}
where bilinear operators $\mathfrak{D}^{[4],k}_{\substack{\epsilon_1,\epsilon_2\\inst}}$ are given by
\begin{multline}
\mathfrak{D}^{[4],1}_{\substack{\epsilon_1,\epsilon_2\\inst}}=D^1_{\epsilon_1,\epsilon_2[\ln t]}+2n(a+n\epsilon),\qquad
\mathfrak{D}^{[4],3}_{\substack{\epsilon_1,\epsilon_2\\inst}}=(1{-}t)D^3_{\epsilon_1,\epsilon_2[\ln t]}-\big(6n(a{+}n\epsilon)(t{-}1)+\epsilon(1{+}t)\big)D^2_{\epsilon_1,\epsilon_2[\ln t]}\\+4n^2(a{+}n\epsilon)^2\big(4n(a{+}n\epsilon)(t{-}1)+\epsilon(1{+}t)\big)D^0_{\epsilon_1,\epsilon_2[\ln t]},
\end{multline}
\begin{multline}
\mathfrak{D}^{[4],4}_{\substack{\epsilon_1,\epsilon_2\\inst}}= (1{-}t)^3 D^4_{\epsilon_1,\epsilon_2[\ln t]}+2(1{+}t)(1{-}t)^2 \left(\epsilon_1\epsilon_2\frac{d}{d\ln t}\right)D^2_{\epsilon_1,\epsilon_2[\ln t]}
+(1{-}t)\left(\sum\limits_{i=0}^2 f_{2,i}t^i\right) D^2_{\epsilon_1,\epsilon_2[\ln t]}
\\ +t(1{-}t)\left(\epsilon_1\epsilon_2\frac{d}{d\ln t}\right)^2D^0_{\epsilon_1,\epsilon_2[\ln t]}+\left(\sum\limits_{i=0}^3f_{1,i}t^i\right)\left(\epsilon_1\epsilon_2\frac{d}{d\ln t}\right)D^0_{\epsilon_1,\epsilon_2[\ln t]}+\left(\sum\limits_{i=0}^3f_{0,i}t^i\right) D^0_{\epsilon_1,\epsilon_2[\ln t]},
\end{multline}
where $f_{k,i}$ are certain (too cumbersome to write them here) polynomials in $e_{1,2,3,4}^{[4]},\, \epsilon,\, \epsilon_1\epsilon_2$ and $a,n$ as well. 
Blowup factors $\mathfrak{l}_n^{[4]}$ are the outcome of the $1$-loop part and are calculated using \eqref{gammablowup}
as
\begin{align}\label{blowup_coeff_def}
\mathfrak{l}_n^{[4]}(a;m_{1,2,3,4};\epsilon_1,\epsilon_2)=C_{\mathfrak{p}(2n)}\mathcal{Z}^{[4]}_{1-loop}(a+2n\epsilon_1;2\epsilon_1,\epsilon_2{-}\epsilon_1)\mathcal{Z}^{[4]}_{1-loop}(a+2n\epsilon_2;\epsilon_1{-}\epsilon_2,2\epsilon_2)=\\ \label{blowup_coeff4}
=\prod\limits_{\pm}\frac{\prod\limits_{f=1}^4 \prod\limits_{\mathrm{reg}(2n)} \Big(m_f\pm \big(a+\mathrm{sgn}(n)(i\epsilon_1+j\epsilon_2)\big)\Big)}{\prod\limits_{\mathrm{reg}(4n)}\Big(\epsilon/2\pm \big(2a+\mathrm{sgn}(n)(i\epsilon_1+j\epsilon_2)\big)\Big)},   
\end{align}
where prefactor $C_{\mathfrak{p}(2n)}$ depends only on the parity $\mathfrak{p}$ of $2n$ (so it could be removed from \eqref{blowup_gen_form_inst}) and the regions of the above products are given by
\begin{equation}
\mathrm{reg}(n)=\Big\{i,j\in \mathbb{Z}_{\geq 0}+\frac12: \quad
i+j\leq |n|-1, \quad i+j\equiv n+1 \mod 2\Big\}.    
\end{equation}

\subsection{\texorpdfstring{$\mathbb{C}^2/\mathbb{Z}_2$}{ℂ²/ℤ₂} blowup relations in terms of quantum tau functions}
\label{ssec:blowuptau}
\paragraph{Classical tau function approach.}
For $\epsilon=0$ let us introduce a tau function as the Zak transform of the $N_f=4$ partition function with parameter $\eta\in{\mathbb C}$
\begin{equation}\label{tau_c=1}
\tau^{[4]}_{\epsilon{=}0}(a,\eta;m_{1,2,3,4}|t)=\sum_{n\in\mathbb{Z}}e^{\ri n \eta } \mathcal{Z}^{[4]}(a+2n \epsilon_2;m_{1,2,3,4};-2\epsilon_2, 2\epsilon_2|t).
\end{equation}
We can rewrite $\mathbb{C}^2/\mathbb{Z}_2$ blowup relations \eqref{blowup_gen_form_split} for $\epsilon=0$ as bilinear differential equations on the above tau functions \eqref{tau_c=1}, namely 
\begin{equation}\label{blowup_gen_form_tau}
 \mathfrak{D}^{[4],k}_{-\epsilon_2,\epsilon_2}\Bigl(\tau^{[4]}_{\epsilon{=}0}(t), \tau^{[4]}_{\epsilon{=}0}(t) \Bigr)=0.
\end{equation}
Indeed, substituting \eqref{tau_c=1} to these equations and collecting terms with the same power of $e^{\ri \eta}$ we obtain
\begin{equation}
\mspace{-11mu}\sum_{N\in\mathbb{Z}}e^{\ri N \eta }\sum_{n\in\mathbb{Z}} \mathfrak{D}^{[4],k}_{-\epsilon_2,\epsilon_2}\Bigl(\mathcal{Z}^{[4]}(a+2(N{-}n) \epsilon_2;m_{1,2,3,4};-2\epsilon_2, 2\epsilon_2|t), \mathcal{Z}^{[4]}(a+2n \epsilon_2;m_{1,2,3,4};-2\epsilon_2, 2\epsilon_2|t)\Bigr)=0.   
\end{equation}
We can redefine $a\mapsto a-N\epsilon_2$ and $n\mapsto n+N/2$ to obtain only two different relations depending on the parity $\mathfrak{p}=N\mod 2$, that is
\begin{equation}
\sum_{n\in\mathbb{Z}+\frac{\mathfrak{p}}2} \mathfrak{D}^{[4],k}_{-\epsilon_2,\epsilon_2}\Bigl(\mathcal{Z}^{[4]}(a-2n \epsilon_2;m_{1,2,3,4};-2\epsilon_2, 2\epsilon_2|t), \mathcal{Z}^{[4]}(a+2n \epsilon_2;m_{1,2,3,4};-2\epsilon_2, 2\epsilon_2|t)\Bigr)=0,
\end{equation}
which is just \eqref{blowup_gen_form_split} for $\epsilon=0$. Note that the 
blowup relations for $k=1$ and $k=3$ in the self-dual case $\epsilon=0$ become trivial identities, as it can be seen from \eqref{blowup_gen_form_tau}.

Such a trick was used in \cite{BS14} to prove that \eqref{tau_c=1} gives a tau function of Painlev\'e VI equation, which was initially claimed in \cite{GIL12}. Indeed, \eqref{blowup_gen_form_tau} for $k=4$ is just
Painlev\'e VI in bilinear form 
\eqref{PVI_tau} while Painlev\'e VI tau function $\tau_{\VI}$ is obtained by the following redefinition and the mass rescaling
\begin{equation}\label{rdf_PVI}
\tau^{[4]}_{\epsilon{=}0}(t)= t^{\frac{\mathrm{w}_2^{[4]}}4} (1{-}t)^{-\frac{\se_1^{[4]2}}{4}}\tau_{\VI}(t), \quad \textrm{where} \quad
\se_1^{[4]}=\sum_{f=1}^4 \sm_f, \quad \mathrm{w}_2^{[4]}=\sum_{f=1}^4 \sm_f^2, \qquad m_f=2\epsilon_2\sm_f.
\end{equation}
Parameters $a$ and $\eta$ are integration constants of the Painlev\'e VI solution through ansatz \eqref{tau_c=1}. More details on Painlev\'e VI equation are given in App.~\ref{ssec:PVI}.

\paragraph{Quantum tau function approach.}
We can rewrite $\mathbb{C}^2/\mathbb{Z}_2$ blowup relations \eqref{blowup_gen_form_split} also for $\epsilon\neq0$ in terms of the Zak transformed partition functions if we canonically quantize variables $a,\eta$ labeling the initial conditions. Specifically, let us set a nontrivial commutation relation
\begin{equation}\label{as_comm_rel}
\ri [a,\eta]=\epsilon \quad \Rightarrow \quad a e^{\ri \eta}=e^{\ri \eta} (a+\epsilon).    
\end{equation}
Introducing the (ordered) Zak transform of a given function $\mathcal{Z}(a;\epsilon_1,\epsilon_2|t)$ with operator $e^{\ri\eta}$
\begin{equation}\label{tau_quantum}
\tau(a,\eta;\epsilon_1,\epsilon_2|t)=\sum_{n\in\mathbb{Z}}e^{\ri n \eta}\mathcal{Z}(a+n\epsilon_2;\epsilon_1,\epsilon_2|t),  \end{equation}
and substituting it into a bilinear relation
\begin{equation}\label{blowup_gen_form_tau_quantum}
\mathfrak{D}\Bigl(\tau(2\epsilon_1,\epsilon_2{-}\epsilon_1|t), \tau(\epsilon_1{-}\epsilon_2,2\epsilon_2|t) \Bigr)=0 
\end{equation}
given in terms of a specific bilinear differential operator $\mathfrak{D}$ in $t$, after collecting powers of $e^{\ri \eta}$, we obtain
\begin{equation}
\sum_{N\in\mathbb{Z}} e^{\ri N \eta} \sum_{n\in\mathbb{Z}} \mathfrak{D}\Bigl(\mathcal{Z}(a+(N{-}n) (\epsilon_2{-}\epsilon_1)+n \epsilon;2\epsilon_1,\epsilon_2{-}\epsilon_1|t),\mathcal{Z}(a+2n \epsilon_2;\epsilon_1{-}\epsilon_2,2\epsilon_2|t)\Bigr)=0.
\end{equation}
Under substitution $a\mapsto a-N\epsilon_2$ and $n\mapsto n+N/2$ the latter becomes
\begin{equation}\label{blowup_tau_subst}
\sum_{n\in\mathbb{Z}+\frac{\mathfrak{p}}2} \mathfrak{D}\Bigl(\mathcal{Z}(a+2n\epsilon_1;2\epsilon_1,\epsilon_2{-}\epsilon_1|t),\mathcal{Z}(a+2n\epsilon_2;\epsilon_1{-}\epsilon_2,2\epsilon_2|t)\Bigr)=0, \qquad \mathfrak{p}=N\mod2,
\end{equation}
which is \eqref{blowup_gen_form_split} for  general $\epsilon$ for $\mathfrak{D}=\mathfrak{D}^{[4],k}_{\epsilon_1,\epsilon_2}$ and $\mathcal{Z}=\mathcal{Z}^{[4]}$.
The above procedure gives a general way to rewrite bilinear relations on partition functions also for other gauge theories. 

Because of the quantization \eqref{as_comm_rel}, we call system of equations \eqref{blowup_gen_form_tau_quantum} for $\mathfrak{D}=\mathfrak{D}^{[4],k}_{\epsilon_1,\epsilon_2}$ {\it quantum} Painlev\'e VI equation (QPVI for brevity)
\begin{equation}\label{PVI_tau_quantum}
\mathfrak{D}^{[4],k}_{\epsilon_1,\epsilon_2}\Bigl(\tau^{(1)}(t), \tau^{(2)}(t) \Bigr)=0,\qquad k=1,3,4,
\end{equation}
where superscripts $^{(1)}$ and $^{(2)}$ abbreviate the shifted $\Omega$-background dependence of the tau functions
\begin{equation}
\tau^{(1)}(t)=\tau(2\epsilon_1,\epsilon_2{-}\epsilon_1|t), \qquad   \tau^{(2)}(t)=\tau(\epsilon_1{-}\epsilon_2,2\epsilon_2|t).
\end{equation}
By construction, Zak transform \eqref{tau_quantum} with $\mathcal{Z}=\mathcal{Z}^{[4]}$
\begin{equation}\label{tau_quantum_4}
\tau^{[4]}(a,\eta;m_{1,\ldots 4};\epsilon_1,\epsilon_2|t)=\sum_{n\in\mathbb{Z}}e^{\ri n \eta} \mathcal{Z}^{[4]}(a+n \epsilon_2;m_{1,\ldots4};\epsilon_1, \epsilon_2|t)
\end{equation}
gives a solution of quantum Painlev\'e VI \eqref{PVI_tau_quantum}.

Notice that the above construction is a simplified variant of the quantization procedure used in \cite{BGM17} to prove an analogous formula for $q$-Painlev\'e III$_3$ equation.

\paragraph{Quantum Painlev\'e VI symmetries.}
Rescaling the tau function with an $\epsilon$-deformed version of \eqref{rdf_PVI} 
\begin{equation}
\tau(\epsilon_1,\epsilon_2|t)= t^{-\frac{w_2^{[4]}{-}2\epsilon^2}{4\epsilon_1\epsilon_2}} (1{-}t)^{\frac{e_1^{[4]}(e_1^{[4]}{+}2\epsilon)}{4\epsilon_1\epsilon_2}}\tau_{s}(\epsilon_1,\epsilon_2|t)    \, ,
\end{equation}
we obtain that \eqref{PVI_tau_quantum} for $k=1,3$ remain unchanged while for $k=4$ becomes an $\epsilon$-deformed version of \eqref{PVI_tau}, that is
\begin{multline}\label{QPVI}
D^4_{\epsilon_1,\epsilon_2[\ln t]}(\tau_s^{(1)},\tau_s^{(2)})+2\,\frac{1+t}{1-t} \left(\epsilon_1\epsilon_2\frac{d}{d\ln t}\right)D^2_{\epsilon_1,\epsilon_2[\ln t]}(\tau_s^{(1)},\tau_s^{(2)})+\frac{t}{(1-t)^2}\left(\epsilon_1\epsilon_2\frac{d}{d\ln t}\right)^2(\tau_s^{(1)}\tau_s^{(2)})\\
-\left(w_2^{[4]}/4{+}\epsilon_1\epsilon_2{+}\epsilon^2/2+(\epsilon_1\epsilon_2{+}6\epsilon^2)\frac{t}{(1{-}t)^2}\right)D^2_{\epsilon_1,\epsilon_2[\ln t]}(\tau_s^{(1)},\tau_s^{(2)})
\\ 
-\frac{t}{16(1{-}t)^2}\left(\frac{\big(w_2^{[4]}{-}2\epsilon^2\big)^2}4+\frac{\big(w_4^{[4]}{+}2e_4^{[4]}{-}\frac14(w_2^{[4]})^2\big)t{-}4e_4^{[4]}}{1-t}\right) \tau_s^{(1)}\tau_s^{(2)}=0.  
\end{multline}
Here the analogs $w_2^{[4]},e_4^{[4]},w_4^{[4]},w_6^{[4]}$ of Painlev\'e VI parameters $\mathrm{w}_2^{[4]},\se_4^{[4]},\mathrm{w}_4^{[4]},\mathrm{w}_6^{[4]}$
in
\eqref{PVI_inv} given by
\begin{multline}\label{QPVI_inv}
w_2^{[4]}=m_1^2+m_2^2+m_3^2+m_4^2, \qquad \
e_4^{[4]}=m_1m_2m_3m_4,\qquad
w_4^{[4]}=m_1^2m_2^2+m_1^2m_3^2+m_1^2m_4^2+m_2^2m_3^2+m_2^2m_4^2+m_3^2m_4^2,\\
w_6^{[4]}=m_1^2m_2^2m_3^2+m_1^2m_2^2m_4^2+m_1^2m_3^2m_4^2+m_2^2m_3^2m_4^2 \qquad \qquad
\end{multline}
are also invariant under permutations of masses and changes of sign of an even number of masses. This corresponds to the action of the Weyl group of root system $D_4$. 
Actually, as it happens for the classical case (see the end of App.~\ref{ssec:PVI}), quantum Painlev\'e VI equation does not depend on $w_6^{[4]}$, but has an additional nontrivial integration constant besides quantum variables $a$ and $\eta$.
Additionally to such obvious symmetries also the Painlev\'e VI crossing symmetries, defined by Table~\ref{tab:PVI_tau_symm} are deformed according to Table~\ref{tab:quant_PVI_tau_symm},
namely there are presented symmetries of \eqref{QPVI} together with \eqref{PVI_tau_quantum} for $k=1,3$.

\begin{table}[ht]
    \centering
    \begin{tabular}{|c||c|c|c|c|c|}
      \hline 
     Generator & $w_2^{[4]}$ & $w_4^{[4]}$ & $e_4^{[4]}$ & $t$ & $\tau_s(\epsilon_1,\epsilon_2)$ \\
     \hline
     $q_{0\infty}$ & $w_2^{[4]}$ & $\frac38(w_2^{[4]})^2{+}3e_4^{[4]}{-}\frac12w_4^{[4]}$ & $\frac12e_4^{[4]}{+}\frac14w_4^{[4]}{-}\frac1{16}(w_2^{[4]})^2$  &   $\frac1{t}$ & $\tau_s$ \\
      \hline
    $q_{01}$ & $w_2^{[4]}$ & $\frac38(w_2^{[4]})^2{-}3e_4^{[4]}{-}\frac12w_4^{[4]}$  & $\frac12e_4^{[4]}{-}\frac14w_4^{[4]}{+}\frac1{16}(w_2^{[4]})^2$ &  $1{-}t$ & $\left(\frac{t}{1{-}t}\right)^{-\frac{w_2^{[4]}{-}2\epsilon^2}{4\epsilon_1\epsilon_2}}\tau_s$ \\
      \hline
    $q_{1\infty}$ & $w_2^{[4]}$ & $w_4^{[4]}$ & $-e_4^{[4]}$ & $\frac{t}{t-1}$ & $(1{-}t)^{\frac{w_2^{[4]}{-}2\epsilon^2}{4\epsilon_1\epsilon_2}}\tau_s$ \\
      \hline
    \end{tabular}
    \caption{Quantum Painlev\'e VI tau form symmetries}
    \label{tab:quant_PVI_tau_symm}
\end{table}
Symmetries $q_{01}, q_{0\infty}$ imply that also around $t=1,\infty$ there exist solutions of the form \eqref{tau_quantum} for $\mathcal{Z}=\mathcal{Z}^{[4]}$ with the mass parameters determined by the above symmetry action. Likely this is a re-expansion around $t=1,\infty$ of the solution \eqref{tau_quantum} found around $t=0$, analogous to that in the  classical case \cite[Thm 1.2]{J82}, but here we do not have a proof of this. For symmetry $q_{1\infty}=q_{0\infty}q_{01}q_{0\infty}$ (that keeps $t=0$) we have the following partition function re-expansion
\begin{equation}
\mathcal{Z}_{inst}^{[4]}\left(a;m_{1,2,3,4};\epsilon_1,\epsilon_2\Big|\frac{t}{t{-}1}\right)=(1{-}t)^{-\frac{a^2{+}e_2^{[4]}{+}(m_2{-}e_1^{[4]})(m_2{-}\epsilon){+}3\epsilon^2/4}{\epsilon_1\epsilon_2}}\mathcal{Z}^{[4]}_{inst}(a;m_1,-m_2,m_3,m_4;\epsilon_1,\epsilon_2|t),
\end{equation}
which is simply the crossing symmetry relation 
for the corresponding Virasoro conformal block \eqref{AGT_block}.
The instanton partition function after  multiplication by a specific $U(1)$ factor
\begin{equation}\label{symm_form}
(1-t)^{-\frac{e_2^{[4]}+e_1^{[4]}\epsilon}{2\epsilon_1\epsilon_2}}\mathcal{Z}^{[4]}_{inst}(a;m_{1,2,3,4};\epsilon_1,\epsilon_2|t)  
\end{equation}
is invariant under the Weyl group of $D_4$ and $(\epsilon_1,\epsilon_2)\mapsto(-\epsilon_1,-\epsilon_2)$ as well. This is also natural from AGT correspondence because the Virasoro conformal block \eqref{AGT_block} depends on the masses through highest weights \eqref{AGT_dictionary_masses}.

\section{Coalescence for quantum Painlev\'e equations}\label{sec:coalesc}

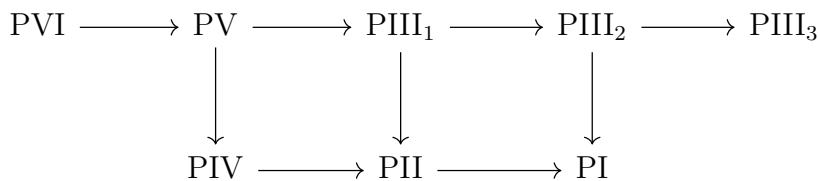
\begin{figure}[ht]

\begin{center}
\begin{tikzcd}[row sep=3em, column sep=3em]
\mathrm{PVI} \arrow[r] & \mathrm{PV}  \arrow[r]  \arrow[d] & \mathrm{PIII}_1 \arrow[r]  \arrow[d] & \mathrm{PIII}_2 \arrow[r]  \arrow[d] & \mathrm{PIII}_3 \\
& \mathrm{PIV} \arrow[r] & \mathrm{PII} \arrow[r] & \mathrm{PI} &
\end{tikzcd}
\end{center}

\caption{Painlev\'e coalescence diagram}
	        
	        \label{fig:Painleve_coalescence}                
\end{figure}

Classical Painlev\'e equations are second order nonlinear differential equations 
involving a single function of one variable
such that their solutions do not have movable singularities except poles.
The most general equation is the so-called Painlev\'e VI equation \eqref{PVI}, which depends on $4$ parameters. The other Painlev\'e equations can be obtained from Painlev\'e VI via successive limiting (coalescence) procedures (Fig.~\ref{fig:Painleve_coalescence}). Each arrow of the diagram denotes a limit which reduces the number of parameters by $1$. 
More details about classical Painlev\'e VI equation and the limiting procedure are presented in App.~\ref{sec:Painleve}. 

\begin{figure}[ht]

\begin{center}
\begin{tikzcd}[row sep=3em, column sep=3em]
N_f=4 \arrow[r] & N_f=3  \arrow[r]  \arrow[d] & N_f=2 \arrow[r]  \arrow[d] & N_f=1 \arrow[r]  \arrow[d] & N_f=0 \\
& H_2 \arrow[r] & H_1 \arrow[r] & H_0 &
\end{tikzcd}
\end{center}

\caption{Gauge theory degeneration diagram}
	        
	        \label{fig:gauge_coalescence}               
\end{figure}
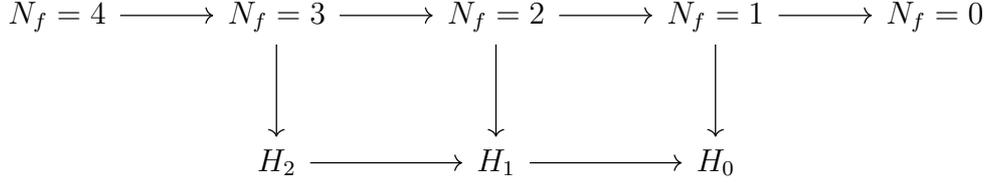

In recent years it was discovered that
tau functions (solutions) of Painlev\'e equations around their critical points can be expressed as the Zak transforms of partition functions of different SUSY gauge theories with self-dual ($\epsilon{=}0$) $\Omega$-background. 
Indeed, the Painlev\'e VI tau function expansions around the critical points $t=0,1,\infty$ are given by the Zak transform of the $N_f=4 $ SUSY $\mathcal{N}=2$ $SU(2)$ $\epsilon=0$ partition function $\mathcal{Z}^{[4]}$ \cite{GIL12}, as we already mentioned in the previous section.
From this result it follows by a straightforward limit (see Sec. \ref{ssec:deg_sol} for general $\epsilon$ as well) that Painlev\'e V, III$_{1,2,3}$ tau functions around $0$ are given by the Zak transform of SUSY $\mathcal{N}=2$ $SU(2)$ $\epsilon=0$ partition functions $\mathcal{Z}^{[N_f]}$ with $N_f=3,2,1,0$ respectively \cite{GIL13}.

\begin{figure}[ht]\label{fig3}
	
	\begin{center}
	\begin{tikzpicture}[scale=1.5]
                \draw[dashed](0,0) circle (1); 
                \draw [blue,line width=1pt] (0,-1) -- (0,1);
                \draw [red,line width=1pt] (-1,0) -- (1,0);
                \filldraw[color=white, fill=white](0,0) circle (0.4);
			\node at (0,0) 	{PV};
                
                \draw[dashed](2.5,0) circle (1); 
                \draw [blue,line width=1pt] (1.5,0) -- (3.5,0);
                \filldraw[color=white, fill=white](2.5,0) circle (0.4);
			\node at (2.5,0) 	{PIII$_1$};

                \draw[dashed](5,0) circle (1); 
                \draw [blue,line width=1pt] (5,-1) -- (5,1);
                \filldraw[color=white, fill=white](5,0) circle (0.4);
			\node at (5,0) 	{PIII$_2$};

                \draw[dashed](7.5,0) circle (1); 
                \draw [blue,line width=1pt] (7.5,0) -- (8.5,0); 
                \filldraw[color=white, fill=white](7.5,0) circle (0.4);
			\node at (7.5,0) 	{PIII$_3$};

                 \draw[dashed](1.25,-2) circle (1); 
                \draw [red,line width=1pt] (1.25,-3) -- (1.25,-1);
                \draw [red,line width=1pt] (0.25,-2) -- (2.25,-2);
                 \draw [blue,line width=1pt] (1.25-0.707,-2-0.707) -- (1.25+0.707,-2+0.707);
                \draw [blue,line width=1pt] (1.25-0.707,-2+0.707) -- (1.25+0.707,-2-0.707);
                \filldraw[color=white, fill=white](1.25,-2) circle (0.4);
			\node at (1.25,-2) 	{PIV};

                 \draw[dashed](3.75,-2) circle (1); 
                \draw [blue,line width=1pt] (3.75,-2) -- (4.75,-2);
                \draw [red,line width=1pt] (2.75,-2) -- (3.75,-2);
                \draw [blue,line width=1pt] (3.25,-2+0.866) -- (3.75,-2);
                \draw [blue,line width=1pt] (3.25,-2-0.866) -- (3.75,-2);
                \draw [red,line width=1pt] (4.25,-2+0.866) -- (3.75,-2);
                \draw [red,line width=1pt] (4.25,-2-0.866) -- (3.75,-2);
                \filldraw[color=white, fill=white](3.75,-2) circle (0.4);
			\node at (3.75,-2) 	{PII};
   
                  \draw[dashed](6.25,-2) circle (1); 
                \draw [blue,line width=1pt] ({6.25+cos(36)},{-2+sin(36)}) -- (6.25,-2);
                \draw [blue,line width=1pt] ({6.25+cos(108)},{-2+sin(108)}) -- (6.25,-2);
                \draw [blue,line width=1pt] ({6.25+cos(180)},{-2+sin(180)}) -- (6.25,-2);
                \draw [blue,line width=1pt] ({6.25+cos(-108)},{-2+sin(-108)}) -- (6.25,-2);
                \draw [blue,line width=1pt] ({6.25+cos(-36)},{-2+sin(-36)}) -- (6.25,-2);
               
                \filldraw[color=white, fill=white](6.25,-2) circle (0.4);
			\node at (6.25,-2) 	{PI};
                
     \end{tikzpicture}
	        \end{center} 
	      
	        \caption{Canonical rays around critical point $\infty$}
	        
	        \label{fig:canonical_rays}
                
		\end{figure}
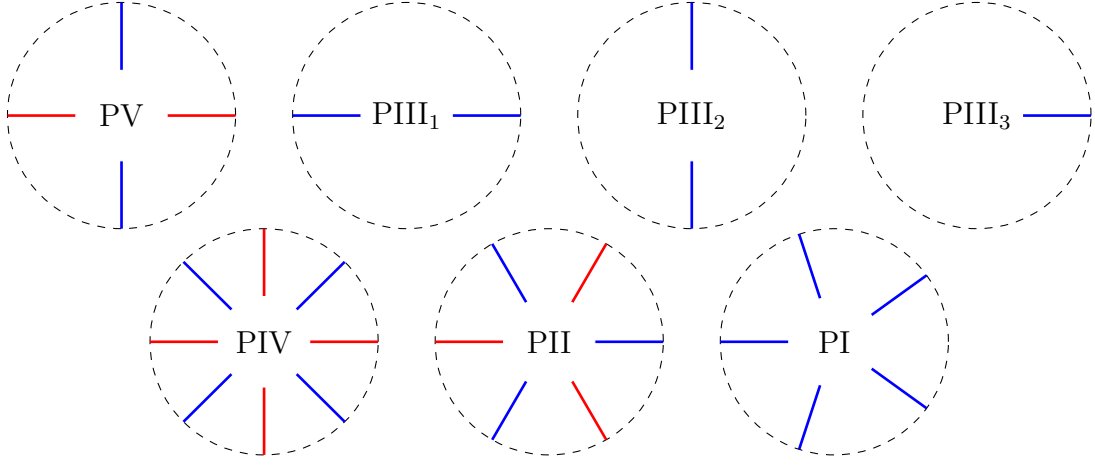

All these expansions are of a regular type, as the corresponding instanton partition functions are power series of the instanton counting parameter with a nonzero radius of convergence (finite for $N_f=4$ and infinite for $N_f<4$ \cite{ABT22}).
For Painlev\'e V, III's as well as for Painlev\'e IV, II, I point $t=\infty$ is an irregular critical point and for the last three ones it is the only one. Near this point the tau function expansions display a nonlinear Stokes phenomenon, the corresponding canonical rays being presented in Fig.~\ref{fig:canonical_rays} whose coloring is explained in the following.
These expansions along the canonical rays as asymptotic series in large $t$ were obtained in \cite{BLMST16}. 
Namely, for the tau function of Painlev\'e V and III's, such expansions are given by the Zak transform of partition functions of the same corresponding gauge theory 
but in the strong coupling regime (see \cite[App. A]{BLMST16}). 
For the tau functions of Painlev\'e IV, II and I they are given by the Zak transform of partition functions of Argyres-Douglas (AD) theories H$_{2,1,0}$ respectively (see \cite[Sec. 3]{BLMST16}).

The critical points of each Painlev\'e equation and the corresponding theory are collected in Table~\ref{tab:theories} (the theories are also presented in Fig.~\ref{fig:gauge_coalescence}). 
In Table~\ref{tab:expansions} one can find the angles of the canonical rays together with the symmetries $\mathbb{Z}_k$ of the equations, which rotate $t\mapsto e^{2\pi \ri/k}\,t$. These groups are presented explicitly in App.~\ref{ssec:D-coal} and \ref{ssec:E-coal}. 
The latter symmetries split the set of canonical rays in orbits, which are distinguished by the colouring in Fig.~\ref{fig:canonical_rays}. 
On each different orbit there is a different  asymptotic expansion, so for Painlev\'e V, IV and II there are two expansions and for the others --- only one.

\begin{table}[ht]
    \centering
    \begin{tabular}{|c|c|c|c|c|c||c|c|c|}
      \hline Equation & PVI & PV & PIII$_1$ & PIII$_2$ & PIII$_3$ & PIV & PII & PI \\ \hline
        Theory & $N_f=4$ & $N_f=3$ & $N_f=2$ & $N_f=1$ & $N_f=0$ & $H_2$ & $H_1$ & $H_0$ \\ \hline\hline
        Reg. exp. & $0,1,\infty$ & 0 & 0 & 0 & 0 & - & - & -  \\ 
        \hline
        Irr. exp. & - & $\infty$ & $\infty$ & $\infty$ & $\infty$ & $\infty$ & $\infty$ & $\infty$  \\ \hline
    \end{tabular}
    \caption{Painlev\'e tau function expansions at the critical points}
    \label{tab:theories}
\end{table}

\begin{table}[ht!]
    \centering
    \begin{tabular}{|c|c|c|c|c||c|c|c|}
      \hline Equation  & PV & PIII$_1$ & PIII$_2$ & PIII$_3$ & PIV & PII & PI \\ \hline 
      Can. rays & $(k{+}\frac12)\pi, k\pi$ & $k\pi$ & $(k{+}\frac12)\pi$ & $0$ & $(\frac{k}2{+}\frac14)\pi, \frac{k}2\pi$ & $\frac{2k}3\pi, \frac{(2k{+}1)}3\pi$ & $\frac{(2k{+}1)}5\pi$
         \\ \hline
          Symmetry &  $\mathbb{Z}_2$ & $\mathbb{Z}_2$ & $\mathbb{Z}_2$ & - & $\mathbb{Z}_4$ & $\mathbb{Z}_3$ & $\mathbb{Z}_5$ \\ 
        \hline    
    \end{tabular}
    \caption{Canonical rays around critical point $\infty$ ($k\in\mathbb{Z}$)}
    \label{tab:expansions}
\end{table}

We expect that tau functions of quantum Painlev\'e equations are then given by the operator-like Zak transforms \eqref{tau_quantum} of the corresponding partition functions $\mathcal{Z}$ with $\epsilon\neq0$, in analogy with the quantum Painlev\'e VI case \eqref{tau_quantum_4} and 
we indeed find proper expansions for these partition functions by solving the quantum Painlev\'e equations, under an appropriate ansatz. 
These equations are degeneracies of quantum Painlev\'e VI equation in bilinear form \eqref{PVI_tau_quantum} and are obtained by the same limit procedure prescribed in terms of the instanton counting parameter and the masses as that for the classical case at self-dual $\Omega$-background $\epsilon=0$ (see App.~\ref{ssec:D-coal} and \ref{ssec:E-coal} for details on the classical case).

In the coming Sec.~\ref{ssec:D-coal_q} we follow the limits along the first row of the coalescence diagram (Fig. \ref{fig:Painleve_coalescence}) while in Sec.~\ref{ssec:E-coal_q} we follow the "vertical" limits from the first to the second row connecting the Lagrangian theories to the Argyres-Douglas ones. Here we also verify the commutativity of the diagram by describing the limits along the second line. Finally, in Sec.~\ref{ssec:deg_sol} we show that, while the expansions around the regular critical points $t=0$ can be easily obtained from the Painlev\'e VI expansion at the level of partition functions, it is on the contrary highly nontrivial to take the limit to the expansions near the irregular critical point $t=\infty$ at the level of partition functions.

\subsection{Limits to quantum Painlev\'e V, III's / \texorpdfstring{$SU(2)$}{SU(2)} theories with \texorpdfstring{$N_f\leq3$}{N\_f≤3}}
\label{ssec:D-coal_q}

In this subsection we follow the first line of the Painlev\'e coalescence diagram (Fig.~\ref{fig:Painleve_coalescence}).  
This corresponds to the standard holomorphic decoupling limits in gauge theory $N_f\to N_f-1$
obtained by rescaling the gauge coupling constant as
$t^{[N_f-1]}=t^{[N_f]}m_{N_f}$ and decoupling the $N_f$-th hypermultiplet
by letting $m_{N_f}\to\infty$ keeping the other masses finite. Let us analyze these limits one by one
from the viewpoint of the quantum Painlev\'e equations.
\begin{itemize} 
 \item {\bf QPVI ($N_f=4$) $\rightarrow$ QPV ($N_f=3$).}
    Rescaling variable  $t^{[4]}=t^{[3]}/m_4$ 
 and letting $m_4\rightarrow \infty$ from \eqref{PVI_tau_quantum} we obtain blowup equations
 \begin{equation}\label{D13irr}
    D^1_{\epsilon_1,\epsilon_2[\ln t]}\left(\tau^{(1)},\tau^{(2)}\right)=0,\qquad \qquad
  D^3_{\epsilon_1,\epsilon_2[\ln t]}\left(\tau^{(1)},\tau^{(2)}\right)=\epsilon D^2_{\epsilon_1,\epsilon_2[\ln t]}\left(\tau^{(1)},\tau^{(2)}\right),
    \end{equation}
    \vspace{-0.7cm}
    \begin{multline}\label{quantum_PV}
    D^4_{\epsilon_1,\epsilon_2[\ln t]}\left(\tau^{(1)},\tau^{(2)}\right)+2\left(\epsilon_1\epsilon_2\frac{d}{d\ln t}\right)D^2_{\epsilon_1,\epsilon_2[\ln t]}\left(\tau^{(1)},\tau^{(2)}\right)-
    \left(\frac{t^2}4{+} (e^{[3]}_1{+}\epsilon)t{+}\epsilon_1\epsilon_2{+}\epsilon^2\right) D^2_{\epsilon_1,\epsilon_2[\ln t]}\left(\tau^{(1)},\tau^{(2)}\right)\\
    -\frac{t}4(t{+}2(e^{[3]}_1{+}\epsilon))\left(\epsilon_1\epsilon_2\frac{d}{d\ln t}\right)\left(\tau^{(1)}\tau^{(2)}\right)+\frac{t}8\left((e_2^{[3]}{+}e_1^{[3]}\epsilon{+}\epsilon^2)t{+}2e_3^{[3]}\right) \tau^{(1)}\tau^{(2)}=0,
    \end{multline}
    where $e^{[3]}_{1,2,3}$ are the elementary symmetric polynomials of three mass parameters $m_1,m_2,m_3$
    \begin{equation}
    e^{[3]}_1=m_1{+}m_2{+}m_3, \qquad e^{[3]}_2=m_1m_2{+}m_2m_3{+}m_3m_1, \qquad  e^{[3]}_3=m_1m_2m_3,
    \end{equation}
    and $t^{[3]}=t$ for notational simplicity.
    
    \item {\bf QPV ($N_f=3$) $\rightarrow$ QPIII$_1$ ($N_f=2$).}
   
     Rescaling variable $t^{[3]}=t^{[2]}/m_3$ we see that blowup equations \eqref{D13irr}  remain unchanged 
     while the order $4$ blowup equation \eqref{quantum_PV} under $m_3\rightarrow\infty$ becomes \begin{multline}\label{quantum_PIII1}
     D^4_{\epsilon_1,\epsilon_2[\ln t]}\left(\tau^{(1)},\tau^{(2)}\right)+2\left(\epsilon_1\epsilon_2\frac{d}{d\ln t}\right)D^2_{\epsilon_1,\epsilon_2[\ln t]}\left(\tau^{(1)},\tau^{(2)}\right)-\left(t{+}\epsilon_1\epsilon_2{+}\epsilon^2\right)D^2_{\epsilon_1,\epsilon_2[\ln t]}\left(\tau^{(1)},\tau^{(2)}\right)\\-\frac{t}2 \left(\epsilon_1\epsilon_2\frac{d}{d\ln t}\right)\left(\tau^{(1)}\tau^{(2)}\right)+\frac{m_1m_2}4 t\,\tau^{(1)}\tau^{(2)}=0,
     \end{multline}
     with $t^{[2]}=t$ for notational simplicity.
     
     \item {\bf QPIII$_1$ ($N_f=2$) $\rightarrow$ QPIII$_2$ ($N_f=1$).}
     
    Rescaling variable $t^{[2]}=t^{[1]}/m_2$ 
    still blowup equations \eqref{D13irr}
    remain unchanged while the order $4$ blowup equation \eqref{quantum_PIII1} under $m_2\rightarrow\infty$ becomes
 \begin{equation}\label{quantum_PIII2}
     D^4_{\epsilon_1,\epsilon_2[\ln t]}\left(\tau^{(1)},\tau^{(2)}\right)+2\left(\epsilon_1\epsilon_2 \frac{d}{d\ln t}\right)D^2_{\epsilon_1,\epsilon_2[\ln t]}\left(\tau^{(1)},\tau^{(2)}\right)-(\epsilon_1\epsilon_2{+}\epsilon^2) D^2_{\epsilon_1,\epsilon_2[\ln t]}\left(\tau^{(1)},\tau^{(2)}\right)+ \frac{m_1}4 t\,\tau^{(1)}\tau^{(2)}=0,
    \end{equation}
    with $t^{[1]}=t$ for notational simplicity.
     \item { \bf QPIII$_2$ ($N_f=1$) $\rightarrow$ QPIII$_3$ ($N_f=0$).}
     
    Rescaling variable $ t^{[1]}=t^{[0]}/m_1$ 
    still blowup equations \eqref{D13irr}
    remain unchanged and the order $4$ blowup equation \eqref{quantum_PIII2} becomes
    \begin{equation}\label{quantum_PIII3}
     D^4_{\epsilon_1,\epsilon_2[\ln t]}\left(\tau^{(1)},\tau^{(2)}\right)+2\left(\epsilon_1\epsilon_2 \frac{d}{d\ln t}\right)D^2_{\epsilon_1,\epsilon_2[\ln t]}\left(\tau^{(1)},\tau^{(2)}\right)-(\epsilon_1\epsilon_2{+}\epsilon^2)D^2_{\epsilon_1,\epsilon_2[\ln t]}\left(\tau^{(1)},\tau^{(2)}\right)+\frac{t}4\tau^{(1)}\tau^{(2)}=0,
    \end{equation}
    with $t^{[0]}=t$ for notational simplicity, even without sending $m_1\rightarrow\infty$.

    Analogously to the case of the Painlev\'e equations, on the quantum Painlev\'e equation level the bilinear forms of PIII$_2$ and PIII$_3$ coincide as the mass parameter is reduced to an integration constant eventually specified by the asymptotics of the solution.
\end{itemize}
Obtained \eqref{quantum_PV}, \eqref{quantum_PIII1}, \eqref{quantum_PIII2}, \eqref{quantum_PIII3} are $\epsilon$-deformations of \eqref{PV_tau}, \eqref{PIII1_tau}, \eqref{PIII2_tau}, \eqref{PIII3_tau} respectively.
    
\subsection{Limits to and among quantum Painlev\'e IV, II, I / AD theories \texorpdfstring{$H_{2,1,0}$}{H₂, H₁, H₀}}
\label{ssec:E-coal_q}

In this subsection we follow the vertical arrows of the Painlev\'e coalescence diagram (Fig.~\ref{fig:Painleve_coalescence}) and display its commutativity by checking the coalescence limits along the lower horizontal line. Differently from the previous section, here we rescale the tau functions before taking the limits. To define these rescalings we $\epsilon$-deform in a minimal way the ones of the classical tau-functions described in Sec. \ref{ssec:E-coal}. In doing this deformation, we respect the mass dimension counting and the permutation symmetries like those of the partition functions of Sec. \ref{ssec:part_funct}.

\begin{itemize}
\item {\bf QPV ($N_f=3$) $\rightarrow$ QPIV ($H_2$)}

In this case we rescale the tau function with the following $\epsilon$-deformed version of \eqref{PV_IV_resc_cl}
\begin{equation}\label{PV_genlimresc}
\tau(\epsilon_1,\epsilon_2|t)=t^{\frac{e_2^{[3]}+\kappa_1 e_1^{[3]}\epsilon+\kappa_2\epsilon^2}{\epsilon_1\epsilon_2}}\tau_r(\epsilon_1,\epsilon_2|t)
\end{equation}
for some $\kappa_1, \kappa_2\in\mathbb{C}$.
Under this rescaling equations \eqref{D13irr} do not change but \eqref{quantum_PV} becomes
\begin{multline}\label{quantum_PVr}
D^4_{\epsilon_1,\epsilon_2[\ln t]}\left(\tau^{(1)},\tau^{(2)}\right)+2\left(\epsilon_1\epsilon_2\frac{d}{d\ln t}\right)D^2_{\epsilon_1,\epsilon_2[\ln t]}\left(\tau^{(1)},\tau^{(2)}\right)\\ \mspace{-10mu}-
    \left(\frac{t^2}4{+} (e^{[3]}_1{+}\epsilon)t{+}\epsilon_1\epsilon_2{+}\epsilon^2-(e_2^{[3]}{+}\kappa_1 e_1^{[3]}\epsilon{+}\kappa_2\epsilon^2)\right) D^2_{\epsilon_1,\epsilon_2[\ln t]}\left(\tau^{(1)},\tau^{(2)}\right)
    -\frac{t}4(t{+}2(e^{[3]}_1{+}\epsilon))\left(\epsilon_1\epsilon_2\frac{d}{d\ln t}\right)\left(\tau^{(1)}\tau^{(2)}\right)\\+\frac{t}8\left((e_2^{[3]}{+}e_1^{[3]}\epsilon{+}\epsilon^2)t{+}2e_3^{[3]}-(2e_1^{[3]}{+}2\epsilon{+}t)(e_2^{[3]}{+}\kappa_1 e_1^{[3]}\epsilon{+}\kappa_2\epsilon^2)\right) \tau^{(1)}\tau^{(2)}=0.   
   \end{multline}
Then, redefining the variable and the parameters as
\begin{equation}\label{resc_3}
t^{[3]}=-\frac1{r^2}-\frac{t^{[H_2]}}{r}, \qquad
m_1=-\frac1{2r^2}, \quad m_2=\frac1{2r^2}+2(2\boldsymbol{m}_1{+}\boldsymbol{m}_2), \quad  m_3=\frac1{2r^2}+2(\boldsymbol{m}_1{+}2\boldsymbol{m}_2)
\end{equation}
and, letting $r\rightarrow0$,
from \eqref{D13irr} we obtain blowup equations
\begin{equation}\label{D13irr_2row}
 D^1_{\epsilon_1,\epsilon_2[t]}\left(\tau^{(1)},\tau^{(2)}\right)=0, \qquad \qquad
     D^3_{\epsilon_1,\epsilon_2[t]}\left(\tau^{(1)},\tau^{(2)}\right)=0,    
\end{equation}
and \eqref{quantum_PVr} goes to an $\epsilon$-deformation of \eqref{PIV_tau} in the leading order iff $\kappa_1=1$, therefore obtaining
\begin{multline}\label{quantum_PIV}
D^4_{\epsilon_1,\epsilon_2[t]}\left(\tau^{(1)},\tau^{(2)}\right)-\left(\frac{t^2}4{-}6(\boldsymbol{m}_1{+}\boldsymbol{m}_2){-}\frac32\epsilon\right)D^2_{\epsilon_1,\epsilon_2[t]}\left(\tau^{(1)},\tau^{(2)}\right)-\frac14t\left(\epsilon_1\epsilon_2\frac{d}{dt}\right)\left(\tau^{(1)}\tau^{(2)}\right)\\+\left(2\boldsymbol{m}_1{+}\boldsymbol{m}_2{+}\frac{\epsilon}2\right)\left(\boldsymbol{m}_1{+}2\boldsymbol{m}_2{+}\frac{\epsilon}2\right)\,\tau^{(1)}\tau^{(2)}=0,
\end{multline}
with $t^{[H_2]}=t$ for notational simplicity. Notice that the parameter $\kappa_2$ is left undetermined.

\item {\bf QPIII$_1$ ($N_f=2$) $\rightarrow$ QPII ($H_1$). }
In this case for the rescaling it is enough to $\epsilon$-deform $\epsilon_2^2$ to $\epsilon_1\epsilon_2$ in the denominator of \eqref{PIII1_PII_resc_cl}, namely
\begin{equation}\label{PIII1_genlimresc}
\tau(\epsilon_1,\epsilon_2|t)=t^{-\frac{m_1m_2}{\epsilon_1\epsilon_2}}e^{-\frac{t}{\epsilon_1\epsilon_2}}\tau_r(\epsilon_1,\epsilon_2|t)\,.
\end{equation}
Under this rescaling equations \eqref{D13irr} (which are still the same for QPV and QPIII's) do not change but \eqref{quantum_PIII1} becomes
\begin{multline}
\label{quantum_PIII1r}
     D^4_{\epsilon_1,\epsilon_2[\ln t]}\left(\tau^{(1)},\tau^{(2)}\right)+2\left(\epsilon_1\epsilon_2\frac{d}{d\ln t}\right)D^2_{\epsilon_1,\epsilon_2[\ln t]}\left(\tau^{(1)},\tau^{(2)}\right)+(t{-}m_1m_2{-}\epsilon_1\epsilon_2{-}\epsilon^2)D^2_{\epsilon_1,\epsilon_2[\ln t]}\left(\tau^{(1)},\tau^{(2)}\right)\\+\frac{t}2 \left(\epsilon_1\epsilon_2\frac{d}{d\ln t}\right)\tau^{(1)}\tau^{(2)}=0.
    \end{multline}    
Then, redefining the variable and the parameters as 
\begin{equation}\label{resc_2}
t^{[2]}=\frac{4}{r^6}+\frac{2t^{[H_1]}}{r^4}, \qquad
m_1=\frac2{r^3}, \quad m_2=\frac2{r^3}+4\boldsymbol{m}
\end{equation}
and, letting $r\rightarrow0$,
from \eqref{D13irr} we obtain blowup equations 
\eqref{D13irr_2row}
and \eqref{quantum_PIII1r} goes to $\epsilon$-deformation of \eqref{PII_tau}
\begin{equation}\label{quantum_PII}
  D^4_{\epsilon_1,\epsilon_2[t]}\left(\tau^{(1)},\tau^{(2)}\right)+\frac{t}2 D^2_{\epsilon_1,\epsilon_2[t]}\left(\tau^{(1)},\tau^{(2)}\right)+\frac14\left(\epsilon_1\epsilon_2\frac{d}{dt}\right)\left(\tau^{(1)}\tau^{(2)}\right)=0,
\end{equation}
with $t^{[H_1]}=t$ for notational simplicity.

\item {\bf QPIII$_2$ ($N_f=1$) $\rightarrow$ QPI ($H_0$).}
In this case for the rescaling it is enough to $\epsilon$-deform $\epsilon_2^2$ to $\epsilon_1\epsilon_2$ in the denominator of \eqref{PIII2_PI_resc_cl}, namely
\begin{equation}\label{PIII2_genlimresc}
\tau(\epsilon_1,\epsilon_2|t)=t^{-\frac{8m_1^2}{9\epsilon_1\epsilon_2}}e^{-\frac{3t}{4m_1\epsilon_1\epsilon_2}}\tau_r(\epsilon_1,\epsilon_2|t)\,.
\end{equation}
Under this rescaling equations \eqref{D13irr} (which are still the same for QPV and QPIII's) do not change but \eqref{quantum_PIII2} becomes
\begin{multline}
\label{quantum_PIII2r}
     D^4_{\epsilon_1,\epsilon_2[\ln t]}\left(\tau^{(1)},\tau^{(2)}\right)+2\left(\epsilon_1\epsilon_2\frac{d}{d\ln t}\right)D^2_{\epsilon_1,\epsilon_2[\ln t]}\left(\tau^{(1)},\tau^{(2)}\right)+\left(\frac{3t}{2m_1}{-}\frac{8m_1^2}9{-}\epsilon_1\epsilon_2{-}\epsilon^2\right)D^2_{\epsilon_1,\epsilon_2[\ln t]}\left(\tau^{(1)},\tau^{(2)}\right)\\+\frac{3t}{4m_1} \left(\epsilon_1\epsilon_2\frac{d}{d\ln t}\right)\left(\tau^{(1)}\tau^{(2)}\right)+t \left(\frac{9t}{64m_1^2}{-}\frac{m_1}{12}\right)\tau^{(1)}\tau^{(2)}=0.    
\end{multline}
Then, redefining the variable and the parameter as 
\begin{equation}\label{resc_1}
t^{[1]}=-\frac{16}{r^{15}}\big(1+r^4t^{[H_0]}/2\big), \qquad
m_1=-\frac3{r^5}
\end{equation}
and, letting $r\rightarrow0$,
from \eqref{D13irr} we obtain blowup equations \eqref{D13irr_2row}
and \eqref{quantum_PIII2r} goes to $\epsilon$-deformation of \eqref{PI_tau}
\begin{equation}\label{quantum_PI}
D^4_{\epsilon_1,\epsilon_2[t]}\left(\tau^{(1)},\tau^{(2)}\right)+\frac{t}8 \tau^{(1)}\tau^{(2)}=0,
\end{equation}
with $t^{[H_0]}=t$ for notational simplicity.
\end{itemize}

For self-consistency we also report the corresponding coalescence limits among the 
Argyres-Douglas theories (lower line on the diagram of Fig.~\ref{fig:Painleve_coalescence}).

\begin{itemize}
    \item {\bf QPIV ($H_2$) $\rightarrow$  QPII ($H_1$).}
In this case we should rescale the tau function with the following $\epsilon$-deformed version of \eqref{PIV_PII_resc_cl}
\begin{equation}\label{PIV_genlimresc}
\tau(\epsilon_1,\epsilon_2|t)=t^{\frac{8(2\boldsymbol{m}_1{+}\boldsymbol{m}_2{+}\epsilon/2)(\boldsymbol{m}_1{+}2\boldsymbol{m}_2{+}\epsilon/2)}{\epsilon_1\epsilon_2}}\tau_r(\epsilon_1,\epsilon_2|t)
\end{equation}
and redefining the variable and parameters as
\begin{equation}\label{resc_H2}
t^{[H_2]}=-\frac1{r^3}+rt^{[H_1]}, \qquad
\boldsymbol{m}_1=-\frac1{24r^6}, \quad \boldsymbol{m}_2=\frac1{12r^6}+2\boldsymbol{m},
\end{equation}
after letting $r\rightarrow0$,
we see that \eqref{D13irr_2row} remain unchanged and \eqref{quantum_PIV} reproduces \eqref{quantum_PII}.

    \item {\bf QPII ($H_1$) $\rightarrow$  QPI ($H_0$).}
In this case for the rescaling it is enough to $\epsilon$-deform $\epsilon_2^2$ to $\epsilon_1\epsilon_2$ in the denominator of \eqref{PII_PI_resc_cl}, namely
\begin{equation}\label{PII_genlimresc}
\tau(\epsilon_1,\epsilon_2|t)=t^{-\frac{12\boldsymbol{m}^2}{\epsilon_1\epsilon_2}}e^{\frac{t^3}{27\epsilon_1\epsilon_2}}\tau_r(\epsilon_1,\epsilon_2|t)
\end{equation}
and redefining the variable and parameter as
\begin{equation}\label{resc_H1}
t^{[H_1]}=-\frac6{r^{10}}+r^2t^{[H_0]}, \qquad
\boldsymbol{m}=\frac1{r^{15}}\,,
\end{equation}
after letting $r\rightarrow0$, we see that \eqref{D13irr_2row} remain unchanged and \eqref{quantum_PII} reproduces \eqref{quantum_PI}.
\end{itemize}

\subsection{A remark on the coalescence of the solutions}
\label{ssec:deg_sol}

The aim of this subsection is to {\it remark a crucial gauge theoretic difference between the features under coalescence of the solutions of the quantum Painlev\'e 
equations at weak and strong coupling}.
While the first naturally follow from the usual 
holomorphic decoupling limit of massive hypermultiplets ---
and are therefore well tamed and understood ---, the second are much more subtle and less straightforward as this 
clear interpretation is not available in terms of weakly coupled degrees of freedom. On the contrary our results
shall match possible proposals for a dual formulation of the gauge theory in the magnetic and AD frames in terms of weakly coupled degrees of freedom.

\paragraph{Coalescence for solutions at weak coupling.}
In the weak coupling 
region, where $t$ is in the neighborhood of $0$, the limits of Sec.~\ref{ssec:D-coal_q} along theories with fundamental matter are realized at the level of partition functions in a very simple way. Indeed, the instanton parts \eqref{Zinst} satisfy 
\begin{equation}
\lim_{m_{N_f}\rightarrow\infty}\mathcal{Z}_{inst}^{[N_f]}(a,m_{1,\ldots N_f};\epsilon_1,\epsilon_2|t/m_f)=\mathcal{Z}_{inst}^{[N_f-1]}(a,m_{1,\ldots N_f-1};\epsilon_1,\epsilon_2|t)    
\end{equation}
because 
\begin{equation}
Z_{fund}(a,m|Y^+,Y^-)\sim m^{|Y^+|+|Y^-|}, \qquad m\rightarrow\infty.   
\end{equation}
For the 1-loop part, according to \eqref{gamma_as}
\begin{equation}
\prod_{\pm}\exp\gamma_{\epsilon_1,\epsilon_2}(m\pm a{-}\epsilon/2)\sim e^{\frac{3m^2}{2\epsilon_1\epsilon_2}}m^{\frac{\epsilon^2/12-m^2-a^2}{\epsilon_1\epsilon_2}-\frac16},\qquad m\rightarrow\infty,
\end{equation}
so we see that the product of classical and 1-loop factors for $N_f$ hypermultiplets in the limit $m_{N_f}\rightarrow\infty$ goes under the above rescalings to the one for $N_f-1$ hypermultiplets up to an $a$- and $t$-independent scaling factor
\begin{multline}
\mathcal{Z}_{cl}^{[N_f]}(a;\epsilon_1,\epsilon_2|t/m_{N_f}) \mathcal{Z}_{1-loop}^{[N_f]}(a,m_{1,\ldots N_f};\epsilon_1,\epsilon_2)\\\sim e^{\frac{3m_{N_f}^2}{2\epsilon_1\epsilon_2}}m_{N_f}^{-\frac{m_{N_f}^2}{\epsilon_1\epsilon_2}-\frac16\left(1+\frac{\epsilon^2}{\epsilon_1\epsilon_2}\right)}\mathcal{Z}_{cl}(a;\epsilon_1,\epsilon_2|t) \mathcal{Z}_{1-loop}^{[N_f-1]}(a,m_{1,\ldots N_f-1};\epsilon_1,\epsilon_2).
\end{multline}
This means that starting from quantum Painlev\'e VI solution \eqref{tau_quantum_4} via the limits of Sec. \ref{ssec:D-coal_q} we obtain solutions for the quantum Painlev\'e V, III$_{1,2,3}$ equations as the operator-like Zak transforms \eqref{tau_quantum} of partition function $\mathcal{Z}^{[N_f]}$ with $N_f=3,2,1,0$ respectively
\begin{equation}\label{tau_quantum_3210}
\tau^{[N_f]}(a,\eta;m_{1,\ldots N_f};\epsilon_1,\epsilon_2|t)=\sum_{n\in\mathbb{Z}}e^{\ri n \eta} \mathcal{Z}^{[N_f]}(a+n \epsilon_2;m_{1,\ldots N_f};\epsilon_1, \epsilon_2|t), \qquad 0\leq N_f\leq3.
\end{equation}

In the classical limit \eqref{tau_quantum_4}, \eqref{tau_quantum_3210} reproduce results of \cite{GIL13}. Namely, for $\epsilon\equiv\epsilon_1{+}\epsilon_2=0$ under dictionary
\begin{equation}
a=\sigma\epsilon_2, \qquad t=\epsilon_2^{4{-}N_f}\st, \qquad m_f=\epsilon_2\sm_f,    
\end{equation}
according to \eqref{gamma_Barnes}, we have 
\begin{equation}\label{pert_clas}
\mathcal{Z}_{cl}^{[N_f]}(a;-\epsilon_2,\epsilon_2|t) \mathcal{Z}_{1-loop}^{[N_f]}(a,m_{1,\ldots N_f};-\epsilon_2,\epsilon_2)=\frac{e^{2(1{-}N_f)\zeta'(-1)}}{(2\pi)^{\se_1^{[N_f]}}}\epsilon_2^{\mathrm{w}_2^{[N_f]}+\frac{1{-}N_f}6}\prod\limits_{\pm}\frac{\prod\limits_{f=1}^{N_f}\mathsf{G}(1+\sm_f\pm \sigma)}{\mathsf{G}(1\pm 2\sigma)}\st^{\sigma^2}
\end{equation}
and
\begin{equation}
\mathcal{Z}_{inst}^{[N_f]}(a,m_{1,\ldots N_f};-\epsilon_2,\epsilon_2|t)=\mathcal{Z}_{inst}^{[N_f]}(\sigma,\sm_{1,\ldots N_f};-1,1|\st)    
\end{equation}
according to \eqref{Zinst_scale}.
The $\sigma-$ and $\st-$independent prefactor before the product of
\eqref{pert_clas} is omitted below. Then, due to \eqref{Zinst} for $N_f=4$, the tau function \eqref{tau_quantum_4} in the classical limit  coincides with GIL tau function \cite[(4.1)]{GIL13} up to an elementary prefactor
\begin{equation}
\tau^{[4]}(a,\eta;m_{1,\ldots 4};-\epsilon_2,\epsilon_2|t)=t^{\theta_0^2+\theta_t^2}(1-t)^{-2\theta_t\theta_1}\tau_{\VI}^{\mathrm{GIL}}(\sigma,s^{\mathrm{GIL}}_{\VI};\theta_{0,t,1,\infty}|\st), \qquad e^{\ri\eta}=s^{\mathrm{GIL}}_{\VI},   
\end{equation}
while adding relations \eqref{masses_theta_4} to the dictionary. Analogously, tau functions \eqref{tau_quantum_3210} for $N_f=3,2,1,0$ in the classical limit coincide with GIL tau functions \cite[(4.14,4.17,4.21,4.25)]{GIL13} respectively
\begin{equation}
\tau^{[N_f]}(a,\eta;m_{1,\ldots N_f};-\epsilon_2,\epsilon_2|t)=\tau_{\mathrm{eq}}^{\mathrm{GIL}}(\sigma,s^{\mathrm{GIL}}_{\mathrm{eq}};\{\theta_\#\}|\st), \qquad e^{\ri\eta}=s^{\mathrm{GIL}}_{\mathrm{eq}},   
\end{equation}
where $\theta$'s and masses are related by the formulas: for $N_f=3$ by \eqref{PV_dct}, for $N_f=2$ by \eqref{PIII1_dct} and for $N_f=1$ by \eqref{PIII2_dct}.

\paragraph{Coalescence for solutions at strong coupling.}
In the coming Sec.~\ref{sec:quant_exp} we are looking for the large time asymptotic expansion of the tau function of quantum Painlev\'e equations around irregular singularity $t=\infty$, which corresponds to the strong coupling expansion for the partition functions. A natural question arises: could we control the limit directly at the level of solutions (or partition functions) in the same way as it is possible in the weak coupling case? From our analysis it seems that such limits are very delicate and that more conceptual work is needed in order to control them in a unified and straightforward way. 

There are two cases in which these limits are tamed. One is the
limit from the regular four-point Virasoro conformal block around the regular singularity at $t=\infty$ to a three point confluent conformal block around the irregular one 
\cite[Sec. 1.3]{LNR18} and the other is its further limit to a double confluent two point conformal block \cite[Sec. 3.4.1]{BIPT22}.
Below we present the results of \cite{LNR18} in terms of the gauge theory language through the AGT dictionary, while those of
\cite{BIPT22} were already discussed in these terms there.

In \cite{LNR18} the holomorphic decoupling limit 
of the strongly coupled partition function for $N_f=4$
is discussed. This exactly gives
the limit {\bf PVI}$\rightarrow${\bf PV} we discussed in
Sec. \ref{ssec:D-coal_q} through the action by $q_{0\infty}$. Namely, the limit of the
instanton partition function reads as
\begin{equation}
(1{-}1/t^{[4]})^{\frac{(a{-}m_4)^2-(e_4^{[4]}{+}\epsilon)^2/4}{\epsilon_1\epsilon_2}}\mathcal{Z}_{inst}^{[4]}(a,q_{0\infty}(m_{1,2,3,4});\epsilon_1,\epsilon_2|1/t^{[4]})\Bigg|_{\substack{t^{[4]}=t/m_4 \\ a=\frac{m_4}2{+}a_{\D}}} \xrightarrow[]{m_4\rightarrow\infty} \mathcal{Z}_{inst}^{[\mathbf{3_L}]}(a_{\D}, m_{1,2,3};\epsilon_1,\epsilon_2|t), 
\end{equation}
where we take $q_{0\infty}(m_i)=\frac12\sum_{f=1}^4m_f-m_i|_ {i=1,2,3,4}$.
It is clear that, in comparison to the weak coupling, this limit is highly nontrivial, because each term of the Nekrasov formula diverges, but its divergence is compensated by a $U(1)$ additional factor
and a proper scaling of the vacuum expectation value $a$. 
We checked this limit up to order $4$, reproducing the linear exponent singularity expansion \eqref{Zinst_3l}, considering the expressions as formal series in $t^{-1}$.

How to systematically tame such limits is indeed a very interesting and nontrivial question, whose study got started already in \cite{GT12}. In the  AGT-dual perspective, more general results are expected at the level of Virasoro irregular vertex operators \cite{NN}.

\section{Quantum expansions}
\label{sec:quant_exp}
In this section we present asymptotic large time expansions of formal local solutions of the quantum Painlev\'e equations obtained in the previous section, which describe via Painlev\'e/gauge correspondence the {\it strong coupling phase} of the gauge theory. Our solutions generalize the ones of \cite{BLMST16} to general $\Omega$-background $\epsilon\neq 0$ by the formal quantization of the Zak transform.

We first review the structure of the ansatz for the solutions in \cite{BLMST16} and illustrate modifications needed to generalize it to the quantum Painlev\'e case.
Then we discuss case by case the concrete forms of the solutions.

\paragraph{Ansatz of \cite{BLMST16}.}
Each solution of \cite{BLMST16} is of form \eqref{tau_quantum} with $\epsilon=0$ and $a=\nu\epsilon_2$. Partition function $\mathcal{Z}$ actually depends on $t$ as fractional power $\mathrm{s}=\kappa \mathrm{t}^d$, $d\in \mathbb{Q}$, where $\mathrm{t}$, $\mathrm{s}$ are staying for the variables $t,s$ of \cite{BLMST16} and $\kappa\in\mathbb{C}$ is a convenient normalization factor. The partition function factorizes into three terms with the following dependence on $\mathrm{s}$ and $\nu$ 
\begin{enumerate}
    \item A monomial factor of form $\mathrm{s}^{\upxi_2-N_p \nu^2/2}\, e^{\upbeta \mathrm{s}^2+\upxi_1 \mathrm{s}+\updelta\nu \mathrm{s}}$,
    where $N_p\in\mathbb{N}$.
    \item A $\mathrm{t}$-independent factor of form $e^{\upchi\nu^2/2}\prod\limits_{i=1}^{N_p} \mathsf{G}(\nu+\upmu_i)$.
    \item An asymptotic power series in $\mathrm{s}^{-1}$ that starts from $1$, the $k$-th coefficient being a polynomial of order $3k$ in $\nu$.
\end{enumerate} 
The parameters 
entering the above ansatz are complex numbers or simple polynomial invariants of the masses fixed case by case. We discuss more on these coefficients in the quantum case.
The terms in the second and third item respectively are interpreted as the 1-loop and nonperturbative part of the quantum field theory describing the low-energy excitations around the expansion point. In particular number $N_p$ of the Barnes $\mathsf{G}$-function factors corresponds to the number of light BPS particles. The first term is needed to provide consistent asymptotic behavior. In the expansion in the electric frame, where a Lagrangian description of the gauge theory is known, it corresponds to the classical term of the partition function. Accordingly, 
we will call the above factors classical, 1-loop and instanton parts of $\mathcal{Z}$.

In order to make sense to each whole solution as an asymptotic expansion in large $\mathrm{t}$, $\updelta \mathrm{s}$ from Item 1 above must be purely imaginary to prevent an exponential growth of the monomial factor under shift $\nu\mapsto\nu+n$, $n\in\mathbb{Z}$. So each asymptotic expansion is forced to be performed along a set of rays
$\arg \delta+\arg \kappa +d\arg \mathrm{t}=\pm\frac{\pi}2$. This set of rays is just a subset of canonical rays described at the beginning of Sec.~\ref{sec:coalesc}.
As already explained there, the expansions of \cite{BLMST16} along the whole set of canonical rays get identified by the action of the rotational symmetry presented in Tab.~\ref{tab:expansions}. Recall that, according to this symmetry, we have two different expansions for PV, PIV and PII, actually with same $d$. The two large $\mathrm{t}$ expansions for same Painlev\'e equation are distinguished by their asymptotic behavior, namely, by the monomial factor, as follows. The expansions 
along red rays of Fig.~\ref{fig:canonical_rays} correspond to $N_p=1$ while the expansions along the blue rays correspond to $N_p$ equal to the number of hypermultiplets plus $1$. From the gauge theory viewpoint they describe two different phases, one with heavy hypermultiplets and the other with light ones respectively. The  expansions along the red and blue rays also have different $\upbeta$, in the gauge of \cite{BLMST16} they have $\upbeta=0$ for the blue rays. For PIII$_{1,2,3}$ and PI we have only one expansion, with $N_p=1$ except PIII$_1$ where $N_p=2$.

\paragraph{Ansatz for quantum Painlev\'e equation.} The ansatz we use to solve quantum Painlev\'e equations is given by the operator-like Zak transform \eqref{tau_quantum} of
\begin{equation}\label{Zstr_q}
\mathcal{Z}(a_{\D};\epsilon_1,\epsilon_2|s)=\mathcal{Z}_{cl}(a_{\D};\epsilon_1,\epsilon_2|s) \mathcal{Z}_{1-loop}(a_{\D};\epsilon_1,\epsilon_2) \mathcal{Z}_{inst}(a_{\D};\epsilon_1,\epsilon_2|s)    
\end{equation}
where we denote by
$a_{\D}$ the Cartan variable in the 
strong coupling expansion/magnetic frame
of the gauge theory and $s=\kappa t^d$ with a convenient $\kappa\in\mathbb{C}$ is an expansion variable. An important novelty with respect to classical Painlev\'e equations is that the monodromy parameters $a_{\D},\eta_{\D}$ obey the nontrivial commutation relations \eqref{as_comm_rel}. 
We expect the following expressions for the factors of $\mathcal{Z}$ as natural deformations of the above classical ones
\begin{enumerate}
    \item A classical term of form (c.f. weak coupling case \eqref{Zcl})
    \begin{equation}\label{Zcl_anz}
    \mathcal{Z}_{cl}(a_{\D};\epsilon_1,\epsilon_2|s)=s^{-\frac{\xi_2-N_p a_{\D}^2/2}{\epsilon_1\epsilon_2}} e^{-\frac{\beta s^2+\xi_1 s+\delta a_{\D} s}{\epsilon_1\epsilon_2}},    
    \end{equation}
    We keep same $N_p\in\mathbb{N}$ because it appears in combination
    \begin{equation}
    \frac{\mathcal{Z}_{cl}(a_{\D}+2n\epsilon_1;2\epsilon_1,\epsilon_2{-}\epsilon_1|s) \mathcal{Z}_{cl}(a_{\D}+2n\epsilon_2;\epsilon_1{-}\epsilon_2,2\epsilon_2|s)}{\mathcal{Z}_{cl}(a_{\D};2\epsilon_1,\epsilon_2{-}\epsilon_1|s) \mathcal{Z}_{cl}(a_{\D};\epsilon_1{-}\epsilon_2,2\epsilon_2|s)}=s^{-N_pn^2}.  
\end{equation}
 Note that we always choose power $d$ in $s=\kappa t^d$ so that $s$ has mass dimension $1$. This implies that $\beta$ and $\delta$ are just complex numbers and that $\xi_1$ and $\xi_2$ are polynomials of mass dimension $1$ and $2$ respectively in the masses and in $\Omega$-background permutation invariants $\epsilon=\epsilon_1+\epsilon_2$ and $\epsilon_1\epsilon_2$.
     \item A 1-loop part of form
     \begin{equation}\label{Z1loop_anz}
     \mathcal{Z}_{1-loop}(a_{\D};\epsilon_1,\epsilon_2)=e^{-\frac{\chi a_{\D}^2}{2\epsilon_1\epsilon_2}}\prod_{i=1}^{N_p} \exp\gamma_{\epsilon_1,\epsilon_2}(a_{\D}+\mu_i-\epsilon/2),
     \end{equation}
    Function $\gamma_{\epsilon_1,\epsilon_2}$ naturally deforms the classical Barnes function according to \eqref{gamma_Barnes} just as in the weak coupling case \eqref{Z1loop}.
    Dimension counting prescribes that $\chi$ is a complex number and $\mu_i$ are mass dimension $1$ polynomials\footnote{Actually it also appears that $\sum_{i=1}^{N_p}\mu_i=0$.} in the masses and $\epsilon$.
    \item An instanton part which we write as a power series in $s^{-1}$ 
     \begin{equation} \label{inst_ansatz}
     \mathcal{Z}_{inst}(a_{\D};\epsilon_1,\epsilon_2|s)=1+\sum_{k=1}^{K} Z_{3k}(a_{\D}) (\epsilon_1\epsilon_2s)^{-k}+O(s^{-K-1}),    
    \end{equation}
    where $Z_{3k}(a_{\D})$ is a polynomial of order $3k$ in $a_{\D}$ and the same mass dimension with coefficients being polynomials in the masses and $\epsilon,\, \epsilon_1\epsilon_2$. As in the  $\epsilon=0$ case, this power series turns out to be an asymptotic one.
\end{enumerate} 
As it was already mentioned for (Q)PVI at the end of Sec. \ref{ssec:blowuptau}, the (quantum) Painlev\'e equations in the tau form have an additional integration constant replacing the dependence on one of the mass variables. In the classical case this integration constant is fixed by the Hamiltonian form of the equation but in the quantum case we do not know an analog of that. Therefore we fix the $\epsilon$-corrections for such dependence indirectly, primarily by comparing the partition functions we obtain with the holomorphic anomaly calculations in Sec. \ref{sec:holom_anom}.

\paragraph{General procedure to determine the ansatz.}
Now we describe how we determine the coefficients entering the above ansatz by imposing that it solves the quantum Painlev\'e equations with a suitable asymptotic behavior.
In each case we substitute the above ansatz to the quantum Painlev\'e equation, i.e. in a system of equations of form \eqref{blowup_tau_subst} obtained in Sec.~\ref{sec:coalesc}. Up to a general factor the relation we obtain is a power series in $s^{-1}$ whose coefficients impose relations on the parameters entering the ansatz, including the coefficients of polynomial $Z_{3k}(a_{\D})$. We will call them $s^{-k}$ relations. Note that a $t$-independent term $\mathcal{Z}_{1-loop}$ appears in these relations as an analog of \eqref{blowup_coeff_def}
\begin{align}\label{l_n_def}
\mathfrak{l}_n=\frac{\mathcal{Z}_{1-loop}(a_{\D}+2n\epsilon_1;2\epsilon_1,\epsilon_2{-}\epsilon_1)\mathcal{Z}_{1-loop}(a_{\D}+2n\epsilon_2;\epsilon_1{-}\epsilon_2,2\epsilon_2)}{\mathcal{Z}_{1-loop}(a_{\D}+\mathfrak{p}(2n)\epsilon_1;2\epsilon_1,\epsilon_2{-}\epsilon_1)\mathcal{Z}_{1-loop}(a_{\D}+\mathfrak{p}(2n)\epsilon_2;\epsilon_1{-}\epsilon_2,2\epsilon_2)}=\\
=e^{\chi (n^2{-}\mathfrak{p}(2n)/4)}\prod_{i=1}^{N_p} \prod\limits_{\mathrm{reg}(2n)} \big(a_{\D}+\mu_i+\mathrm{sgn}(n) (i \epsilon_1+j \epsilon_2)\big). \label{l_n_fact}
\end{align} 

In each case to find $\mathcal{Z}$ we follow the general procedure below
\begin{enumerate}
    \item The very first factor we set is the classical one. We find it from the first few $s^{-k}$ relations.
    \item Then, following the $s^{-k}$ relations step by step, we obtain blowup factors $\mathfrak{l}_n$ together with the coefficients of polynomials $Z_{3k}(a_{\D})$. 
    It will appear that the coefficients of $Z_{3k}(a_{\D})$ and $\mathfrak{l}_n$ depend on mass invariants, corresponding to those of the classical Painlev\'e equations, and on the additional integration constant replacing the highest dimension mass invariant. 
    When available, we can use the factorization \eqref{l_n_fact} to fix this freedom completely.
    \item For $\mathcal{Z}_{inst}$ we present the first few terms of the power series in $s^{-1}$ of $\ln \mathcal{Z}_{inst}$. Dimension counting prescribes that taking the logarithm drastically reduces the power in $a_{\D}$ of the $s^{-k}$-coefficients, namely
    \begin{equation}\label{inst_ansatz_ln}
    -\epsilon_1\epsilon_2 \ln \mathcal{Z}_{inst}=\sum_{k=1}^K P_{k+2}(a_{\D}) s^{-k}+O(s^{-K-1}),   
    \end{equation}
    where $P_{k+2}(a_{\D})$ is a polynomial in $a_{\D}$ of order $k+2$ and the same mass dimension.
    \item Finally, we compare our results with the undeformed, classical ones by checking their self-consistency and fixing the integration constant freedom up to $\epsilon$-corrections. These last are completely fixed in Sec.~\ref{sec:holom_anom}.
      
\end{enumerate}
The final answer we obtain in the quantum case is the Zak transform in noncommutative with $a_{\D}$ operator $e^{\ri\eta_{\D}}$,   with coefficients given by the asymptotic expansion of \eqref{Zstr_q}\footnote{We could consider that $\eta_{\D}=\ri\epsilon\frac{\partial}{\partial a_{\D}}$, however, in the given setting we do not need to specify the nature of operator $\eta_{\D}$.}. Note that for asymptotic expansions of individual functions $\mathcal{Z}$ it is not necessary to restrict $t$ along the canonical rays. Though, unlike the classical Painlev\'e equations, we, strictly speaking, cannot even guarantee the existence of solutions $\mathcal{Z}$ to the considered equations in the form \eqref{blowup_tau_subst}.

\subsection{QPV (linear exp singularity) / \texorpdfstring{$N_f=3$}{N\_f=3} with light hypers}
\label{ssec:exp_3l}
As already mentioned, for Painlev\'e V we have two different large time expansions. Here we present the quantum deformation of the one 
on the imaginary canonical rays from \cite[Sec. A.4]{BLMST16}, called there "expansion~2". For the (quantum) Painlev\'e V we have $d=1$ and we set $s=t$ for the both expansions.

\paragraph{Classical part.}
For this expansion we have $N_p=4$. The first few leading $t^{-k}$ relations for \eqref{D13irr}, \eqref{quantum_PV} imply that the classical part can be taken with $\beta=0$ as
\begin{align}
&\mathcal{Z}_{cl}^{[\mathbf{3_L}]}(a_{\D}, e_{1,2,3};\epsilon_1,\epsilon_2|t)=\mathcal{Z}^{[\mathbf{3_L}]}_{cl-m}(e_{1,2};\epsilon_1,\epsilon_2|t) \, \mathcal{Z}^{[\mathbf{3_L}]}_{cl-a}(a_{\D};\epsilon_1,\epsilon_2|t) \qquad \textrm{with}\nonumber \\  \label{Zcl_V_2}
&\mathcal{Z}^{[\mathbf{3_L}]}_{cl-m}(e_{1,2};\epsilon_1,\epsilon_2|t)=e^{-\frac{(e_1+\epsilon)t}{2\epsilon_1\epsilon_2}} \, t^{\frac{\epsilon^2-w_2}{2\epsilon_1\epsilon_2}}, \qquad  \quad \mathcal{Z}^{[\mathbf{3_L}]}_{cl-a}(a_{\D};\epsilon_1,\epsilon_2|t)=e^{-\frac{a_{\D}t}{\epsilon_1\epsilon_2}} \, t^{\frac{2a_{\D}^2}{\epsilon_1\epsilon_2}}.
\end{align}
The superscript for $\mathcal{Z}$ here and below will be in boldface for the large time expansions and subscript {\bf L} inside it denotes the Linear behavior in $s$ of the exponent in the classical part. 
On the contrary, in each special expansion we omit the superscript on the mass invariants for brevity, i.e. here we write $w_2$ instead of $w_2^{[3]}=m_1^2{+}m_2^2{+}m_3^2$.
By normalizing the tau function of QPV with $a_{\D}$-independent factor $\mathcal{Z}_{cl-m}^{[\mathbf{3_L}]}$ 
\begin{equation}\label{pref_rem_2}
\tau_r(\epsilon_1,\epsilon_2|t)=\frac{\tau(\epsilon_1,\epsilon_2|t)}{\mathcal{Z}^{[\mathbf{3_L}]}_{cl-m}(e_{1,2};\epsilon_1,\epsilon_2|t)},
\end{equation}
we obtain from \eqref{D13irr},  \eqref{quantum_PV}
\begin{equation}\label{D13irr_r}
D^1_{\epsilon_1,\epsilon_2[\ln t]}(\tau_r^{(1)},\tau_r^{(2)})=0,\qquad \qquad
   D^3_{\epsilon_1,\epsilon_2[\ln t]}(\tau_r^{(1)},\tau_r^{(2)})=\epsilon D^2_{\epsilon_1,\epsilon_2[\ln t]}(\tau^{(1)}_r,\tau^{(2)}_r),
   \end{equation}
   \vspace{-0.7cm}
    \begin{multline}\label{quantum_PV_r}
    D^4_{\epsilon_1,\epsilon_2[\ln t]}(\tau_r^{(1)},\tau_r^{(2)})+2\left(\epsilon_1\epsilon_2\frac{d}{d\ln t}\right)D^2_{\epsilon_1,\epsilon_2[\ln t]}(\tau_r^{(1)},\tau_r^{(2)})-
    \left(\frac{t^2}4+\frac{w_2}2{+}\epsilon_1\epsilon_2{+}\frac{\epsilon^2}2\right) D^2_{\epsilon_1,\epsilon_2[\ln t]}(\tau_r^{(1)},\tau_r^{(2)})\\
    -\frac{t^2}4\left(\epsilon_1\epsilon_2\frac{d}{d\ln t}\right)D^0_{\epsilon_1,\epsilon_2[\ln t]}\left(\tau^{(1)}_r\tau^{(2)}_r\right)+\frac{e_3}4 t\,\tau^{(1)}_r\tau^{(2)}_r=0,
    \end{multline}
    so we see that the quantum Painlev\'e V actually depends only on two mass parameters, namely $w_2$ and $e_3$. The third one appears as an integration constant as for the classical Painlev\'e V (App.~\ref{ssec:D-coal}).

\paragraph{Expansion.}
We then solve the successive $s^{-k}$ relations, obtaining step by step the leading terms of $\mathcal{Z}_{inst}^{[\mathbf{3_L}]}$ together with blowup factors $\mathfrak{l}_n^{[\mathbf{3_L}]}$. In this way we computed $\mathfrak{l}_n^{[\mathbf{3_L}]}$ for $|n|\leq\frac32$ 
\begin{multline}
\label{l_n_V_2}
\mathfrak{l}_0^{[\mathbf{3_L}]}=\mathfrak{l}_{\pm \frac12}^{[\mathbf{3_L}]}=1, \qquad \mathfrak{l}_{\pm 1}^{[\mathbf{3_L}]}=
\prod\limits_{\lambda,\lambda'=\pm1}  \left(a_{\D}+\frac{\lambda \tilde{m}_1{+}\lambda' \tilde{m}_2{+}\lambda \lambda' \tilde{m}_3}2 \pm\frac{\epsilon}2\right), \\ \mathfrak{l}_{\pm \frac32}^{[\mathbf{3_L}]}=
\prod\limits_{\lambda,\lambda'=\pm1} \prod\limits_{\substack{(i,j)=\\ \scriptscriptstyle(1,3),(3,1)}}\left(a_{\D}+\frac{\lambda \tilde{m}_1{+}\lambda' \tilde{m}_2{+}\lambda \lambda' \tilde{m}_3}2 \pm \frac{i \epsilon_1{+}j\epsilon_2}2\right)
\end{multline}
together with the terms of $\mathcal{Z}_{inst}^{[\mathbf{3_L}]}$ up to $t^{-8}$, up to $t^{-4}$ in the logarithmic expansion they are
\begin{multline}\label{Zinst_3l}
-\epsilon_1\epsilon_2 \ln \mathcal{Z}_{inst}^{[\mathbf{3_L}]}(a_{\D},w_2,e_3,\tilde{w}_4;\epsilon_1,\epsilon_2|t)=\left(4a_{\D}^3-(w_2{-}\epsilon^2)a_{\D}+e_3\right)\cdot\frac1{t}\\+\left(10 a_{\D}^4-(3w_2{-}5\epsilon^2)a_{\D}^2+4e_3 a_{\D} +\frac{(w_2{-}\epsilon^2)^2}8-\frac{\tilde{w}_4}2\right)\cdot\frac1{t^2}\\ +\left(44a_{\D}^5-2\frac{23w_2{+}2\epsilon_1\epsilon_2{-}57\epsilon^2}3a_{\D}^3+22e_3a_{\D}^2-\frac{c_1}{12} a_{\D}-\frac{e_3}6(w_2{+}2\epsilon_1\epsilon_2{-}13\epsilon^2)\right)\cdot\frac1{t^3}\\
+\Bigg(252a_{\D}^6-2(50w_2{+}13\epsilon_1\epsilon_2{-}170\epsilon^2)a_{\D}^4+146e_3 a_{\D}^3-\frac{c_2}4a_{\D}^2-\frac{e_3(9w_2{+}22\epsilon_1\epsilon_2{-}93\epsilon^2)}2a_{\D} \\ +\frac{w_2{+}3\epsilon_1\epsilon_2{-}7\epsilon^2}2\left(\tilde{w}_4{-}\frac{(w_2{-}\epsilon^2)^2}4\right){+}\frac{e_3^2}2\Bigg)\cdot\frac1{t^4}+\ldots+O\left(\frac1{t^9}\right), 
\end{multline}
where
\begin{align}
&c_1=44\tilde{w}_4-13w_2^2-2(2\epsilon_1\epsilon_2{-}25\epsilon^2)w_2+4\epsilon_1\epsilon_2\epsilon^2-37\epsilon^4,\\
&c_2=108\tilde{w}_4-39w_2^2-2(16\epsilon_1\epsilon_2{-}115\epsilon^2)w_2+56\epsilon_1\epsilon_2\epsilon^2-239\epsilon^4.
\end{align}
This expansion has the expected dependence on integration constant $\tilde{w}_4$, for $\mathfrak{l}_n^{[\mathbf{3_L}]}$
it is parametrized as
\begin{equation}\label{mtilda_def}
w_2=\tilde{m}_1^2+\tilde{m}_2^2+\tilde{m}_3^2, \qquad e_3=\tilde{m}_1\tilde{m}_2\tilde{m}_3, \qquad \tilde{w}_4=\tilde{m}_1^2 \tilde{m}_2^2+\tilde{m}_1^2 \tilde{m}_3^2+\tilde{m}_2^2 \tilde{m}_3^2\, .
\end{equation}
Note that when we substitute the ansatz into the equations we should control the dependence on $\epsilon_{1,2}$ of the variable $\tilde{w}_4$. The substitution gives  $\tilde{w}_4(2\epsilon_1,\epsilon_2{-}\epsilon_1)=\tilde{w}_4(\epsilon_1{-}\epsilon_2,2\epsilon_2)$, and then it is natural to consider that $\tilde{w}_4$ depends on $\epsilon$ but not on $\epsilon_1\epsilon_2$, so the substitution into the equation does not affect $\tilde{w}_4$.
Above $\mathfrak{l}_n^{[\mathbf{3_L}]}$ are of form \eqref{l_n_fact}, so our ansatz \eqref{Z1loop_anz} gives
\begin{equation}\label{Z1loop_V_2}
\mathcal{Z}_{1-loop}^{[\mathbf{3_L}]}(a_{\D},\tilde{m}_{1,2,3};\epsilon_1,\epsilon_2)= \prod\limits_{\lambda,\lambda'=\pm1}  \exp \gamma_{\epsilon_1,\epsilon_2} \left(a_{\D}{+}\frac12(\lambda \tilde{m}_1{+}\lambda' \tilde{m}_2{+}\lambda \lambda' \tilde{m}_3{-}\epsilon)\right). 
\end{equation}

\paragraph{Comparison with the classical limit ($\epsilon=0$).}
Let us compare our result in the  $\epsilon=0$ limit with expansion BLMST \cite[(A.49)]{BLMST16}. According to \eqref{PV_tau_cl_app} and \eqref{PV_tau_lit} we check that 
\begin{equation}\label{PV_tau_cl}
\tau(a_{\D},\eta_{\D},m_{1,2,3};-2\epsilon_2,2\epsilon_2|t)=e^{-\theta_* \mathrm{t}_{\sV}/2}e^{\se_1 \mathrm{t}_{\sV}/2}\tau_{\V}^{\mathrm{BLMST}}(\nu,\rho;\theta_0,\theta_t,\theta_*|\mathrm{t}_{\V}).
\end{equation}
The dictionary is given by $\eta_{\D}=\rho$ and standard scalings (see \eqref{PV_tau_cl_app} for the all except the first one)
\begin{equation}
\label{PV_par_sc}
a_{\D}=2\epsilon_2\nu, \qquad t=2\epsilon_2 \mathrm{t}_{\V}, \qquad m_f=2\epsilon_2\sm_f|_{f=1,2,3},
\end{equation}
where masses $\sm_{1,2,3}$ are given by \eqref{PV_dct}.
Then, classical part \eqref{Zcl_V_2} becomes
\begin{equation}
\mathcal{Z}_{cl}^{[\mathbf{3_L}]}(-2\epsilon_2,2\epsilon_2)=e^{(\nu+\se_1/2)\mathrm{t}_{\sV}}\, (2\epsilon_2 \mathrm{t}_{\V})^{\theta_0^2+\theta_t^2+\theta_*^2/2-2\nu^2}.  
\end{equation} 
1-loop part \eqref{Z1loop_V_2}, by using \eqref{gamma_Barnes}, becomes 
\begin{equation}
\mathcal{Z}_{1-loop}^{[\mathbf{3_L}]}(-2\epsilon_2,2\epsilon_2)=\frac{(2\epsilon_2)^{2\nu^2{+}\theta_0^2{+}\theta_t^2{+}\frac{\theta_*^2}2{-}\frac13}}{e^{4\zeta'(-1)}(2\pi)^{2\nu}}\prod_{\pm} \mathsf{G}\left(1+\nu\pm\tilde{\theta}_t+\frac{\tilde{\theta}_*}2\right) \mathsf{G}\left(1+\nu\pm\tilde{\theta}_0-\frac{\tilde{\theta}_*}2\right),
\end{equation}
where the tilded thetas are connected with the tilded masses by same formulas \eqref{PV_par_sc}, \eqref{PV_dct} as for the non-tilded ones. The Barnes $\mathsf{G}$-function factor of the above expression coincides with the counterpart from \cite[(A.49)]{BLMST16} iff in the self-dual case the tilded and the non-tilded masses are equal up to discrete group $W(A_3)=S_4$, which acts by  transpositions of the masses and changing signs for an even number of the masses. This means that $\tilde{w}_4|_{\epsilon=0}=w_4$, where $w_4=m_1^2m_2^2{+}m_1^2m_3^2{+}m_2^2m_3^2$. 
Then, up to the factor in \eqref{PV_tau_cl} and a constant factor in $\mathrm{t}_{\V}$ and $\nu$, $\mathcal{Z}_{cl}^{[\mathbf{3_L}]}\mathcal{Z}_{1-loop}^{[\mathbf{3_L}]}$ coincides with the prefactor of the $\mathrm{t}_{\V}^{-1}$ series in \cite[(A.49)]{BLMST16}\footnote{In \cite{BLMST16} there is additional factor $2^{-2\nu^2}$, it is a typo there.}. Finally, for the first two terms of the instanton expansion we obtain \cite[(A.50)]{BLMST16} 
\begin{multline}
\mathcal{Z}_{inst}^{[\mathbf{3_L}]}(-2\epsilon_2,2\epsilon_2)=1+\left(4\nu^3-\mathrm{w}_2 \nu+\se_3\right)\cdot\frac1{\mathrm{t}_{\V}}\\+\left(8\nu^6-2(2\mathrm{w}_2{-}5)\nu^4+4\se_3\nu^3+\frac{\mathrm{w}_2 (\mathrm{w}_2{-}6)}2\nu^2-\se_3(\mathrm{w}_2{-}4)\nu +\frac{\se_3^2}2{+}\frac{\mathrm{w}_2^2}8{-}\frac{\mathrm{w}_4}2\right)\cdot\frac1{\mathrm{t}_{\V}^2}+O\left(\frac1{\mathrm{t}_{\V}^3}\right), 
\end{multline}
where $\mathrm{w}_2, \se_3, \mathrm{w}_4$ are defined in \eqref{PV_inv}. In terms of $\theta$-variables \eqref{PV_dct} of \cite[(A.50)]{BLMST16} they read 
\begin{equation}\label{PV_theta}
\mathrm{w}_2=2\theta_0^2+2\theta_t^2+\theta_*^2, \qquad \se_3=\theta_*(\theta_t^2-\theta_0^2), \qquad \mathrm{w}_4=(\theta_t^2-\theta_0^2)^2+2\theta_*^2(\theta_0^2{+}\theta_t^2).    
\end{equation}

Let us comment more on the connection between the tilded and the non-tilded masses, the latter being the gauge theory ones. The factorization \eqref{l_n_fact} prescribes that, without loss of generality, the differences $\tilde{m}_f-m_f|_{f=1,2,3}$ are proportional to $\epsilon$ up to a numerical coefficient. The first two relations of \eqref{mtilda_def} imply that such coefficients vanish. So, under the assumptions of the ansatz, we have $\tilde{w}_4=w_4$, as it is further confirmed in Sec.~\ref{ssec:holom_anom_3}.

\subsection{QPV (square exp singularity) / \texorpdfstring{$N_f=3$}{N\_f=3} with heavy hypers}
\label{ssec:exp_3s}
Here we present the quantum deformation of the other Painlev\'e V large time expansion, namely the one on the real canonical rays from \cite[Sec. A.4]{BLMST16}, called there "expansion~1".
Recall that for QPV we set $s=t$.

\paragraph{Classical part.}
For this expansion we have $N_p=1$. The first few leading $t^{-k}$ relations for \eqref{D13irr}, \eqref{quantum_PV} imply that the classical part can be taken with $\beta\neq0$ as
\begin{align} 
&\mathcal{Z}_{cl}^{[\mathbf{3_S}]}(a_{\D},e_{1,2,3};\epsilon_1,\epsilon_2|t)=\mathcal{Z}_{cl-m}^{[\mathbf{3_S}]}(e_{1,2};\epsilon_1,\epsilon_2|t) \mathcal{Z}_{cl-a}^{[\mathbf{3_S}]}(a_{\D};\epsilon_1,\epsilon_2|t) \qquad \textrm{with}\nonumber \\ \label{Zcl_V_1}
&\mathcal{Z}_{cl-m}^{[\mathbf{3_S}]}(e_{1,2};\epsilon_1,\epsilon_2|t)=e^{-\frac{t^2}{32\epsilon_1\epsilon_2}}e^{-\frac{(e_1+\epsilon)t}{2\epsilon_1\epsilon_2}} \, t^{\frac{5\epsilon^2/8-w_2}{\epsilon_1\epsilon_2}-\frac14}, \qquad \quad \mathcal{Z}_{cl-a}^{[\mathbf{3_S}]}(a_{\D};\epsilon_1,\epsilon_2|t)=e^{-\frac{\ri a_{\D}t}{2\epsilon_1\epsilon_2}} \, t^{\frac{a_{\D}^2}{2\epsilon_1\epsilon_2}},
\end{align}
where subscript {\bf S} denotes the Square behavior in $s$ of the exponent in the classical part. 
By normalizing the tau function of QPV with $a_{\D}$-independent factor $\mathcal{Z}_{cl-m}^{[\mathbf{3_S}]}$ 
\begin{equation}\label{pref_rem_1}
\tau_r(\epsilon_1,\epsilon_2|t)=\frac{\tau(\epsilon_1,\epsilon_2|t)}{\mathcal{Z}^{[\mathbf{3_S}]}_{cl-m}(e_{1,2};\epsilon_1,\epsilon_2|t)}
\end{equation}
we obtain from \eqref{D13irr},  \eqref{quantum_PV}
\begin{equation}\label{D13irr_rr}
   D^1_{\epsilon_1,\epsilon_2[\ln t]}(\tau_r^{(1)},\tau_r^{(2)})=\frac{\epsilon}4\tau_r^{(1)}\tau_r^{(2)},\qquad
   D^3_{\epsilon_1,\epsilon_2[\ln t]}(\tau_r^{(1)},\tau_r^{(2)})=\frac74\epsilon D^2_{\epsilon_1,\epsilon_2[\ln t]}(\tau^{(1)}_r,\tau^{(2)}_r)-\frac{\epsilon}{16}\left(t^2{+}\frac{3\epsilon^2}2\right)\tau^{(1)}_r\tau^{(2)}_r,
    \end{equation}
    \vspace{-0.7cm}
    \begin{multline}\label{quantum_PV_rr}
    D^4_{\epsilon_1,\epsilon_2[\ln t]}(\tau_r^{(1)},\tau_r^{(2)})+2\left(\epsilon_1\epsilon_2\frac{d}{d\ln t}\right)D^2_{\epsilon_1,\epsilon_2[\ln t]}(\tau_r^{(1)},\tau_r^{(2)})+
    \left(\frac{t^2}{16}{-}w_2{-}2\epsilon_1\epsilon_2{-}\frac{7\epsilon^2}4\right) D^2_{\epsilon_1,\epsilon_2[\ln t]}(\tau_r^{(1)},\tau_r^{(2)})\\
    -\frac18 (t^2{+}\epsilon^2)\left(\epsilon_1\epsilon_2\frac{d}{d\ln t}\right)D^0_{\epsilon_1,\epsilon_2[\ln t]}\left(\tau^{(1)}_r\tau^{(2)}_r\right)+\left(\frac{69\epsilon^2}{256}t^2{+}\frac{e_3}4t{+}\frac{\epsilon^2}{16}\left(w_2{+}2\epsilon_1\epsilon_2{+}\frac{27\epsilon^2}{16}\right)\right)\tau^{(1)}_r\tau^{(2)}_r=0,
    \end{multline}
    so again the quantum Painlev\'e V depends only on two mass parameters $w_2$ and $e_3$. And the third mass parameter appears as an integration constant as well.

\paragraph{Expansion.}
We then solve the successive $s^{-k}$ relations, obtaining step by step the leading terms of $\mathcal{Z}_{inst}^{[\mathbf{3_S}]}$ together with blowup factors $\mathfrak{l}_n^{[\mathbf{3_S}]}$.
In this way we computed  $\mathfrak{l}_n^{[\mathbf{3_S}]}$ for $|n|\leq \frac52$ 
\begin{multline}\label{l_n_V_1}
\mathfrak{l}_0^{[\mathbf{3_S}]}=\mathfrak{l}_{\pm \frac12}^{[\mathbf{3_S}]}=1, \qquad \mathfrak{l}_{\pm 1}^{[\mathbf{3_S}]}=
\frac1{2\ri}\left(a_{\D}\pm\frac{\epsilon}2\right), \qquad \mathfrak{l}_{\pm \frac32}^{[\mathbf{3_S}]}=
-\frac14\prod\limits_{\substack{(i,j)=\\ \scriptscriptstyle (1,3),(3,1)}}\left(a_{\D}\pm \frac{i \epsilon_1{+}j\epsilon_2}2\right), \\
\mathfrak{l}_{\pm2}^{[\mathbf{3_S}]}=\frac1{16}\prod\limits_{\substack{(i,j)={\scriptscriptstyle(1,1),}\\ \scriptscriptstyle(5,1),(3,3),(1,5)}}\left(a_{\D}\pm\frac{i \epsilon_1{+}j\epsilon_2}2\right), \qquad 
\mathfrak{l}_{\pm\frac52}^{[\mathbf{3_S}]}=-\frac1{64}\prod\limits_{\substack{(i,j)={\scriptscriptstyle(1,3),(3,1),}\\ \scriptscriptstyle(7,1),(3,5),(5,3)(1,7)}}\left(a_{\D}\pm\frac{i \epsilon_1{+}j\epsilon_2}2\right) 
\end{multline}  
together with the terms of $\mathcal{Z}_{inst}^{[\mathbf{3_S}]}$ up to $t^{-5}$, up to $t^{-4}$ in the logarithmic expansion with $\alpha_{\D}=\ri a_{\D}$ they are 
{\allowdisplaybreaks\begin{multline}\label{Zinst_3s}
-\epsilon_1\epsilon_2 \ln \mathcal{Z}_{inst}^{[\mathbf{3_S}]}(a_{\D},w_2,e_3,\check{w}_4;\epsilon_1,\epsilon_2|t)=\Bigg(\frac{\alpha_{\D}^3}2+\left(4w_2{+}\frac74\epsilon_1\epsilon_2{-}\frac{11}8\epsilon^2\right)\alpha_{\D}+8e_3\Bigg)\cdot\frac1{t}\\+\Bigg(-\frac58\alpha_{\D}^4-\left(12w_2{+}\frac{45}8\epsilon_1\epsilon_2{-}\frac{65}{16}\epsilon^2\right)\alpha_{\D}^2-64e_3\alpha_{\D}-8\check{w}_4{-}2w_2\epsilon_1\epsilon_2{-}\frac12(\epsilon_1\epsilon_2)^2{+}\frac{61}{32}\epsilon_1\epsilon_2\epsilon^2\Bigg)\cdot\frac1{t^2}\\+\Bigg(\frac{11}8\alpha_{\D}^5+\left(\frac{148}3w_2{+}\frac{191}8\epsilon_1\epsilon_2{-}\frac{797}{48}\epsilon^2\right)\alpha_{\D}^3+448e_3\alpha_{\D}^2 +\frac{c_1}3\alpha_{\D}+\frac{16e_3}{3}(8w_2{+}10\epsilon_1\epsilon_2{-}17\epsilon^2)\Bigg)\cdot\frac1{t^3}
\\ +\Bigg(-\frac{63}{16}\alpha_{\D}^6-\left(230w_2{+}\frac{227}2\epsilon_1\epsilon_2{-}\frac{1235}{16}\epsilon^2\right)\alpha_{\D}^4-3008e_3\alpha_{\D}^3-c_2 \alpha_{\D}^2 \\  -16e_3(96w_2{+}146\epsilon_1\epsilon_2{-}165\epsilon^2)\alpha_{\D}+c_3  \Bigg)\cdot\frac1{t^4}+\ldots+O\left(\frac1{t^6}\right),
\end{multline}}
where 
\begin{align}
&c_1=448\check{w}_4+16w_2^2+(166\epsilon_1\epsilon_2{+}\epsilon^2)w_2+\frac{813}{16}(\epsilon_1\epsilon_2)^2-\frac{3797}{32}\epsilon_1\epsilon_2\epsilon^2-\frac{259}{128}\epsilon^4,\\
&c_2=1728\check{w}_4+96w_2^2+7(106\epsilon_1\epsilon_2{-}\epsilon^2)w_2+\frac{3811}{16}(\epsilon_1\epsilon_2)^2-\frac{31067}{64}\epsilon_1\epsilon_2\epsilon^2-\frac{2033}{256}\epsilon^4, \\
&c_3=-16(4w_2{+}10\epsilon_1\epsilon_2{-}19\epsilon^2)\check{w}_4-320e_3^2-16\epsilon_1\epsilon_2w_2^2-\left(44(\epsilon_1\epsilon_2)^2{-}\frac{259}2\epsilon_1\epsilon_2\epsilon^2{-}\frac{459}{16}\epsilon^4\right)w_2 \nonumber\\
&-10(\epsilon_1\epsilon_2)^3+\frac{4961}{64}(\epsilon_1\epsilon_2)^2\epsilon^2-\frac{17779}{256}\epsilon_1\epsilon_2\epsilon^4-\frac{4725}{512}\epsilon^6.
\end{align}
This expansion has the expected dependence on integration constant $\check{w}_4$, and  $\check{w}_4(2\epsilon_1,\epsilon_2{-}\epsilon_1)=\check{w}_4(\epsilon_1{-}\epsilon_2,2\epsilon_2)$ so we consider that $\check{w}_4$ depends on $\epsilon$ but not on $\epsilon_1\epsilon_2$ as in the linear exp case. Above $\mathfrak{l}_n^{[\mathbf{3_S}]}$ are of form \eqref{l_n_fact}, so our ansatz \eqref{Z1loop_anz} gives
\begin{equation}\label{Z1loop_V_1}
\mathcal{Z}_{1-loop}^{[\mathbf{3_S}]}(a_{\D};\epsilon_1,\epsilon_2)=(2\ri)^{\frac{a_{\D}^2}{2\epsilon_1\epsilon_2}}  \exp \gamma_{\epsilon_1,\epsilon_2} (a_{\D}-\epsilon/2). 
\end{equation}

\paragraph{Comparison with the classical limit ($\epsilon=0$).}
Let us compare our result in the $\epsilon=0$ limit with expansion \cite[(A.45)]{BLMST16}. Namely, we check \eqref{PV_tau_cl} under $\eta_{\D}=\rho$ and scaling \eqref{PV_par_sc} with  
\eqref{PV_dct}. Then, classical part \eqref{Zcl_V_1} becomes 
\begin{equation}
\mathcal{Z}_{cl}^{[\mathbf{3_S}]}(-2\epsilon_2,2\epsilon_2)=e^{\mathrm{t}_{\sV}^2/32}e^{(\ri\nu/2+\theta_t+\theta_*/2)\mathrm{t}_{\sV}}\, (2\epsilon_2 \mathrm{t}_{\V})^{2\theta_0^2+2\theta_t^2+\theta_*^2-1/4-\nu^2/2}.  
\end{equation} 
1-loop part \eqref{Z1loop_V_1}, by using \eqref{gamma_Barnes}, becomes 
\begin{equation}
\mathcal{Z}_{1-loop}^{[\mathbf{3_S}]}(-2\epsilon_2,2\epsilon_2)=(2\ri)^{-\frac{\nu^2}2}\frac{(2\epsilon_2)^{\frac{\nu^2}2-\frac1{12}}}{e^{\zeta'(-1)}(2\pi)^{\nu/2}}\mathsf{G}(1+\nu).
\end{equation}
Then, up to the factor in \eqref{PV_tau_cl} and a constant factor in $\mathrm{t}_{\V}$ and $\nu$, $\mathcal{Z}_{cl}^{[\mathbf{3_S}]}\mathcal{Z}_{1-loop}^{[\mathbf{3_S}]}$ coincides with the prefactor of the $\mathrm{t}_{\V}^{-1}$ series in \cite[(A.45)]{BLMST16}. Finally, for the first two terms of the instanton expansion we obtain \cite[(A.46)]{BLMST16} with  \eqref{PV_theta}
\begin{multline}
\mathcal{Z}_{inst}^{[\mathbf{3_S}]}(-2\epsilon_2,2\epsilon_2)=1+\Bigg(\frac{\nu^3}{4\ri}+\left(2\mathrm{w}_2{-}\frac78\right) \ri\nu+4\se_3\Bigg)\cdot\frac2{\mathrm{t}_{\V}}+\Bigg(-\frac{\nu^6}{32}+\frac12\left(\mathrm{w}_2{-}\frac34\right)\nu^4-\se_3\ri\nu^3\\ -\left(2\mathrm{w}_2^2{-}\frac{19}4\mathrm{w}_2{+}\frac{229}{128}\right)\nu^2+\se_3\left(8\mathrm{w}_2{-}\frac{39}2\right)\ri\nu+8\se_3^2{-}2\mathrm{w}_4{+}\frac{\mathrm{w}_2}2{-}\frac18\Bigg)\cdot\frac4{\mathrm{t}_{\V}^2}+O\left(\frac8{\mathrm{t}_{\V}^3}\right),
\end{multline}    
where, to match the $\nu$-independent part of $\mathrm{t}^{-2}_{\V}$ in \cite[(A.46)]{BLMST16} we imposed condition $\check{w}_4|_{\epsilon=0}=w_4$.   

\subsection{\texorpdfstring{QPIII$_1$}{QPIII₁} / \texorpdfstring{$N_f=2$}{N\_f=2}}
\label{ssec:exp_2}
Here we present the quantum deformation of the Painlev\'e III$_1$ large time expansion on the real positive canonical ray from \cite[Sec. A.3]{BLMST16}.  For the (quantum) Painlev\'e III$_1$ we have $d=1/2$ and we set $s=8\ri\, t^{1/2}$.

\paragraph{Classical part.}
For this expansion we have $N_p=2$. The first few leading $s^{-k}$ relations for \eqref{D13irr}, \eqref{quantum_PIII1} imply that the classical part can be taken as
\begin{align}
&\mathcal{Z}_{cl}^{[\mathbf{2}]}(a_{\D}, \phi;\epsilon_1,\epsilon_2|s)=\mathcal{Z}^{[\mathbf{2}]}_{cl-m}(\phi;\epsilon_1,\epsilon_2|s) \mathcal{Z}^{[\mathbf{2}]}_{cl-a}(a_{\D};\epsilon_1,\epsilon_2|s) \qquad \textrm{with}\nonumber \\ \label{Zcl_III1} &\mathcal{Z}^{[\mathbf{2}]}_{cl-m}(\phi;\epsilon_1,\epsilon_2|s)=e^{\frac{s^2}{64\epsilon_1\epsilon_2}} \, s^{\frac{\phi}{\epsilon_1\epsilon_2}}, \qquad \quad \mathcal{Z}^{[\mathbf{2}]}_{cl-a}(a_{\D};\epsilon_1,\epsilon_2|s)=e^{-\frac{a_{\D} s}{2\epsilon_1\epsilon_2}} \, s^{\frac{a_{\D}^2}{\epsilon_1\epsilon_2}},
\end{align}
where $\phi$ is a dimension $2$ integration constant that is not fixed by the further $s^{-k}$ relations, such that  $\phi(2\epsilon_1,\epsilon_2{-}\epsilon_1)=\phi(\epsilon_1{-}\epsilon_2,2\epsilon_2)$. So we consider that $\phi$ depends on $\epsilon$ but not on $\epsilon_1\epsilon_2$ as in the previous cases. Note that the quantum Painlev\'e III$_1$ \eqref{D13irr}, \eqref{quantum_PIII1} depends only on $e_2=m_1m_2$ but not on $w_2=m_1^2{+}m_2^2$ (c.f. App.~ \ref{ssec:D-coal} for the classical case).

\paragraph{Expansion.}
We then solve the successive $s^{-k}$ relations, obtaining step by step the leading terms of $\mathcal{Z}_{inst}^{[\mathbf{2}]}$ together with blowup factors $\mathfrak{l}_n^{[\mathbf{2}]}$. 
In this way we computed  $\mathfrak{l}_n^{[\mathbf{2}]}$ for $|n|\leq2$ 
\begin{multline} \label{l_n_III1}
\mspace{-20mu}\mathfrak{l}_0^{[\mathbf{2}]}=\mathfrak{l}_{\pm \frac12}^{[\mathbf{2}]}=1, \quad \mathfrak{l}_{\pm 1}^{[\mathbf{2}]}=
\prod\limits_{\lambda=\pm1}  \left(a_{\D}+\lambda\frac{\tilde{m}_1{-}\tilde{m}_2}2\pm\frac{\epsilon}2\right), \quad \mathfrak{l}_{\pm \frac32}^{[\mathbf{2}]}=
\prod\limits_{\lambda=\pm1} \prod\limits_{\substack{(i,j)=\\ \scriptscriptstyle(1,3),(3,1)}}\left(a_{\D}+\lambda \frac{\tilde{m}_1{-}\tilde{m}_2}2\pm\frac{i \epsilon_1{+}j\epsilon_2}2\right)\\
\mathfrak{l}_{\pm 2}^{[\mathbf{2}]}=
\prod\limits_{\lambda=\pm1} \prod\limits_{\substack{(i,j)={\scriptscriptstyle(1,1),}\\ \scriptscriptstyle(5,1),(3,3),(1,5)}}\left(a_{\D}+\lambda\frac{\tilde{m}_1{-}\tilde{m}_2}2\pm\frac{i \epsilon_1{+}j\epsilon_2}2\right),
\end{multline}
\begin{equation}
\textrm{with reparametrization} \qquad \phi(e_2,\tilde{w}_2)=\frac34(\epsilon^2{-}\tilde{w}_2)-\frac{e_2}2, \qquad e_2=\tilde{m}_1\tilde{m}_2, \quad \tilde{w}_2=\tilde{m}_1^2+\tilde{m}_2^2 \, .
\end{equation}
We computed the terms of $\mathcal{Z}_{inst}^{[\mathbf{2}]}$ up to $s^{-7}$, up to $s^{-4}$ in the logarithmic expansion they are
\begin{multline}\label{Zinst_2}
-\epsilon_1\epsilon_2 \ln \mathcal{Z}_{inst}^{[\mathbf{2}]}(a_{\D},\tilde{w}_2,e_2;\epsilon_1,\epsilon_2|s)=\left(2a_{\D}^3-\frac{14e_2{+}9\tilde{w}_2{+}2\epsilon_1\epsilon_2{-}3\epsilon^2}2a_{\D}\right)\cdot\frac1{s}\\+\left(5 a_{\D}^4-\frac{90e_2{+}51\tilde{w}_2{+}14\epsilon_1\epsilon_2{-}17\epsilon^2}2 a_{\D}^2-\frac{62e_2{+}33\tilde{w}_2{+}12\epsilon_1\epsilon_2{-}9\epsilon^2}{16}(2e_2{-}\tilde{w}_2{+}\epsilon^2)\right)\cdot\frac1{s^2}\\+\left(22a_{\D}^5-\frac{1138e_2{+}615\tilde{w}_2{+}186\epsilon_1\epsilon_2{-}205\epsilon^2}3a_{\D}^3+c_1 a_{\D}\right)\cdot\frac1{s^3}
\\+\bigg(126a_{\D}^6-2(1790e_2{+}945\tilde{w}_2{+}303\epsilon_1\epsilon_2{-}315\epsilon^2)a_{\D}^4+c_2 a_{\D}^2+c_3(2e_2{-}\tilde{w}_2{+}\epsilon^2)
\bigg)\cdot\frac1{s^4}+\ldots+O\left(\frac1{s^8}\right),
\end{multline}
where
\begin{align}
&c_1=\frac{271}8\tilde{w}_2^2+\frac{230e_2{+}282\epsilon_1\epsilon_2{-}861\epsilon^2}{12}\tilde{w}_2-\frac{177}2e_2^2+\frac{38\epsilon_1\epsilon_2{-}643\epsilon^2}6e_2+\frac{24(\epsilon_1\epsilon_2)^2{-}220\epsilon_1\epsilon_2\epsilon^2{+}135\epsilon^4}8,\\
&c_2=\frac{4455}8\tilde{w}_2^2+\frac{3(998e_2{+}878\epsilon_1\epsilon_2{-}2025\epsilon^2)}4\tilde{w}_2-\frac{1305}2e_2^2+\frac{5(318\epsilon_1\epsilon_2{-}1025\epsilon^2)}2e_2 \nonumber\\&+\frac{1056(\epsilon_1\epsilon_2)^2{-}5428\epsilon_1\epsilon_2\epsilon^2{+}2943\epsilon^4}8,\\
&c_3=\frac{369}{16}\tilde{w}_2^2+\frac{3(85e_2{+}73\epsilon_1\epsilon_2{-}87\epsilon^2)}4\tilde{w}_2+\frac{145}4e_2^2+\frac{197\epsilon_1\epsilon_2{-}243\epsilon^2}2e_2\nonumber\\&+\frac{3(80(\epsilon_1\epsilon_2)^2{-}204\epsilon_1\epsilon_2\epsilon^2{+}81\epsilon^4)}{16}.
\end{align}
Above $\mathfrak{l}_n^{[\mathbf{2}]}$ are of form \eqref{l_n_fact}, so our ansatz \eqref{Z1loop_anz} gives
\begin{equation}\label{Z1loop_III1}
\mathcal{Z}_{1-loop}^{[\mathbf{2}]}(a_{\D},\tilde{m}_{1,2};\epsilon_1,\epsilon_2)=\prod\limits_{\lambda=\pm1}  \exp \gamma_{\epsilon_1,\epsilon_2} \left(a_{\D}{+}\frac{\lambda}2(\tilde{m}_1{-}\tilde{m}_2)-\epsilon/2\right). 
\end{equation}

Note that the expansion on the real {\it negative} canonical ray from \cite[Sec. A.3]{BLMST16} was obtained by action of symmetry $\mathbb{Z}_2$ that identifies the canonical rays (see App.~\ref{ssec:D-coal}). Analogously, the quantum deformation of this expansion is obtained via the  $\mathbb{Z}_2$ symmetry of \eqref{D13irr}, \eqref{quantum_PIII1} that acts by $t\mapsto -t, e_2\mapsto -e_2$, $\tau\mapsto e^{\frac{t}{\epsilon_1\epsilon_2}}\tau$. 
{\color{red}At the level of the answers this leads to }

\paragraph{Comparison with the classical limit ($\epsilon=0$).}
Let us compare our result in the $\epsilon=0$ limit with expansion \cite[(A.30)]{BLMST16}. According to \eqref{PIII1_tau_cl_app} we check that 
\begin{equation}\label{PIII1_tau_cl}
\tau(a_{\D},\eta_{\D},m_{1,2};-2\epsilon_2,2\epsilon_2|t)=e^{\mathrm{t}_{\sIII_1}/2}\tau_{\III_1}(\nu,\rho;\theta_{\star},\theta_*|\mathrm{t}_{\III_1}).  
\end{equation}
The dictionary is given by $\eta_{\D}=\rho$ and standard scalings (see \eqref{PIII1_tau_cl_app} for the all except the first one)
\begin{equation}\label{PIII1_par_sc}
a_{\D}=2\epsilon_2\nu, \qquad t=(2\epsilon_2)^2\mathrm{t}_{\III_1}, \qquad m_f=2\epsilon_2\sm_f|_{f=1,2}, 
\end{equation}
where masses $\sm_{1,2}$ are given by \eqref{PIII1_dct}.
Then, classical part \eqref{Zcl_III1} becomes 
\begin{equation}
\mathcal{Z}_{cl}^{[\mathbf{2}]}(-2\epsilon_2,2\epsilon_2)=e^{\mathrm{t}_{\sIII_1}}e^{4\ri\nu \mathrm{t}_{\sIII_1}^{1/2}}\, (2\epsilon_2 {\cdot} 8\ri \mathrm{t}_{\III_1}^{1/2})^{\frac34(\tilde{\theta}_{\star}^2{+}\tilde{\theta}_*^2)+\frac12\theta_{\star}\theta_*-\nu^2}.  
\end{equation} 
1-loop part \eqref{Z1loop_III1}, by using \eqref{gamma_Barnes}, becomes  
\begin{equation}
\mathcal{Z}_{1-loop}^{[\mathbf{2}]}(-2\epsilon_2,2\epsilon_2)=\frac{(2\epsilon_2)^{\nu^2{+}\frac14(\tilde{\theta}_{\star}{-}\tilde{\theta}_*)^2{-}\frac16}}{e^{2\zeta'(-1)}(2\pi)^{\nu}}\prod_{\pm} \mathsf{G}\left(1+\nu\pm \frac{\tilde{\theta}_{\star}{-}\tilde{\theta}_*}2\right), 
\end{equation}
where the tilded thetas are connected with the tilded masses by same formulas \eqref{PIII1_par_sc}, \eqref{PIII1_dct} as for the non-tilded ones. The Barnes $\mathsf{G}$-function factor of the above expression coincides with the counterpart from \cite[(A.30)]{BLMST16} iff in the self-dual case the tilded and the non-tilded masses are equal up to discrete group $W(2A_1)=\mathbb{Z}_2^2$, which acts by the transpositions of the masses and changing signs of an even number of masses. This means that $\tilde{w}_2|_{\epsilon=0}=w_2$, where $w_2=m_1^2{+}m_2^2$.
Then, up to the factor in \eqref{PIII1_tau_cl} and a constant factor in $\mathrm{t}_{\III_1}$ and $\nu$, $\mathcal{Z}_{cl}^{[\mathbf{2}]}\mathcal{Z}_{1-loop}^{[\mathbf{2}]}$ coincides with the prefactor of the $\mathrm{t}_{\III_1}^{-1/2}$ series in \cite[(A.30)]{BLMST16}. Finally, for the first two terms of the instanton expansion we obtain \cite[(A.31)]{BLMST16}
\begin{multline}
\mathcal{Z}_{inst}^{[\mathbf{2}]}(-2\epsilon_2,2\epsilon_2)=1+\left(\nu^3{-}\frac{9\mathrm{w}_2{+}14\se_2{-}2}4 \nu\right)\cdot\frac1{4\ri \mathrm{t}_{\III_1}^{1/2}}+\Bigg(\frac12\nu^6-\frac{9\mathrm{w}_2{+}14\se_2{-}7}4\nu^4\\ +\left(\frac{(9\mathrm{w}_2{+}14\se_2)^2}{32}{-}\frac{15\mathrm{w}_2{+}26\se_2}2{+}\frac{15}8\right)\nu^2+\frac{(\mathrm{w}_2{-}2\se_2)(33\mathrm{w}_2{+}62\se_2{-}12)}{64}\Bigg)\cdot\frac1{(4\ri \mathrm{t}_{\III_1}^{1/2})^2}+O\left(\frac1{\mathrm{t}_{\III_1}^{3/2}}\right), 
\end{multline}    
where $\mathrm{w}_2, \se_2$ are defined in \eqref{PIII1_inv}. In terms of $\theta$-variables \eqref{PIII1_dct} of \cite[(A.31)]{BLMST16} they read 
\begin{equation}
\mathrm{w}_2=\theta_{\star}^2+\theta_*^2,\qquad \se_2=\theta_{\star}\theta_*.
\end{equation}

As in Sec. \ref{ssec:exp_3l}, factorization \eqref{l_n_fact} prescribes that, without loss of generality, differences $\tilde{m}_f-m_f|_{f=1,2}$ are proportional to $\epsilon$ up to a numerical coefficient. Relation $\tilde{m}_1\tilde{m}_2=m_1m_2\equiv e_2$ implies that such coefficient vanishes. So, under the assumptions of the ansatz, we have $\tilde{w}_2=w_2$, as it is further confirmed in Sec.~\ref{ssec:holom_anom_2}.

\subsection{\texorpdfstring{QPIII$_2$}{QPIII₂} / \texorpdfstring{$N_f=1$}{N\_f=1}}
\label{ssec:exp_1}
Here we present the quantum deformation of the Painlev\'e III$_2$ large time expansion on the imaginary rays from \cite[Sec. A.2]{BLMST16}. For the (quantum) Painlev\'e III$_2$ we have $d=1/3$ and we set $s=(54t)^{1/3}$.

\paragraph{Classical part.}
For this expansion we have $N_p=1$. The first few leading $s^{-k}$ relations for \eqref{D13irr}, \eqref{quantum_PIII2} imply that the classical part can be taken as
\begin{align}
&\mathcal{Z}_{cl}^{[\mathbf{1}]}(a_{\D},m_1;\epsilon_1,\epsilon_2|s)=\mathcal{Z}_{cl-m}^{[\mathbf{1}]}(m_1;\epsilon_1,\epsilon_2|t) \mathcal{Z}_{cl-a}^{[\mathbf{1}]}(a_{\D};\epsilon_1,\epsilon_2|s) \qquad \textrm{with}\nonumber \\ &\mathcal{Z}_{cl-m}^{[\mathbf{1}]}(m_1;\epsilon_1,\epsilon_2|s)=e^{\frac{\frac{s^2}{8\psi^2}+m_1\psi^2 s}{\epsilon_1\epsilon_2}} s^{\frac{23\epsilon^2/24-m_1^2\psi^6}{\epsilon_1\epsilon_2}+\frac1{12}}, \qquad \quad \mathcal{Z}_{cl-a}^{[\mathbf{1}]}(a_{\D};\epsilon_1,\epsilon_2|s)=e^{-\frac{\sqrt3a_{\D} s}{\psi\epsilon_1\epsilon_2}} s^{\frac{a_{\D}^2}{2\epsilon_1\epsilon_2}},
\end{align}
where extra dimensionless parameter $\psi$, such that $\psi(2\epsilon_1,\epsilon_2{-}\epsilon_1)=\psi(\epsilon_1{-}\epsilon_2,2\epsilon_2)$, corresponds to obvious scaling symmetry $t\mapsto t\psi^3 ,m_1\mapsto m_1 \psi^{-3}$ of quantum Painlev\'e III$_2$ \eqref{D13irr},  \eqref{quantum_PIII2}. As in the previous cases we consider that $\psi$ depends on $\epsilon$ but not on $\epsilon_1\epsilon_2$. Let us introduce for convenience 
$\tilde{m}_1=m_1 \psi^3,\, \tilde{s}=s/\psi$
then, up to a power of $\psi$ that actually redefines the 1-loop term, we get
\begin{equation}\label{Zcl_III2}
\mathcal{Z}_{cl-m}^{[\mathbf{1}]}(m_1;\epsilon_1,\epsilon_2|\tilde{s})=e^{\frac{\tilde{s}^2/8+\tilde{m}_1 \tilde{s}}{\epsilon_1\epsilon_2}} \tilde{s}^{\frac{23\epsilon^2/24-\tilde{m}_1^2}{\epsilon_1\epsilon_2}+\frac1{12}}, \qquad \mathcal{Z}_{cl-a}^{[\mathbf{1}]}(a_{\D};\epsilon_1,\epsilon_2|\tilde{s})=e^{-\frac{\sqrt3 a_{\D} \tilde{s}}{\epsilon_1\epsilon_2}} \tilde{s}^{\frac{a_{\D}^2}{2\epsilon_1\epsilon_2}}.    
\end{equation}

\paragraph{Expansion.}
We then solve the successive $s^{-k}$ relations, obtaining step by step the leading terms of $\mathcal{Z}_{inst}^{[\mathbf{1}]}$ together with blowup factors $\mathfrak{l}_n^{[\mathbf{1}]}$. 
In this way we computed $\mathfrak{l}_n^{[\mathbf{1}]}$ for $|n|\leq2$ 
\begin{multline}
\mathfrak{l}_0^{[\mathbf{1}]}=\mathfrak{l}_{\pm \frac12}^{[\mathbf{1}]}=1, \qquad \mathfrak{l}_{\pm 1}^{[\mathbf{1}]}=
-\frac1{12\sqrt3}\left(a_{\D}\pm\frac{\epsilon}2\right), \qquad \mathfrak{l}_{\pm \frac32}^{[\mathbf{1}]}=
\frac1{{(12\sqrt3)^2}}\prod\limits_{\substack{(i,j)=\\ \scriptscriptstyle(1,3),(3,1)}}\left(a_{\D}\pm\frac{i \epsilon_1{+}j\epsilon_2}2\right), \\
\mathfrak{l}_{\pm2}^{[\mathbf{1}]}=\frac1{(12\sqrt3)^4}\prod\limits_{\substack{(i,j)={\scriptscriptstyle(1,1)},\\ \scriptscriptstyle(5,1),(3,3),(1,5)}}\left(a_{\D}\pm\frac{i \epsilon_1{+}j\epsilon_2}2\right) ,
\end{multline}  
together with the terms of $\mathcal{Z}_{inst}^{[\mathbf{1}]}$ up to $\tilde{s}^{-4}$, in the logarithmic expansion with $\alpha_{\D}=a_{\D}/\sqrt3$ they are
\begin{multline}\label{Zinst_1}
-\epsilon_1\epsilon_2 \ln \mathcal{Z}_{inst}^{[\mathbf{1}]}(a_{\D},\tilde{m}_1;\epsilon_1,\epsilon_2|\tilde{s})=\Bigg(-\frac54\alpha_{\D}^3+6\tilde{m}_1\alpha_{\D}^2-\left(6\tilde{m}_1^2{+}\frac{13\epsilon_1\epsilon_2}{24}{-}\frac{25\epsilon^2}{48}\right)\alpha_{\D}+\frac{\tilde{m}_1(8\tilde{m}_1^2{+}2\epsilon_1\epsilon_2{-}\epsilon^2)}6\Bigg)\cdot\frac1{\tilde{s}}\\
+\Bigg(-\frac{515}{96}\alpha_{\D}^4+19\tilde{m}_1\alpha_{\D}^3-\left(21\tilde{m}_1^2{+}\frac{533\epsilon_1\epsilon_2}{288}{-}\frac{425\epsilon^2}{576}\right)\alpha_{\D}^2+\frac{\tilde{m}_1(96\tilde{m}_1^2{+}22\epsilon_1\epsilon_2{+}\epsilon^2)}{12}\alpha_{\D}-c_1\Bigg)\cdot\frac1{\tilde{s}^2}\\+\Bigg(-\frac{10759}{576}\alpha_{\D}^5+\frac{805\tilde{m}_1}{12}\alpha_{\D}^4-\left(\frac{165\tilde{m}_1^2}2{+}\frac{9877\epsilon_1\epsilon_2}{1728}{+}\frac{3275\epsilon^2}{3456}\right)\alpha_{\D}^3 \\ +\frac{\tilde{m}_1(2880\tilde{m}_1^2{+}422\epsilon_1\epsilon_2{+}545\epsilon^2)}{72}\alpha_{\D}^2-c_2\alpha_{\D}-\tilde{m}_1\left(\frac{2\tilde{m}_1^2(2\epsilon_1\epsilon_2{-}7\epsilon^2){+}(\epsilon_1\epsilon_2)^2}9{-}\frac{649\epsilon_1\epsilon_2\epsilon^2}{1296}{+}\frac{613\epsilon^4}{5184}\right)\Bigg)\cdot\frac1{\tilde{s}^3}\\+\Bigg(-\frac{144683}{2304}\alpha_{\D}^6+\frac{11057\tilde{m}_1}{48}\alpha_{\D}^5-\left(\frac{7055\tilde{m}_1^2}{24}{+}\frac{2879\epsilon_1\epsilon_2}{288}{+}\frac{50645\epsilon^2}{2304}\right)\alpha_{\D}^4\\ +\tilde{m}_1\left(\frac{430\tilde{m}_1^2}3{-}\frac{1261\epsilon_1\epsilon_2}{144}{+}\frac{23245\epsilon^2}{288}\right)\alpha_{\D}^3-c_3\alpha_{\D}^2-c_4\tilde{m}_1\alpha_{\D}+c_5\Bigg)\cdot\frac1{\tilde{s}^4}+O\left(\frac1{\tilde{s}^5}\right),
\end{multline}
where
{\allowdisplaybreaks\begin{align}
&c_1= \frac{2\tilde{m}_1^4}3+\frac{\tilde{m}_1^2(2\epsilon_1\epsilon_2{+}5\epsilon^2)}{12}+\frac{671\epsilon_1\epsilon_2\epsilon^2}{10368}-\frac{1571\epsilon^4}{41472},\\
&c_2=6\tilde{m}_1^4-\frac{\tilde{m}_1^2(2\epsilon_1\epsilon_2{-}61\epsilon^2)}8-\frac{221(\epsilon_1\epsilon_2)^2}{1152}+\frac{68263\epsilon_1\epsilon_2\epsilon^2}{62208}-\frac{104263\epsilon^4}{248832},\\
&c_3=10\tilde{m}_1^4-\frac{\tilde{m}_1^2(5150\epsilon_1\epsilon_2{-}12187\epsilon^2)}{144}-\frac{76289(\epsilon_1\epsilon_2)^2}{20736}+\frac{836233\epsilon_1\epsilon_2\epsilon^2}{82944}-\frac{757333\epsilon^4}{331776},\\
&c_4=8\tilde{m}_1^4+\frac{\tilde{m}_1^2(322\epsilon_1\epsilon_2{-}509\epsilon^2)}{18}+\frac{1057(\epsilon_1\epsilon_2)^2}{288}-\frac{35921\epsilon_1\epsilon_2\epsilon^2}{5184}-\frac{5263\epsilon^4}{20736},\\
&c_5=\frac{8\tilde{m}_1^6}9+\frac{\tilde{m}_1^4(26\epsilon_1\epsilon_2{-}31\epsilon^2)}{18}+\left(\frac{11(\epsilon_1\epsilon_2)^2}{36}{+}\frac{997\epsilon_1\epsilon_2\epsilon^2}{2592}{-}\frac{11329\epsilon^4}{10368}\right)\tilde{m}_1^2\nonumber\\ &+\left(\frac{109057(\epsilon_1\epsilon_2)^2}{746496}{-}\frac{267689\epsilon_1\epsilon_2\epsilon^2}{995328}{+}\frac{483715\epsilon^4}{5971968}\right)\epsilon^2.
\end{align}}
Above $\mathfrak{l}_n^{[\mathbf{1}]}$ are of form \eqref{l_n_fact}, so our ansatz \eqref{Z1loop_anz} gives
\begin{equation}\label{Z1loop_III2}
\mathcal{Z}_{1-loop}^{[\mathbf{1}]}(a_{\D};\epsilon_1,\epsilon_2)=(-12\sqrt3)^{\frac{a_{\D}^2}{2\epsilon_1\epsilon_2}}  \exp \gamma_{\epsilon_1,\epsilon_2} (a_{\D}-\epsilon/2). 
\end{equation}

\paragraph{Comparison with the classical limit ($\epsilon=0$).}
Let us compare our result in the $\epsilon=0$ limit with expansion \cite[(A.19)]{BLMST16}. According to \eqref{PIII2_tau_cl_app} we check that
\begin{equation}\label{PIII2_tau_cl}
\tau(a_{\D},\eta_{\D},m_1;-2\epsilon_2,2\epsilon_2|t)=\tau_{\III_2}(\nu,\rho;\theta_*|\mathrm{t}_{\III_2}),
\end{equation}
The dictionary is given by $\eta_{\D}=\rho$ and scalings
\begin{equation}
a_{\D}=2\epsilon_2\nu/\sqrt3, \qquad t=(2\epsilon_2)^3 \mathrm{t}_{\III_2}\quad(s=2\epsilon_2 \mathrm{s}_{\III_2}), \qquad m_1=2\epsilon_2 \mathrm{m}_1= 2\epsilon_2\theta_{\star},    
\end{equation}
where $\sqrt3$ appears due to the corresponding noninteger shift of the Zak transform for the tau function in \cite[(A.19)]{BLMST16} and the latter equality is \eqref{PIII2_dct} . 
Then, classical part \eqref{Zcl_III2} becomes
\begin{equation}
\mathcal{Z}_{cl}^{[\mathbf{1}]}(-2\epsilon_2,2\epsilon_2)=e^{-\mathrm{s}_{\sIII_2}^2/8+(\nu-m_1) \mathrm{s}_{\sIII_2}} (2\epsilon_2\mathrm{s}_{\III_2})^{m_1^2+1/12-\nu^2/6},   
\end{equation} 
where we have already put $\psi|_{\epsilon=0}=1$ (i.e. $\tilde{s}=s, \, \tilde{m}_1=m_1$), that is the only choice to match with \cite[(A.19)]{BLMST16}.  
1-loop part \eqref{Z1loop_III2}, by using \eqref{gamma_Barnes}, becomes  
\begin{equation}
\mathcal{Z}_{1-loop}^{[\mathbf{1}]}(-2\epsilon_2,2\epsilon_2)=(-12\sqrt3)^{-\frac{\nu^2}6}\frac{(2\epsilon_2)^{\frac{\nu^2}6{-}\frac1{12}}}{e^{\zeta'(-1)}(2\pi)^{\frac{\nu}{2\sqrt3}}} \mathsf{G}(1+\nu/\sqrt3).
\end{equation}
Then, up to a constant factor in $\mathrm{t}_{\III_2}$ and $\nu$, $\mathcal{Z}_{cl}^{[\mathbf{1}]}\mathcal{Z}_{1-loop}^{[\mathbf{1}]}$ coincides with the prefactor of the $\mathrm{s}_{\III_2}^{-1}$ series in \cite[(A.19)]{BLMST16}. Finally, for the first two terms of the instanton part we obtain \cite[(A.20)]{BLMST16} 
\begin{multline}
\mathcal{Z}_{inst}^{[\mathbf{1}]}(-2\epsilon_2,2\epsilon_2)=1+\Bigg(-\frac{5\nu^3}{108}+\frac{2\theta_{\star}}3\nu^2-\left(2\theta_{\star}^2{-}\frac{13}{72}\right)\nu+\frac{\theta_{\star}(4\theta_{\star}^2{-}1)}3\Bigg)\cdot\frac1{\mathrm{s}_{\III_2}}\\+\Bigg(\frac{25\nu^6}{23328}-\frac{5\theta_{\star}}{162}\nu^5+\frac{612\theta_{\star}^2{-}145}{1944}\nu^4-\frac{113\theta_{\star}^2{-}68}{81}\theta_{\star}\nu^3+\left(\frac{26\theta_{\star}^4}9{-}\frac{35\theta_{\star}^2}{12}{+}\frac{767}{3456}\right)\nu^2\\ -\frac{576\theta_{\star}^4{-}772\theta_{\star}^2{+}145}{216}\theta_{\star}\nu+\frac{2\theta_{\star}^2(\theta_{\star}^2{-}1)(4\theta_{\star}^2{-}1)}9\Bigg)\cdot\frac1{\mathrm{s}_{\III_2}^2}+O\left(\frac1{\mathrm{s}_{\III_2}^3}\right). 
\end{multline}

\subsection{\texorpdfstring{QPIII$_3$}{QPIII₃} / \texorpdfstring{$N_f=0$}{N\_f=0}}
\label{ssec:exp_0}
Here we present the quantum deformation of the Painlev\'e III$_3$ large time expansion on the real positive canonical ray from \cite[Sec. A.1]{BLMST16}. For the (quantum) Painlev\'e III$_3$ we have $d=1/4$ and we set $s=-32\ri\, t^{1/4}$.
\paragraph{Classical part.}
For this expansion we have $N_p=1$. The first few leading $s^{-k}$ relations for \eqref{D13irr}, \eqref{quantum_PIII3} imply that the classical part can be taken as
\begin{align}
&\mathcal{Z}_{cl}^{[\mathbf{0}]}(a_{\D};\epsilon_1,\epsilon_2|s)=\mathcal{Z}_{cl-m}^{[\mathbf{0}]}(\epsilon_1,\epsilon_2|s) \mathcal{Z}_{cl-a}^{[\mathbf{0}]}(a_{\D};\epsilon_1,\epsilon_2|s) \qquad \textrm{with}\nonumber \\ \label{Zcl_III3} &\mathcal{Z}_{cl-m}^{[\mathbf{0}]}(\epsilon_1,\epsilon_2|s)=e^{\frac{s^2}{256\epsilon_1\epsilon_2}} \, s^{\frac{9\epsilon^2}{8\epsilon_1\epsilon_2}+\frac14}, \qquad \quad \mathcal{Z}_{cl-a}^{[\mathbf{0}]}(a_{\D};\epsilon_1,\epsilon_2|s)=e^{\frac{a_{\D} s}{4\epsilon_1\epsilon_2}} \, s^{\frac{a_{\D}^2}{2\epsilon_1\epsilon_2}}.
\end{align}

\paragraph{Expansion.}
We then solve the successive $s^{-k}$ relations, obtaining step by step the leading terms of $\mathcal{Z}_{inst}^{[\mathbf{0}]}$ together with blowup factors $\mathfrak{l}_n^{[\mathbf{0}]}$. 
In this way we computed $\mathfrak{l}_n^{[\mathbf{0}]}$ for $|n|\leq3$ 
\begin{multline}
\mathfrak{l}_0^{[\mathbf{0}]}=\mathfrak{l}_{\pm \frac12}^{[\mathbf{0}]}=1, \qquad \mathfrak{l}_{\pm 1}^{[\mathbf{0}]}=
a_{\D}\pm\frac{\epsilon}2, \qquad \mathfrak{l}_{\pm \frac32}^{[\mathbf{0}]}=
\prod\limits_{\substack{(i,j)=\\ \scriptscriptstyle(1,3),(3,1)}}\left(a_{\D}\pm\frac{i \epsilon_1{+}j\epsilon_2}2\right), \qquad
\mathfrak{l}_{\pm2}^{[\mathbf{0}]}=\prod\limits_{\substack{(i,j)={\scriptscriptstyle(1,1)},\\ \scriptscriptstyle(5,1),(3,3),(1,5)}}\left(a_{\D}\pm\frac{i \epsilon_1{+}j\epsilon_2}2\right),\\
\mathfrak{l}_{\pm\frac52}^{[\mathbf{0}]}=\prod\limits_{\substack{(i,j)={\scriptscriptstyle(1,3),(3,1)},\\ \scriptscriptstyle(7,1),(3,5),(5,3)(1,7)}}\left(a_{\D}\pm\frac{i \epsilon_1{+}j\epsilon_2}2\right), \qquad \mathfrak{l}_{\pm3}^{[\mathbf{0}]}=\prod\limits_{\substack{(i,j)={\scriptscriptstyle(1,1)},\\ \scriptscriptstyle(5,1),(3,3),(1,5),\\
\scriptscriptstyle(9,1),(7,3),(5,5),(3,7),(1,9)}}\left(a_{\D}\pm\frac{i \epsilon_1{+}j\epsilon_2}2\right)
\end{multline}   
together with the terms of $\mathcal{Z}_{inst}^{[\mathbf{0}]}$ up to $s^{-8}$, up to $s^{-4}$ in the logarithmic expansion they are
\begin{multline}\label{Zinst_0}
-\epsilon_1\epsilon_2 \ln \mathcal{Z}_{inst}^{[\mathbf{0}]}(a_{\D};\epsilon_1,\epsilon_2|s)=-\left(a_{\D}^3+\frac{2\epsilon_1
\epsilon_2{+}3\epsilon^2}4a_{\D}\right)\cdot\frac{1}{s}+\Bigg(\frac52a_{\D}^4+\left(\frac32\epsilon_1\epsilon_2{+}\frac{17}4\epsilon^2\right)a^2_{\D}+\frac{3\epsilon^2(4\epsilon_1\epsilon_2{+}3\epsilon^2)}{32}\Bigg)\cdot\frac{1}{s^2}\\-\left(11a_{\D}^5+\frac{38\epsilon_1\epsilon_2{+}205\epsilon^2}6a_{\D}^3-\frac{8(\epsilon_1\epsilon_2)^2{-}100\epsilon_1\epsilon_2\epsilon^2{-}135\epsilon^4}{16}a_{\D}\right)\cdot\frac{1}{ s^3}+\Bigg(63a_{\D}^6+3(8\epsilon_1\epsilon_2{+}105\epsilon^2)a_{\D}^4\\  -\left(13(\epsilon_1\epsilon_2)^2{-}\frac{229}4\epsilon_1\epsilon_2\epsilon^2{-}\frac{2943}{16}\epsilon^4\right)a_{\D}^2-\frac{3\epsilon^2(10\epsilon_1\epsilon_2{-}27\epsilon^2)(4\epsilon_1
\epsilon_2{+}3\epsilon^2)}{32}\Bigg)\cdot\frac{1}{s^4}+\ldots+O\left(\frac{1}{s^9}\right),
\end{multline}
Above $\mathfrak{l}_n^{[\mathbf{0}]}$ are in form \eqref{l_n_fact}, so our ansatz \eqref{Z1loop_anz} gives
\begin{equation}\label{Z1loop_III3}
\mathcal{Z}_{1-loop}^{[\mathbf{0}]}(a_{\D};\epsilon_1,\epsilon_2)=\exp \gamma_{\epsilon_1,\epsilon_2} (a_{\D}-\epsilon/2). 
\end{equation}

\paragraph{Comparison with the classical limit ($\epsilon=0$).}
Let us compare our result in the $\epsilon=0$ limit with expansion \cite[(A.8)]{BLMST16}. According to \eqref{PIII3_tau_cl_app} we check that 
\begin{equation}\label{PIII3_tau_cl}
\tau(a_{\D},\eta_{\D};-2\epsilon_2,2\epsilon_2|t)=\tau_{\III_3}(\nu,\rho|\mathrm{t}_{\III_3}).   
\end{equation}
The dictionary is given by $\eta_{\D}=\rho$ and standard scalings (see \eqref{PIII3_tau_cl_app} for the second one)
\begin{equation}
a_{\D}=2\epsilon_2\nu, \qquad t=(2\epsilon_2)^4 \mathrm{t}_{\III_3}.    
\end{equation}  
Then, classical part \eqref{Zcl_III3} becomes 
\begin{equation}
\mathcal{Z}_{cl}^{[\mathbf{0}]}(-2\epsilon_2,2\epsilon_2)=e^{4\mathrm{t}_{\sIII_3}^{1/2}+8\ri\nu \mathrm{t}_{\sIII_3}^{1/4}}\, (2\epsilon_2 (-32\ri) \mathrm{t}_{\III_3}^{1/4})^{1/4-\nu^2/2}.   
\end{equation} 
1-loop part \eqref{Z1loop_III3}, by using \eqref{gamma_Barnes}, becomes  
\begin{equation}
\mathcal{Z}_{1-loop}^{[\mathbf{0}]}(-2\epsilon_2,2\epsilon_2)=\frac{(2\epsilon_2)^{\frac{\nu^2}2-\frac1{12}}}{e^{\zeta'(-1)}(2\pi)^{\nu/2}} \mathsf{G}(1+\nu).
\end{equation}
Up to a constant factor in $\mathrm{t}_{\III_3}$ and $\nu$, $\mathcal{Z}_{cl}^{[\mathbf{0}]}\mathcal{Z}_{1-loop}^{[\mathbf{0}]}$ coincides with the prefactor of the $\mathrm{t}_{\III_3}^{-1/4}$ series in \cite[(A.8)]{BLMST16}. Finally, for the first two terms of the instanton expansion we obtain \cite[(A.9)]{BLMST16} 
\begin{equation}
\mathcal{Z}_{inst}^{[\mathbf{0}]}(-2\epsilon_2,2\epsilon_2)=1+\left(\frac{\nu^3}{4\ri}+\frac{\ri\nu}8\right)\cdot\frac1{8\mathrm{t}_{\III_3}^{1/4}}+\left(-\frac{\nu^6}{32}-\frac{\nu^4}8+\frac{11\nu^2}{128}\right)\cdot\frac1{(8\mathrm{t}_{\III_3}^{1/4})^2}+O\left(\frac1{(8\mathrm{t}_{\III_3}^{1/4})^3}\right). 
\end{equation}

\paragraph{Comparison with \cite{GMS20}.}
As it was already mentioned, the approach we use in this paper was initially introduced in \cite[Sec. 7]{GMS20}. Namely, they computed the first $7$ orders of the large time expansion for the quantum Painlev\'e III$_3$ in the so-called Toda-like form, that is instead of considering the single fourth-order equation \eqref{quantum_PIII3} they use two second order differential equations, which in the autonomous limit become the equations of the two-particle Toda system. Similarly to us they use it together with the first equation of \eqref{D13irr}.

For completeness we compare their result with ours. Namely, we check the equality between above instanton expansions \eqref{Zinst_0} and \cite[App. D]{GMS20} up to the power $7$ in $s$ (accordingly $r$)
\begin{equation}
\mathsf{B}^{\infty}(\nu,\epsilon_1,\epsilon_2|r)=\mathcal{Z}_{inst}^{[\mathbf{0}]}(a_{\D};\epsilon_1,\epsilon_2|s)    
\end{equation}
by using dictionary
\begin{equation}
\nu=-\ri a_{\D}, \qquad r=\frac{s}{4\ri}.
\end{equation}
For the classical part this dictionary leads to equality
\begin{equation}
C_{cl}^{\infty}(\nu,\epsilon_1,\epsilon_2|r)=(4\ri)^{-\frac{\epsilon^2/4+a_{\D}^2}{2\epsilon_1\epsilon_2}-\frac14}s^{-\frac{\epsilon^2}{\epsilon_1\epsilon_2}}\mathcal{Z}_{cl}^{[\mathbf{0}]}(a_{\D};\epsilon_1,\epsilon_2|s).
\end{equation}
For the 1-loop part (called "perturbative" in \cite{GMS20}) it leads to
\begin{equation}
C_{pert}^{\infty}(\nu,\epsilon_1,\epsilon_2)=(4\ri)^{\frac{a_{\D}^2}{2\epsilon_1\epsilon_2}}\mathcal{Z}_{1-loop}^{[\mathbf{0}]}(a_{\D};\epsilon_1,\epsilon_2),
\end{equation}
if we identify double Gamma function $\mathsf{G}$ of \cite[Sec. 7.2]{GMS20} with $\Gamma_2^{-1}$ of App.~\ref{sec:gamma}. So $C_{cl}^{\infty}C_{pert}^{\infty}$ coincides with $\mathcal{Z}_{cl}^{[\mathbf{0}]}\mathcal{Z}_{1-loop}^{[\mathbf{0}]}$ up to a constant factor in $a_{\D}$. The latter is proportional to $s^{\frac{\phi(\epsilon)}{\epsilon_1\epsilon_2}}$ that is actually a gauge symmetry of the Toda-like equations and equations \eqref{D13irr}.

Notice that Toda-like blowup relations should exist also for $N_f>0$ and they possibly can help to connect the extra integration constants with the corresponding mass variables of the theory. Studying such equations seem to be a natural part of understanding the notion of quantum Painlev\'e equations (see also Sec.~\ref{sec:further}). 

\subsection{QPIV (linear exp singularity) / \texorpdfstring{$H_2$}{H₂} with light hypers}
\label{ssec:exp_H2l}
As already mentioned, for Painlev\'e IV we have two different large time expansions. Here we present the quantum  deformation of the one on canonical rays $\arg\, t=\pm \frac{\pi}4,\, \pm\frac{3\pi}4$ from \cite[Sec. 3.3]{BLMST16}, called there "expansion 2". For the (quantum) Painlev\'e IV we have $d=2$ and we set $s=t^2$ for the both expansions.

\paragraph{Classical part.}
For this expansion we have $N_p=3$. The first few leading $s^{-k}$ relations for \eqref{D13irr_2row}, \eqref{quantum_PIV} imply that the classical part can be taken with $\beta=0$ as
\begin{align}
&\mathcal{Z}_{cl}^{[\mathbf{H_{2,L}}]}(a_{\D}, \boldsymbol{m}_{1,2,3};\epsilon_1,\epsilon_2|s)=\mathcal{Z}^{[\mathbf{H_{2,L}}]}_{cl-m}(\boldsymbol{m}_{1,2,3};\epsilon_1,\epsilon_2|s) \, \mathcal{Z}^{[\mathbf{H_{2,L}}]}_{cl-a}(a_{\D};\epsilon_1,\epsilon_2|s) \qquad \textrm{with}\nonumber \\ \label{Zcl_IV(p)} &\mathcal{Z}^{[\mathbf{H_{2,L}}]}_{cl-m}(\boldsymbol{m}_{1,2,3};\epsilon_1,\epsilon_2|s)=e^{\frac{(\epsilon{-}4\boldsymbol{m}_3)s}{4\epsilon_1\epsilon_2}} \, s^{\frac{4\boldsymbol{e}_2{+}\epsilon^2/4}{2\epsilon_1\epsilon_2}}, \qquad \quad \mathcal{Z}^{[\mathbf{H_{2,L}}]}_{cl-a}(a_{\D};\epsilon_1,\epsilon_2|s)=e^{-\frac{a_{\D}s}{2\epsilon_1\epsilon_2}} \, s^{\frac{3a_{\D}^2}{2\epsilon_1\epsilon_2}},   
\end{align}
where $\boldsymbol{e}_2=\boldsymbol{m}_1\boldsymbol{m}_2{+}\boldsymbol{m}_1\boldsymbol{m}_3{+}\boldsymbol{m}_2\boldsymbol{m}_3$.
By normalizing the tau functions of QPIV with 
$a_{\D}$-independent factor $\mathcal{Z}_{cl-m}^{[\mathbf{H_{2,L}}]}$ 
\begin{equation}\label{rem_IV_2}
\tau_r(\epsilon_1,\epsilon_2|t)=\frac{\tau(\epsilon_1,\epsilon_2|t)}{\mathcal{Z}^{[\mathbf{H_{2,L}}]}_{cl-m}(\boldsymbol{m}_{1,2,3};\epsilon_1,\epsilon_2|t)},
\end{equation}
we obtain from \eqref{D13irr_2row},  \eqref{quantum_PIV}
\begin{equation}\label{D13irr_2row_r}
D^1_{\epsilon_1,\epsilon_2[t]}(\tau_r^{(1)},\tau_r^{(2)})=0,\qquad 
   D^3_{\epsilon_ 1,\epsilon_2[t]}(\tau_r^{(1)},\tau_r^{(2)})=\epsilon \left(4\boldsymbol{e}_2{+}\frac{\epsilon^2}4\right) t^{-3}(\tau^{(1)}_r,\tau^{(2)}_r),
   \end{equation}
   \vspace{-0.7cm}
    \begin{multline}\label{quantum_PIV_r}
    D^4_{\epsilon_1,\epsilon_2[t]}(\tau_r^{(1)},\tau_r^{(2)})+
    \left(3\left(4\boldsymbol{e}_2{+}\frac{\epsilon^2}4\right)-\frac{t^4}4\right) D^2_{\epsilon_1,\epsilon_2[t]}(\tau_r^{(1)},\tau_r^{(2)})\\
    -\frac{t}4\left(\epsilon_1\epsilon_2\frac{d}{dt}\right)\tau^{(1)}_r\tau^{(2)}_r+3\left(\boldsymbol{e}_2{-}\epsilon_1\epsilon_2{+}\frac{17\epsilon^2}{16}\right)\left(4\boldsymbol{e}_2{+}\frac{\epsilon^2}4\right)t^{-4}\tau^{(1)}_r\tau^{(2)}_r=0,
    \end{multline}
    so we see that the quantum Painlev\'e IV actually depends only on one mass parameter  $\boldsymbol{e}_2$. The other mass parameter appears as an integration constant as for the classical Painlev\'e IV (App.~\ref{ssec:E-coal}).

\paragraph{Expansion.}
We then solve the successive $s^{-k}$ relations, obtaining step by step the leading terms of $\mathcal{Z}_{inst}^{[\mathbf{H_{2,L}}]}$ together with blowup factors $\mathfrak{l}_n^{[\mathbf{H_{2,L}}]}$. In this way we computed $\mathfrak{l}_n^{[\mathbf{H_{2,L}}]}$ for $|n|\leq\frac32$ 
\begin{equation}\label{l_n_IV(p)}
\mathfrak{l}_0^{[\mathbf{H_{2,L}}]}=\mathfrak{l}_{\pm \frac12}^{[\mathbf{H_{2,L}}]}=1, \qquad \mathfrak{l}_{\pm 1}^{[\mathbf{H_{2,L}}]}=
\prod\limits_{i=1}^3  \left(a_{\D}+2\tilde{\boldsymbol{m}}_i \pm\frac{\epsilon}2\right), \qquad \mathfrak{l}_{\pm \frac32}^{[\mathbf{H_{2,L}}]}=
\prod\limits_{i=1}^3 \prod\limits_{\substack{(i,j)=\\ \scriptscriptstyle(1,3),(3,1)}}\left(a_{\D}+2\tilde{\boldsymbol{m}}_i\pm\frac{i\epsilon_1{+}j\epsilon_2}2\right)
\end{equation}
together with the terms of $\mathcal{Z}_{inst}^{[\mathbf{H_{2,L}}]}$ up to $s^{-6}$, up to $s^{-4}$ in the logarithmic expansion they are
\begin{multline}\label{Zinst_IVl}
-\frac13\epsilon_1\epsilon_2 \ln \mathcal{Z}_{inst}^{[\mathbf{H_{2,L}}]}(a_{\D},\boldsymbol{e}_2,\tilde{\boldsymbol{e}}_3;\epsilon_1,\epsilon_2|s)=\Bigg(2a_{\D}^3+\left(4\boldsymbol{e}_2{+}\frac{\epsilon^2}2\right)a_{\D}+4\tilde{\boldsymbol{e}}_3\Bigg)\cdot\frac1{s}\\+\Bigg(\frac{35}4 a_{\D}^4+\left(22\boldsymbol{e}_2{-}\frac{\epsilon_1\epsilon_2}4{+}\frac{37\epsilon^2}8\right)a_{\D}^2+28\tilde{\boldsymbol{e}}_3 a_{\D} +\frac{\left(16\boldsymbol{e}_2{+}\epsilon^2\right) \left(16\boldsymbol{e}_2{-}4\epsilon_1\epsilon_2{+}25\epsilon^2\right)}{192}\Bigg)\cdot\frac1{s^2} \\+\Bigg(67a_{\D}^5+\left(200\boldsymbol{e}_2{-}7\epsilon_1\epsilon_2{+}\frac{125\epsilon^2}2\right)a_{\D}^3+272\tilde{\boldsymbol{e}}_3a_{\D}^2+\frac{c_1}3 a_{\D}
+4\tilde{\boldsymbol{e}}_3(12\boldsymbol{e}_2{-}4\epsilon_1\epsilon_2{+}9\epsilon^2)\Bigg)\cdot\frac1{s^3}
\\+\Bigg(\frac{2667}4a_{\D}^6+3\left(765\boldsymbol{e}_2{-}\frac{427\epsilon_1\epsilon_2}8{+}\frac{5235\epsilon^2}{16}\right)a_{\D}^4 +3180\tilde{\boldsymbol{e}}_3 a_{\D}^3{+}3c_2a_{\D}^2+3\tilde{\boldsymbol{e}}_3(504\boldsymbol{e}_2{-}190\epsilon_1\epsilon_2{+}423\epsilon^2)a_{\D}\\ +(16\boldsymbol{e}_2{+}\epsilon^2)c_3{+}264\tilde{\boldsymbol{e}}_3^2\Bigg)\cdot\frac1{s^4}+\ldots+O\left(\frac1{s^7}\right), 
\end{multline}
where
\begin{align}
&c_1=128\boldsymbol{e}_2^2-2(22\epsilon_1\epsilon_2{-}95\epsilon^2)\boldsymbol{e}_2-\frac{23}4\epsilon_1\epsilon_2\epsilon^2+\frac{281\epsilon^4}{16},\\
&c_2=316\boldsymbol{e}_2^2-\left(139\epsilon_1\epsilon_2{-}\frac{1011\epsilon^2}2\right)\boldsymbol{e}_2+\frac{7(\epsilon_1\epsilon_2)^2}8-\frac{519\epsilon_1\epsilon_2\epsilon^2}{16}+\frac{4303\epsilon^4}{64},\\
&c_3=\boldsymbol{e}_2^2-\frac38(3\epsilon_1\epsilon_2{-}16\epsilon^2)\boldsymbol{e}_2+\frac{7(\epsilon_1\epsilon_2)^2}{32}-\frac{389\epsilon_1\epsilon_2\epsilon^2}{128}+\frac{911\epsilon^4}{256}.
\end{align}
This expansion has the expected dependence on integration constant $\tilde{\boldsymbol{e}}_3$, for $\mathfrak{l}_n^{[\mathbf{H_{2,L}}]}$ it is parametrized as
\begin{equation}\label{mbtilda_def}
\boldsymbol{e}_1=\tilde{\boldsymbol{m}}_1+\tilde{\boldsymbol{m}}_2+\tilde{\boldsymbol{m}}_3=0, \quad \boldsymbol{e}_2=\tilde{\boldsymbol{m}}_1 \tilde{\boldsymbol{m}}_2+\tilde{\boldsymbol{m}}_1 \tilde{\boldsymbol{m}}_3+\tilde{\boldsymbol{m}}_2 \tilde{\boldsymbol{m}}_3, \quad \tilde{\boldsymbol{e}}_3=\tilde{\boldsymbol{m}}_1\tilde{\boldsymbol{m}}_2\tilde{\boldsymbol{m}}_3 
\end{equation}
As in the previous cases,  $\tilde{\boldsymbol{e}}_3(2\epsilon_1,\epsilon_2{-}\epsilon_1)=\tilde{\boldsymbol{e}}_3(\epsilon_1{-}\epsilon_2,2\epsilon_2)$, so we consider that $\tilde{\boldsymbol{e}}_3$ depends on $\epsilon$ but not on $\epsilon_1\epsilon_2$.
Above $\mathfrak{l}_n^{[\mathbf{H_{2,L}}]}$ are of form \eqref{l_n_fact}, so our ansatz \eqref{Z1loop_anz} gives
\begin{equation}\label{Z1loop_IV(p)}
\mathcal{Z}_{1-loop}^{[\mathbf{H_{2,L}}]}(a_{\D},\tilde{\boldsymbol{m}}_{1,2,3};\epsilon_1,\epsilon_2)= \prod\limits_{i=1}^3  \exp \gamma_{\epsilon_1,\epsilon_2} \left(a_{\D}+2\tilde{\boldsymbol{m}}_i-\epsilon/2\right). 
\end{equation}

\paragraph{Comparison with the classical limit ($\epsilon=0$).}
Let us compare our result in the $\epsilon=0$ limit with expansion BLMST \cite[(3.48)]{BLMST16}. According to \eqref{PIV_tau_cl_app} and \eqref{PIV_tau_lit} we check that 
\begin{equation}\label{PIV_tau_cl}
\tau(a_{\D},\eta_{\D},\boldsymbol{m}_{1,2,3};-2\epsilon_2,2\epsilon_2|t)=\tau_{\IV}^{\mathrm{BLMST}}(\nu,\rho;\theta_t,\theta_s|\mathrm{t}_{\IV}).
\end{equation}
The dictionary is given by $\eta_{\D}=\rho$ and standard scalings (see \eqref{PIV_tau_cl_app} for the all except the first one)
\begin{equation}\label{PIV_par_sc}
a_{\D}=2\epsilon_2\nu, \qquad t=(2\epsilon_2)^{1/2} \mathrm{t}_{\IV}, \qquad \boldsymbol{m}_k=2\epsilon_2\mathfrak{m}_k|_{k=1,2,3}, 
\end{equation}
where masses $\mathfrak{m}_{1,2,3}$ are given by \eqref{PIV_dct}.
Then, classical part \eqref{Zcl_IV(p)} becomes
\begin{equation}
\mathcal{Z}_{cl}^{[\mathbf{H_{2,L}}]}(-2\epsilon_2,2\epsilon_2)=e^{(\nu+2(\theta_s{+}\theta_t)/3)\mathrm{t}_{\sIV}^2/2}\, (2\epsilon_2 \mathrm{t}_{\IV}^2)^{\frac23(\theta_s^2{-}\theta_s\theta_t{+}\theta_t^2)-3\nu^2/2}.  
\end{equation} 
1-loop part \eqref{Z1loop_IV(p)}, by using \eqref{gamma_Barnes}, becomes  
\begin{equation}
\mathcal{Z}_{1-loop}^{[\mathbf{H_{2,L}}]}(-2\epsilon_2,2\epsilon_2)=\frac{(2\epsilon_2)^{\frac{3\nu^2}2{+}\frac43(\theta_s^2{-}\theta_s\theta_t{+}\theta_t^2){-}\frac14}}{e^{3\zeta'(-1)}(2\pi)^{3\nu/2}} \mathsf{G}\left(1{+}\nu{+}\frac{2\tilde{\theta}_t{-}4\tilde{\theta}_s}3\right) \mathsf{G}\left(1{+}\nu{+}\frac{2\tilde{\theta}_s{-}4\tilde{\theta}_t}3\right)\mathsf{G}\left(1{+}\nu{+}\frac{2\tilde{\theta}_s{+}2\tilde{\theta}_t}3\right),
\end{equation}
where the tilded thetas are connected with the tilded masses by same formulas \eqref{PIV_par_sc}, \eqref{PIV_dct} as for the non-tilded ones. The Barnes $\mathsf{G}$-function factor of the above expression coincides with the counterpart from \cite[(3.48)]{BLMST16} iff in the self-dual case the tilded and the non-tilded masses are equal up to discrete group $W(A_2)=S_3$, which acts by  the transpositions of the masses. This means that $\tilde{\boldsymbol{e}}_3|_{\epsilon=0}=\boldsymbol{e}_3$, where $\boldsymbol{e}_3=\boldsymbol{m}_1\boldsymbol{m}_2\boldsymbol{m}_3$.  
Then, up to the factor in \eqref{PIV_tau_cl} and a constant factor in $\mathrm{t}_{\IV}$ and $\nu$, $\mathcal{Z}_{cl}^{[\mathbf{H_{2,L}}]}\mathcal{Z}_{1-loop}^{[\mathbf{H_{2,L}}]}$ coincides with the prefactor of the $\mathrm{t}_{\IV}^{-2}$ series in \cite[(3.48)]{BLMST16}. Finally, for the first two terms of the instanton expansion we obtain \cite[(3.49)]{BLMST16} 
\begin{multline}
\mathcal{Z}_{inst}^{[\mathbf{H_{2,L}}]}(-2\epsilon_2,2\epsilon_2)=1+3\left(\nu^3+2\mathfrak{e}_2 \nu+2\mathfrak{e}_3\right)\cdot\frac2{\mathrm{t}_{\IV}^2}\\+9\Bigg(\frac{\nu^6}2+\left(2\mathfrak{e}_2{+}\frac{35}{48}\right)\nu^4+2\mathfrak{e}_3\nu^3+\left(2\mathfrak{e}_2^2{+}\frac{11\mathfrak{e}_2}6{+}\frac1{48}\right)\nu^2+\mathfrak{e}_3\left(4\mathfrak{e}_2{+}\frac73\right)\nu+2\mathfrak{e}_3^2{+}\frac{\mathfrak{e}_2(4\mathfrak{e}_2{+}1)}{36}\Bigg)\cdot\frac4{\mathrm{t}_{\IV}^4}+O\left(\frac8{\mathrm{t}_{\IV}^6}\right), 
\end{multline}
where parameters $\mathfrak{e}_2,\mathfrak{e}_3$ are defined in \eqref{PIV_inv}. In terms of $\theta$-variables \eqref{PIV_dct} of \cite[(3.49)]{BLMST16} they are 
\begin{equation}\label{PIV_theta}
\mathfrak{e}_2=-\frac{\theta_s^2{-}\theta_s\theta_t{+}\theta_t^2}3, \qquad \mathfrak{e}_3=\frac{(\theta_s{+}\theta_t) (\theta_s{-}2\theta_t) (\theta_t{-}2\theta_s)}{27}.
\end{equation}

As in Secs. \ref{ssec:exp_3l} and \eqref{ssec:exp_2}, factorization \eqref{l_n_fact} prescribes that, without loss of generality, differences $\tilde{\boldsymbol{m}}_k-\boldsymbol{m}_k|_{k=1,2,3}$ are proportional to $\epsilon$ up to some numerical coefficients. The first two relations of \eqref{mbtilda_def} imply that such coefficients vanish. So, under the assumptions of the ansatz, we have $\tilde{\boldsymbol{e}}_3=\boldsymbol{e}_3$, which is further confirmed in Sec.~\ref{ssec:holom_anom_H2}.

\subsection{QPIV (square exp singularity) / \texorpdfstring{$H_2$}{H₂} with heavy hypers}
\label{ssec:exp_H2s}
Here we present the quantum deformation of the other Painlev\'e IV large time expansion, namely the one on the real and imaginary canonical rays from \cite[Sec. 3.3]{BLMST16}, called there "expansion 1".
Recall that for QPIV we set $s=t^2$

\paragraph{Classical part.}
For this expansion we have $N_p=1$. The first few leading $s^{-k}$ relations for \eqref{D13irr_2row}, \eqref{quantum_PIV} imply that the classical part can be taken with $\beta\neq0$ as
\begin{align}
&\mathcal{Z}_{cl}^{[\mathbf{H_{2,S}}]}(a_{\D},\boldsymbol{m}_{1,2,3};\epsilon_1,\epsilon_2|t)=\mathcal{Z}_{cl-m}^{[\mathbf{H_{2,S}}]}(\boldsymbol{m}_{1,2,3};\epsilon_1,\epsilon_2|t) \mathcal{Z}_{cl-a}^{[\mathbf{H_{2,S}}]}(a_{\D};\epsilon_1,\epsilon_2|t) \qquad \textrm{with} \nonumber \\ 
\label{Zcl_IV(e)} &\mathcal{Z}_{cl-m}^{[\mathbf{H_{2,S}}]}(\boldsymbol{m}_{1,2,3};\epsilon_1,\epsilon_2|t)=e^{-\frac{s^2}{108\epsilon_1\epsilon_2}}e^{\frac{(\epsilon{-}4\boldsymbol{m}_3)s}{4\epsilon_1\epsilon_2}} \, s^{\frac{5\epsilon^2/24+6\boldsymbol{e}_2}{\epsilon_1\epsilon_2}-\frac16}, \qquad  \mathcal{Z}_{cl-a}^{[\mathbf{H_{2,S}}]}(a_{\D};\epsilon_1,\epsilon_2|t)=e^{-\frac{\ri a_{\D} s}{2\sqrt3\epsilon_1\epsilon_2}} \, s^{\frac{a_{\D}^2}{2\epsilon_1\epsilon_2}}.
\end{align}
By normalizing the 
tau function of QPIV with $a_{\D}$-independent factor $\mathcal{Z}_{cl-m}^{[\mathbf{H_{2,S}}]}$ 
\begin{equation}
\tau_r(\epsilon_1,\epsilon_2|t)=\frac{\tau(\epsilon_1,\epsilon_2|t)}{\mathcal{Z}^{[\mathbf{H_{2,S}}]}_{cl-m}(\boldsymbol{m}_{1,2,3};\epsilon_1,\epsilon_2|t)}
\end{equation}
we obtain from \eqref{D13irr_2row},  \eqref{quantum_PIV}
\begin{multline}\label{D13irr_2row_rr}
D^1_{\epsilon_1,\epsilon_2[t]}(\tau_r^{(1)},\tau_r^{(2)})=\frac{\epsilon}3t^{-1}\tau_r^{(1)}\tau_r^{(2)},\\
   D^3_{\epsilon_1,\epsilon_2[t]}(\tau_r^{(1)},\tau_r^{(2)})=\epsilon t^{-1} D^2_{\epsilon_1,\epsilon_2[t]}(\tau^{(1)}_r,\tau^{(2)}_r)-\epsilon \left(\frac{t^4}9{-}12\boldsymbol{e}_2{+}2\epsilon_1\epsilon_2{-}\frac{109\epsilon^2}{108}\right)t^{-3}\tau^{(1)}_r\tau^{(2)}_r,
   \end{multline}
    \vspace{-0.7cm}
    \begin{multline}\label{quantum_PIV_rr}
    D^4_{\epsilon_1,\epsilon_2[t]}(\tau_r^{(1)},\tau_r^{(2)})+
    \left(\frac{t^4}{12}{+}36\boldsymbol{e}_2{-}4\epsilon_1\epsilon_2{+}\frac{31\epsilon^2}{12}\right) t^{-2}D^2_{\epsilon_1,\epsilon_2[t]}(\tau_r^{(1)},\tau_r^{(2)})
    -\frac14 \epsilon_1\epsilon_2 t\frac{d}{dt}\left(\tau^{(1)}_r\tau^{(2)}_r\right)\\+\left(\frac{11\epsilon^2}{36}t^4{+}108\boldsymbol{e}_2^2{+}\frac{71\epsilon^2{-}120\epsilon_1\epsilon_2}2\boldsymbol{e}_2{+}\frac{16(\epsilon_1\epsilon_2)^2}3{-}\frac{299\epsilon_1\epsilon_2\epsilon^2}{36}{+}\frac{12371\epsilon^4}{5184}\right)t^{-4}\tau^{(1)}_r\tau^{(2)}_r=0,
    \end{multline}
    so again the quantum Painlev\'e IV depends only on one mass parameter $\boldsymbol{e}_2$. And the other mass parameter appears as an integration constant as well.

\paragraph{Expansion.}
We then solve the successive $s^{-k}$ relations, obtaining step by step the leading terms of $\mathcal{Z}_{inst}^{[\mathbf{H_{2,S}}]}$ together with blowup factors $\mathfrak{l}_n^{[\mathbf{H_{2,S}}]}$.
In this way we computed $\mathfrak{l}_n^{[\mathbf{H_{2,S}}]}$ for $|n|\leq\frac52$ 
\begin{multline}\label{l_n_IV(e)}
\mathfrak{l}_0^{[\mathbf{H_{2,S}}]}=\mathfrak{l}_{\pm \frac12}^{[\mathbf{H_{2,S}}]}=1, \qquad \mathfrak{l}_{\pm 1}^{[\mathbf{H_{2,S}}]}=
\frac1{\sqrt3\ri}\left(a_{\D}\pm\frac{\epsilon}2\right), \qquad \mathfrak{l}_{\pm \frac32}^{[\mathbf{H_{2,S}}]}=
-\frac13 \prod\limits_{\substack{(i,j)=\\
\scriptscriptstyle (1,3),(3,1)}}\left(a_{\D}\pm\frac{i \epsilon_1{+}j\epsilon_2}2\right), \\
\mathfrak{l}_{\pm2}^{[\mathbf{H_{2,S}}]}=\frac19\prod\limits_{\substack{(i,j)={\scriptscriptstyle(1,1),}\\ \scriptscriptstyle(5,1),(3,3),(1,5)}}\left(a_{\D}\pm\frac{i \epsilon_1{+}j\epsilon_2}2\right), \qquad 
\mathfrak{l}_{\pm\frac52}^{[\mathbf{H_{2,S}}]}=-\frac1{27}\prod\limits_{\substack{(i,j)={\scriptscriptstyle(1,3),(3,1),}\\ \scriptscriptstyle(7,1),(3,5),(5,3)(1,7)}}\left(a_{\D}\pm\frac{i \epsilon_1{+}j\epsilon_2}2\right)
\end{multline}   
together with the terms of $\mathcal{Z}_{inst}^{[\mathbf{H_{2,S}}]}$ up to $s^{-5}$, up to $s^{-4}$ in the logarithmic expansion with $\alpha_{\D}=\ri a_{\D}/\sqrt3$ they are
\begin{multline}\label{Zinst_IVs}
-\frac13\epsilon_1\epsilon_2 \ln \mathcal{Z}_{inst}^{[\mathbf{H_{2,S}}]}(a_{\D},\boldsymbol{e}_2,\check{\boldsymbol{e}}_3;\epsilon_1,\epsilon_2|s)=\Bigg(2\alpha_{\D}^3-\left(36\boldsymbol{e}_2{-}\frac{4\epsilon_1\epsilon_2}3{+}\frac{7\epsilon^2}6\right) \alpha_{\D}+36\check{\boldsymbol{e}}_3\Bigg)\cdot\frac1{s}\\-\Bigg(\frac{35}4\alpha_{\D}^4-\left(378\boldsymbol{e}_2{-}\frac{181}{12}\epsilon_1\epsilon_2{+}\frac{289}{24}\epsilon^2\right)\alpha_{\D}^2+972 \check{\boldsymbol{e}}_3 \alpha_{\D}+c_1\Bigg)\cdot\frac1{s^2}\\+\Bigg(67\alpha_{\D}^5-\left(5400\boldsymbol{e}_2{-}\frac{671\epsilon_1\epsilon_2}3{+}\frac{1025\epsilon^2}6\right)\alpha_{\D}^3+23328 \check{\boldsymbol{e}}_3 \alpha_{\D}^2 +c_2\alpha_{\D}-648\check{\boldsymbol{e}}_3(6\boldsymbol{e}_2{-}\epsilon_1\epsilon_2{+}2\epsilon^2)\Bigg)\cdot\frac1{s^3}\\-\Bigg(\frac{2667}4\alpha_{\D}^6-3\left(29115\boldsymbol{e}_2{-}\frac{9877}8\epsilon_1\epsilon_2{+}\frac{14685}{16}\epsilon^2\right)\alpha_{\D}^4+539460 \check{\boldsymbol{e}}_3 \alpha_{\D}^3+c_3 \alpha_{\D}^2\\ -243\check{\boldsymbol{e}}_3(1944\boldsymbol{e}_2{-}382\epsilon_1\epsilon_2{+}503\epsilon^2)\alpha_{\D}+52488\check{\boldsymbol{e}}_3^2-c_4\Bigg)\cdot\frac1{s^4}+\ldots+O\left(\frac1{s^6}\right),
\end{multline}
where 
\begin{align}
&c_1=108 \boldsymbol{e}_2^2-\frac12(30\epsilon_1\epsilon_2{-}51\epsilon^2)  \boldsymbol{e}_2+\frac13(\epsilon_1\epsilon_2)^2-\frac{601}{432}\epsilon_1\epsilon_2\epsilon^2+\frac{1105}{1728}\epsilon^4,\\
&c_2=7776\boldsymbol{e}_2^2-18\left(74\epsilon_1\epsilon_2{-}85\epsilon^2\right)\boldsymbol{e}_2+36(\epsilon_1\epsilon_2)^2-\frac{10091}{108}\epsilon_1\epsilon_2\epsilon^2+\frac{16049}{432}\epsilon^4,\\
&c_3=329508\boldsymbol{e}_2^2-\frac92(13402\epsilon_1\epsilon_2{-}13397\epsilon^2)\boldsymbol{e}_2+\frac{41167}{24}(\epsilon_1\epsilon_2)^2-\frac{185311}{48}\epsilon_1\epsilon_2\epsilon^2+\frac{277015}{192}\epsilon^4, \\
&c_4=19440\boldsymbol{e}_2^3-81(110\epsilon_1\epsilon_2{-}223\epsilon^2)\boldsymbol{e}_2^2+\frac12\left(1845(\epsilon_1\epsilon_2)^2{-}\frac{17657}2\epsilon_1\epsilon_2\epsilon^2{+}\frac{37025}8\epsilon^4\right)\boldsymbol{e}_2\nonumber\\ &-\frac{115}6(\epsilon_1\epsilon_2)^3+\frac{139307}{864}(\epsilon_1\epsilon_2)^2\epsilon^2-\frac{218659}{1152}\epsilon_1\epsilon_2\epsilon^4+\frac{351755}{6912}\epsilon^6.
\end{align}
This expansion has the expected dependence on integration constant $\check{\boldsymbol{e}}_3$ and  $\check{\boldsymbol{e}}_3(2\epsilon_1,\epsilon_2{-}\epsilon_1)=\check{\boldsymbol{e}}_3(\epsilon_1{-}\epsilon_2,2\epsilon_2)$ so we consider that $\check{\boldsymbol{e}}_3$ depends on $\epsilon$ but not on $\epsilon_1\epsilon_2$ as in the previous cases. Above $\mathfrak{l}_n^{[\mathbf{H_{2,S}}]}$ are of form \eqref{l_n_fact}, so our ansatz \eqref{Z1loop_anz} gives
\begin{equation}\label{Z1loop_IV(e)}
\mathcal{Z}_{1-loop}^{[\mathbf{H_{2,S}}]}(a_{\D};\epsilon_1,\epsilon_2)=(\sqrt3\ri)^{\frac{a_{\D}^2}{2\epsilon_1\epsilon_2}}  \exp \gamma_{\epsilon_1,\epsilon_2} (a_{\D}-\epsilon/2). 
\end{equation}

\paragraph{Comparison with the classical limit ($\epsilon=0$).}
Let us compare our result in the $\epsilon=0$ limit with expansion \cite[(3.44)]{BLMST16}.
Namely, we check \eqref{PIV_tau_cl} under $\eta_{\D}=\rho$ and scaling \eqref{PIV_par_sc} with  
\eqref{PIV_dct}. Then, classical part \eqref{Zcl_IV(e)} becomes
\begin{equation}
\mathcal{Z}_{cl}^{[\mathbf{H_{2,S}}]}(-2\epsilon_2,2\epsilon_2)=e^{\mathrm{t}_{\sIV}^4/108}e^{\left(\ri\nu/\sqrt3+2(\theta_s{+}\theta_t)/3)\right)\mathrm{t}^2_{\sIV}/2}\, (2\epsilon_2 \mathrm{t}_{\IV}^2)^{2(\theta_s^2{-}\theta_s\theta_t{+}\theta_t^2)-1/6-\nu^2/2}. 
\end{equation} 
1-loop part \eqref{Z1loop_IV(e)}, by using \eqref{gamma_Barnes}, becomes  
\begin{equation}
\mathcal{Z}_{1-loop}^{[\mathbf{H_{2,S}}]}(-2\epsilon_2,2\epsilon_2)=(\sqrt3\ri)^{-\frac{\nu^2}2}\frac{(2\epsilon_2)^{\frac{\nu^2}2{-}\frac1{12}}}{e^{\zeta'(-1)}(2\pi)^{\nu/2}}\mathsf{G}(1+\nu) .
\end{equation}
Then, up to the factor in \eqref{PV_tau_cl} and a constant factor in $\mathrm{t}_{\IV}$ and $\nu$, $\mathcal{Z}_{cl}^{[\mathbf{H_{2,S}}]}\mathcal{Z}_{1-loop}^{[\mathbf{H_{2,S}}]}$ coincides with the prefactor of the $\mathrm{t}_{\IV}^{-2}$ series in \cite[(3.44)]{BLMST16}. Finally, for the first two terms of the instanton expansion we obtain \cite[(3.45)]{BLMST16} with \eqref{PIV_theta}
\begin{multline}
\mathcal{Z}_{inst}^{[\mathbf{H_{2,S}}]}(-2\epsilon_2,2\epsilon_2)=1+\Bigg(\frac{\nu^3}{3\ri}-\left(18\mathfrak{e}_2{+}\frac23\right) \ri\nu+18\sqrt3\mathfrak{e}_3\Bigg)\cdot\frac{2\sqrt3}{\mathrm{t}_{\IV}^2}\\-\Bigg(\frac{\nu^6}{18}+\left(6\mathfrak{e}_2{+}\frac{67}{144}\right)\nu^4+6\sqrt3\mathfrak{e}_3\ri\nu^3+3\left(54\mathfrak{e}_2^2{+}\frac{29}2\mathfrak{e}_2{+}\frac{71}{144}\right)\nu^2+3\sqrt3\mathfrak{e}_3\left(108\mathfrak{e}_2{+}31\right)\ri\nu\\ -486\mathfrak{e}_3^2{+}27\mathfrak{e}_2^2{+}\frac{15\mathfrak{e}_2}4{+}\frac1{12}\Bigg)\cdot\frac{12}{\mathrm{t}_{\IV}^4}+O\left(\frac1{\mathrm{t}_{\IV}^6}\right),
\end{multline}    
where, to match the $\nu$-independent part of $\mathrm{t}^{-2}_{\IV}$ in \cite[(3.45)]{BLMST16}, we imposed condition $\check{\boldsymbol{e}}_3|_{\epsilon=0}=\boldsymbol{e}_3$. 

\subsection{QPII (linear exp singularity) / \texorpdfstring{$H_1$}{H₁} with light hyper}
\label{ssec:exp_H1l}
As already mentioned, for Painlev\'e II we have two different large time expansions. Here we present the quantum deformation of the one on canonical rays $\arg\, t=0, \, \pm \frac{2\pi}3$ from \cite[Sec. 3.2]{BLMST16}, called there "expansion 2".
For the (quantum) Painlev\'e II we have $d=3/2$ and we set $s=2\sqrt2\ri\,t^{3/2}$ for the both expansions.

\paragraph{Classical part.}
For this expansion we have 
$N_p=2$. The first few leading $s^{-k}$ relations for \eqref{D13irr_2row}, \eqref{quantum_PII} imply that the classical part can be taken with $\beta=0$ as 
\begin{align}
&\mathcal{Z}_{cl}^{[\mathbf{H_{1,L}}]}(a_{\D}, \phi;\epsilon_1,\epsilon_2|s)=\mathcal{Z}^{[\mathbf{H_{1,L}}]}_{cl-m}(\phi;\epsilon_1,\epsilon_2|s) \, \mathcal{Z}^{[\mathbf{H_{1,L}}]}_{cl-a}(a_{\D};\epsilon_1,\epsilon_2|s) \qquad \textrm{with}\nonumber \\ \label{Zcl_II(p)} &\mathcal{Z}^{[\mathbf{H_{1,L}}]}_{cl-m}(\phi;\epsilon_1,\epsilon_2|s)=s^{\frac{\phi}{\epsilon_1\epsilon_2}}, \qquad \quad \mathcal{Z}^{[\mathbf{H_{1,L}}]}_{cl-a}(a_{\D};\epsilon_1,\epsilon_2|s)=e^{-\frac{a_{\D} s}{3\epsilon_1\epsilon_2}} \, s^{\frac{a_{\D}^2}{\epsilon_1\epsilon_2}}, 
\end{align}
where $\phi$ is a dimension $2$ integration constant that is not fixed by the further $s^{-k}$ relations, such that  $\phi(2\epsilon_1,\epsilon_2{-}\epsilon_1)=\phi(\epsilon_1{-}\epsilon_2,2\epsilon_2)$. So we consider that $\phi$ depends on $\epsilon$ but not on $\epsilon_1\epsilon_2$ as in the previous cases. Note that the quantum Painlev\'e II \eqref{D13irr_2row}, \eqref{quantum_PII} does not depend on any mass variables (c.f. App~\ref{ssec:E-coal} for the classical case).

\paragraph{Expansion.}
We then solve the successive $s^{-k}$ relations, obtaining step by step the leading terms of $\mathcal{Z}_{inst}^{[\mathbf{H_{1,L}}]}$ together with blowup factors $\mathfrak{l}_n^{[\mathbf{H_{1,L}}]}$. In this way we computed $\mathfrak{l}_n^{[\mathbf{H_{1,L}}]}$ for $|n|\leq2$ 
\begin{multline}\label{l_n_II(p)}
\mathfrak{l}_0^{[\mathbf{H_{1,L}}]}{=}\mathfrak{l}_{\pm \frac12}^{[\mathbf{H_{1,L}}]}=1, \qquad \mathfrak{l}_{\pm 1}^{[\mathbf{H_{1,L}}]}=\frac14
\prod\limits_{\lambda=\pm1}  \left(a_{\D}{+}2\lambda\tilde{\boldsymbol{m}} {\pm}\frac{\epsilon}2\right), \qquad \mathfrak{l}_{\pm \frac32}^{[\mathbf{H_{1,L}}]}=
\frac1{16}\prod\limits_{\lambda=\pm1} \prod\limits_{\substack{(i,j)=\\ \scriptscriptstyle(1,3),(3,1)}}\left(a_{\D}{+}2\lambda\tilde{\boldsymbol{m}}{\pm}\frac{i \epsilon_1{+}j\epsilon_2}2\right), \\
\mathfrak{l}_{\pm 2}^{[\mathbf{H_{1,L}}]}=
\frac1{256}\prod\limits_{\lambda=\pm1} \prod\limits_{\substack{(i,j)={\scriptscriptstyle(1,1),}\\ \scriptscriptstyle(5,1) (3,3) (1,5)}}\left(a_{\D}{+}2\lambda\tilde{\boldsymbol{m}}{\pm}\frac{i \epsilon_1{+}j\epsilon_2}2\right), \quad \textrm{with reparametrization} \quad \phi=\frac{\epsilon^2}{12}-\frac43\tilde{\boldsymbol{m}}^2.
\end{multline}
We computed the terms of $\mathcal{Z}_{inst}^{[\mathbf{H_{1,L}}]}$ up to $s^{-7}$, up to $s^{-4}$
in the logarithmic expansion they are 
{\allowdisplaybreaks\begin{multline}\label{Zinst_IIl}
-\epsilon_1\epsilon_2 \ln \mathcal{Z}_{inst}^{[\mathbf{H_{1,L}}]}(a_{\D},\tilde{\boldsymbol{m}}^2;\epsilon_1,\epsilon_2|s)=\Bigg(\frac{17}3a_{\D}^3-\left(12\tilde{\boldsymbol{m}}^2{+}\frac{\epsilon_1\epsilon_2}6{-}\frac{19\epsilon^2}{12}\right)a_{\D}\Bigg)\cdot\frac1{s}\\+\Bigg(\frac{125}4 a_{\D}^4-\left(86\tilde{\boldsymbol{m}}^2{+}\frac{15\epsilon_1\epsilon_2}4{-}\frac{153\epsilon^2}8\right)a_{\D}^2+(16\tilde{\boldsymbol{m}}^2{-}\epsilon^2)\left(\frac{11}{12}\tilde{\boldsymbol{m}}^2{+}\frac{17\epsilon_1\epsilon_2}{48}{-}\frac{131\epsilon^2}{192}\right)\Bigg)\cdot\frac1{s^2}\\ +\Bigg(\frac{3563}{12}a_{\D}^5-\left(990\tilde{\boldsymbol{m}}^2{+}\frac{3101\epsilon_1\epsilon_2}{36}{-}\frac{23405\epsilon^2}{72}\right)a_{\D}^3+c_1a_{\D}\Bigg)\cdot\frac1{s^3}\\
+\Bigg(\frac{29183}8a_{\D}^6-\left(14210\tilde{\boldsymbol{m}}^2{+}\frac{16143\epsilon_1\epsilon_2}8{-}\frac{50715\epsilon^2}8\right)a_{\D}^4+c_2a_{\D}^2-(16\tilde{\boldsymbol{m}}^2{-}\epsilon^2)c_3\Bigg)\cdot\frac1{s^4}+\ldots+O\left(\frac1{s^8}\right), 
\end{multline}}
where
\begin{align}
&c_1=404\tilde{\boldsymbol{m}}^4+\left(197\epsilon_1\epsilon_2{-}\frac{821\epsilon^2}2\right)\tilde{\boldsymbol{m}}^2+\frac{35(\epsilon_1\epsilon_2)^2}{24}-\frac{4133\epsilon_1\epsilon_2\epsilon^2}{144}+\frac{22709\epsilon^4}{576},\\
&c_2=9550\tilde{\boldsymbol{m}}^4+\frac12\left(11621\epsilon_1\epsilon_2{-}\frac{47675\epsilon^2}2\right)\tilde{\boldsymbol{m}}^2+\frac{577(\epsilon_1\epsilon_2)^2}4-\frac{48589\epsilon_1\epsilon_2\epsilon^2}{32}+\frac{217663\epsilon^4}{128},\\
&c_3=\frac{115}3\tilde{\boldsymbol{m}}^4+\frac1{12}(673\epsilon_1\epsilon_2{-}1111\epsilon^2)\tilde{\boldsymbol{m}}^2+\frac{239(\epsilon_1\epsilon_2)^2}{16}-\frac{4009\epsilon_1\epsilon_2\epsilon^2}{64}+\frac{10483\epsilon^4}{256}.
\end{align}
Above $\mathfrak{l}_n^{[\mathbf{H_{1,L}}]}$ are of form \eqref{l_n_fact}, so our ansatz \eqref{Z1loop_anz} gives
\begin{equation}\label{Z1loop_II(p)}
\mathcal{Z}_{1-loop}^{[\mathbf{H_{1,L}}]}(a_{\D},\tilde{\boldsymbol{m}};\epsilon_1,\epsilon_2)= 2^{\frac{a_{\D}^2}{\epsilon_1\epsilon_2}}\prod\limits_{\lambda=\pm1}  \exp \gamma_{\epsilon_1,\epsilon_2} \left(a_{\D}+2\lambda\tilde{\boldsymbol{m}}-\epsilon/2\right). 
\end{equation}

\paragraph{Comparison with the classical limit ($\epsilon=0$).}
Let us compare our result in the $\epsilon=0$ limit with expansion \cite[(3.32)]{BLMST16}. According to \eqref{PII_tau_cl_app} we check that 
\begin{equation}\label{PII_tau_cl}
\tau(a_{\D},\eta_{\D},\boldsymbol{m};-2\epsilon_2,2\epsilon_2|t)=\tau_{\II}(\nu,\rho;\theta|\mathrm{t}_{\II}).   
\end{equation}
The dictionary is given by $\eta_{\D}=\rho$ and standard scalings (see \eqref{PII_tau_cl_app} for the all except the first one)
\begin{equation}\label{PII_par_sc}
a_{\D}=2\epsilon_2\nu,  \qquad t=(2\epsilon_2)^{2/3} \mathrm{t}_{\II}, \qquad \boldsymbol{m}=2\epsilon_2\mathfrak{m}, 
\end{equation}
where mass $\mathfrak{m}$ is given by \eqref{PII_dct}.
Then, classical part \eqref{Zcl_II(p)} becomes 
\begin{equation}
\mathcal{Z}_{cl}^{[\mathbf{H_{1,L}}]}(-2\epsilon_2,2\epsilon_2)=e^{\frac{2\sqrt2\ri}3\nu \mathrm{t}_{\sII}^{3/2}}\, (2\epsilon_2 (2\sqrt2\ri \mathrm{t}_{\II}^{3/2}))^{\frac{\tilde{\theta}^2}{12}-\nu^2}.   
\end{equation} 
1-loop part \eqref{Z1loop_II(p)}, by using \eqref{gamma_Barnes}, becomes  
\begin{equation}
\mathcal{Z}_{1-loop}^{[\mathbf{H_{1,L}}]}(-2\epsilon_2,2\epsilon_2)=2^{-\nu^2}\frac{(2\epsilon_2)^{\nu^2+\tilde{\theta}^2/4-\frac16}}{e^{2\zeta'(-1)}(2\pi)^{\nu}}\prod_{\lambda=\pm1}\mathsf{G}\left(1+\nu+\lambda\tilde{\theta}/2\right),
\end{equation}
where the tilded theta is connected with the the tilded mass by same formula $\tilde{\boldsymbol{m}}=(2\epsilon_2)\tilde{\theta}/4$ as for the non-tilded one. The Barnes $\mathsf{G}$-function factor of the above expression coincides with the counterpart in \cite[(3.32)]{BLMST16} iff in the self-dual case the tilded and the non-tilded mass are equal up to discrete group $W(A_1)=\mathbb{Z}_2$, which acts by changing the sign of the mass. This means that $\tilde{\boldsymbol{m}}^2|_{\epsilon=0}=\boldsymbol{m}^2$.
Then, up to a constant factor in $\mathrm{t}_{\II}$ and $\nu$, $\mathcal{Z}_{cl}^{[\mathbf{H_{1,L}}]}\mathcal{Z}_{1-loop}^{[\mathbf{H_{1,L}}]}$ coincides with the prefactor of the $\mathrm{t}_{\II}^{-3/2}$ series in \cite[(3.32)]{BLMST16}. Finally, for the first two terms of the instanton expansion we obtain \cite[(3.33)]{BLMST16} 
\begin{multline}
\mathcal{Z}_{inst}^{[\mathbf{H_{1,L}}]}(-2\epsilon_2,2\epsilon_2)=1+\frac1{9\ri}\left(17\nu^3-\frac{9\theta^2{-}2}4 \nu\right)\cdot\frac3{2\sqrt2 \mathrm{t}_{\II}^{3/2}}\\-\frac1{81}\Bigg(\frac{17^2}2\nu^6-\frac{153\theta^2{-}1159}4\nu^4+\left(\frac{81\theta^4}{32}{-}\frac{99\theta^2}2{+}\frac{271}8\right)\nu^2+\frac{3(11\theta^2{-}68)}{64}\Bigg)\cdot\frac9{8 \mathrm{t}_{\II}^3}+O\left(\frac1{\mathrm{t}_{\II}^{9/2}}\right). 
\end{multline}

 As in Secs. \ref{ssec:exp_3l}, \ref{ssec:exp_2}, and \ref{ssec:exp_H2l}, factorization \eqref{l_n_fact} prescribes that, without loss of generality, difference $\tilde{\boldsymbol{m}}-\boldsymbol{m}$ is proportional to $\epsilon$ up to a numerical coefficient. {\color{red}Even if, contrary to the other cases, we have no relation to fix this shift, we see that it is actually zero in Sec.~\ref{ssec:holom_anom_H1}.}

\subsection{QPII (square exp singularity) / \texorpdfstring{$H_1$}{H₁} with heavy hyper}
\label{ssec:exp_H1s}
Here we present the quantum deformation of the other Painlev\'e II large time expansion, namely the one on canonical rays $\arg\, t=\pi,\pm \frac{\pi}3$ from \cite[Sec. 3.3]{BLMST16}, called there "expansion 1". Recall that for QPII we set $s=2\sqrt2\ri \,t^{3/2}$.

\paragraph{Classical part.}
For this expansion we have $N_p=1$. The first few leading $s^{-k}$ relations for \eqref{D13irr_2row}, \eqref{quantum_PII} imply that the classical part can be taken with $\beta\neq0$ as
\begin{align}
&\mathcal{Z}_{cl}^{[\mathbf{H_{1,S}}]}(a_{\D},\phi;\epsilon_1,\epsilon_2|s)=\mathcal{Z}_{cl-m}^{[\mathbf{H_{1,S}}]}(\phi;\epsilon_1,\epsilon_2|s) \mathcal{Z}_{cl-a}^{[\mathbf{H_{1,S}}]}(a_{\D};\epsilon_1,\epsilon_2|s) \qquad \textrm{with}\nonumber \\ 
\label{Zcl_II(e)} &\mathcal{Z}_{cl-m}^{[\mathbf{H_{1,S}}]}(\phi;\epsilon_1,\epsilon_2|s)=e^{-\frac{s^2}{192\epsilon_1\epsilon_2}} \, s^{\frac{\phi}{48\epsilon_1\epsilon_2}-\frac1{12}}, \qquad \quad \mathcal{Z}_{cl-a}^{[\mathbf{H_{1,S}}]}(a_{\D};\epsilon_1,\epsilon_2|t)=e^{\frac{\ri\sqrt2 a_{\D} s}{6\epsilon_1\epsilon_2}} \, s^{\frac{a_{\D}^2}{2\epsilon_1\epsilon_2}},
\end{align}
where $\phi$ is again a dimension $2$ integration constant that is not fixed by the further $s^{-k}$ relations, such that  $\phi(2\epsilon_1,\epsilon_2{-}\epsilon_1)=\phi(\epsilon_1{-}\epsilon_2,2\epsilon_2)$. So we again consider that $\phi$ depends on $\epsilon$ but not on $\epsilon_1\epsilon_2$.

\paragraph{Expansion.}
We then solve the successive $s^{-k}$ relations, obtaining step by step the leading terms of $\mathcal{Z}_{inst}^{[\mathbf{H_{1,S}}]}$ together with blowup factors $\mathfrak{l}_n^{[\mathbf{H_{1,S}}]}$.
In this way we computed $\mathfrak{l}_n^{[\mathbf{H_{1,S}}]}$ for $|n|\leq\frac52$ 
\begin{multline}\label{l_n_II(e)}
\mathfrak{l}_0^{[\mathbf{H_{1,S}}]}=\mathfrak{l}_{\pm \frac12}^{[\mathbf{H_{1,S}}]}=1, \qquad \mathfrak{l}_{\pm 1}^{[\mathbf{H_{1,S}}]}=
\frac{\ri}{2\sqrt2}\left(a\pm\frac{\epsilon}2\right), \qquad \mathfrak{l}_{\pm \frac32}^{[\mathbf{H_{1,S}}]}=
-\frac18\prod\limits_{\substack{(i,j)=\\ \scriptscriptstyle(1,3),(3,1)}}\left(a\pm\frac{i \epsilon_1{+}j\epsilon_2}2\right), \\
\mathfrak{l}_{\pm2}^{[\mathbf{H_{1,S}}]}=\frac1{64}\prod\limits_{\substack{(i,j)={\scriptscriptstyle(1,1),}\\ \scriptscriptstyle(5,1),(3,3),(1,5)}}\left(a\pm\frac{i \epsilon_1{+}j\epsilon_2}2\right), \qquad 
\mathfrak{l}_{\pm\frac52}^{[\mathbf{H_{1,S}}]}=-\frac1{512}\prod\limits_{\substack{(i,j)={\scriptscriptstyle(1,3),(3,1),}\\ \scriptscriptstyle(7,1),(3,5),(5,3)(1,7)}}\left(a\pm\frac{i \epsilon_1{+}j\epsilon_2}2\right)
\end{multline}   
together with the terms of $\mathcal{Z}_{inst}^{[\mathbf{H_{1,S}}]}$ up to $s^{-5}$, up to $s^{-4}$
in the logarithmic expansion with  $\alpha_{\D}=\ri a_{\D}/\sqrt2$ and under reparametrization $\phi=-256\check{\boldsymbol{m}}^2$ they are
\begin{multline}\label{Zinst_IIe}
\epsilon_1\epsilon_2 \ln \mathcal{Z}_{inst}^{[\mathbf{H_{1,S}}]}(a_{\D},\check{\boldsymbol{m}}^2;\epsilon_1,\epsilon_2|s)=\Bigg(\frac{17}3\alpha_{\D}^3+\left(2^7\check{\boldsymbol{m}}^2{+}\frac{31}{12}\epsilon_1\epsilon_2{+}\frac5{24}\epsilon^2\right) \alpha_{\D}\Bigg)\cdot\frac1{ s}\\+\Bigg(\frac{125}4\alpha_{\D}^4+\left(13 {\cdot} 2^7\check{\boldsymbol{m}}^2{+}\frac{293}8\epsilon_1\epsilon_2{+}\frac{55}{16}\epsilon^2\right)\alpha_{\D}^2+\frac{2^{10}\check{\boldsymbol{m}}^4}3{+}\frac{160}3(\epsilon_1\epsilon_2{-}2\epsilon^2)\check{\boldsymbol{m}}^2{+}\frac34(\epsilon_1\epsilon_2)^2{-}\frac{165}{256}\epsilon_1\epsilon_2\epsilon^2{-}\frac{175}{32}\epsilon^4\Bigg)\cdot\frac1{s^2}\\+\frac13\Bigg(\frac{3563}4 \alpha_{\D}^5+\left(685{\cdot}2^7\check{\boldsymbol{m}}^2{+}\frac{48557}{24}\epsilon_1\epsilon_2{+}\frac{9475}{48}\epsilon^2\right) \alpha_{\D}^3+c_1 \alpha_{\D}\Bigg)\cdot\frac1{ s^3}\\+\Bigg(\frac{29183}8\alpha_{\D}^6+\left(9065{\cdot}2^6\check{\boldsymbol{m}}^2\phi{+}\frac{27617}2\epsilon_1\epsilon_2{+}\frac{21805}{16}\epsilon^2\right)\alpha_{\D}^4+c_2 \alpha_{\D}^2+\frac{c_3}3\Bigg)\cdot\frac1{ s^4}+\ldots+O\left(\frac1{s^6}\right),
\end{multline}
where 
\begin{align}
&c_1=11{\cdot}2^{13}\check{\boldsymbol{m}}^4+80(206\epsilon_1\epsilon_2{-}259\epsilon^2)\check{\boldsymbol{m}}^2+\frac{8831}{32}(\epsilon_1\epsilon_2)^2-\frac{93605}{192}\epsilon_1\epsilon_2\epsilon^2-\frac{105835}{768}\epsilon^4,\\
&c_2=95{\cdot}2^{14} \check{\boldsymbol{m}}^4+11(1702\epsilon_1\epsilon_2{-}1775\epsilon^2)2^4\check{\boldsymbol{m}}^2+\frac{168659}{32}(\epsilon_1\epsilon_2)^2-\frac{926843}{128}\epsilon_1\epsilon_2\epsilon^2-\frac{1080577}{512}\epsilon^4, \\
&c_3=2^{18}\check{\boldsymbol{m}}^6+(31\epsilon_1\epsilon_2{-}74\epsilon^2)2^{12}\check{\boldsymbol{m}}^4+(14016(\epsilon_1\epsilon_2)^2{-}74956\epsilon_1\epsilon_2\epsilon^2{+}34855\epsilon^4)\check{\boldsymbol{m}}^2 \nonumber\\ &+189(\epsilon_1\epsilon_2)^3-\frac{387315}{256}(\epsilon_1\epsilon_2)^2\epsilon^2+\frac{614625}{1024}\epsilon_1\epsilon_2\epsilon^4+\frac{564375}{2048}\epsilon^6.
\end{align}
Above $\mathfrak{l}_n^{[\mathbf{H_{1,S}}]}$ are of form \eqref{l_n_fact}, so our ansatz \eqref{Z1loop_anz} gives
\begin{equation}\label{Z1loop_II(e)}
\mathcal{Z}_{1-loop}^{[\mathbf{H_{1,S}}]}(a_{\D};\epsilon_1,\epsilon_2)=(\sqrt2\ri)^{\frac{3a_{\D}^2}{2\epsilon_1\epsilon_2}}\exp \gamma_{\epsilon_1,\epsilon_2} (a_{\D}-\epsilon/2). 
\end{equation}

\paragraph{Comparison with the classical limit ($\epsilon=0$).}
Let us compare our result in the $\epsilon=0$ limit with expansion \cite[(3.28)]{BLMST16}. Namely, we check \eqref{PII_tau_cl} under $\eta_{\D}=\rho$ 
and scaling \eqref{PII_par_sc} with \eqref{PII_dct}.
Then, classical part \eqref{Zcl_II(e)} becomes
\begin{equation}
\mathcal{Z}_{cl}^{[\mathbf{H_{1,S}}]}(-2\epsilon_2,2\epsilon_2)=e^{-\mathrm{t}_{\sII}^3/24+2\nu \mathrm{t}_{\sII}^{3/2}/3}\, (2\epsilon_2 (2\sqrt2\ri \mathrm{t}_{\II}^{3/2}))^{16\mathfrak{m}^2/3-1/12-\nu^2/2}.   
\end{equation} 
1-loop part \eqref{Z1loop_II(e)}, by using \eqref{gamma_Barnes}, becomes  
\begin{equation}
\mathcal{Z}_{1-loop}^{[\mathbf{H_{1,S}}]}(-2\epsilon_2,2\epsilon_2)=(\sqrt2\ri)^{-\frac{3\nu^2}2}\frac{(2\epsilon_2)^{\frac{\nu^2}2-\frac1{12}}}{e^{\zeta'(-1)}(2\pi)^{\nu/2}}\mathsf{G}(1+\nu).
\end{equation}
Then, up to a constant factor in $\mathrm{t}_{\II}$ and $\nu$, $\mathcal{Z}_{cl}^{[\mathbf{H_{1,S}}]}\mathcal{Z}_{1-loop}^{[\mathbf{H_{1,S}}]}$ coincides with the prefactor of the $\mathrm{t}_{\II}^{-3/2}$ series in \cite[(3.28)]{BLMST16} iff $\check{\boldsymbol{m}}^2|_{\epsilon=0}=\boldsymbol{m}^2$. Finally, for the first two terms of the instanton expansion we obtain \cite[(3.29)]{BLMST16} 
\begin{multline}
\mathcal{Z}_{inst}^{[\mathbf{H_{1,S}}]}(-2\epsilon_2,2\epsilon_2)=1+\left(\frac{17}{36}\nu^3-\frac{96\theta^2{-}31}{72}\nu\right)\cdot\frac3{2\mathrm{t}_{\II}^{3/2}}\\+\Bigg(\frac{289}{2592}\nu^6-\frac{408\theta^2{-}413}{648}\nu^4+\left(\frac89\theta^4{-}\frac{187}{54}\theta^2{+}\frac{11509}{10368}\right)\nu^2+\frac{(4\theta^2{-}1) (4\theta^2{-}9)}{216}\Bigg)\cdot\frac{9}{4\mathrm{t}_{\II}^3}+O\left(\frac1{\mathrm{t}_{\II}^{9/2}}\right).
\end{multline}    

\subsection{QPI / \texorpdfstring{$H_0$}{H₀}}
\label{ssec:exp_H0}
Here we present the quantum deformation of the Painlev\'e I large time expansion on canonical rays $\arg\, t=\frac{(2k+1)\pi}5,\, k=0\ldots 4$ from \cite[Sec. 3.1]{BLMST16}. For the (quantum) Painlev\'e I we have $d=5/4$ and we set $s=-48(\ri{+}1)6^{1/4}t^{5/4}$.
\paragraph{Classical part.}
For this expansion we have $N_p=1$. The first few leading $s^{-k}$ relations for \eqref{D13irr_2row}, \eqref{quantum_PI} imply that the classical part can be taken as
\begin{align}
&\mathcal{Z}_{cl}^{[\mathbf{H_0}]}(a_{\D};\epsilon_1,\epsilon_2|s)=\mathcal{Z}_{cl-m}^{[\mathbf{H_0}]}(\epsilon_1,\epsilon_2|s) \mathcal{Z}_{cl-a}^{[\mathbf{H_0}]}(a_{\D};\epsilon_1,\epsilon_2|s) \qquad \textrm{with}\nonumber \\ \label{Zcl_I} &\mathcal{Z}_{cl-m}^{[\mathbf{H_0}]}(\epsilon_1,\epsilon_2|s)=e^{\frac{s^2}{103680\epsilon_1\epsilon_2}} \, s^{\frac{7\epsilon^2}{120\epsilon_1\epsilon_2}-\frac1{60}}, \qquad \quad \mathcal{Z}_{cl-a}^{[\mathbf{H_0}]}(a_{\D};\epsilon_1,\epsilon_2|s)=e^{-\frac{a_{\D} s}{60\epsilon_1\epsilon_2}} \, s^{\frac{a_{\D}^2}{2\epsilon_1\epsilon_2}}.
\end{align}

\paragraph{Expansion.}
We then solve the successive $s^{-k}$ relations, obtaining step by step the leading terms of $\mathcal{Z}_{inst}^{[\mathbf{H_0}]}$ together with blowup factors $\mathfrak{l}_n^{[\mathbf{H_0}]}$. 
In this way we computed $\mathfrak{l}_n^{[\mathbf{H_0}]}$ for $|n|\leq\frac52$ 
\begin{multline}
\mathfrak{l}_0^{[\mathbf{H_0}]}=\mathfrak{l}_{\pm \frac12}^{[\mathbf{H_0}]}=1, \qquad \mathfrak{l}_{\pm 1}^{[\mathbf{H_0}]}=
a_{\D}\pm\frac{\epsilon}2, \qquad \mathfrak{l}_{\pm \frac32}^{[\mathbf{H_0}]}=
\prod\limits_{\substack{(i,j)=\\ \scriptscriptstyle(1,3),(3,1)}}\left(a_{\D}\pm\frac{i \epsilon_1{+}j\epsilon_2}2\right), \\
\mathfrak{l}_{\pm2}^{[\mathbf{H_0}]}=\prod\limits_{\substack{(i,j)={\scriptscriptstyle(1,1),}\\ \scriptscriptstyle(5,1),(3,3),(1,5)}}\left(a_{\D}\pm\frac{i \epsilon_1{+}j\epsilon_2}2\right), \qquad \mathfrak{l}_{\pm\frac52}^{[\mathbf{H_0}]}=\prod\limits_{\substack{(i,j)={\scriptscriptstyle(1,3),(3,1),}\\ \scriptscriptstyle(7,1),(3,5),(5,3),(1,7)}}\left(a_{\D}\pm\frac{i \epsilon_1{+}j\epsilon_2}2\right)
\end{multline}   
together with the terms of $\mathcal{Z}_{inst}^{[\mathbf{H_0}]}$ up to $s^{-6}$, up to $s^{-4}$ in the logarithmic expansion they are
{\allowdisplaybreaks\begin{multline}\label{Zinst_I}
-\epsilon_1\epsilon_2 \ln \mathcal{Z}_{inst}^{[\mathbf{H_0}]}(a_{\D};\epsilon_1,\epsilon_2|s)=\left(47a_{\D}^3-\frac{34\epsilon_1
\epsilon_2{-}77\epsilon^2}4a_{\D}\right)\cdot\frac{1}{s}\\+\Bigg(\frac{7717}2a_{\D}^4-\frac{7354\epsilon_1\epsilon_2{-}13937\epsilon^2}4a_{\D}^2+7\left(\frac{24}5(\epsilon_1\epsilon_2)^2{-}\frac{4597}{120}\epsilon_1\epsilon_2\epsilon^2{+}\frac{14497}{480}\epsilon^4\right)\Bigg)\cdot\frac{1}{s^2}\\+\Bigg(\frac{2663129}5a_{\D}^5-\frac{1003906\epsilon_1\epsilon_2{-}1717793\epsilon^2)}2a_{\D}^3+\left(\frac{541269}{10}(\epsilon_1\epsilon_2)^2{-}\frac{15982183}{60}\epsilon_1\epsilon_2\epsilon^2{+}\frac{43147783}{240}\epsilon^4\right)a_{\D}\Bigg)\cdot\frac{1}{s^3}\\+\Bigg(94160703a_{\D}^6-9(16692776\epsilon_1\epsilon_2{-}26678883\epsilon^2)a_{\D}^4 +3\left(14438609(\epsilon_1\epsilon_2)^2{-}\frac{237879817}4\epsilon_1\epsilon_2\epsilon^2{+}\frac{589817557}{16}\epsilon^4\right)a_{\D}^2\\  -7\left(80640(\epsilon_1\epsilon_2)^3-\frac{21788387}{20}(\epsilon_1\epsilon_2)^2\epsilon^2+\frac{150631761}{80}\epsilon_1\epsilon_2\epsilon^4-\frac{113120777}{160}\epsilon^6\right)\Bigg)\cdot\frac{1}{s^4}+\ldots+O\left(\frac{1}{s^7}\right).
\end{multline}}
Above $\mathfrak{l}_n^{[\mathbf{H_0}]}$ are of form \eqref{l_n_fact}, so our ansatz \eqref{Z1loop_anz} gives
\begin{equation}\label{Z1loop_I}
\mathcal{Z}_{1-loop}^{[\mathbf{H_0}]}(a_{\D};\epsilon_1,\epsilon_2)=\exp \gamma_{\epsilon_1,\epsilon_2} (a_{\D}-\epsilon/2). 
\end{equation}

\paragraph{Comparison with the classical limit ($\epsilon=0$).}
Let us compare our result in the $\epsilon=0$ limit with expansion BLMST \cite[(3.9)]{BLMST16}. According to \eqref{PI_tau_cl_app} we check that 
\begin{equation}\label{PI_tau_cl}
\tau(a_{\D},\eta_{\D};-2\epsilon_2,2\epsilon_2|t)=\tau_{\I}(\nu,\rho|\mathrm{t}_{\I}).   
\end{equation}
The dictionary is given by $\eta_{\D}=\rho$ and standard scalings (see \eqref{PI_tau_cl_app} for the second one)
\begin{equation}
a_{\D}=2\epsilon_2\nu, \qquad  t=(2\epsilon_2)^{4/5} \mathrm{t}_{\I}\quad (s=2\epsilon_2 \mathrm{s}_{\I}). 
\end{equation}
Then, classical part \eqref{Zcl_I} becomes
\begin{equation}
\mathcal{Z}_{cl}^{[\mathbf{H_0}]}(-2\epsilon_2,2\epsilon_2)=e^{-\frac{\mathrm{s}_{\sI}^2}{45\cdot 48^2}+\frac{\nu \mathrm{s}_{\sI}}{60}}\, (2\epsilon_2 \mathrm{s}_{\I})^{-1/60-\nu^2/2}.   
\end{equation} 
1-loop part \eqref{Z1loop_I}, by using \eqref{gamma_Barnes}, becomes 
\begin{equation}
\mathcal{Z}_{1-loop}^{[\mathbf{H_0}]}(-2\epsilon_2,2\epsilon_2)=\frac{(2\epsilon_2)^{\frac{\nu^2}2-\frac1{12}}}{e^{\zeta'(-1)}(2\pi)^{\nu/2}}\mathsf{G}(1+\nu).
\end{equation}
Up to a constant factor in $\mathrm{t}_{\I}$ and $\nu$, $\mathcal{Z}_{cl}^{[\mathbf{H_0}]}\mathcal{Z}_{1-loop}^{[\mathbf{H_0}]}$ coincides with the prefactor of the $\mathrm{s}_{\I}^{-1}$ series in \cite[(3.9)]{BLMST16}, where BLMST expansion variable differs from our convention, namely $\mathrm{s}_{\I}=48\ri\,\mathrm{s}_{\I}^{\mathrm{BLMST}}$. Finally, for the first two terms of the instanton expansion we obtain \cite[(3.10)]{BLMST16} 
\begin{equation}
\mathcal{Z}_{inst}^{[\mathbf{H_0}]}(-2\epsilon_2,2\epsilon_2)=1+\left(\frac{47\nu^3}{48\ri}+\frac{17\nu}{96\ri}\right)\cdot\frac{48\ri}{\mathrm{s}_{\I}}-\left(\frac{2209\nu^6}{4608}+\frac{2129\nu^4}{1152}+\frac{4999\nu^2}{6144}+\frac7{480}\right)\cdot\frac{(48\ri)^2}{\mathrm{s}_{\I}^2}+O\left(\frac1{\mathrm{s}_{\I}^3}\right). 
\end{equation}

\section{Holomorphic anomaly calculations}
\label{sec:holom_anom}
In this section we explicitly connect the expansions of the quantum Painlev\'e tau functions of the previous section with the partition functions of the corresponding gauge theories. Actually, since the expansions we consider are in the large time regime, from Painlev\'e/gauge theory correspondence it follows that we have to study the expansion of gauge theory partition functions in the  magnetic frame, where the light degrees of freedom are given by particles with the magnetic charge. To access this regime in the full $\Omega$-background we use the refined topological string partition function, which geometrically engineers the relevant gauge theory. More precisely, we compare our quantum Painlev\'e tau function expansion coefficients $\mathcal{Z}$ with refined topological string $\epsilon_{1,2}$-expansion
\begin{equation}
\epsilon_1\epsilon_2\ln \mathcal{Z}=\sum_{n,m=0}^{+\infty}\epsilon^{2n}(\epsilon_1\epsilon_2)^m\mathcal{F}_{n,m}=\sum_{g=0}^{\infty}(\epsilon_1\epsilon_2)^g \mathcal{F}_g, \qquad \mathcal{F}_g=\sum_{n=0}^g Q^{2n} \mathcal{F}_{n,g-n}, \quad \textrm{for} \quad Q^2=\frac{\epsilon^2}{\epsilon_1\epsilon_2},
\end{equation}
where we introduced CFT notation $Q$ according to \eqref{AGT_dictionary}. Now we very briefly recall how to calculate $\mathcal{F}_g$ using the refined holomorphic anomaly equations. We partially follow papers \cite{HKK11,PP23}.

\paragraph{Seiberg-Witten curve.} Let us consider a Seiberg-Witten (SW) theory \cite{SW94} governed by an elliptic curve, in the Weierstrass form it is
\begin{equation}\label{Welc}
y^2=4z^3-g_2z-g_3,    
\end{equation}
where $g_2$ and $g_3$ are polynomials in global Coulomb modulus parameter $u$. In a given framing this torus has periods
\begin{equation}
\omega_1=\int\limits_{A-cycle}\frac{dz}{\ri\pi y}, \qquad \omega_2=\int\limits_{B-cycle}\frac{dz}{\ri\pi y}    
\end{equation}
and torus parameter $\tau=\omega_2/\omega_1$ is identified with the infrared coupling of the effective $U(1)$ gauge theory. These periods can be expressed in terms of two solutions of hypergeometric equation
\begin{align}\label{periods_hyph}
\omega_1&=(3g_2/4)^{-1/4}{}_2F_1\Big(\frac1{12},\frac5{12},1\Big|\frac{\Delta}{g_2^3}\Big),\\
2\pi\ri\omega_2&=(3g_2/4)^{-1/4} \left(\frac{d}{dv}\Bigg|_{v=0}{}_2F_1\Big(\frac1{12}{+}v,\frac5{12}{+}v,1{+}2v\Big|\frac{\Delta}{g_2^3}\Big)+\ln\left(\frac{\Delta}{(12g_2)^3}\right){}_2F_1\Big(\frac1{12},\frac5{12},1\Big|\frac{\Delta}{g_2^3}\Big)\right),
\end{align}
where
\begin{equation}
\Delta(u)=g_2^3-27 g_3^2    
\end{equation}
is the modular discriminant of the curve.

\paragraph{Flat coordinates.}
The so-called "flat" coordinates $a_{\D}$ and $a$ are introduced by
\begin{equation}\label{dada}
\omega_1=\frac{da_{\D}}{du}, \qquad  \omega_2=\frac{da}{du},
\end{equation}
where the periods are interchanged because we study partition functions in the dual, magnetic frame.
SW prepotential $\mathcal{F}_0$ 
is defined such that its second derivative obeys
\begin{equation}\label{F0}
\frac{d^2\mathcal{F}_0}{da_{\D}^2}=-\ln q^2, \qquad  q^2\equiv e^{2\pi\ri\tau}.    
\end{equation}

Flat coordinates $a_{\D}$ and $a$ in \eqref{dada} are defined up to a shift. Moreover, note that in \eqref{periods_hyph} we also have a branch cut due to factor $(3g_2/4)^{-1/4}$. Up to the overall sign $(a, a_{\D})\mapsto-(a, a_{\D})$, this branching can be  
fixed by standard formula regarding elliptic curves
\begin{equation}\label{omega_1_ell}
\omega_1^2=\frac{2g_2E_6(q)}{9g_3E_4(q)}, 
\end{equation}
where Eisenstein series $E_{2n}(q)$ are defined by
\begin{equation}
E_{2n}(q)=1+\frac{2}{\zeta(1{-}2n)}\sum_{k=1}^{+\infty}\frac{k^{2n-1}q^{2n}}{1-q^{2n}}, \qquad n\in\mathbb{N}.
\end{equation}

\paragraph{Holomorphic anomaly equation.}
First correction $\mathcal{F}_1$ is given by formula
\begin{equation}\label{F1}
\mathcal{F}_1=\frac{Q^2-2}{24}\ln \Delta-\frac14\ln \omega_1^2.
\end{equation}
Higher terms $\mathcal{F}_{g>1}$ can be calculated inductively from $\mathcal{F}_0$ and $\mathcal{F}_1$ by using the so-called refined holomorphic anomaly equation
\begin{equation}\label{ha}
\partial_X \mathcal{F}_g(X(u),u)=\frac3{16}\left(\frac{d^2 \mathcal{F}_{g-1}}{du^2}+\left(\frac{9w}{4\Delta}X+\frac1{12}\frac{d}{du}\ln\Delta\right)\frac{d\mathcal{F}_{g-1}}{du}+\sum_{g'=1}^{g-1}\frac{d\mathcal{F}_{g'}}{du} \frac{d\mathcal{F}_{g-g'}}{du}\right),
\end{equation}
where 
\begin{equation}
X=\frac{2E_2}{9\omega_1^2}, \qquad    w=2g_2\frac{d}{du}g_3-3g_3\frac{d}{du}g_2.
\end{equation}
We are looking for solutions of $\mathcal{F}_{g>1}$ which are polynomials in $X(u)$ with coefficients depending on $u$ and on the mass parameters as well.
Partial derivative $\partial_X$ is understood as applied on such polynomials. 

Then term $\mathcal{F}_2$ for given $g_2$ and $g_3$ reads as follows
\begin{multline}\label{F2}
\frac{2^{10}\Delta^2}9\mathcal{F}_2=45 
w^2 X^3+\big(12\Delta w'{+}(2Q^2{-}11)\Delta' w\big)X^2+\left(2g_2w^2{+}\frac{8(Q^2{-}1)}9\Delta''\Delta{+}\frac{(Q^2{-}1) (Q^2{-}23)}{27}\Delta'^2\right)X\\
+\frac4{45}\left((9{-}18Q^2{+}7Q^4)\Delta w' g_2-\frac{71{-}142Q^2{+}53Q^4}{12}\Delta' w g_2+\frac32(43{-}96Q^2{+}39Q^4)w^2 g_3\right).
\end{multline}
While the $X$-dependent terms, appearing in the first row, are fixed by the holomorphic anomaly equation \eqref{ha}, the $X$-independent terms, given by the second row, are not determined by it. This holomorphic ambiguity should be fixed by the so-called gap condition. This condition describes the behavior of $\mathcal{F}_g$ around the zeroes of modular discriminant $\Delta(u)$. For $g=2$ the gap condition around zero $u_*$ reads as follows
\begin{equation}\label{gap_cond}
\mathcal{F}_2(u\rightarrow u_*)=\frac{24-28Q^2+7Q^4}{5760}\tilde{a}_{\D}^{-2}+O(\tilde{a}_{\D}^0), \qquad \tilde{a}_{\D}=a_{\D}-a_{\D}(u_*).    
\end{equation}
It is claimed in \cite{HKK11} that the gap condition around all zeroes of $\Delta(u)$ fixes the $X$-independent part uniquely.
The general form we find for the $X$-independent part, given by the second row of \eqref{F2}, is conjectural. Indeed, we find that \eqref{F2}, upon suitable specification of polynomials $g_2$ and $g_3$ to the physical ones appearing in the given SW curve, satisfies gap condition \eqref{gap_cond} for all the ten cases of the expansions that we consider below. However, we have no proof of this general formula.

\paragraph{General comments.}
In the following subsections, using the holomorphic anomaly approach, we obtain $\epsilon_1,\epsilon_2$-expansions of  the topological strings partition functions in the magnetic frame of the SUSY gauge theories corresponding to the Painlev\'e equations (Table \ref{tab:theories}). We compare them 
with partition functions obtained in Sec.~\ref{sec:quant_exp} as the Zak transform input of the quantum Painlev\'e tau functions. To perform the comparison for each corresponding SW curve, we follow these steps:
\begin{enumerate} 
    \item We find the zeroes of $\Delta(u)$.
    \item In the vicinity of each zero we find the relation between $u$ and $\tilde{a}_{\D}$.
    \item We calculate $\frac{d^2\mathcal{F}_0}{da_{\D}^2}$, $\mathcal{F}_1$, $\mathcal{F}_2$.
    \item We check gap condition \eqref{gap_cond} for $\mathcal{F}_2$.
    \item We compare $\mathcal{F}_{0,1,2}$ with the $\epsilon_{1,2}$-expanded partition functions of Sec.~\ref{sec:quant_exp}, choosing the consistent sign of $a_{\D}$, the shift $\tilde{a}_{\D}-a_{\D}$ and also fixing the mass parameter freedom in the quantum Painlev\'e solution. Note that shift $\tilde{a}_{\D}-a_{\D}\neq0$ is possible only when allowed by the dimension counting.
\end{enumerate}
All the calculations we are going to perform are in terms of strong coupling asymptotic series for $t\rightarrow\infty$. 
In the following for brevity we sometimes report only the first four terms of these asymptotic expansions, being the ones necessary to reproduce $\mathcal{F}_{0,1,2}$ up to $t^{-4}$, denoting with $\ldots$ the omitted ones.

For the explicit expressions of the required SW curves we take \cite{BLMST16} as a source, but use Painlev\'e notations of the previous section from the very beginning. In the case of the quartic curve we bring it to the Weierstrass form \eqref{Welc} by using standard formulas and choosing  convenient normalizations for $g_2$ and $g_3$ (see e.g. \cite[(3.11-3.13)]{HKK11})
\begin{equation}\label{quart_to_W}
\!y^2=x^4{+}c_1 x^3{+}c_2 x^2{+}c_3 x{+}c_4\,\Rightarrow \quad R^2g_2=c_4{-}\frac{c_1c_3}4{+}\frac{c_2^2}{12},\quad R^3g_3=\frac{c_2c_4}6{+}\frac{c_1c_2c_3}{48}{-}\frac{c_3^2}{16}{-}\frac{c_1^2c_4}{16}{-}\frac{c_2^3}{216}.   
\end{equation}

In parallel to this comparison we also follow the renormalization group flow at the level of SW curves that corresponds to the Painlev\'e coalescence of Secs.~\ref{ssec:D-coal}, \ref{ssec:E-coal} (presented in Fig.~\ref{fig:gauge_coalescence}).
We also support this correspondence explicitly presenting the known  relation between the SW curves and the Painlev\'e Hamiltonians of \cite{KMNOY04}, further discussed in \cite[Sec. 2.3]{BLMST16}. As a starting point for the renormalization group flow we start from presenting the SW curve for $N_f=4$ and its connection with Painlev\'e VI Hamiltonian formalism (Sec.~\ref{ssec:PVI}).

\paragraph{SW curve for $N_f=4$.}
SW curve \cite[(17.58)]{SW94} can be easily converted into the Weierstrass form with
\begin{equation}\label{g2_4}
\frac{g_2}{16}=\frac43u^2(1{-}t{+}t^2)-\frac23uw_2t^2(2{-}t)+\left(\frac{w_2^2}{12}t^3+w_4t(1{-}t)-2e_4(1{-}t)(2{-}t)\right)t,
\end{equation}
\vspace{-0.7cm}
\begin{multline}\label{g3_4}
\mspace{70mu}-\frac{g_3}{64}=-\frac4{27}u^3(1{+}t)(2{-}t)(1{-}2t)-\frac{u^2}9w_2t^2(5{-}5t{+}2t^2)\\+\frac{u}3\left(\frac{w_2^2}6t^3(2{-}t)+w_4t(1{-}t)(2{-}t)+2e_4(1{-}t)(2{-}2t{-}t^2)\right)t\\-\left(\frac{w_2^3}{216}t^4+\frac{w_2w_4}{12}t^2(1{-}t)-\frac{w_2e_4}6t(1{-}t)(2{-}t)+w_6(1{-}t)^2\right)t^2,
\end{multline}
where, instead of SW notations \cite{SW94}, we introduced
\begin{equation}
t=1-\frac{\beta^{\mathrm{SW}}}{\alpha^{\mathrm{SW}}}, \qquad u=-\frac{u^{\mathrm{SW}}}{4\alpha^{\mathrm{SW}}}.    
\end{equation}
This dictionary is determined by a match with the instanton expansion \eqref{Zinst} via the $\epsilon_{1,2}$-expansion. Following \cite{KMNOY04} we see that this SW curve is equivalent to the curve defined by PVI zeta function \eqref{zeta_VI}
\begin{equation}\label{zeta_SW_VI}
\zeta_{\VI}(p,q|t)+\mathrm{u}+\frac{\mathrm{w}_2}4t=0.   
\end{equation}
To see this, in the latter we introduce variable $\hat{p}$ instead of momentum $p$  
\begin{equation}
p=\frac{\theta_{\infty}\hat{p}}{q(q{-}t)(q{-}1)}+\frac{\theta_0}{q}+\frac{\theta_t}{q{-}t}+\frac{\theta_1}{q{-}1},    
\end{equation}
obtaining a quartic curve $\hat{p}^2=q^4+\ldots$, which in the Weierstrass form \eqref{Welc} is just the above SW curve with \eqref{g2_4}, \eqref{g3_4} after restoring the dimensions by $u=(2\epsilon_2)^2\mathrm{u}$ and $m_f=2\epsilon_2\sm_f|_{f=1,2,3,4}$. 

\subsection{\texorpdfstring{$N_f=3$}{N\_f=3}}
\label{ssec:holom_anom_3}
\paragraph{SW curve.} We bring quartic curve \cite[(B.31)]{BLMST16} with $\Lambda=4t$ in Weierstrass form \eqref{Welc} using \eqref{quart_to_W} with $R=-\frac14$
\begin{align}\label{g2_3}
\frac{g_2}{16}&=\frac43u(u-t^2)+\left(\frac{t^3}{12}+w_2t-4e_3\right)t,\\ \label{g3_3}
-\frac{g_3}{64}&=-\frac8{27}u^3-\frac59u^2t^2+\frac13u\left(\frac{t^3}3+2w_2 t+4e_3\right)t-\left(\frac{t^4}{216}+\frac{w_2t^2}{12}-\frac{e_3t}3+w_4\right)t^2.
\end{align}
This is just limit $m_4\rightarrow\infty$ of \eqref{g2_4}, \eqref{g3_4} after standard rescaling $t^{[4]}=t^{[3]}/m_4$ (Sec.~\ref{ssec:D-coal_q}). Analogously to $N_f=4$ / PVI case \eqref{zeta_SW_VI} (and by \cite[App. A]{KMNOY04}, \cite[App. A.4]{BLMST16}), this $N_f=3$ SW curve is equivalent to level curve $\zeta_{\V}(p,q|t)=\mathrm{u}$ of PV zeta function \eqref{zeta_V} with dimension rescalings $u=(2\epsilon_2)^2\mathrm{u}$, $t=2\epsilon_2\st$, $m_f=2\epsilon_2\sm_f|_{f=1,2,3}$.

Modular discriminant $\Delta(u)$ has power $5$. It has zero
\begin{equation}
u_s=\frac{t^2}{16}+w_2-\frac{8e_3}t+\frac{16w_4}{t^2}-\frac{128w_2e_3}{t^3}+\frac{256(w_2w_4+5e_3^2)}{t^4}+O\left(\frac1{t^5}\right)    
\end{equation}
that, as we will see, corresponds to the square exponent singularity expansion in the Painlev\'e setting. The remaining $4$ zeroes correspond instead to the linear exponent singularity expansion
\begin{multline}
u_l^{\vec\lambda}=-\frac{e_1^{\vec\lambda}}2t-\frac{(e_1^{\vec\lambda})^2{-}w_2}2+\frac{(e_1^{\vec\lambda})^3{-}w_2e_1^{\vec\lambda}{-}2e_3}{2t}\left(1-\frac{2e_1^{\vec\lambda}}t+\frac{11(e_1^{\vec\lambda})^2{-}w_2}{2t^2}-2\frac{9(e_1^{\vec\lambda})^3{-}2w_2e_1^{\vec\lambda}{-}e_3}{t^3}\right.\\ \left.+\frac{107w_2(e_1^{\vec\lambda})^2{+}506e_3e_1^{\vec\lambda}{+}(262w_4{-}65w_2^2)}{t^4}\right.\\ \left.-2\frac{190w_2(e_1^{\vec\lambda})^3{+}965e_3(e_1^{\vec\lambda})^2{+}4(128w_4{-}31w_2^2)e_1^{\vec\lambda}{+}3e_3w_2}{t^5}+\ldots+O\left(\frac1{t^{10}}\right)\right).    
\end{multline}
Here the zeroes are enumerated by elements $\vec\lambda$ of subgroup $(\mathbb{Z}_2)^2 \subset W(A_3)$, which acts by changing signs of an even number of the masses. We denote by $e_1^{\vec\lambda}$ the action of $\vec\lambda$ on $e_1$. Other mass parameters $w_2,e_3,w_4$ are, of course, invariant under this subgroup.

\paragraph{Heavy hypers expansion.}
Now in the vicinity of $u_s$ we pass to the flat coordinate (with $\alpha_{\D}=\ri a_{\D}$)
\begin{multline}
\frac{2 (u{-}u_s)}{\alpha_{\D} t}=1+\frac{\alpha_{\D}}{t}-\frac{\alpha_{\D}^2+8w_2}{t^2}+\left(\frac52\alpha_{\D}^3+48w_2 \alpha_{\D}+256 e_3\right)\cdot\frac1{t^3}\\-\left(\frac{33}4\alpha_{\D}^4+296 w_2\alpha_{\D}^2+2688e_3\alpha_{\D}+32(w_2^2{+}28w_4)\right)\cdot\frac1{t^4}\\+\left(\frac{63}2\alpha_{\D}^5+1840w_2\alpha_{\D}^3+24064e_3\alpha_{\D}^2+768(w_2^2{+}18w_4)\alpha_{\D}+12288e_3w_2\right)\cdot\frac1{t^5}+O\left(\frac1{t^6}\right),    
\end{multline}
where we have already chosen $(3g_2/4)^{-1/4}\sim -2\ri/t$ in \eqref{periods_hyph}, which is consistent with \eqref{omega_1_ell}.
By comparing the $t^{-1}$ expansion parts of obtained $\mathcal{F}_{0,1,2}$ with the corresponding $\epsilon_{1,2}$ terms of $\mathcal{Z}_{inst}^{[\mathbf{3_S}]}$ (of Sec. \ref{ssec:exp_3s}) up to $t^{-5}$ we obtain that they coincide iff
\begin{equation}
\check{w}_4=w_4-\frac38w_2\epsilon^2+\frac{105}{1024}\epsilon^4.
\end{equation}
In the assumption that the instanton expansion of the tau function has polynomial coefficients, this is the exact relation and it is not necessary to compute $\mathcal{F}_{g>2}$. Besides the instanton part we have also perturbative parts of $\mathcal{F}_g$
\begin{equation}\label{F012_pert_V_1}
\frac{d^2\mathcal{F}_{0,pert}}{da_{\D}^2}=\ln\left(\frac{2\ri t}{a_{\D}}\right), \quad \mathcal{F}_{1,pert}=\frac{Q^2}{24}\ln \left(\frac{a_{\D} t^{11}}{2\ri}\right)-\frac1{12}\ln((2\ri t)^5a_{\D}), \quad  \mathcal{F}_{2,pert}=\frac{24{-}28Q^2{+}7Q^4}{5760a_{\D}^2},
\end{equation}
so gap condition \eqref{gap_cond} is satisfied without any higher orders $O(a_{\D}^0)$. From \eqref{Zcl_V_1} and \eqref{Z1loop_V_1} using \eqref{gamma_as} we get
\begin{equation}
\epsilon_1
\epsilon_2\ln\mathcal{Z}_{cl-a}^{[\mathbf{3_S}]}\mathcal{Z}_{1-loop}^{[\mathbf{3_S}]}=\frac{a_{\D}^2}2\ln\frac{2\ri t}{a_{\D}}{+}\frac{3a_{\D}^2}4{-}\frac{\ri a_{\D} t}2+\frac{Q^2{-}2}{24}\ln a_{\D} \cdot (\epsilon_1\epsilon_2)+\frac{24{-}28Q^2{+}7Q^4}{5760a_{\D}^2}(\epsilon_1\epsilon_2)^2+O\big((\epsilon_1\epsilon_2)^3\big),
\end{equation}
which gives \eqref{F012_pert_V_1} up to $a_{\D}$-independent terms in $\mathcal{F}_1$.

\paragraph{Light hypers expansion.} Now in the vicinity of $u_l^{\vec\lambda}$ we pass to the flat coordinate
{\allowdisplaybreaks
\begin{multline}
\frac{u{-}u_l^{\vec\lambda}}{\tilde{a}_{\D}t}=1-2\frac{\tilde{a}_{\D}{-}e_1^{\vec\lambda}}t-\frac{4\tilde{a}_{\D}^2{-}6e_1^{\vec\lambda} \tilde{a}_{\D}{+}(3(e_1^{\vec\lambda})^2{-}w_2)}{t^2}-2\frac{10\tilde{a}_{\D}^3{-}20e_1^{\vec\lambda}\tilde{a}_{\D}^2{+}3(5(e_1^{\vec\lambda})^2{-}w_2)\tilde{a}_{\D}{-}(5(e_1^{\vec\lambda})^3{-}3w_2e_1^{\vec\lambda}{-}4e_3)}{t^3}\\
-\frac{132\tilde{a}_{\D}^4{-}330e_1^{\vec\lambda}\tilde{a}_{\D}^3{+}2(165(e_1^{\vec\lambda})^2{-}23w_2)\tilde{a}_{\D}^2{-}3(55(e_1^{\vec\lambda})^3{-}23w_2e_1^{\vec\lambda}{-}22e_3)\tilde{a}_{\D}{+}2(24w_2(e_1^{\vec\lambda})^2{+}132e_3e_1^{\vec\lambda}{+}77w_4{-}19w_2^2)}{t^4}\\
-16\left(63\tilde{a}_{\D}^5-189e_1^{\vec\lambda}\tilde{a}_{\D}^4+\frac54(189(e_1^{\vec\lambda})^2{-}20w_2)\tilde{a}_{\D}^3-\frac{315(e_1^{\vec\lambda})^3{-}100w_2e_1^{\vec\lambda}{-}73e_3}2\tilde{a}_{\D}^2\right.\\ \left.+\frac38(215w_2(e_1^{\vec\lambda})^2{+}1114e_3e_1^{\vec\lambda}{+}612w_4{-}151w_2^2)\tilde{a}_{\D}-\frac{89w_2(e_1^{\vec\lambda})^3{+}537e_3(e_1^{\vec\lambda})^2{+}3(108w_4{-}25w_2^2)e_1^{\vec\lambda}{+}9e_3w_2}8\right)\cdot\frac1{t^5}\\-16\Bigg(527\tilde{a}_{\D}^6-\frac{3689}2e_1^{\vec\lambda}\tilde{a}_{\D}^5+\frac34\left(3689(e_1^{\vec\lambda})^2{-}313w_2\right)\tilde{a}_{\D}^4-\frac52\left(\frac{3689}4(e_1^{\vec\lambda})^3{-}\frac{939}4w_2e_1^{\vec\lambda}{-}137e_3\right)\tilde{a}_{\D}^3\\ +\frac52\left(\frac{1375}2w_2(e_1^{\vec\lambda})^2{+}3415e_3e_1^{\vec\lambda}{+}1818w_4{-}\frac{899}2w_2^2\right)\tilde{a}_{\D}^2-\frac32c_1\tilde{a}_{\D}+\frac{c_2}8\Bigg)\cdot\frac1{t^6}\\ -24\Bigg(3129\tilde{a}_{\D}^7-12516e_1^{\vec\lambda}\tilde{a}_{\D}^6+7(3129(e_1^{\vec\lambda})^2{-}221w_2)\tilde{a}_{\D}^5-(21903(e_1^{\vec\lambda})^3{-}4641w_2e_1^{\vec\lambda}{-}2242e_3)\tilde{a}_{\D}^4\\ +5\left(\frac{8631}2w_2(e_1^{\vec\lambda})^2{+}20782e_3e_1^{\vec\lambda}{+}10863w_4{-}\frac{5383}2w_2^2\right)\tilde{a}_{\D}^3-c_3\tilde{a}_{\D}^2+c_4\tilde{a}_{\D}-c_5\Bigg)\cdot\frac1{t^7}+\ldots+O\left(\frac1{t^{12}}\right),
\end{multline}}
where
{\allowdisplaybreaks\begin{align}
&c_1=\frac{531}2w_2(e_1^{\vec\lambda})^3+1502e_3(e_1^{\vec\lambda})^2+\frac{1712w_4{-}403w_2^2}2e_1^{\vec\lambda}+15e_3w_2,\\   
&c_2=2319e_3(e_1^{\vec\lambda})^3+(1447w_4{+}384w_2^2)(e_1^{\vec\lambda})^2+2863e_3w_2e_1^{\vec\lambda}+1367w_4w_2{+}42e_3^2{-}342w_2^3,\\
&c_3=7084w_2(e_1^{\vec\lambda})^3+38201e_3(e_1^{\vec\lambda})^2+2(10509w_4{-}2506w_2^2)e_1^{\vec\lambda}+\frac{725}3e_3w_2,\\
&c_4=8149e_3(e_1^{\vec\lambda})^3+6(802w_4{+}259w_2^2)(e_1^{\vec\lambda})^2+10663e_3w_2e_1^{\vec\lambda}+2(2594w_4w_2{+}39e_3^2{-}649w_2^3),\\
&c_5=\left(561w_4{+}\frac{245}2w_2^2\right) (e_1^{\vec\lambda})^3+
\frac{5435}2e_3w_2(e_1^{\vec\lambda})^2+\left(442 w_4w_2{+}6989e_3^2{-}\frac{223}2w_2^3\right)e_1^{\vec\lambda}\nonumber\\ &+
\left(\frac{10415}3w_4{-}\frac{1739}2 w_2^2\right)e_3.
\end{align}}
Here we have already chosen $(3g_2/4)^{-1/4}\sim 1/t$ in \eqref{periods_hyph}, which is consistent with \eqref{omega_1_ell}.
By comparing the $t^{-1}$ expansion parts of obtained $\mathcal{F}_{0,1,2}$ with the corresponding $\epsilon_{1,2}$ terms of $\mathcal{Z}_{inst}^{[\mathbf{3_L}]}$ (of Sec. \ref{ssec:exp_3l}) up to $t^{-8}$ we obtain that they coincide iff $\tilde{a}_{\D}=a_{\D}+e_1^{\vec\lambda}/2$ and $\tilde{w}_4=w_4$.
In the assumption of polynomial coefficients for the instanton expansion, the latter is the exact relation and it is not necessary to compute $\mathcal{F}_{g>2}$. Besides the instanton part, we have also perturbative parts of $\mathcal{F}_g$
\begin{multline}\label{F012_pert_V_2} 
\frac{d^2\mathcal{F}_{0,pert}}{da^2}=4\ln t-\ln \prod_{\lambda,\lambda'=\pm1}a_{\lambda,\lambda'}, \quad \mathcal{F}_{1,pert}=\frac{2Q^2{-}1}6 \ln t+\frac{Q^2{-}2}{24}\ln \left( 2^{12}\prod_{\lambda,\lambda'=\pm1}a_{\lambda,\lambda'} \right),\\
\mathcal{F}_{2,pert}=-\frac{24{-}28Q^2{+}7Q^4}{5760}\frac{d^2}{da_{\D}^2}\left(\ln\prod_{\lambda,\lambda'=\pm1}a_{\lambda,\lambda'}\right)=\frac{24{-}28Q^2{+}7Q^4}{5760\tilde{a}_{\D}^2}+O(\tilde{a}_{\D}^0),
\end{multline}
where
\begin{equation}
a_{\lambda,\lambda'}=a_{\D}+\frac12(\lambda m_1{+}\lambda' m_2{+}\lambda\lambda'm_3).    
\end{equation}
So we see that gap condition \eqref{gap_cond} is satisfied. From \eqref{Zcl_V_2} and \eqref{Z1loop_V_2} using \eqref{gamma_as} we get 
\begin{multline}
\epsilon_1
\epsilon_2\ln\mathcal{Z}_{cl-a}^{[\mathbf{3_L}]}\mathcal{Z}_{1-loop}^{[\mathbf{3_L}]}=2a_{\D}^2\ln t{-}a_{\D}t+\sum_{\lambda,\lambda'=\pm1}\frac{3{-}2\ln a_{\lambda,\lambda'}}4 a_{\lambda,\lambda'}^2+\frac{Q^2{-}2}{24}\ln \left(\prod_{\lambda,\lambda'=\pm1}a_{\lambda,\lambda'}\right) \cdot (\epsilon_1\epsilon_2)\\-\frac{24{-}28Q^2{+}7Q^4}{5760}\frac{d^2}{da_{\D}^2}\left(\ln\prod_{\lambda,\lambda'=\pm1}a_{\lambda,\lambda'}\right)(\epsilon_1\epsilon_2)^2+O\big((\epsilon_1\epsilon_2)^3\big),
\end{multline}
which gives \eqref{F012_pert_V_2} up to $a_{\D}$-independent terms in $\mathcal{F}_1$.

\subsection{\texorpdfstring{$N_f=2$}{N\_f=2}}
\label{ssec:holom_anom_2}
\paragraph{SW curve.} We bring quartic curve \cite[(B.20)]{BLMST16} with $\Lambda^2=4t$ in Weierstrass form \eqref{Welc} using \eqref{quart_to_W} with $R=-\frac14$
\begin{align}\label{g2_2}
\frac{g_2}{16}&=\frac43u^2-4e_2t+t^2,\\ \label{g3_2}
-\frac{g_3}{64}&=-\frac8{27}u^3+\frac23u(2e_2{+}t)t-w_2t^2.
\end{align}
This is just limit $m_3\rightarrow\infty$ of \eqref{g2_3}, \eqref{g3_3} after standard rescaling $t^{[3]}=t^{[2]}/m_3$ (Sec.~\ref{ssec:D-coal_q}). Analogously to $N_f=4$ / PVI case \eqref{zeta_SW_VI} (and by \cite[App. A]{KMNOY04}, \cite[App. A.3]{BLMST16}), this $N_f=2$ SW curve is equivalent to level curve $\zeta_{\III_1}(p,q|t)=\mathrm{u}$ of PIII$_1$ zeta function \eqref{zeta_III1} with dimension rescalings $u=(2\epsilon_2)^2\mathrm{u}$, $t=(2\epsilon_2)^2\st$, $m_f=2\epsilon_2\sm_f|_{f=1,2}$.

Modular discriminant $\Delta(u)$ has power $4$. Its zeroes can be enumerated by roots of unity $\iota$, $\iota^4=1$
\begin{multline}
u_*^\iota=-\frac{s_{\iota}^2}{128}+\frac{m_1{-}\iota^2m_2}8s_{\iota}+\frac{w_2{+}2\iota^2e_2}4\left(1+4\frac{m_1{-}\iota^2m_2}{s_{\iota}}+16\frac{w_2{-}2\iota^2e_2}{s_{\iota}^2}+16\frac{(m_1{-}\iota^2m_2)(3w_2{-}10\iota^2e_2)}{s_{\iota}^3}\right.\\ \left.-2^{10}\frac{\iota^2e_2(w_2{-}2\iota^2e_2)}{s_{\iota}^4}-2^7\frac{(m_1{-}\iota^2m_2)(11w_2^2{+}28w_2(\iota^2e_2){-}116(\iota^2e_2)^2)}{s_{\iota}^5}-2^{12}\frac{(w_2{-}2\iota^2e_2)(3w_2^2{-}28(\iota^2e_2)^2)}{s_{\iota}^6}\right.\\ \left.-2^8 \frac{(m_1{-}\iota^2m_2)(221w_2^3{-}1266w_2^2(\iota^2e_2){-}1828w_2(\iota^2e_2)^2{+}7272(\iota^2e_2)^3)}{s_{\iota}^7}+O\left(\frac1{s_{\iota}^8}\right)\right),
\end{multline}
where $s_{\iota}=\iota s$. These $4$ solutions correspond to the root branches of
$s=8\ri t^{1/2}$ and to the $t\mapsto -t, e_2\mapsto -e_2$ symmetry of the SW curve already appeared in Sec. \ref{ssec:exp_2} as a symmetry of QPIII$_1$.
We discuss the sense of these solutions below in more details.

\paragraph{Expansion.} Now in the vicinity of $u_*^{\iota}$ we pass to the flat coordinate
{\allowdisplaybreaks\begin{multline}
\frac{4(u{-}u_*^{\iota})}{\tilde{a}_{\D}s_{\iota}}=1-2\frac{\tilde{a}_{\D}+(m_1{-}\iota^2m_2)}{s_{\iota}}-2\frac{2\tilde{a}_{\D}^2+3(m_1{-}\iota^2m_2)\tilde{a}_{\D}-(10\iota^2e_2{+}3w_2)}{s_{\iota}^2}\\-4\frac{5\tilde{a}_{\D}^3+10(m_1{-}\iota^2m_2)\tilde{a}_{\D}^2-6(10\iota^2e_2{+}3w_2)\tilde{a}_{\D}-(m_1{-}\iota^2m_2)(50\iota^2e_2{+}23w_2)}{s_{\iota}^3}\\
-2\Big(66\tilde{a}_{\D}^4+165(m_1{-}\iota^2m_2)\tilde{a}_{\D}^3-2(734\iota^2e_2{+}225w_2) \tilde{a}_{\D}^2-24 (m_1{-}\iota^2m_2)(78\iota^2e_2{+}35w_2)\tilde{a}_{\D} \\ +(1524(\iota^2e_2)^2){+}44w_2(\iota^2e_2){-}339w_2^2\Big)\cdot\frac1{s_{\iota}^4}\\-4\Big(252\tilde{a}_{\D}^5+756(m_1{-}\iota^2m_2)\tilde{a}_{\D}^4-5(1810\iota^2e_2{+}567w_2)\tilde{a}_{\D}^3-10(m_1{-}\iota^2m_2)(1558\iota^2e_2{+}693w_2)\tilde{a}_{\D}^2 \\ +96(220e_2^2{+}12w_2(\iota^2e_2){-}45w_2^2)\tilde{a}_{\D}+(m_1{-}\iota^2m_2)(6044e_2^2{+}1508w_2(\iota^2e_2){-}729w_2^2)\Big)\cdot\frac1{s_{\iota}^5}
\\-4\Big(2108\tilde{a}_{\D}^6+7378(m_1{-}\iota^2m_2)\tilde{a}_{\D}^5-24(4670\iota^2e_2{+}1491w_2)\tilde{a}_{\D}^4-5(m_1{-}\iota^2m_2)(48662\iota^2e_2{+}21581w_2)\tilde{a}_{\D}^3\\ +20(23428e_2^2{+}1796w_2(\iota^2e_2){-}4483w_2^2)\tilde{a}_{\D}^2+48(m_1{-}\iota^2m_2)(4812e_2^2{+}1388w_2(\iota^2e_2){-}477w_2^2)\tilde{a}_{\D} \\ -(106760(\iota^2e_2)^3{+}3156w_2(\iota^2e_2)^2{-}26010w_2^2(\iota^2e_2){-}617w_2^2)\Big)\cdot\frac1{s_{\iota}^6}+\ldots+O\left(\frac1{s_{\iota}^{10}}\right),
\end{multline}}
where we have already chosen $(3g_2/4)^{-1/4}\sim 4/s_{\iota}$ in \eqref{periods_hyph}, which is consistent with \eqref{omega_1_ell}.
By comparing the $s^{-1}$ expansion parts of obtained $\mathcal{F}_{0,1,2}$ with the corresponding $\epsilon_{1,2}$ terms of $\mathcal{Z}_{inst}^{[\mathbf{2}]}$ (of Sec. \ref{ssec:exp_2}) up to $s^{-7}$ we obtain that they coincide iff $\tilde{a}_{\D}=a_{\D}-\frac12(m_1{-}\iota^2m_2)$ and $\tilde{w}_2=w_2$, under the additional substitutions $m_2\mapsto \iota^2m_2$ and $s\mapsto s_{\iota}$.
So the $\iota=1$ answer literally coincides with \eqref{Zinst_2} and the $\iota=\ri$ one give us the instanton series after the above mentioned symmetry transformation. We observe that changing of the sign $\iota\mapsto-\iota$ corresponds just to the changing of the root branch of $s=8\ri t^{1/2}$.
In the assumption of polynomial coefficients for the instanton expansion, $\tilde{w}_2=w_2$ is the exact relation and it is not necessary to compute $\mathcal{F}_{g>2}$.
Besides the instanton part we have also perturbative parts of $\mathcal{F}_g$
\begin{multline}\label{F012_pert_III1} 
\frac{d^2\mathcal{F}_{0,pert}}{da^2}=\ln \frac{4s_{\iota}^2}{4a^2{-}w_2{+}2\iota^2e_2}, \quad \mathcal{F}_{1,pert}=\frac{Q^2{-}2}{24}\ln \frac{(4a^2{-}w_2{+}2\iota^2e_2)s_{\iota}^{10}}{2^{14}}+\frac14\ln\frac{s_{\iota}^2}{16},\\
\mathcal{F}_{2,pert}=-\frac{24{-}28Q^2{+}7Q^4}{5760}\frac{d^2}{da_{\D}^2}\ln(4a^2{-}w_2{+}2\iota^2e_2)=\frac{24{-}28Q^2{+}7Q^4}{5760\tilde{a}_{\D}^2}+O(\tilde{a}_{\D}^0),    
\end{multline}
so gap condition \eqref{gap_cond} is satisfied. From \eqref{Zcl_III1} and \eqref{Z1loop_III1} using \eqref{gamma_as} we get
\begin{multline}
\epsilon_1
\epsilon_2\ln\mathcal{Z}_{cl-a}^{[\mathbf{2}]}\mathcal{Z}_{1-loop}^{[\mathbf{2}]}=a_{\D}^2\ln s{-}\frac{a_{\D}s}2+\sum_{\lambda=\pm1}\frac{3{-}2\ln \left(a_{\D}{+}\lambda\frac{m_1{-}m_2}2\right)}4 \left(a_{\D}{+}\lambda\frac{m_1{-}m_2}2\right)^2\\+\frac{Q^2{-}2}{24}\ln\left(a^2{-}\frac{w_2}4{+}\frac{\iota^2e_2}2\right) \cdot (\epsilon_1\epsilon_2)-\frac{24{-}28Q^2{+}7Q^4}{5760}\frac{d^2}{da_{\D}^2}\ln \left(a^2{-}\frac{w_2}4{+}\frac{\iota^2e_2}2\right) \cdot(\epsilon_1\epsilon_2)^2+O\big((\epsilon_1\epsilon_2)^3\big),
\end{multline}
which gives \eqref{F012_pert_III1} up to $a_{\D}$-independent terms in $\mathcal{F}_1$ and the substitutions related to the different value of $\iota$ discussed above.

\subsection{\texorpdfstring{$N_f=1$}{N\_f=1}}
\label{ssec:holom_anom_1}
\paragraph{SW curve.} We bring quartic curve \cite[(B.10)]{BLMST16} with $\Lambda^3=4t$ and $m=m_1$ in Weierstrass form \eqref{Welc} using \eqref{quart_to_W} with $R=-\frac14$
\begin{align}\label{g2_1}
\frac{g_2}{16}&=\frac43u^2-4m_1t, \\ \label{g3_1}
-\frac{g_3}{64}&=-\frac8{27}u^3+\frac43m_1ut-t^2.
\end{align} 
This is just limit $m_2\rightarrow\infty$ of \eqref{g2_2}, \eqref{g3_2} after standard rescaling $t^{[2]}=t^{[1]}/m_2$ (Sec.~\ref{ssec:D-coal_q}). Analogously to $N_f=4$ / PVI case \eqref{zeta_SW_VI} (and by \cite[App. A]{KMNOY04}, \cite[App. A.2]{BLMST16}), this $N_f=1$ SW curve is equivalent to level curve $\zeta_{\III_2}(p,q|t)=\mathrm{u}$ of PIII$_2$ zeta function \eqref{zeta_III2} with dimension rescalings $u=(2\epsilon_2)^2\mathrm{u}$, $t=(2\epsilon_2)^3\st$, $m_1=2\epsilon_2\sm_1$.

Modular discriminant $\Delta(u)$ has power $3$. Its $3$ zeroes
correspond to a choice of the root branch of $s=(54t)^{1/3}$ and can be enumerated by roots of unity $\varsigma$, $\varsigma^3=1$
\begin{equation}
3u^{\varsigma}_*=-\frac{s_{\varsigma}^2}4-m_1s_{\varsigma}+m_1^2-\frac{4m_1^3}{3s_{\varsigma}}+\frac{4m_1^4}{3s_{\varsigma}^2}+O\left(\frac1{s_{\varsigma}^4}\right), \qquad s_{\varsigma}=\varsigma s.
\end{equation}

\paragraph{Expansion.} Now in the vicinity of $u^{\varsigma}_*$ we pass to the flat coordinate (with $\tilde{\alpha}_{\D}=\tilde{a}_{\D}/\sqrt3$)
\begin{multline}
\frac{u{-}u^{\varsigma}_*}{\tilde\alpha_{\D} s_{\varsigma}}=1-\frac{\tilde\alpha_{\D}}{2s_{\varsigma}}+\left(\frac5{12}\tilde\alpha_{\D}^2{-}2m_1\tilde\alpha_{\D}{+}2m_1^2\right)\cdot\frac1{s_{\varsigma}^2}+\left(\frac{515}{144}\tilde\alpha_{\D}^3{-}\frac{38}3m_1\tilde\alpha_{\D}^2{+}14m_1^2\tilde\alpha_{\D}{-}\frac{16}3m_1^3\right)\cdot\frac1{s_{\varsigma}^3}\\
+\left(\frac{10759}{576}\tilde\alpha_{\D}^4{-}\frac{805}{12}m_1\tilde\alpha_{\D}^3{+}\frac{165}2m_1^2\tilde\alpha_{\D}^2{-}40m_1^3\tilde\alpha_{\D}{+}6m_1^4\right)\cdot\frac1{s_{\varsigma}^4}+O\left(\frac1{s_{\varsigma}^5}\right),
\end{multline}
where we have already chosen $(3g_2/4)^{-1/4}\sim \sqrt3/s_{\varsigma}$ in \eqref{periods_hyph}, which is consistent with \eqref{omega_1_ell}. 
By comparing the $s^{-1}$ expansion parts of obtained $\mathcal{F}_{0,1,2}$ with the corresponding $\epsilon_{1,2}$ terms of $\mathcal{Z}_{inst}^{[\mathbf{1}]}$ (of Sec. \ref{ssec:exp_1}) up to $s^{-4}$ we obtain that they coincide iff $\tilde{a}_{\D}=a_{\D}$ and $\psi=1$. More precisely, for $s_1$ we obtain literally \eqref{Zinst_1} and for $s_{e^{\pm2\pi\ri/3}}$ we obtain other branches of the expansion. In the assumption of polynomial coefficients for the instanton expansion, $\psi=1$ is the exact relation and it is not necessary to compute $\mathcal{F}_{g>2}$.  Besides the instanton part we have also perturbative parts of $\mathcal{F}_g$
\begin{equation}\label{F012_pert_1}
\frac{d^2\mathcal{F}_{0,pert}}{da_{\D}^2}=\ln\left({-}\frac{12s_{\varsigma}}{\alpha_{\D}}\right), \quad \mathcal{F}_{1,pert}=\frac{Q^2{-}2}{24}\ln \left({-}\frac{2^{10}\alpha_{\D} s_{\varsigma}^{11}}{3^7}\right)+\frac14\ln\left(\frac{s_{\varsigma}^2}3\right), \quad  \mathcal{F}_{2,pert}=\frac{24{-}28Q^2{+}7Q^4}{5760\cdot 3\alpha_{\D}^2}
\end{equation}
so gap condition \eqref{gap_cond} is satisfied without any higher orders $O(a_{\D}^0)$. From \eqref{Zcl_III2} and \eqref{Z1loop_III2} using \eqref{gamma_as} we get 
\begin{equation}
\epsilon_1
\epsilon_2\ln\mathcal{Z}_{cl-a}^{[\mathbf{1}]}\mathcal{Z}_{1-loop}^{[\mathbf{1}]}=\frac{a_{\D}^2}2\ln\frac{-12\sqrt3s}{a_{\D}}{+}\frac{3a_{\D}^2}4{-}\sqrt3a_{\D} s+\frac{Q^2{-}2}{24}\ln a_{\D} \cdot (\epsilon_1\epsilon_2)+\frac{24{-}28Q^2{+}7Q^4}{5760a_{\D}^2}(\epsilon_1\epsilon_2)^2+O\big((\epsilon_1\epsilon_2)^3\big),
\end{equation}
which for the suitable root branch of $s$ gives \eqref{F012_pert_1} up to $a_{\D}$-independent terms in $\mathcal{F}_1$.

\subsection{\texorpdfstring{$N_f=0$}{N\_f=0}}
\label{ssec:holom_anom_0}
\paragraph{SW curve.} We bring quartic curve \cite[(B.2)]{BLMST16} with $\Lambda^4=4t$ in Weierstrass form \eqref{Welc}  using \eqref{quart_to_W} with $R=-\frac14$
\begin{align}
\frac{g_2}{16}&=\frac43u^2-4t, \\
-\frac{g_3}{64}&=-\frac8{27}u^3+\frac43ut.
\end{align}
This is just limit $m_1\rightarrow\infty$ of \eqref{g2_1}, \eqref{g3_1} after standard rescaling $t^{[1]}=t^{[0]}/m_1$ (Sec.~\ref{ssec:D-coal_q}). Analogously to $N_f=4$ / PVI case \eqref{zeta_SW_VI} (and by \cite[App. A]{KMNOY04}, \cite[App. A.1]{BLMST16}), this $N_f=0$ SW curve is equivalent to level curve $\zeta_{\III_3}(p,q|t)=\mathrm{u}$ of PIII$_3$ zeta function \eqref{zeta_III3} with dimension rescalings $u=(2\epsilon_2)^2\mathrm{u}$, $t=(2\epsilon_2)^4\st$.

Modular discriminant $\Delta(u)$ has power $2$. It has zeroes
\begin{equation}
u^{\pm}_*=\mp s^2/2^9\equiv-s^2_{\pm}/2^9,
\end{equation}
where the sign actually corresponds to a choice of the root branch of $s=-32\ri t^{1/4}$ (i.e. $s_+=s, s_-=\ri s$).

\paragraph{Expansion.} Now in the vicinity of $u^{\pm}_*$ we pass to the flat coordinate 
\begin{multline}
-\frac{16(u{-}u^{\pm}_*)}{a_{\D}s_{\pm}}=1+2(a_{\D}/s_{\pm})-4(a_{\D}/s_{\pm})^2+20(a_{\D}/s_{\pm})^3-132(a_{\D}/s_{\pm})^4+1008(a_{\D}/s_{\pm})^5\\-8432(a_{\D}/s_{\pm})^6+75096(a_{\D}/s_{\pm})^7-700180(a_{\D}/s_{\pm})^8+O\left(\frac1{s_{\pm}^9}\right),
\end{multline}
where we have already chosen $(3g_2/4)^{-1/4}\sim -16/s_{\pm}$ in \eqref{periods_hyph}, which is consistent with \eqref{omega_1_ell}. 
Then the $s^{-1}$ expansion parts of obtained $\mathcal{F}_{0,1,2}$ coincide with the corresponding $\epsilon_{1,2}$ terms of $\mathcal{Z}_{inst}^{[\mathbf{0}]}$ (of Sec. \ref{ssec:exp_0}) up to $s^{-8}$. More precisely, for $s_+$ we obtain literally \eqref{Zinst_0} and for $s_-$ we obtain the other root branch of the expansion.  Besides the instanton part we have also perturbative parts of $\mathcal{F}_g$
\begin{equation}\label{F012_pert_0}
\frac{d^2\mathcal{F}_{0,pert}}{da_{\D}^2}=\ln\left(\frac{s_{\pm}}{a_{\D}}\right), \quad \mathcal{F}_{1,pert}=\frac{Q^2{-}2}{24}\ln \left(\frac{a_{\D} s_{\pm}^{11}}{2^{36}}\right)+\frac14\ln\left(\frac{s_{\pm}^2}{2^8}\right), \quad  \mathcal{F}_{2,pert}=\frac{24{-}28Q^2{+}7Q^4}{5760 a_{\D}^2}
\end{equation}
so gap condition \eqref{gap_cond} is satisfied without any higher orders $O(a_{\D}^0)$. From \eqref{Zcl_III3} and \eqref{Z1loop_III3} using \eqref{gamma_as} we get
\begin{equation}
\epsilon_1
\epsilon_2\ln\mathcal{Z}_{cl-a}^{[\mathbf{0}]}\mathcal{Z}_{1-loop}^{[\mathbf{0}]}=\frac{a_{\D}^2}2\ln\frac{s}{a_{\D}}{+}\frac{3a_{\D}^2}4{+}\frac{ a_{\D} s}4+\frac{Q^2{-}2}{24}\ln a_{\D} \cdot (\epsilon_1\epsilon_2)+\frac{24{-}28Q^2{+}7Q^4}{5760a_{\D}^2}(\epsilon_1\epsilon_2)^2+O\big((\epsilon_1\epsilon_2)^3\big),
\end{equation}
which for the suitable root branch of $s$ gives \eqref{F012_pert_0} up to $a_{\D}$-independent terms in $\mathcal{F}_1$.

\subsection{\texorpdfstring{$H_2$}{H₂}}
\label{ssec:holom_anom_H2}
\paragraph{SW curve.} We bring curve \cite[(4.26)]{BLMST16} in Weierstrass form \eqref{Welc} by substituting $y\mapsto y/z$ and then using \eqref{quart_to_W} with $R=-1$. Finally, shifting global modulus parameter $u\mapsto u-\frac43 c \tilde{m}$ and using dictionary
\begin{equation}
t=2\sqrt2c,\qquad \boldsymbol{m}_1=\frac23(\tilde{m}+3m_-), \quad \boldsymbol{m}_2=\frac23(\tilde{m}-3m_-),\quad \boldsymbol{m}_3=-\frac43\tilde{m}    
\end{equation}
we obtain following polynomials $g_2,g_3$
\begin{align}\label{g2_H2}
4g_2&=-\frac{t u}{\sqrt2}+\frac{t^4}{3{\cdot}2^6}-\boldsymbol{e}_2,\\ \label{g3_H2}
16g_3&=u^2-\frac{t^3 u}{3{\cdot} 2^{7/2}}+\frac{t^6}{3^3{\cdot} 2^8}-\frac{\boldsymbol{e}_2t^2}{3{\cdot} 2^3}-\boldsymbol{e}_3.
\end{align}
These polynomials are obtained from \eqref{g2_3}, \eqref{g3_3}
by limit $r\rightarrow0$
after rescaling \eqref{resc_3} and (c.f. \eqref{zeta_IV})
\begin{equation}
u^{[3]}=\frac1{2r^4}+\frac{t^{[H_2]}}{4r^3}-\frac{3\boldsymbol{m}_3}{r^2}+\frac{4\sqrt2u^{[H_2]}{-}\boldsymbol{m}_3t^{[H_2]}}{r}, \qquad  z^{[3]}=2^5 z^{[H_2]}/r^2, \quad y^{[3]}=2^{15/2}y^{[H_2]}/r^3,
\end{equation}
note that this limit actually resembles the one of \cite[Sec. 3.3]{APSW95}.
Analogously to $N_f=4$ / PVI case \eqref{zeta_SW_VI} (and by \cite[App. A]{KMNOY04}, \cite[Sec. 3.3]{BLMST16}), this $H_2$ SW curve is equivalent to level curve $\zeta_{\IV}(p,q|t)=4\sqrt2\mathrm{u}$ of PIV zeta function \eqref{zeta_IV} with dimension rescalings $u=(2\epsilon_2)^{3/2}\mathrm{u}$, $t=(2\epsilon_2)^{1/2}\st$, $\boldsymbol{m}_k=2\epsilon_2\mathfrak{m}_k|_{k=1,2,3}$.

Modular discriminant $\Delta(u)$ has power $4$. It has zero
\begin{equation}
u_s=6\sqrt2t\left(\frac{t^2}{6^4}-\frac{\boldsymbol{e}_2}{4t^2}-\frac{9\boldsymbol{e}_3}{2t^4}+\frac{27\boldsymbol{e}_2^2}{t^6}+\frac{2 {\cdot}3^6\boldsymbol{e}_2\boldsymbol{e}_3}{t^8}+\frac{4{\cdot} 3^5(27\boldsymbol{e}_3^2{-}10\boldsymbol{e}_2^3)}{t^{10}}+O\left(\frac1{t^{12}}\right) \right)   
\end{equation}
that, as we will see, corresponds to the square exponent singularity expansion in the Painlev\'e setting. Remaining $3$ zeroes $u_l^{(i)}, \, i=1,2,3$ correspond to the linear exponent expansion
\begin{multline}
u_l^{(i)}=-\frac{\boldsymbol{m}_i t}{2\sqrt2}-\frac{\boldsymbol{e}_2{+}3\boldsymbol{m}_i^2}{\sqrt2 t}\left(1-\frac{6\boldsymbol{m}_i}{t^2}+\frac{4(\boldsymbol{e}_2{+}21\boldsymbol{m}_i^2)}{t^4}+\frac{24(59\boldsymbol{m}_i\boldsymbol{e}_2{-}66\boldsymbol{e}_3)}{t^6}+\frac{48(2\boldsymbol{e}_2^2{+}729\boldsymbol{m}_i\boldsymbol{e}_3{-}606\boldsymbol{m}_i^2\boldsymbol{e}_2)}{t^8}\right.\\ \left.+\frac{3{\cdot} 2^6(3408\boldsymbol{e}_2\boldsymbol{e}_3{-}3449\boldsymbol{m}_i\boldsymbol{e}_2^2{-}4446\boldsymbol{m}_i^2\boldsymbol{e}_3)}{t^{10}}+\frac{3{\cdot} 2^6(116127\boldsymbol{e}_3^2{+}19\boldsymbol{e}_2^3{-}197487\boldsymbol{m}_i\boldsymbol{e}_2\boldsymbol{e}_3 {+}83667\boldsymbol{m}_i^2\boldsymbol{e}_2^2}{t^{12}}\right.\\ \left.-\frac{9{\cdot} 2^7(355458\boldsymbol{e}_2^2\boldsymbol{e}_3{+}(531441\boldsymbol{e}_3^2{-}355069\boldsymbol{e}_2^3)\boldsymbol{m}_i{-}868356\boldsymbol{m}_i^2\boldsymbol{e}_2\boldsymbol{e}_3)}{t^{14}}\right)+O\left(\frac1{t^{17}}\right). 
\end{multline}
\paragraph{Heavy hypers expansion.}
Now in the vicinity of $u_s$ we pass to the flat coordinate (with $\tilde\alpha_{\D}=\frac{\ri \tilde{a}_{\D}}{\sqrt3}$)
\begin{multline}
\frac{4\sqrt2 (u{-}u_s)}{\tilde\alpha_{\D}t}=1+\frac{3\tilde\alpha_{\D}}{t^2}-\frac{12(\tilde\alpha_{\D}^2{-}18\boldsymbol{e}_2)}{t^4}+\frac{3(35\tilde\alpha_{\D}^3{-}7{\cdot} 6^3 \boldsymbol{e}_2 \tilde\alpha_{\D}{+}3{\cdot}6^4 \boldsymbol{e}_3)}{t^6}-\frac{18(67\tilde\alpha_{\D}^4{-}25{\cdot}6^3 \boldsymbol{e}_2\tilde\alpha_{\D}^2{+}3{\cdot}6^5 \boldsymbol{e}_3\tilde\alpha_{\D}{+}6^5\boldsymbol{e}_2^2)}{t^8}\\+\frac{18(889\tilde\alpha_{\D}^5{-}3235{\cdot} 6^2 \boldsymbol{e}_2\tilde\alpha_{\D}^3{+}555{\cdot} 6^4 \boldsymbol{e}_3\tilde\alpha_{\D}^2{+}339{\cdot} 6^4\boldsymbol{e}_2^2\tilde\alpha_{\D}{-}3^4{\cdot} 6^5\boldsymbol{e}_2\boldsymbol{e}_3)}{t^{10}}+O\left(\frac1{t^{12}}\right),\end{multline}
where we have already chosen $(3g_2/4)^{-1/4}\sim -4\sqrt6\ri/t$ in \eqref{periods_hyph}, which is consistent with \eqref{omega_1_ell}.
By comparing the $s^{-1}$ expansion parts of obtained $\mathcal{F}_{0,1,2}$ with the corresponding $\epsilon_{1,2}$ terms of $\mathcal{Z}_{inst}^{[\mathbf{H_{2,S}}]}$ (of Sec. \ref{ssec:exp_H2s}) up to $s^{-5}$ we obtain that they coincide iff $\tilde{a}_{\D}=a_{\D}$ and $\check{\boldsymbol{e}}_3=\boldsymbol{e}_3$.
In the assumption of polynomial coefficients for the instanton expansion the latter is the exact relation and it is not necessary to compute $\mathcal{F}_{g>2}$. Besides the instanton part we have also perturbative parts of $\mathcal{F}_g$
\begin{equation}\label{F012_pert_IV_1}
\mspace{-12mu}
\frac{d^2\mathcal{F}_{0,pert}}{da_{\D}^2}=\ln\left({-}\frac{t^2}{\alpha_{\D}}\right), \quad \mathcal{F}_{1,pert}=\frac{Q^2{-}2}{24}\ln \left({-}\frac{\alpha_{\D} t^{10}}{24^6}\right)+\frac14\ln\left({-}\frac{t^2}{96}\right), \quad  \mathcal{F}_{2,pert}=\frac{24{-}28Q^2{+}7Q^4}{5760\cdot (-3\alpha_{\D}^2)}
\end{equation}
so gap condition \eqref{gap_cond} is satisfied  without any higher orders $O(a_{\D}^0)$. From \eqref{Zcl_IV(e)} and \eqref{Z1loop_IV(e)} using \eqref{gamma_as} we get 
\begin{equation}
\epsilon_1
\epsilon_2\ln\mathcal{Z}_{cl-a}^{[\mathbf{H_{2,S}}]}\mathcal{Z}_{1-loop}^{[\mathbf{H_{2,S}}]}=\frac{a_{\D}^2}2\ln\frac{\sqrt3\ri s}{a_{\D}}{+}\frac{3a_{\D}^2}4{-}\frac{\ri a_{\D} s}{2\sqrt3}+\frac{Q^2{-}2}{24}\ln a_{\D} \cdot (\epsilon_1\epsilon_2)+\frac{24{-}28Q^2{+}7Q^4}{5760a_{\D}^2}(\epsilon_1\epsilon_2)^2+O\big((\epsilon_1\epsilon_2)^3\big),
\end{equation}
which gives \eqref{F012_pert_IV_1} up to $a_{\D}$-independent terms in $\mathcal{F}_1$.

\paragraph{Light hypers expansion.}
Now in the vicinity of $u_l^{(i)}$ we pass to the flat coordinate
\begin{multline}
\frac{4\sqrt2(u{-}u_l^{(i)})}{\tilde{a}_{\D}}=1-3\frac{\tilde{a}_{\D}{-}4\boldsymbol{m}_i}{t^2}-12\frac{\tilde{a}_{\D}^2{-}6\boldsymbol{m}_i \tilde{a}_{\D}{+}2(\boldsymbol{e}_2{+}6\boldsymbol{m}_i^2)}{t^4}\\-24\left(\frac{35}8\tilde{a}_{\D}^3{-}280\boldsymbol{m}_i\tilde{a}_{\D}^2{+}8(11\boldsymbol{e}_2{+}105\boldsymbol{m}_i^2)\tilde{a}_{\D}{-}48(21\boldsymbol{e}_3{-}16\boldsymbol{m}_i\boldsymbol{e}_2)\right)\cdot\frac1{t^6}\\
-36\Bigg(\frac{67}2\tilde{a}_{\D}^4{-}335\boldsymbol{m}_i\tilde{a}_{\D}^3{+}20(5\boldsymbol{e}_2{+}67\boldsymbol{m}_i^2)\tilde{a}_{\D}^2{-}16(159\boldsymbol{e}_3{-}130\boldsymbol{m}_i\boldsymbol{e}_2)\tilde{a}_{\D}{+}8\left(\frac83\boldsymbol{e}_2^2{+}267\boldsymbol{m}_i\boldsymbol{e}_3{-}185\boldsymbol{m}_i^2\boldsymbol{e}_2\right)\Bigg)\cdot\frac1{t^8}\\
-72\Bigg(\frac{889}4\tilde{a}_{\D}^5{-}2667\boldsymbol{m}_i\tilde{a}_{\D}^4{+}15(51\boldsymbol{e}_2{+}889\boldsymbol{m}_i^2)\tilde{a}_{\D}^3{-}460(75\boldsymbol{e}_3{-}64\boldsymbol{m}_i\boldsymbol{e}_2)\tilde{a}_{\D}^2{+}4(79\boldsymbol{e}_2^2{+}11745\boldsymbol{m}_i\boldsymbol{e}_3{-}8745\boldsymbol{m}_i^2\boldsymbol{e}_2)\tilde{a}_{\D}\\ {+}8(2337\boldsymbol{e}_3\boldsymbol{e}_2{-}2432\boldsymbol{m}_i\boldsymbol{e}_2^2{-}3744\boldsymbol{m}_i^2\boldsymbol{e}_3)\Bigg)\cdot\frac1{t^{10}}\\
-288\Bigg(807\tilde{a}_{\D}^6{-}11298\boldsymbol{m}_i\tilde{a}_{\D}^5{+}7(449\boldsymbol{e}_2{+}9684\boldsymbol{m}_i^2)\tilde{a}_{\D}^4{-}5(44319\boldsymbol{e}_3{-}38906\boldsymbol{m}_i\boldsymbol{e}_2)\tilde{a}_{\D}^3\\ {+}20\left(\frac{299}3\boldsymbol{e}_2^2{+}20850\boldsymbol{m}_i\boldsymbol{e}_3{-}16310\boldsymbol{m}_i^2\boldsymbol{e}_2\right)\tilde{a}_{\D}^2{+}4(73641\boldsymbol{e}_3\boldsymbol{e}_2{-}75706\boldsymbol{m}_i\boldsymbol{e}_2^2{-}109386\boldsymbol{m}_i^2\boldsymbol{e}_3)\tilde{a}_{\D}\\{+}192 \left(\frac{17}{24}\boldsymbol{e}_2^3{+}1161\boldsymbol{e}_3^2{-}1806 \boldsymbol{m}_i\boldsymbol{e}_3\boldsymbol{e}_2{+}698\boldsymbol{m}_i^2\boldsymbol{e}_2^2\right)\Bigg)\cdot\frac1{t^{12}}+\ldots+O\left(\frac1{t^{18}}\right),
\end{multline}
where we have already chosen $(3g_2/4)^{-1/4}\sim 4\sqrt2/t$ in \eqref{periods_hyph}, which is consistent with \eqref{omega_1_ell}. 
By comparing the $s^{-1}$ expansion parts of obtained $\mathcal{F}_{0,1,2}$ with the corresponding $\epsilon_{1,2}$ terms of $\mathcal{Z}_{inst}^{[\mathbf{H_{2,L}}]}$ (of Sec. \ref{ssec:exp_H2l}) up to $s^{-6}$ we obtain that they coincide iff $\tilde{a}_{\D}=a_{\D}+2\boldsymbol{m}_i$ and $\tilde{\boldsymbol{e}}_3=\boldsymbol{e}_3$.
In the assumption of polynomial coefficients for the instanton expansion, the latter is the exact relation and it is not necessary to compute $\mathcal{F}_{g>2}$. Besides the instanton part we have also perturbative parts of $\mathcal{F}_g$
\begin{multline}\label{F012_pert_IV_2}
\frac{d^2\mathcal{F}_{0,pert}}{da_{\D}^2}=6\ln t-\ln \prod_{i=1}^3 (a_{\D}{+}2\boldsymbol{m}_i) , \quad \mathcal{F}_{1,pert}=\frac{Q^2{-}2}{24}\ln \left(\frac{t^6}{2^{18}}\prod_{i=1}^3 (a_{\D}{+}2\boldsymbol{m}_i)\right)+\frac14\ln \frac{t^2}{2^5},\\ \mathcal{F}_{2,pert}=-\frac{24{-}28Q^2{+}7Q^4}{5760}\frac{d^2}{da_{\D}^2}\left(\ln\prod_{i=1}^3 (a_{\D}{+}2\boldsymbol{m}_i)\right)=\frac{24{-}28Q^2{+}7Q^4}{5760\tilde{a}_{\D}^2}+O(\tilde{a}_{\D}^0), 
\end{multline}
so gap condition \eqref{gap_cond} is satisfied. From \eqref{Zcl_IV(p)} and \eqref{Z1loop_IV(p)} using \eqref{gamma_as} we get
\begin{multline}
\epsilon_1
\epsilon_2\ln\mathcal{Z}_{cl-a}^{[\mathbf{H_{2,L}}]}\mathcal{Z}_{1-loop}^{[\mathbf{H_{2,L}}]}=\frac32a_{\D}^2\ln s{-}\frac{a_{\D}s}2+\sum_{\lambda,\lambda'=\pm1}\frac{3{-}2\ln (a_{\D}{+}2\boldsymbol{m}_i)}4 a_{\lambda,\lambda'}^2+\frac{Q^2{-}2}{24}\ln \left(\prod_{i=1}^3(a_{\D}{+}2\boldsymbol{m}_i)\right) \cdot (\epsilon_1\epsilon_2)\\-\frac{24{-}28Q^2{+}7Q^4}{5760}\frac{d^2}{da_{\D}^2}\left(\ln\prod_{i=1}^3(a_{\D}{+}2\boldsymbol{m}_i)\right)(\epsilon_1\epsilon_2)^2+O\big((\epsilon_1\epsilon_2)^3\big),
\end{multline}
which gives \eqref{F012_pert_IV_2} up to $a_{\D}$-independent terms in $\mathcal{F}_1$.

\subsection{\texorpdfstring{$H_1$}{H₁}}
\label{ssec:holom_anom_H1}
\paragraph{SW curve.} We bring quartic curve \cite[(4.17)]{BLMST16} with $c=2^{-10/3}t$ in Weierstrass form \eqref{Welc} using \eqref{quart_to_W} with $R=-1$
\begin{align}\label{g2_H1}
g_2&=u+\frac{t^2}{3{\cdot} 2^{14/3}},\\ \label{g3_H1}
g_3&=-\frac{tu}{3{\cdot} 2^{7/3}}+\frac{t^3}{3^3{\cdot} 2^7}+\frac{\boldsymbol{m}^2}4.
\end{align}
These polynomials are obtained from \eqref{g2_2}, \eqref{g3_2}
by limit $r\rightarrow0$
after rescaling \eqref{resc_2} and (c.f. \eqref{zeta_II})
\begin{equation}
u^{[2]}=\frac6{r^6}+\frac{t^{[H_1]}}{r^4}+\frac{8\boldsymbol{m}}{r^3}+\frac{2^{8/3}u^{[H_1]}{-}t^2/4}{r^2}, \qquad  z^{[2]}=2^{16/3} z^{[H_1]}/r^4, \quad y^{[2]}=2^8y^{[H_1]}/r^6,
\end{equation}
note that this limit actually resembles the one of \cite[Sec. 3.2]{APSW95}. They are also obtained from \eqref{g2_H2}, \eqref{g3_H2}
by limit $r\rightarrow0$
after rescaling \eqref{resc_H2} and (c.f. \eqref{zeta_tau_IV_II})
\begin{equation}
4\sqrt2u^{[H_2]}=-\frac1{12r^9}+\frac{t^{[H_1]}}{12r^5}-\frac{2\boldsymbol{m}}{r^3}+\frac{2^{5/3}u^{[H_1]}{-}t^2/8}{r}, \qquad  z^{[H_2]}=2^{-5/3} z^{[H_1]}/r^2, \quad y^{[H_2]}=2^{-5/2}y^{[H_1]}/r^3.
\end{equation}
Analogously to $N_f=4$ / PVI case \eqref{zeta_SW_VI} (and by \cite[App. A]{KMNOY04}, \cite[Sec. 3.2]{BLMST16}), this $H_1$ SW curve is equivalent to level curve $\zeta_{\II}(p,q|t)=2^{5/3}\mathrm{u}-\st^2/8$ of PII zeta function \eqref{zeta_II} with dimension rescalings $u=(2\epsilon_2)^{4/3}\mathrm{u}$, $t=(2\epsilon_2)^{2/3}\st$, $\boldsymbol{m}=2\epsilon_2\mathfrak{m}$.

Modular discriminant $\Delta(u)$ has power $3$. It has zero
\begin{equation}
u_s=\frac{2^{4/3}\boldsymbol{m}^2}{t}\left(1-\frac{2^4\boldsymbol{m}^2}{t^3}+\frac{2^{10}\boldsymbol{m}^4}{t^6}+O\left(\frac1{t^9}\right)\right)   
\end{equation}
that, as we will see, corresponds to the square exponent singularity expansion. Remaining $2$ zeroes correspond to the linear exponent singularity expansion ($s=\ri(2t)^{3/2}$)
\begin{multline}
u_l^{\pm}=\frac{t^2}{2^{14/3}}\left(1\pm\frac{2^6\boldsymbol{m}}{s}+\frac{2^8\boldsymbol{m}^2}{s^2}\mp\frac{2^{11}\boldsymbol{m}^3}{s^3}+\frac{2^{15}\boldsymbol{m}^4}{s^4}\mp\frac{21{\cdot}2^{15}\boldsymbol{m}^5}{s^5}+\frac{2^{24}\boldsymbol{m}^6}{s^6}\mp\frac{429{\cdot}2^{20}\boldsymbol{m}^7}{s^7}\right.\\ \left.+\frac{3{\cdot}2^{32}\boldsymbol{m}^8}{s^8}\mp\frac{46189{\cdot}2^{23}\boldsymbol{m}^9}{s^9}+O\left(\frac1{s^{10}}\right)\right)    
\end{multline}
that differ by trivial symmetry of the theory $\boldsymbol{m}\mapsto-\boldsymbol{m}$.

\paragraph{Heavy hypers expansion.}
Now in the vicinity of $u_s$ we pass to the flat coordinate ($\alpha_{\D}=\frac{\ri a_{\D}}{\sqrt2}$)
\begin{multline}
\frac{2^{7/6}\ri(u{-}u_s)}{\alpha_{\D}t^{1/2}}=1-\frac{3\alpha_{\D}}{s}-\frac{17\alpha_{\D}^2{+}3{\cdot}2^7\boldsymbol{m}^2}{s^2}-\left(\frac{125}2\alpha_{\D}^3{+}13{\cdot}2^8\boldsymbol{m}^2\alpha_{\D}\right)\cdot\frac3{s^3}\\-\left(\frac{3563}4\alpha_{\D}^4{+}685{\cdot}2^7\boldsymbol{m}^2\alpha_{\D}^2{+}11{\cdot}2^{13}\boldsymbol{m}^4\right)\cdot\frac3{s^4}-\left(\frac{29183}2\alpha_{\D}^5{+}9065{\cdot}2^8\boldsymbol{m}^2\alpha_{\D}^3{+}95{\cdot}2^{16}\boldsymbol{m}^4\alpha_{\D}\right)\cdot\frac3{s^5}+O\left(\frac1{s^6}\right),
\end{multline}
where we have already chosen $(3g_2/4)^{-1/4}\sim 2^{5/3}/t^{1/2}$ in \eqref{periods_hyph}, which is consistent with \eqref{omega_1_ell}. 
By comparing the $s^{-1}$ expansion parts of obtained $\mathcal{F}_{0,1,2}$ with the corresponding $\epsilon_{1,2}$ terms of $\mathcal{Z}_{inst}^{[\mathbf{H_{1,S}}]}$ (of Sec. \ref{ssec:exp_H1s}) up to $s^{-5}$ we obtain that they coincide iff
\begin{equation}
\check{\boldsymbol{m}}^2=\boldsymbol{m}^2-\frac3{128}\epsilon^2.   
\end{equation}
In the assumption of polynomial coefficients of the instanton expansion this is the exact relation and it is not necessary to compute $\mathcal{F}_{g>2}$. Besides the instanton part we have also perturbative parts of $\mathcal{F}_g$
\begin{equation}\label{F012_pert_II_1}
\frac{d^2\mathcal{F}_{0,pert}}{da_{\D}^2}=\ln\left(\frac{2s}{\alpha_{\D}}\right), \quad \mathcal{F}_{1,pert}=\frac{Q^2{-}2}{24}\ln \left(\frac{\alpha_{\D} s^3}{2^{15}}\right)+\frac14\ln\frac{t}{2^{10/3}}, \quad  \mathcal{F}_{2,pert}=\frac{24{-}28Q^2{+}7Q^4}{5760(-2\alpha_{\D}^2)},
\end{equation}
so gap condition \eqref{gap_cond} is satisfied without any higher orders $O(a_{\D}^0)$. From \eqref{Zcl_II(e)} and \eqref{Z1loop_II(e)} using \eqref{gamma_as} we get
\begin{equation}
\epsilon_1
\epsilon_2\ln\mathcal{Z}_{cl-a}^{[\mathbf{H_{1,S}}]}\mathcal{Z}_{1-loop}^{[\mathbf{H_{1,S}}]}=\frac{a_{\D}^2}2\ln\frac{(\sqrt2\ri)^3 s}{a_{\D}}{+}\frac{3a_{\D}^2}4{-}\frac{\ri a_{\D} s}{3\sqrt2}+\frac{Q^2{-}2}{24}\ln a_{\D} \cdot (\epsilon_1\epsilon_2)+\frac{24{-}28Q^2{+}7Q^4}{5760a_{\D}^2}(\epsilon_1\epsilon_2)^2+O\big((\epsilon_1\epsilon_2)^3\big),
\end{equation}
which gives \eqref{F012_pert_II_1} up to $a_{\D}$-independent terms in $\mathcal{F}_1$.

\paragraph{Light hypers expansion.} Now in the vicinity of $u_l^{\pm}$ we pass to the flat coordinate
\begin{multline}
\frac{2^{7/6}(u{-}u_l^{\pm})}{\ri\tilde{a}_{\D}t^{1/2}}=1-3\frac{\tilde{a}_{\D}{\mp}4\boldsymbol{m}}{s}-\frac{17\tilde{a}_{\D}^2{\mp}102\boldsymbol{m}\tilde{a}_{\D}{+}168\boldsymbol{m}^2}{s^2}-3\left(\frac{125}{2}\tilde{a}_{\D}^3{\mp}500\boldsymbol{m}\tilde{a}_{\D}^2{+}83{\cdot}2^4\boldsymbol{m}^2\tilde{a}_{\D}{\mp}41{\cdot}2^5\boldsymbol{m}^3\right)\cdot\frac1{s^3}\\
-3\left(\frac{3563}4\tilde{a}_{\D}^4{\mp}\frac{17815}2\boldsymbol{m}\tilde{a}_{\D}^3{+}32660\boldsymbol{m}^2\tilde{a}_{\D}^2{\mp}835{\cdot}2^6\boldsymbol{m}^3\tilde{a}_{\D}{+}1151{\cdot}2^5\boldsymbol{m}^4\right)\cdot\frac1{s^4}\\-3\left(\frac{29183}2\tilde{a}_{\D}^5{\mp}175098\boldsymbol{m}\tilde{a}_{\D}^4{+}818650\boldsymbol{m}^2\tilde{a}_{\D}^3{\mp}117495{\cdot}2^4\boldsymbol{m}^3\tilde{a}_{\D}^2{+}2125{\cdot}2^{10}\boldsymbol{m}^4\tilde{a}_{\D}{\mp}8871{\cdot}2^7\boldsymbol{m}^5\right)\cdot\frac1{s^5}+\ldots+O\left(\frac1{s^9}\right),
\end{multline}
where we have already chosen $(3g_2/4)^{-1/4}\sim-2^{7/6}\ri/t^{1/2}$ in \eqref{periods_hyph}, which is consistent with \eqref{omega_1_ell}.
By comparing the $s^{-1}$ expansion parts of obtained $\mathcal{F}_{0,1,2}$ with the corresponding $\epsilon_{1,2}$ terms of $\mathcal{Z}_{inst}^{[\mathbf{H_{1,L}}]}$ (of Sec. \ref{ssec:exp_H1l}) up to $s^{-7}$ we obtain that they coincide iff $\tilde{a}_{\D}=a_{\D}\pm 2\boldsymbol{m}$ and $\tilde{\boldsymbol{m}}^2=\boldsymbol{m}^2$.
In the assumption of polynomial coefficients for the instanton expansion, the latter is the exact relation and it is not necessary to compute $\mathcal{F}_{g>2}$. Besides the instanton part we have also perturbative parts of $\mathcal{F}_g$
\begin{multline}\label{F012_pert_II_2}
\frac{d^2\mathcal{F}_{0,pert}}{da_{\D}^2}=\ln \frac{32t^3}{4\boldsymbol{m}^2{-}a_{\D}^2}, \quad \mathcal{F}_{1,pert}=\frac{Q^2{-}2}{24}\ln \frac{(4\boldsymbol{m}^2{-}a_{\D}^2)t^3}{2^7}+\frac14\ln\left(-\frac{t}{2^{7/3}}\right),\\
\mathcal{F}_{2,pert}=-\frac{24{-}28Q^2{+}7Q^4}{5760}\frac{d^2}{da_{\D}^2}\ln(4\boldsymbol{m}^2{-}a_{\D}^2)=\frac{24{-}28Q^2{+}7Q^4}{5760\tilde{a}_{\D}^2}+O(\tilde{a}_{\D}^0),    
\end{multline}
so gap condition \eqref{gap_cond} is satisfied. From \eqref{Zcl_II(p)} and \eqref{Z1loop_II(p)} using \eqref{gamma_as} we get
\begin{multline}
\epsilon_1
\epsilon_2\ln\mathcal{Z}_{cl-a}^{[\mathbf{H_{1,L}}]}\mathcal{Z}_{1-loop}^{[\mathbf{H_{1,L}}]}=a_{\D}^2\ln (2s){-}\frac{a_{\D}s}3+\sum_{\lambda=\pm1}\frac{3{-}2\ln (a_{\D}{+}2\lambda\boldsymbol{m})}4 (a_{\D}{+}2\lambda\boldsymbol{m})^2+\frac{Q^2{-}2}{24}\ln (a_{\D}^2{-}4\boldsymbol{m}^2) \cdot (\epsilon_1\epsilon_2)\\-\frac{24{-}28Q^2{+}7Q^4}{5760}\frac{d^2}{da_{\D}^2}\ln (a_{\D}^2{-}4\boldsymbol{m}^2)\cdot(\epsilon_1\epsilon_2)^2+O\big((\epsilon_1\epsilon_2)^3\big),
\end{multline}
which gives \eqref{F012_pert_II_2} up to $a_{\D}$-independent terms in $\mathcal{F}_1$.

\subsection{\texorpdfstring{$H_0$}{H₀}}
\label{ssec:holom_anom_H0}
\paragraph{SW curve.} Curve \cite[(4.10)]{BLMST16} is almost in Weierstrass form \eqref{Welc}: we should just rescale $y\mapsto \frac12 e^{\pi\ri/4}y, z\mapsto -\ri z$. Then, substituting $6c=t$, we obtain
\begin{equation}
g_2=-2t, \qquad
g_3=4\ri u.
\end{equation}
These polynomials are obtained from \eqref{g2_1}, \eqref{g3_1}
by limit $r\rightarrow0$
after rescaling \eqref{resc_1} and (c.f. \eqref{zeta_I})
\begin{equation}
u^{[1]}=\frac{12}{r^{10}}+\frac{2t^{[H_0]}}{r^6}+\frac{4\ri u^{[H_0]}}{r^4}, \qquad  z^{[1]}=16z^{[H_0]}/r^8, \quad y^{[1]}=64y^{[H_0]}/r^{12},
\end{equation}
note that this limit actually resembles the one of \cite[Sec. 3.1]{APSW95}. They are also obtained from \eqref{g2_H1}, \eqref{g3_H1}
by limit $r\rightarrow0$
after rescaling \eqref{resc_H1} and (c.f. \eqref{zeta_tau_II_I})
\begin{equation}
2^{8/3}u^{[H_1]}=-\frac3{r^{20}}-\frac{t^{[H_0]}}{r^8}+\frac{4\ri u^{[H_0]}}{r^2}, \qquad  z^{[H_1]}=2^{-4/3} z^{[H_0]}/r^4, \quad y^{[H_1]}=2^{-2}y^{[H_0]}/r^6.
\end{equation}
Analogously to $N_f=4$ / PVI case \eqref{zeta_SW_VI} (and by \cite[App. A]{KMNOY04}, \cite[Sec. 3.1]{BLMST16}), this $H_0$ SW curve is equivalent to level curve $\zeta_{\I}(p,q|t)=-2\ri\mathrm{u}$ of PI zeta function \eqref{zeta_I} with dimension rescalings $u=(2\epsilon_2)^{6/5}\mathrm{u}$, $t=(2\epsilon_2)^{4/5}\st$.

Modular discriminant $\Delta(u)$ is of order $2$. It has zeroes
\begin{equation}
u_*^{\pm}=\pm\frac{t^{3/2}}{3\sqrt6}=\frac{s_{\pm}^2}{3^4{\cdot}2^{10}\ri t},
\end{equation}
where the sign actually corresponds to a choice of the root branch of $s=2^{7/2}(6e^{\pi\ri}t)^{5/4}$ (i.e. $s_+{=}s, s_-{=}\ri s$).

\paragraph{Expansion.}Now in the vicinity of $u_*^{\pm}$ we pass to the flat coordinate 
\begin{multline}
\frac{3{\cdot}2^5t(u{-}u_*^{\pm})}{\ri s_{\pm}a_{\D}}=1-30(a/s_{\pm})-2820(a/s_{\pm})^2-463020(a/s_{\pm})^3-95872644(a/s_{\pm})^4\\-22598568720(a/s_{\pm})^5-5788947486960(a/s_{\pm})^6+O\left(\frac1{s_{\pm}^7}\right),
\end{multline}
where we have already chosen $(3g_2/4)^{-1/4}\sim \frac{96t}{\ri s_{\pm}}$ in \eqref{periods_hyph}, which is consistent with \eqref{omega_1_ell}. 
The $s^{-1}$ expansion parts of obtained $\mathcal{F}_{0,1,2}$ coincide with the corresponding $\epsilon_{1,2}$ terms of $\mathcal{Z}_{inst}^{[\mathbf{H_0}]}$ (of Sec. \ref{ssec:exp_H0}) up to $s^{-6}$. More precisely, for $s_+$ we obtain literally \eqref{Zinst_I} and for $s_-$ we obtain the other root branch of the Painlev\'e expansion.  Besides the instanton part we have also perturbative parts of $\mathcal{F}_g$
\begin{equation}\label{F012_pert_I}
\frac{d^2\mathcal{F}_{0,pert}}{da_{\D}^2}=\ln\left(\frac{s_{\pm}}{a_{\D}}\right), \quad \mathcal{F}_{1,pert}=\frac{Q^2{-}2}{24}\ln \left(\frac{a_{\D}(-24t)^3}{s_{\pm}}\right)+\frac14\ln\left(-\frac{s_{\pm}^2}{(96t)^2}\right), \quad  \mathcal{F}_{2,pert}=\frac{24{-}28Q^2{+}7Q^4}{5760 a_{\D}^2}
\end{equation}
so gap condition \eqref{gap_cond} is satisfied without any higher orders $O(a_{\D}^0)$. From \eqref{Zcl_I} and \eqref{Z1loop_I} using \eqref{gamma_as} we get
\begin{equation}
\epsilon_1
\epsilon_2\ln\mathcal{Z}_{cl-a}^{[\mathbf{H_0}]}\mathcal{Z}_{1-loop}^{[\mathbf{H_0}]}=\frac{a_{\D}^2}2\ln\frac{s}{a_{\D}}{+}\frac{3a_{\D}^2}4{-}\frac{ a_{\D} s}{60}+\frac{Q^2{-}2}{24}\ln a_{\D} \cdot (\epsilon_1\epsilon_2)+\frac{24{-}28Q^2{+}7Q^4}{5760a_{\D}^2}(\epsilon_1\epsilon_2)^2+O\big((\epsilon_1\epsilon_2)^3\big),
\end{equation}
which for the suitable root branch of $s$ gives \eqref{F012_pert_I} up to $a_{\D}$-independent terms in $\mathcal{F}_1$.

\section{Comparison with CFT approaches}
\label{sec:comparison}
\subsection{Partition functions from irregular states in Liouville theory}
\label{ssec:GT}
In this subsection we compare our results with those in \cite{NU19} and \cite{PP23, IILZ25}.
In \cite{NU19} the asymptotic expansions of the irregular conformal blocks corresponding to the partition functions of $(A_1,D_4)$ (which is $H_2$) and $(A_1, A_3)$ (which is $H_1$)
Argyres-Douglas theories were evaluated 
using rank-2 and rank-3 irregular states in Liouville theory.
In the following we compare these results with the linear exponent expansions \eqref{Zinst_IVl}, \eqref{Zinst_IIl}, while the analog for the square exponent expansion seems to be unknown. Notice that the above irregular conformal block asymptotic expansion for $H_1$ was also obtained using a half-integer rank-3/2 irregular state \cite{HNNT24} (see also \cite{PP25}).

In \cite{PP23}, by making use of 
rank-5/2 Liouville irregular state, the asymptotic expansion of the irregular conformal block corresponding to the partition function of $H_0$ Argyres-Douglas theory was evaluated. Further an alternative algebraic approach to the rank-5/2 irregular state was developed in \cite[Sec. 3.4]{IILZ25}. We compare these two results with our expansion \eqref{Zinst_I}.

\paragraph{$H_2$ (linear exp singularity).} 
The general form of the expansion is given by formula \cite[(3.24)]{NU19}, where the analog of the instanton counting parameter is $c_2/c_1^2$.
They found this expansion up to power $4$, corresponding coefficients $D_k$ are given by \cite[(3.19),(B.3)-(B.6)]{NU19}\footnote{It seems that $D_3$ in (B.5) and $D_4$ in (B.6) have extra overall minus. Also it seems that there is a typo in (B.5) inside $Q^2$ bracket: $\nu^2$-term should be multiplied by $s_3$ instead of $\nu^1$-term. For the comparison below we use these changes.}. We compare our instanton expansion
\eqref{Zinst_IVl} (recall notation \eqref{inst_ansatz}, adding $Z_0=1$)
with their coefficients $D_k$. We find that, by setting $\tilde{\boldsymbol{e}}_3=\boldsymbol{e}_3$, we match the expansions up to a scaling of the instanton counting parameter
\begin{equation}
2^k D_k(\nu,s_{2,3};Q^2)=(\epsilon_1\epsilon_2)^{-k}Z_{3k}^{[\mathbf{H_{2,L}}]}(a_{\D},\boldsymbol{m}_{1,2,3};\epsilon_1,\epsilon_2), \quad k=0\ldots 4    
\end{equation}
under dictionaries (c.f. $a_{\D}=2\epsilon_2\nu$ and \eqref{AGT_dictionary})
\begin{align}\label{nuQ_dct}
&\nu=\frac{a_{\D}}{\sqrt{-\epsilon_1\epsilon_2}}, \qquad Q^2=\frac{\epsilon^2}{\epsilon_1\epsilon_2},\\
&s_2=\frac94\frac{\boldsymbol{e}_2}{-\epsilon_1\epsilon_2}, \qquad s_3=\frac{27}8\frac{\boldsymbol{e}_3}{(-\epsilon_1\epsilon_2)^{3/2}}.
\end{align}

\paragraph{$H_1$ (linear exp singularity).}
The general form of the expansion is given by formula \cite[(3.50)]{NU19}, where the analog of the instanton counting parameter is $c_3^2/c_2^3$. They found this expansion up to power $3$, corresponding coefficients $D_k$ are given by \cite[(3.44),(4.13)-(4.15)]{NU19}.
We compare our instanton expansion \eqref{Zinst_IIl} (recall notation \eqref{inst_ansatz}, adding $Z_0=1$) up to $s^{-3}$ with their coefficients $D_k$. We find that, by setting $\tilde{\boldsymbol{m}}^2=\boldsymbol{m}^2$, we match the expansions up to a scaling of the instanton counting parameter
\begin{equation}
(3\ri)^k D_k(\nu,\theta;Q^2)=(\epsilon_1\epsilon_2)^{-k}Z_{3k}^{[\mathbf{H_{1,L}}]}(a_{\D},\boldsymbol{m};\epsilon_1,\epsilon_2), \quad k=0\ldots 3    
\end{equation} under dictionary \eqref{nuQ_dct} and 
\begin{equation}
\theta^2=\frac{16\boldsymbol{m}^2}{-\epsilon_1\epsilon_2}.    
\end{equation}

\paragraph{$H_0$ (\cite{PP23}).} 
The full partition function in \cite{PP23} is divided in a tree and an instanton part. The instanton part is given by \cite[(A.4)]{PP23} and we compare it with our \eqref{Zinst_I}. Namely, taking the latter power expansion for $\mathcal{Z}_{inst}^{[\mathbf{H_0}]}$ up to $s^{-2}$, we have
\begin{equation}
\ln \mathcal{Z}_{inst}^{[\mathbf{H_0}]}(a_{\D};\epsilon_1,\epsilon_2|s)=\ln Z_{\mathcal{H}_0 \mathrm{inst}}(a^{\mathrm{PP}};Q|c_1,c_2,\Lambda_5)-\frac{60(a^{\mathrm{PP}})^2{-}2{+}7Q^2}{96}\ln \left(1-\frac{6c_1\Lambda_5}{c_2^3}\right)+o(\Lambda_5^5)
\end{equation}
under the dictionary that connects instanton counting parameter $s$ with eigenvalues $c_1,c_2,\Lambda_5$ of the irregular state (c.f. \cite[(5.6)]{PP23})  and $a_{\D}$ with $a$ in \cite{PP23} (we denote it $a^{\mathrm{PP}}$)
\begin{equation}
a_{\D}=a^{\mathrm{PP}}\sqrt{\epsilon_1\epsilon_2}, \qquad s=-\frac{32 (c_2^4-6c_1c_2\Lambda_5)^{5/4}}{\Lambda_5^2} \sqrt{\epsilon_1\epsilon_2}.   
\end{equation}
Together with tree part 
\cite[(2.20)]{PP23}, we obtain
\begin{multline}
\ln \mathcal{Z}_{cl}^{[\mathbf{H_0}]}(a_{\D};\epsilon_1,\epsilon_2|s)+\ln \mathcal{Z}_{inst}^{[\mathbf{H_0}]}(a_{\D};\epsilon_1,\epsilon_2|s)=\ln Z_{\mathcal{H}_0 \mathrm{tree}}(a^{\mathrm{PP}};Q|c_1,c_2,\Lambda_5)+\ln Z_{\mathcal{H}_0 \mathrm{inst}}(a^{\mathrm{PP}};Q|c_1,c_2,\Lambda_5)\\+C+\frac12\left((a^{\mathrm{PP}})^2{-}\frac1{30}+\frac7{60}Q^2\right)\ln (-32\sqrt{\epsilon_1\epsilon_2}),   
\end{multline}
where
\begin{equation}
C=\frac{c_2^{10}}{270\Lambda_5^4}-\frac{c_2^6 \left(-\frac{s}{32\sqrt{\epsilon_1\epsilon_2}}\right)^{4/5}}{81\Lambda_5^{12/5}}+\frac{c_2^2 \left(-\frac{s}{32\sqrt{\epsilon_1\epsilon_2}}\right)^{8/5}}{54\Lambda_5^{4/5}}.     
\end{equation}

\paragraph{$H_0$ (\cite{IILZ25}).}
Irregular conformal block expansion \cite[(3.57)]{IILZ25} is explicitly presented as logarithmic expansion \cite[(3.59)]{IILZ25}, which coefficients $\mathcal{U}_k|_{k=1,\ldots 5}$ is presented in \cite[(3.61b)-(3.61f)]{IILZ25}. We compare our instanton expansion \eqref{Zinst_I} (recall notation \eqref{inst_ansatz_ln}) up to $s^{-5}$ with their coefficients $\mathcal{U}_k$. We find, under dictionary \eqref{nuQ_dct} and central charge $c$ given by first relation of \eqref{AGT_dictionary}, that
\begin{equation}
(-\epsilon_1\epsilon_2)^{\frac{k{+}2}2}\,\mathcal{U}_k(\nu;c)=P_{k+2}^{[\mathbf{H_0]}}(a_{\D};\epsilon_1,\epsilon_2), \quad k=1\ldots 5,    
\end{equation}
which implies the relation
$s^{-1}=(\varepsilon^{\mathrm{IILZ}})^2\sqrt{-\epsilon_1\epsilon_2}$ with IILZ expansion variable $\varepsilon^{\mathrm{IILZ}}$. Under this relation the logarithmic part of \cite[(3.59)]{IILZ25} with coefficient $\mathcal{U}_{\ln}$, given by \cite[(3.61a)]{IILZ25}, reproduces $\ln s$ coefficient of $\ln\mathcal{Z}_{cl}^{[\mathbf{H_0]}}$ obtained from \eqref{Zcl_I}.

\subsection{Partition functions from irregular vertex operators}
\label{ssec:Nagoya}
The Liouville irregular states approach to irregular conformal blocks started with \cite{GT12} (see also \cite{G09, BMT1112}). Another, representation theoretic approach, based on introducing irregular vertex operators of the Virasoro algebra acting on irregular Verma modules was initiated by Nagoya in paper \cite{N15}. While these two approaches are believed to be intimately connected by the operator-state correspondence, explicit formulas needed to connect them seem to be absent in the literature. 

The Nagoya approach was explicitly used to obtain the Zak transform ansatz expansions near $t=\infty$ for some Painlev\'e equations.
Namely, in paper \cite{N15} tau functions of Painlev\'e V, IV near $t=\infty$ are given by the Zak transform of an expansion of a matrix element of Virasoro $c=1$ rank-0 irregular vertex operator. Further, in paper \cite{N18}, the tau functions of Painlev\'e III$_1$, II near $t=\infty$ limit are given by the Zak transform of an expansion of a matrix element of Virasoro $c=1$ rank-0 ramified irregular vertex operator.

These calculations reproduce the expansions of \cite{BLMST16} at least for the first few orders of the instanton part. More precisely, in \cite{N15,N18} the matrix elements reproduce $\mathcal{Z}_{inst}$ and $\mathcal{Z}_{cl-a}$ for $\epsilon=0$, but the rest of the classical and 1-loop part is introduced there by hands. Note that for all the cases which are analyzed -- except Painlev\'e III$_1$, where there is a single expansion -- only the linear exp singularity one is reproduced, as the representation-theoretic approach for the other expansion is not known just as it is for the Liouville irregular states approach.

It is natural to expect that the quantized expansions are given by the same conformal block but with general $c=1+\frac{6\epsilon^2}{\epsilon_1\epsilon_2}$, according to AGT dictionary \eqref{AGT_dictionary}. That is actually true as it is explicitly shown in the rest of the subsection.

\paragraph{QPV (linear exp singularity).}
According to \cite[Conj. 4.1]{N15} the Painlev\'e V tau function near $t=\infty$ is given by the Zak transform of (the expansion of) the  matrix element of the $c=1$ Virasoro rank-0 irregular vertex operator acting on the rank-1 irregular Verma module. The matrix element is taken between the highest weight vector of the regular Verma module and the rank-1 irregular (Whittaker) vector. 
This tau function coincides with \cite[(A.49)]{BLMST16} up to a constant redefinition of $\eta_{\D}$. This is due to a different parametrization of the arguments in the Barnes $\mathsf{G}$-functions. 

After the quantum ($\epsilon$-) deformation and the corresponding refining of the central charge by \eqref{AGT_dictionary} we checked up to the power $8$ in $t^{-1}$ of $\mathcal{Z}_{inst}^{[\mathbf{3_L}]}$ that for $\tilde{w}_4=w_4$
\begin{multline}
\mathrm{t}_{\V}^{-2\Delta_t{-}\frac{\theta_*^2}2} e^{-\frac{\theta_*\mathrm{t}_{\sV}}2 }\Big\langle \Delta_0\Big|\Phi^{\Delta_t}_{\left(\frac{\theta_*}2-\nu,\frac14\right), \left(\theta_*,\frac14\right)}\big(\mathrm{t}_{\V}^{-1}\big)\Big|(\theta_*,1/4)\Big\rangle\\=(-\epsilon_1\epsilon_2)^{-\frac{a_{\D}^2}{\epsilon_1\epsilon_2}}\mathcal{Z}_{cl-a}^{[\mathbf{3_L}]}(a_{\D};\epsilon_1,\epsilon_2|t)\mathcal{Z}_{inst}^{[\mathbf{3_L}]}(a_{\D},m_{1,2,3};\epsilon_1,\epsilon_2|t) 
    \end{multline}
    for $\Delta_{0,t}=\frac{c-1}{24}+\theta^2_{0,t}$
    under dictionary
    \begin{equation}
\nu=\frac{a_{\D}}{\sqrt{-\epsilon_1\epsilon_2}}, \qquad \mathrm{t}_{\V}=\frac{t}{\sqrt{-\epsilon_1\epsilon_2}}, \qquad \theta_0=\frac{m_1-m_3}{2\sqrt{-\epsilon_1\epsilon_2}}, \quad  \theta_t=\frac{m_1+m_3}{2\sqrt{-\epsilon_1\epsilon_2}}, \quad \theta_*=\frac{m_2}{\sqrt{-\epsilon_1\epsilon_2}}.
\end{equation}
This dictionary is a natural quantum deformation of scalings $a_{\D}=2\epsilon_2\nu$, $t=2\epsilon_2 \mathrm{t}_{\V}$, $m_f=2\epsilon_2\sm_f|_{f=1,2,3}$, according to \eqref{PV_dct}.

\paragraph{QPIV (linear exp singularity).}
According to \cite[Conj. 4.2]{N15} the Painlev\'e IV tau function near $t=\infty$ is given by the Zak transform of (the expansion of) the matrix element of the $c=1$ Virasoro rank-0 irregular vertex operator acting on the rank-2 irregular Verma module. The matrix element is taken between the vacuum vector and the rank-2 irregular (Whittaker) vector. 
This tau function coincides with \cite[(3.48)]{BLMST16} up to a constant redefinition of $\eta_{\D}$
as in the previous case.

After the quantum ($\epsilon$-) deformation and the corresponding refining of the central charge by \eqref{AGT_dictionary} we checked up to the power $4$ in $s^{-1}$ of $\mathcal{Z}_{inst}^{[\mathbf{H_{2,L}}]}$ that for 
$\tilde{\boldsymbol{e}}_3=\boldsymbol{e}_3$
\begin{multline}
    \mathrm{t}_{\IV}^{-3\Delta_t-\frac{\theta^2}3}e^{-\frac{\theta \mathrm{t}_{\sIV}^2}6}\Big\langle 0\Big|\Phi^{\Delta_t}_{\left(\frac{2\theta}3-\nu,0,\frac14\right), \left(\theta,0,\frac14\right)}\big(\mathrm{t}_{\IV}^{-1}\big)\Big|(\theta,0,1/4)\Big\rangle\\=(-\epsilon_1\epsilon_2)^{-\frac{3a_{\D}^2}{4\epsilon_1\epsilon_2}}\mathcal{Z}_{cl-a}^{[\mathbf{H_{2,L}}]}(a_{\D};\epsilon_1,\epsilon_2|s)\mathcal{Z}_{inst}^{[\mathbf{H_{2,L}}]}(a_{\D},\boldsymbol{m}_{1,2,3};\epsilon_1,\epsilon_2|s) 
    \end{multline}
    for $\Delta_{t}=\frac{c-1}{24}+\theta^2_{t}$ and $\theta=2\theta_s{-}\theta_t$ under dictionary
    \begin{equation}\label{PIVdct}
    \nu=\frac{a_{\D}}{\sqrt{-\epsilon_1\epsilon_2}}, \qquad \mathrm{t}_{\IV}=\frac{t}{(-\epsilon_1\epsilon_2)^{1/4}}, \qquad \theta_t=-\frac{2\boldsymbol{m}_1+\boldsymbol{m_2}}{\sqrt{-\epsilon_1\epsilon_2}}, \quad  \theta_s=-\frac{\boldsymbol{m}_1+2\boldsymbol{m}_2}{\sqrt{-\epsilon_1\epsilon_2}}.
\end{equation}
This dictionary is a natural quantum deformation of scalings $a_{\D}=2\epsilon_2\nu$, $t=(2\epsilon_2)^{1/2}\mathrm{t}_{\IV}$, $\boldsymbol{m}_k=2\epsilon_2\mathfrak{m}_k|_{k=1,2,3}$, according to \eqref{PIV_dct}.

\paragraph{QPIII$_1$.}
According to \cite[Conj. 3.10]{N18} the Painlev\'e III$_1$ tau function near $t=\infty$ is given by the Zak transform of (the expansion of) the matrix element of the $c=1$ Virasoro rank-0 ramified irregular vertex operator acting on the rank-1/2 irregular Verma module. The matrix element is taken between the highest weight vector in the Verma module and the rank-1/2 irregular (Whittaker) vector. 
This tau function coincides with \cite[(A.30)]{BLMST16}.

After the quantum ($\epsilon$-) deformation and the corresponding refining of the central charge by\eqref{AGT_dictionary} we checked up to the power $5$ in $s^{-1}$ of $\mathcal{Z}_{inst}^{[\mathbf{2}]}$ that for $\tilde{w}_2=w_2$
\begin{equation}
    \big(-\mathrm{t}_{\III_1}^{-1}\big)^{3\Delta_-/2}\Big\langle \Delta_+\Big|\Phi^{\Delta_-,4\nu}_{(1,0), (1,0)}\big(-\mathrm{t}_{\III_1}^{-1}\big)\Big|(1,0)\Big\rangle=(-64\epsilon_1\epsilon_2)^{-\frac{a_{\D}^2}{2\epsilon_1\epsilon_2}}\mathcal{Z}_{cl-a}^{[\mathbf{2}]}(a_{\D};\epsilon_1,\epsilon_2|t)\mathcal{Z}_{inst}^{[\mathbf{2}]}(a_{\D},m_{1,2};\epsilon_1,\epsilon_2|t) 
    \end{equation}
    with root branch choice $(-\mathrm{t}_{\III_1}^{-1})^{1/2}=(\ri \mathrm{t}_{\III_1}^{1/2})^{-1}$ for $\Delta_{\pm}=\frac{c-1}{24}+\frac{(\theta_*\pm\theta_{\star})^2}4$  under dictionary
    \begin{equation}
     \nu=\frac{a_{\D}}{\sqrt{-\epsilon_1\epsilon_2}},  \qquad \mathrm{t}_{\III_1}^{1/2}=\frac{t^{1/2}}{\sqrt{-\epsilon_1\epsilon_2}}, \qquad \theta_{\star}=\frac{m_1}{\sqrt{-\epsilon_1\epsilon_2}}, \quad  \theta_*=\frac{m_2}{\sqrt{-\epsilon_1\epsilon_2}}.
\end{equation}
This dictionary is a natural quantum deformation of scalings $a_{\D}=2\epsilon_2\nu$, $t=(2\epsilon_2)^2\mathrm{t}_{\III_1}$, $m_f=2\epsilon_2\sm_f|_{f=1,2}$,  according to \eqref{PIII1_dct}.

\paragraph{QPII (linear exp singularity).}
According to \cite[Conj. 3.12]{N18} the Painlev\'e II tau function near $t=\infty$ is given by the Zak transform of (the expansion of) the matrix element of the $c=1$ Virasoro rank-0 ramified irregular vertex operator acting on the rank-3/2 irregular Verma module. The matrix element is taken between the vacuum vector and the rank-3/2 irregular (Whittaker) vector. 
This tau function coincides with \cite[(3.32)]{BLMST16}.

After the quantum ($\epsilon$-) deformation and the corresponding refining of the central charge by \eqref{AGT_dictionary}, we checked up to the power $5$ in $s^{-1}$ of $\mathcal{Z}_{inst}^{[\mathbf{H_{1,L}}]}$ that for $\tilde{\boldsymbol{m}}^2=\boldsymbol{m}^2$
\begin{multline}
   \big(-2^{1/3}\mathrm{t}_{\II}^{-1}\big)^{5\Delta/2}\Big\langle 0\Big|\Phi^{\Delta,4\nu/3}_{(0,1,0), (0,1,0)}\big(-2^{1/3}\mathrm{t}_{\II}^{-1}\big)\Big|(0,1,0)\Big\rangle\\=(-16\epsilon_1\epsilon_2)^{-\frac{a_{\D}^2}{2\epsilon_1\epsilon_2}}\mathcal{Z}_{cl-p}^{[\mathbf{H_{1,L}}]}(a_{\D};\epsilon_1,\epsilon_2|t)\mathcal{Z}_{inst}^{[\mathbf{H_{1,L}}]}(a_{\D},\boldsymbol{m};\epsilon_1,\epsilon_2|t) 
    \end{multline}
    with root branch choice $(-\mathrm{t}_{\II}^{-1})^{3/2}=(\ri \mathrm{t}_{\II}^{3/2})^{-1}$ for $\Delta=\frac{c-1}{24}+\frac{\theta^2}4$
    under dictionary
    \begin{equation}\label{PIIdct}
     \nu=\frac{a_{\D}}{\sqrt{-\epsilon_1\epsilon_2}}, \qquad \mathrm{t}_{\II}^{3/2}=\frac{t^{3/2}}{\sqrt{-\epsilon_1\epsilon_2}}, \qquad \theta^2=\frac{16\boldsymbol{m}^2}{-\epsilon_1\epsilon_2}.
     \end{equation}
     This dictionary is a natural quantum deformation of scalings $a_{\D}=2\epsilon_2\nu$, $t=(2\epsilon_2)^{2/3} \mathrm{t}_{\II}$, $\boldsymbol{m}^2=(2\epsilon_2)^2\theta^2/16$, according to \eqref{PII_dct}.

\section{Further questions}
\label{sec:further}

This paper left many open questions to the authors. Let us list some of them.

\begin{itemize}

\item
We would like to understand better the nature of the quantum deformations of the Painlev\'e equations introduced by passing from self-dual to general $\Omega$-background. The very structure of the solutions we found indicates that the quantization follows from that of the monodromy (Stokes) parameter space of the associated linear system. There are actually two approaches to the quantization: matrix Painlev\'e equations \cite{K15} and canonical quantized Painlev\'e equations \cite{N04,N09}. We could expect that the latter is directly connected with our quantum Painlev\'e equations in the tau form.
Also notice that KZB equations on affine $SL(2)$ conformal blocks have the form of time dependent quantum mechanical wave equations and are a natural quantization of the Painlev\'e equations, as deautonomized quantum Calogero-type two particle systems
\cite{AT10,N17}.
\item
    In the comparison with the holomorphic anomaly equations approach we found a novel compact formula \eqref{F2} for the genus two topological string amplitude on elliptic geometries. It would be interesting to have a direct proof that the chosen $X$-independent term leads to gap condition \eqref{gap_cond} being satisfied. Also it is interesting to obtain a generalization of this compact formula to the higher genera. 
    
\item
    As we discussed in Sec.~\ref{sec:comparison}, interpretations from the viewpoint of the Virasoro algebra representation theory of the strong coupling expansions we found is known only for some cases. Namely, the corresponding irregular states of Liouville field theory or irregular vertex operators are not known yet. 
    The study of irregular Virasoro representations
    is an interesting subject in CFT$_2$ to be further developed.
    
    \item The specific properties of our solutions in the anti-self-dual and Nekrasov-Shatashvili limits are worth to be studied. For example, the anti-self-dual case could shed light on the correlation functions of AD theories on $S^4$. See \cite{BFMMS21,FGMS23} on this topic.

    \item As was presented in Sec.~\ref{ssec:deg_sol}, unlike for the small time expansion, it is hard to follow the renormalization group flow for the partition functions in the large time expansion, thus some new ideas for this analysis are necessary. It is natural to hope that explicit combinatorial formulas exist for the large time expansions as well. These are still unknown and should arise from a proper formulation of the gauge theory in the strong coupling phase.

\item A natural generalization would be to study gauge theories with higher rank simple groups. For self-dual $\Omega$-background these are related to suitable tau forms of nonautonomous Toda system as discussed in \cite{BFT21,BFT22}. The formulation of their quantum version would allow to access the description of the gauge theories in general $\Omega$-background and refined topological strings.

     \item   
In \cite{BGMT24} Hurwitz expansions for the tau functions of Painlev\'e equations around their zeros have been analyzed and proposed to provide
a nonperturbative completion of the related topological string partition function. We do believe that analogous expansions can be worked out for the
quantum Painlev\'e equations and the corresponding refined topological strings. 

\item In \cite{IM23} the resurgent structure of topological string has been analyzed both in general and on the example of the Painlev\'e I case
by giving a closed formula for the action of the Stokes automorphisms on the topological string partition function. 
It would be interesting to explore how to elaborate the extension of such an analysis to the refined topological string case, by considering the quantum Painlev\'e tau functions and their expansions, see also the item below.

 \item As it is well known, a multiplicative (q-) difference version of classical Painlev\'e equations  \cite{S01} is related to five-dimensional supersymmetric gauge theories in the self-dual $\Omega$-background \cite{BS16,BGT16,BGM17}. An obvious question is to promote also this set to the quantum q-difference Painlev\'e equations by studying blowup equations on $\mathbb{C}^2/\mathbb{Z}_2\times S^1$ analogously to the case of $\mathbb{C}^2\times S^1$ \cite{NY05} (which is also connected to $q$-Painlev\'e, see e.g. \cite{BS18,BGMT25}). In this context a comparison with the results of \cite{AMP23} on noncommutative wall-crossing formulae and quantum wave functions, see for example their (4.31), would be welcome.
    
\end{itemize}

\appendix

\section{Notations}
\label{sec:notations}
 Here we collect some notations and abbreviations which we used through the main text. Of course
 this is not the full list, but rather an explanatory note.
 
 \paragraph{$\Omega$-background and vacuum expectation values.}
 \begin{itemize}
  \item $\Omega$-background parameters $\epsilon_1$ and $\epsilon_2$ have mass dimension $1$. Their permutation invariant combinations are $\epsilon=\epsilon_1{+}\epsilon_2$ and $\epsilon_1\epsilon_2$.
  We denote their dimensionless ratio by CFT notation $Q^2=\frac{\epsilon^2}{\epsilon_1\epsilon_2}$.
  \item Superscripts $^{(1)}$ and $^{(2)}$ denote the (tau) function dependence on the $\Omega$-background parameters in the two patches as prescribed by the $\mathbb{C}^2/\mathbb{Z}_2$ blowup relations. In the two toric patches the equivariant weights of the $(\mathbb{C}^*)^2$-action are given respectively by $(2\epsilon_1,\epsilon_2{-}\epsilon_1)$ and $(\epsilon_1{-}\epsilon_2,2\epsilon_2)$. 
  \item In the electric frame the vacuum expectation value is $(a_1, a_2)=(a, -a)$. For the dual, magnetic frame, the corresponding value is $a_{\D}$. Variables $a$ and $a_{\D}$ have mass dimension $1$ and are flat local coordinates on the Coulomb moduli space, the global gauge invariant one being called $u$. Up to a constant shift and the sign, $a$ and $a_{\D}$ are expressed in terms of $u$ via Seiberg-Witten (SW) periods $\omega_{1,2}$ by \eqref{dada} and \eqref{periods_hyph}. We sometimes use notation $\tilde{a}_{\D}=a_{\D}-a_{\D}(u_*)$ referring to zero $u_*$ of SW modular discriminant $\Delta(u)$. 
  \item Dimensionless variable $\nu$ of paper \cite{BLMST16} is related to $a_{\D}$ as $\nu=a_{\D}/\epsilon_2$. Note that for the blowup equations in the self-dual case $\epsilon=0$, the $\Omega$-background parameters are $(-2\epsilon_2,2\epsilon_2)$ both for $\tau^{(1)}$ and $\tau^{(2)}$, so when comparing with \cite{BLMST16} for these tau functions we have rescaled relation $a_{\D}=2\epsilon_2\nu$. 
  We perform the same rescaling also for the masses, see the next paragraph.
 \end{itemize}

  \paragraph{Masses and their invariants}
 \begin{itemize}
  \item We use the standard elementary symmetric polynomials of masses $m_j$, namely
  \begin{equation}
      e_k(m_1,\ldots m_N)=\sum_{1\leq j_1< \ldots < j_k\leq N}m_{j_1} \ldots m_{j_k}, \quad k=1,\ldots N. 
  \end{equation}
  For the Lagrangian theories we use superscript $^{[N_f]}$ to specify the number of masses.
  We also use the elementary symmetric polynomials of the mass squares
   \begin{equation}
  w_{2k}^{[N_f]}=e_k(m_1^2,\ldots m_{N_f}^2), \quad k=1,\ldots N_f,   
  \end{equation}   
  which are insensitive to the sign of any mass.
  \item For the classical Painlev\'e VI, V, III's we use dimensionless mass parameters $\mathrm{m}_f$ and the corresponding invariants
  \begin{multline}
   m_f=2\epsilon_2\mathrm{m}_f,\, f=1,\ldots N_f \qquad \se_k^{[N_f]}=e_k(\sm_1,\ldots,\sm_{N_f}), \quad \mathrm{w}_{2k}^{[N_f]}=w_{2k}(\sm_1,\ldots,\sm_{N_f})\\  \Rightarrow \quad e_k=(2\epsilon_2)^k\se_k, \quad w_{2k}=(2\epsilon_2)^{2k}\mathrm{w}_{2k}.  
  \end{multline}
  \item To parametrize the two mass parameters of Argyres-Douglas (AD) theory $H_2$ we use three linearly dependent masses $\boldsymbol{m}_j$ and their elementary symmetric polynomials
  \begin{equation}
\boldsymbol{e}_k=e_k(\boldsymbol{m}_1,\boldsymbol{m}_2,\boldsymbol{m}_3)|_{k=1,2,3} \quad \textrm{with relation} \quad \boldsymbol{e}_1\equiv0.  
  \end{equation}
For the classical Painlev\'e IV dimensionless mass parameters $\mathfrak{m}_j$ and related invariants $\mathfrak{e}_k$ are  
\begin{equation}
\boldsymbol{m}_j=2\epsilon_2\mathfrak{m}_j,\qquad \mathfrak{e}_k=e_k(\mathfrak{m}_1,\mathfrak{m}_2,\mathfrak{m}_3)|_{k=1,2,3}.
\end{equation}
Analogously, for AD theory $H_1$ (Painlev\'e II) we use masses $\boldsymbol{m}$ ($\mathfrak{m}$) connected by $\boldsymbol{m}=2\epsilon_2\mathfrak{m}$.   
\end{itemize}

 \paragraph{Partition functions and tau functions.}
 \begin{itemize}
  \item When it is necessary to specify the theory we are referring to, we use superscripts $^{[N_f]}$ for instanton counting parameters $t$, partition functions $\mathcal{Z}$, tau functions $\tau$. For the same purposes for the AD theories we use superscripts $^{[H_{2,1,0}]}$. For the strong coupling/large time expansions of the partition functions we use the boldface versions of these superscripts, like $^{\mathbf{[2]}}$ or $^{\mathbf{[H_0]}}$. To distinguish the large time expansions with different exponent singularity asymptotics we add marks $\mathbf{L}$ (linear) and $\mathbf{S}$ (square).
  \item We factorize each partition function as $\mathcal{Z}=\mathcal{Z}_{cl}\mathcal{Z}_{1-loop}\mathcal{Z}_{inst}$.  Factor $\mathcal{Z}_{cl}$ is a product of powers and, possibly, exponents of powers of $t$. Factor $\mathcal{Z}_{1-loop}$ is independent of $t$ and, up to a simple exponent prefactor, is a product of double Gamma functions $\gamma_{\epsilon_1,\epsilon_2}$ (see App.~\ref{sec:gamma}). Factor $\mathcal{Z}_{inst}$ is an (asymptotic) expansion in powers of $t$. 
  \item The asymptotic large time expansions of $\mathcal{Z}_{inst}$ are in $s^{-1}$ for $s=\kappa t^d$, $\kappa\in\mathbb{C}$.  This variable $s$ has mass dimension $1$. Then $s$ and $t$ are connected with corresponding Painlev\'e dimensionless variables $\mathrm{s}_{\mathrm{eq}}, \mathrm{t}_{\mathrm{eq}},\,\, \mathrm{eq}=\mathrm{V},\mathrm{III}_{1,2,3},\mathrm{IV},\mathrm{II},\mathrm{I}$ by 
  \begin{equation}
  s^{[th]}=2\epsilon_2 \mathrm{s}_{\mathrm{eq}},\quad t^{[th]}=(2\epsilon_2)^{1/d}\mathrm{t}_{\mathrm{eq}},
  \end{equation}
  where $\mathrm{eq}$ is related to $th$ via Table \ref{tab:theories}.
  There is no scaling in the (quantum) PVI case, i.e. $t^{[4]}=\mathrm{t}_{\VI}$. 
  \item We use notation $\tau_{\mathrm{eq}}$ with above $\mathrm{eq}$ to refer to the Painlev\'e tau functions of App.~\ref{sec:Painleve}.
 \end{itemize}
 \paragraph{Other notations}
 \begin{itemize}
     \item Generalized Hirota derivative $D^n_{\epsilon_1,\epsilon_2[x]}$ is given by \eqref{Hirota}, its standard version is $D^n_{[x]}= D^n_{1,-1[x]}$.
     \item For the blowup relations and $X_2$ partition functions we use a label for the parity $\mathfrak{p}=0,1$ as well as the corresponding parity function $\mathfrak{p}(x)=x\mod2$.
 \end{itemize}
\section{Double gamma function \texorpdfstring{$\gamma_{\epsilon_1,\epsilon_2}$}{γε₁ε₂}}
\label{sec:gamma}
\paragraph{Definition.} 
According to \cite{NY03L}, double gamma function $\gamma_{\epsilon_1,\epsilon_2}(x)$ is formally defined as 
\begin{equation} \label{gammaNYdef}
\gamma_{\epsilon_1,\epsilon_2}(x):=
   \frac{d}{ds}\Bigg|_{s=0}\frac1{\Gamma(s)}\int\limits_0^{+\infty}\frac{dz}{z} z^s \frac{e^{-xz}}{(e^{\epsilon_1z}-1)(e^{\epsilon_2z}-1)},\qquad \mathrm{Re}(\epsilon_{1,2})\neq0, \quad \mathrm{Re}(x)>0,
\end{equation}
where the integral should be understood as an analytic continuation in $s$ from region $\mathrm{Re}(s)>2$, where it (absolutely) converges at $0$ and gives an analytic function in $s$, to the vicinity of $s=0$.   
The standard analytic continuation of the integral to region $\mathrm{Re}(s)>1-K, \, K\in\mathbb{N}$ is as follows 
\begin{equation}
\int\limits_0^{+\infty}\frac{dz}{z} z^s \frac{e^{-xz}}{(e^{\epsilon_1z}{-}1)(e^{\epsilon_2z}{-}1)}=\int\limits_0^{+\infty}\frac{dz}{z} z^s e^{-xz}\left(\frac{1}{(e^{\epsilon_1z}{-}1)(e^{\epsilon_2z}{-}1)}-\sum_{n=0}^K \frac{c_n}{n!} z^{n-2} \right)+\sum_{n=0}^K \frac{c_n}{n!} \frac{\Gamma(n{+}s{-}2)}{x^{n+s-2}}, 
\end{equation}
where 
\begin{equation}\label{den_Taylor}
\frac1{(e^{\epsilon_1z}-1)(e^{\epsilon_2z}-1)}=\sum_{n=0}^{K}\frac{c_n}{n!} z^{n-2}+O(z^{K-1}), \quad \textrm{with} \quad c_n=\sum_{i=0}^n C^i_n B_i B_{n-i}\epsilon_1^{i-1}\epsilon_2^{n-i-1}.    
\end{equation}
in terms of the Bernoulli numbers $B_i$.
For the analytic continuation of the integral to the vicinity of $s=0$ we should take $K\geq2$. Then we can take the derivative in $s$ explicitly,
using expansion $\Gamma^{-1}(s)=s+O(s^2)$
\begin{equation}\label{gamma_s_cont}
\gamma_{\epsilon_1,\epsilon_2}(x)=\frac{3{-}2\ln x}{4\epsilon_1\epsilon_2}x^2+\epsilon\frac{1{-}\ln x}{2\epsilon_1\epsilon_2}x-\frac{\epsilon_1\epsilon_2{+}\epsilon^2}{12\epsilon_1\epsilon_2}\ln x+\sum_{n=3}^K \frac{c_n}{n(n{-}1)(n{-}2)}x^{2-n}+R_K,
\end{equation}
\begin{equation}
\label{R_K}
\textrm{where}\qquad R_K:=\int\limits_0^{+\infty}\frac{dz}{z} e^{-xz}\left(\frac{1}{(e^{\epsilon_1z}{-}1)(e^{\epsilon_2z}{-}1)}-\sum_{n=0}^K \frac{c_n}{n!} z^{n-2} \right).
\end{equation}
Finally we can analytically continue $R_K$ (and $\gamma_{\epsilon_1,\epsilon_2}(x)$) in $x$. For $\mathrm{Re}(x)>0$ we substitute $z\mapsto z/x$ and rotate the contour back to the real nonnegative half-line
\begin{equation}\label{R_K_x_cont}
R_K=\int\limits_0^{+\infty}\frac{dz}{z} e^{-z}\left(\frac{1}{(e^{\frac{\epsilon_1}xz}{-}1)(e^{\frac{\epsilon_2}xz}{-}1)}-\sum_{n=0}^K \frac{c_n}{n! x^{n-2}} z^{n-2} \right),    
\end{equation}
due to a Jordan lemma like calculation
\begin{equation}\label{Jordan}
R^{K-2}\left|\int\limits_0^{\arg(x)}d\phi\, e^{-R\cos\phi}\right|\leq \frac{\pi R^{K-3}}{2\sin|\arg(x)/2|} e^{-R \cos\arg(x)}\rightarrow0. 
\end{equation}
Now term $R_K$ is defined for $\mathrm{Re}\left(\frac{x}{\epsilon_{1,2}}\right)\neq0$. Note also that there should be a cut for the logarithms in \eqref{gamma_s_cont}, for that we can choose the real negative axis in $x$.

\paragraph{Homogeneity.}
Term $R_K$ given by \eqref{R_K_x_cont} is manifestly invariant under scaling $x\mapsto\Lambda x,\, (\epsilon_1,\epsilon_2)\mapsto(\Lambda\epsilon_1,\Lambda\epsilon_2)$ as well as the power sum in $x$ in \eqref{gamma_s_cont}. Then, for the whole $\gamma_{\epsilon_1,\epsilon_2}$ given by \eqref{gamma_s_cont} we have
\begin{equation}\label{gamma_homogen}
\exp\gamma_{\Lambda\epsilon_1,\Lambda\epsilon_2}(\Lambda x)=\Lambda^{-\frac{(x{+}\epsilon/2)^2+\epsilon_1\epsilon_2/6{-}\epsilon^2/12}{2\epsilon_1\epsilon_2}}\exp\gamma_{\epsilon_1,\epsilon_2}(x).    
\end{equation}
This gives the mass dimension scaling for the 1-loop terms \eqref{Z1loop_scale}.

\paragraph{Difference relation.}
From (the analytic continuation of)  definition \eqref{gammaNYdef} it follows difference relation
\begin{equation}
\label{gammablowup}
\frac{\exp \gamma_{2\epsilon_1,\epsilon_2{-}\epsilon_1}(x+2n\epsilon_1) \exp\gamma_{\epsilon_1{-}\epsilon_2,2\epsilon_2}(x+2n\epsilon_2)}{\exp\gamma_{2\epsilon_1,\epsilon_2{-}\epsilon_1}(x{+}\mathfrak{p}(2n)\epsilon_1)\exp\gamma_{\epsilon_1{-}\epsilon_2,2\epsilon_2}(x{+}\mathfrak{p}(2n)\epsilon_2)}=\prod\limits_{\mathrm{reg}(2n)} \big(x{+}\epsilon/2+\mathrm{sgn}(n) (i \epsilon_1{+}j \epsilon_2)\big),  \quad 2n\in\mathbb{Z}, 
\end{equation}
where $\mathfrak{p}(2n)=2n \mod2$ is the parity of $n$ and the region of the product is
\begin{equation}
\mathrm{reg}(2n)=\Big\{i,j\in \mathbb{Z}_{\geq 0}{+}\frac12: \quad
i+j\leq 2|n|{-}1, \quad i+j\equiv 2n{+}1 \mod 2\Big\}.
\end{equation}

\paragraph{Asymptotic series.} 
Formula \eqref{gamma_s_cont} describing the
analytic continuation actually gives us also the asymptotic series of $\gamma_{\epsilon_1,\epsilon_2}(x)$ in  $\epsilon_{1,2}/x\rightarrow0$, while keeping $\epsilon_1/\epsilon_2$ finite.
Indeed, below we prove that remainder $R_K$ given by \eqref{R_K_x_cont} is $O\left(\left(\frac{\epsilon_{1,2}}x\right)^{K-1}\right)$. To bound remainder $R_K$ we use \eqref{den_Taylor} for any $K\geq2$
\begin{equation}\label{Taylor_rem}
\exists C,r>0  \quad \forall z,\, |z|<r\,: \left|\frac1{(e^{\frac{\epsilon_1}{\epsilon_2}z}-1)(e^z-1)}-\sum_{n=0}^{K}\frac{cp_n(\epsilon_1/\epsilon_2)}{n!} z^{n-2}\right|\leq C(\epsilon_1/\epsilon_2)|z|^{K-1},
\end{equation}
where $cp_n(\epsilon_1/\epsilon_2)=c_n(\epsilon_1,\epsilon_2)/\epsilon_2^{n-2}$ is a polynomial in $\epsilon_1/\epsilon_2$ with numerical coefficients.
We rescale the remainder $R_K$ integral 
\begin{equation}
R_K=\int\limits_0^{+\infty}\frac{dz}{z} e^{-\left|\frac{x}{\epsilon_2}\right|z}\left(\frac{1}{(e^{\frac{\epsilon_1}{\epsilon_2}\frac{\epsilon_2|x|}{|\epsilon_2|x}z}{-}1)(e^{\frac{\epsilon_2|x|}{|\epsilon_2|x}z}{-}1)}-\sum_{n=0}^K \frac{cp_n(\epsilon_1/\epsilon_2)}{n!} \left(\frac{\epsilon_2|x|}{|\epsilon_2|x}z\right)^{n-2} \right)
\end{equation}
and split it in two parts according to \eqref{Taylor_rem}. These are  separately bounded as
\begin{multline}
|R_K|\leq C(\epsilon_1/\epsilon_2)\int\limits_0^r dz\, z^{K-2} e^{-\left|\frac{x}{\epsilon_2}\right|z}+\int\limits_r^{+\infty}\frac{dz}{z} \frac{e^{-\left|\frac{x}{\epsilon_2}\right|z}}{\left|e^{\frac{\epsilon_1}{\epsilon_2}\frac{\epsilon_2|x|}{|\epsilon_2|x}z}{-}1\right|\cdot\left|e^{\frac{\epsilon_2|x|}{|\epsilon_2|x}z}{-}1\right|}+\sum_{n=0}^K\frac{|cp_n(\epsilon_1/\epsilon_2)|}{n!}\int\limits_r^{+\infty}dz\, z^{n-3} e^{-\left|\frac{x}{\epsilon_2}\right|z}  \\
\leq C(\epsilon_1/\epsilon_2)(K{-}2)!\left|\frac{\epsilon_2}{x}\right|^{K-1}+\left(\frac{\left|\frac{\epsilon_2}{x}\right|^3/r^3}{\mathrm{Re}\left(\frac{\epsilon_1}x\right)\mathrm{Re}\left(\frac{\epsilon_2}x\right)}+\sum_{n=0}^K\frac{|cp_n(\epsilon_1/\epsilon_2)|}{n!r^{3-n}}\left|\frac{\epsilon_2}{x}\right|\mathrm{Pl}_{(n{-}3)}\left(\left|\frac{\epsilon_2}{x}\right|/r\right)\right)e^{-\left|\frac{x}{\epsilon_2}\right|r},
\end{multline}
where $\mathrm{Pl}_{(n-3)}$ is a polynomial of degree $n{-}3$ for $n\geq3$ and a constant otherwise.
The second term is exponentially suppressed, so $R_K$ is $O\left(\left(\frac{\epsilon_{1,2}}x\right)^{K-1}\right)$ as expected. Then from \eqref{gamma_s_cont} for $K=4$, we obtain a shifted asymptotical formula which we found useful in this paper
\begin{equation}\label{gamma_as}
\gamma_{\epsilon_1,\epsilon_2}(a-\epsilon/2)=\frac{3-2\ln a}{4\epsilon_1\epsilon_2} a^2+\frac{\epsilon^2{-}2\epsilon_1\epsilon_2}{24\epsilon_1\epsilon_2}\ln a+\frac{7\epsilon^4{-}28\epsilon_1\epsilon_2\epsilon^2{+}24(\epsilon_1\epsilon_2)^2}{5760\epsilon_1\epsilon_2\cdot a^2}+O\left(\left(\frac{\epsilon_{1,2}}{a}\right)^4\right).
\end{equation}

\paragraph{Relation to double gamma function $\Gamma_2$.}
Double gamma function $\gamma_{\epsilon_1,\epsilon_2}(x)$ as a function of $\epsilon_1,\epsilon_2$ is defined by (the analytic continuation of) \eqref{gammaNYdef} on four path-components determined by the signs of $\mathrm{Re}(\epsilon_{1,2})$. Its values on these components are related by 
\begin{equation}\label{gamma_sectors}
\gamma_{\epsilon_1,\epsilon_2}(x)=-\gamma_{-\epsilon_1,\epsilon_2}(x+\epsilon_1)=-\gamma_{\epsilon_1,-\epsilon_2}(x+\epsilon_2)=\gamma_{-\epsilon_1,-\epsilon_2}(x+\epsilon).
\end{equation}
There are an alternative double sum representation for the integral of \eqref{gammaNYdef}. Namely, for $\mathrm{Re} (s)>2$ and, without loss of generality, in path-component $\mathrm{Re}(\epsilon_{1,2})<0$
\begin{equation}
\frac1{\Gamma(s)}\int\limits_0^{+\infty}\frac{dz}{z} z^s \frac{e^{-xz}}{(e^{\epsilon_1z}{-}1)(e^{\epsilon_2z}{-}1)}=\frac1{\Gamma(s)}\int\limits_0^{+\infty}\frac{dz}{z} z^s \sum_{n_1,n_2=0}^{+\infty} e^{-(x{-}n_1\epsilon_1{-}n_2\epsilon_2)z}=\sum_{n_1,n_2=0}^{+\infty}\frac1{(x{-}n_1\epsilon_1{-}n_2\epsilon_2)^s}.
\end{equation}
For this sum, called the Barnes zeta function, condition $\mathrm{Re}(x)>0$ is no longer important except possible branch issues. This function can be analytically continued to $s\in\mathbb{C}\setminus\{1,2\}$ and gives the usual relation between $\gamma_{\epsilon_1,\epsilon_2}$ and standard double gamma function $\Gamma_2$
\begin{equation}\label{gamma_Gamma_2}
\exp\gamma_{\epsilon_1,\epsilon_2}(x)=
\exp\left(\frac{d}{ds}\Bigg|_{s=0} \sum_{n_1,n_2=0}^{+\infty} \frac1{(x-n_1\epsilon_1-n_2\epsilon_2)^s}\right)=:\Gamma_2(x;-\epsilon_1,-\epsilon_2).
\end{equation}
Function $\Gamma_2(x;-\epsilon_1,-\epsilon_2)$ is defined for general $x\in\mathbb{C}$ excluding simple poles $x=n_1\epsilon_1+n_2\epsilon_2,\, n_1,n_2\in\mathbb{Z}_{\geq0}$ but, unlike $\gamma_{\epsilon_1,\epsilon_2}$, only in region $\mathrm{Re}(\epsilon_{1,2})<0$.

\paragraph{Relation to Barnes $\mathsf{G}$-function for $\epsilon=0$.}
Integrating by parts integral \eqref{R_K} in formula \eqref{gamma_s_cont} for $\epsilon=0$ with $\mathrm{Re}(\epsilon_2)>0$ and $K=2$ we obtain the Binet integral representation of the Barnes $\mathsf{G}$-function \cite[Prop. 4.1]{A03}\footnote{Formula  \cite[(25)]{A03} seems to have a wrong sign in front of the integral, this sign seems missed in the proof of Prop. 4.1 while calculating $\zeta'_s(s,z)$ at point $s=-1$.} for $\mathrm{Re}(x)>0$
\begin{multline}
\gamma_{-\epsilon_2,\epsilon_2}(x)=\int\limits_0^{+\infty}\frac{dz}{z} e^{-\frac{x}{\epsilon_2}z}\left(\frac{1}{(e^{-z}{-}1)(e^{z}{-}1)}+\frac1{z^2}{-}\frac1{12}\right)-\frac{3{-}2\ln x}4\left(\frac{x}{\epsilon_2}\right)^2-\frac{\ln x}{12}\\
=\int\limits_0^{+\infty}\frac{dz}{z} e^{-\frac{x}{\epsilon_2}z}\left(\frac1{z}+\frac{x}{\epsilon_2}\right)\left(\frac1{1-e^{-z}}-\frac1{z}-\frac12-\frac{z}{12}\right)-\frac{3{-}2\ln x}4\left(\frac{x}{\epsilon_2}\right)^2-\frac{\ln x}{12}\\
=\ln\mathsf{G}\left(1+\frac{x}{\epsilon_2}\right)+\frac12\left(\left(\frac{x}{\epsilon_2}\right)^2{-}\frac16\right)\ln\epsilon_2-\frac{x}{2\epsilon_2}\ln(2\pi)-\zeta'(-1),
\end{multline}
where we used the classical Binet first integral formula for the gamma function and \eqref{Jordan}. This relation can be continued to the whole complex plain in $x$ (except the negative real axis cut) and, due to \eqref{gamma_Gamma_2}, \eqref{gamma_sectors}, reproduces the well-known relation between $\Gamma_2(x;\epsilon_2,\epsilon_2)$ and the Barnes $\mathsf{G}$-function
\begin{equation}\label{gamma_Barnes}
\Gamma_2^{-1}(x{+}\epsilon_2;\epsilon_2,\epsilon_2)=
\exp \big(-\gamma_{-\epsilon_2,-\epsilon_2}(x{+}\epsilon_2)\big)=\exp \gamma_{-\epsilon_2,\epsilon_2}(x)=\frac{\epsilon_2^{\frac12\left(\frac{x}{\epsilon_2}\right)^2{-}\frac1{12}}}{e^{\zeta'(-1)}(2\pi)^{\frac{x}{2\epsilon_2}}}\,\mathsf{G}\left(1+\frac{x}{\epsilon_2}\right).    
\end{equation}

\section{\texorpdfstring{$\mathbb{C}^2/\mathbb{Z}_2$}{ℂ²/ℤ₂} blowup relations from representation theory}
\label{sec:BS14}
In this appendix we start by reviewing necessary results of 
\cite{BS14}, namely the relations that express $4$-point
$\mathcal{N}=1$ Super Virasoro conformal block as a bilinear combination of $4$-point Virasoro conformal blocks. According to the famous AGT relation, the $4$-point Virasoro conformal blocks are equal to the $N_f=4$ SUSY $\mathcal{N}=2$ $SU(2)$ partition functions \cite{AGT09}. Analogously, the $4$-point
$\mathcal{N}=1$ Super Virasoro conformal blocks are equal to  the $\mathbb{C}^2/\mathbb{Z}_2$ (ALE space) partition functions \cite{BF11, BMT1106,BMT1107,BBB11}. So we finish this appendix by using these equalities to write the $\mathbb{C}^2/\mathbb{Z}_2$ blowup relations on the partition functions that we use in the main text.

\subsection{Relations on conformal blocks from Super Virasoro representations}
\label{ssec:CFT}
\paragraph{Virasoro and Super Virasoro algebras and their Verma modules.}
The Virasoro algebra ($\mathsf{Vir}$) is an infinite dimensional algebra defined by OPE
\begin{equation}\label{Vir}
L(z)L(w)=\frac{c/2}{(z-w)^4}+\frac{2L(w)}{(z-w)^2}+\frac{\partial L(w)}{z-w}+reg 
\end{equation}
for current $L(z)=\sum_{n\in\mathbb{Z}}L_n z^{-n-2}$. The Super Virasoro or Neweu-Schwarz-Ramond ($\mathsf{NSR}$) algebra
is the $\mathcal{N}=1$ superextension of the Virasoro algebra 
by odd current $G(z)=\sum_{r\in\mathbb{Z}+\frac12}G_{r}z^{-r-\frac32}$ with OPE
\begin{eqnarray}
G(z)G(w)=\frac{c^{\NS}}{(z-w)^3}+\frac{2L(w)}{z-w}+reg,\\
L(z)G(w)=\frac{3/2 G(w)}{(z-w)^2}+\frac{\partial G(w)}{z-w}+reg,
\end{eqnarray}
where for the extension we take $c^{\NS}=3c/2$. Notation
$\mathsf{NS}$ here and below means that we consider the so-called Neweu-Schwarz sector of the algebra, where the indices of $G_r$ are half-integer, while in the Ramond sector they are integer. 

Verma module $\pi^{\Delta}_{\mathsf{Vir}}$ of $\mathsf{Vir}$ with highest weight $\Delta$ is generated by highest weight vector $|\Delta\rangle$ such that
\begin{equation}
L_0|\Delta\rangle=\Delta|\Delta\rangle, \quad L_n|\Delta\rangle=0, \quad n>0.
\end{equation}
Analogously, $\mathsf{NSR}$ algebra Verma module $\pi^{\Delta^{\NS}}_{\mathsf{NSR}}$ with highest weight $\Delta^{\NS}$ is generated by highest weight vector 
$|\Delta^{\NS}\rangle$ such that
\begin{equation}
L_0|\Delta^{\NS}\rangle=\Delta^{\NS}|\Delta^{\NS}\rangle; \qquad L_n|\Delta^{\NS}\rangle=0, \quad G_r|\Delta^{\NS}\rangle=0, \quad n,r>0.
\end{equation}

\paragraph{Conformal blocks of Virasoro and Super Virasoro algebras.}
We define the scalar product (Shapovalov form) on $\pi^{\Delta}_{\mathsf{Vir}}$ by  conjugation $L_n^+=L_{-n}$ normalized as $\langle\Delta|\Delta\rangle=1$. Then we can define the $4$-point $\mathsf{Vir}$ conformal blocks as
\begin{equation}\label{Vir_conf_block}
\mathcal{F}_c(\Delta_{0,t,1,\infty},\Delta|t)=t^{\Delta_0+\Delta_t}\big\langle\Delta_{\infty}\big|V^{\Delta_1}_{\Delta_{\infty}, \Delta}(1) V^{\Delta_t}_{\Delta, \Delta_0}(t)\big|\Delta_0\big\rangle,    
\end{equation}
which are the matrix elements of vertex operators $V^{\Delta}_{\Delta_1,\Delta_2}\colon \pi^{\Delta_2}_{\mathsf{Vir}} \mapsto \pi^{\Delta_1}_{\mathsf{Vir}}$, defined by
\begin{equation}\label{vertcommrel}
[L_k,V^{\Delta}_{\Delta_1,\Delta_2}(z)]=z^k\big(z\partial_z +(k{+}1)\Delta\big)V^{\Delta}_{\Delta_1,\Delta_2}(z) 
\end{equation}
normalized as $\big\langle \Delta_2\big|V^{\Delta}_{\Delta_2,\Delta_1}(1)\big| \Delta_1\big\rangle=1$.

For the $\mathsf{NSR}$ superextension to define the scalar product we specify conjugation $G_r^+=G_{-r}$ and  analogous normalization $\langle\Delta^{\NS}|\Delta^{\NS}\rangle=1$. In this case there are two vertex operators: even $\Phi^{\Delta^{\NS}}_{\Delta_1^{\NS},\Delta_2^{\NS}}(z)$ and  odd $\Psi^{\Delta^{\NS}}_{\Delta_1^{\NS},\Delta_2^{\NS}}(z)=\big[G_{-1/2},\Phi^{\Delta^{\NS}}_{\Delta_1^{\NS},\Delta_2^{\NS}}(z)\big]$ acting from $\pi^{\Delta_2^{\NS}}_{\mathsf{NSR}}$ to $\pi^{\Delta_1^{\NS}}_{\mathsf{NSR}}$. They are defined by 
\begin{equation}
\big[L_k,\Phi^{\Delta^{\NS}}_{\Delta_1^{\NS},\Delta_2^{\NS}}(z)\big]=z^k\big(z\partial_z +(k{+}1)\Delta^{\NS}\big)\Phi^{\Delta^{\NS}}_{\Delta_1^{\NS},\Delta_2^{\NS}}(z), \qquad \big[G_r,\Phi^{\Delta^{\NS}}_{\Delta_1^{\NS},\Delta_2^{\NS}}(z)\big]=z^{r+\frac12}\Psi^{\Delta^{\NS}}_{\Delta_1^{\NS},\Delta_2^{\NS}}(z)     
\end{equation}
with normalizations $\big\langle \Delta_2^{\NS}\big|\Phi^{\Delta^{\NS}}_{\Delta_2^{\NS},\Delta_1^{\NS}}(1)\big| \Delta_1^{\NS}\big\rangle=\big\langle \Delta_2^{\NS}\big|\Psi^{\Delta^{\NS}}_{\Delta_2^{\NS},\Delta_1^{\NS}}(1)\big| \Delta_1^{\NS}\big\rangle=1$. Using these vertex operators, we define two $4$-point $\mathsf{NSR}$ conformal blocks
\begin{align}
\mathcal{F}_{c^{\NS}}(\Delta^{\NS}_{0,t,1,\infty},\Delta^{\NS}|t)&=t^{\Delta_0^{\NS}+\Delta_t^{\NS}}\Big\langle\Delta_{\infty}\Big|\Phi^{\Delta_1^{\NS}}_{\Delta_{\infty}^{\NS}, \Delta^{\NS}}(1) \Phi^{\Delta_t^{\NS}}_{\Delta^{\NS}, \Delta_0^{\NS}}(t)\Big|\Delta_0^{\NS}\Big\rangle,\\
\widetilde{\mathcal{F}}_{c^{\NS}}(\Delta^{\NS}_{0,t,1,\infty},\Delta^{\NS}|t)&=t^{\Delta_0^{\NS}+\Delta_t^{\NS}+\frac12}\Big\langle\Delta_{\infty}\Big|\Psi^{\Delta_1^{\NS}}_{\Delta_{\infty}^{\NS}, \Delta^{\NS}}(1) \Psi^{\Delta_t^{\NS}}_{\Delta^{\NS}, \Delta_0^{\NS}}(t)\Big|\Delta_0^{\NS}\Big\rangle.
\end{align}

\paragraph{Relations on the conformal blocks.}
Let us extend the $\mathsf{NSR}$ algebra by a direct sum with free Majorana fermion $\mathsf{F}:f(z)=\sum_{r\in\mathbb{Z}+\frac12}f_r z^{-r-\frac12}$ with OPE
\begin{equation}
f(z)f(w)=\frac1{z-w}+reg.    
\end{equation}
Then, according to \cite{CPSS90}, we have an embedding of two commuting copies of $\mathsf{Vir}$ into a certain completion of universal enveloping algebra $\mathsf{F}\oplus\mathsf{NSR}$. These $\mathsf{Vir}$ are explicitly given by 
\begin{equation}
\left(1{-}(b^2)^{(-1)^{s{+}1}}\right)L^{\scriptscriptstyle{(s)}}(z)=L(z)-\left(1/2{+}(b^2)^{(-1)^{s{+}1}}\right) :f'(z)f(z):+b^{(-1)^{s{+}1}}f(z)G(z), \quad s=1,2, 
\end{equation}
where 
\begin{equation}
c^{\NS}=1+2(b+b^{-1})^2.
\end{equation}
The central charges of these $s=1,2$ $\mathsf{Vir}$ are given by 
\begin{equation}
  c^{(s)}=1+6\big(b^{(s)}{+}1/b^{(s)}\big)^2 \quad 
{\rm{with}}\qquad (b^{(s)})^2=\frac12\left((b^2)^{(-1)^s}-1\right).
\end{equation}
Below we also use the momentum ($P$) parametrization of the $\mathsf{NSR}$ Verma module highest weights
\begin{equation}
\Delta^{\NS}=\frac{(b+b^{-1})^2}8-\frac{P^2}2, \qquad  \Delta^{\NS}_p=\frac{(b+b^{-1})^2}8-\frac{P_p^2}2, \quad p=0,t,1,\infty.   
\end{equation}

Using this embedding and the induced Verma module decomposition, found in \cite{BBFLT11}, in \cite{BS14} certain relations between bilinear differential combinations of $\mathsf{Vir}$ conformal blocks and $\mathsf{NSR}$ conformal blocks were obtained. 
To present them here, for given $b$ and $\Delta^{\NS}$ let us define the following bilinear expression
\begin{equation}
\label{FhatHirota}
\widehat{\mathcal{F}_k}=\sum\limits_{2n\in\mathbb{Z}}\left(\mathfrak{l}_n^{\mathsf{CFT}}  D^k_{b,b^{-1}[\ln t]}(\mathcal{F}^{(1)}_n(t),\mathcal{F}^{(2)}_n(t))\right),
\end{equation}
where $\mathfrak{l}_n^{\mathsf{CFT}}$ are 
given by
\begin{equation}\label{l_n_CFT}
\mathfrak{l}_n^{\mathsf{CFT}}=(-2)^{\mathfrak{p}(2n)}\prod_{\pm}\frac{\prod\limits_{\lambda=\pm1}\prod\limits_{\mathrm{reg}(2n)} \Big(P_t{+}\lambda P_0\pm(P+\mathrm{sgn}(n) (i b{+}j b^{-1}))\Big) \Big(P_1{+}\lambda P_{\infty}\pm(P+\mathrm{sgn}(n) (i b{+}j b^{-1}))\Big)}{\prod\limits_{\mathrm{reg}(4n)} \Big(2P\pm (b{+}b^{-1})/2+\mathrm{sgn}(n) (i b+j b^{-1})\Big)}    
\end{equation}
and $\mathcal{F}^{(s)}_n$ are the conformal blocks of the commuting $s=1,2$ $\mathsf{Vir}$ with the following parameters
\begin{equation}
\mathcal{F}^{(s)}_n=\mathcal{F}_{c^{(s)}}\left(\frac{\Delta_{0,t,1,\infty}^{\NS}}{1-(b^2)^{(-1)^{s{+}1}}},\frac{\Delta^{\NS}(P+2nb^{(-1)^{s{+}1}})}{1-(b^2)^{(-1)^{s{+}1}}}\Big|t\right).
\end{equation}
For these bilinear combinations in \cite[App. A]{BS14} it was proved that
\begin{equation}\label{Fhat0123}
\widehat{\mathcal{F}_0}=\mathcal{F}_{\NS}, \qquad
\widehat{\mathcal{F}_1}=0, \qquad
\widehat{\mathcal{F}_2}=-\frac1{1-t}\,t^{\frac12}\widetilde{\mathcal{F}_{\NS}}, \qquad
\widehat{\mathcal{F}_3}=-(b{+}b^{-1})\frac{1+t}{(1-t)^2}\,t^{\frac12}\widetilde{\mathcal{F}_{\NS}}
\end{equation}
as well as
\begin{multline}\label{Fhat4}
\widehat{\mathcal{F}_4}=2\frac{1{+}t}{(1{-}t)^2}\frac{d}{d\ln t}\left(t^{\frac12}\widetilde{\mathcal{F}_{\NS}}\right)-\frac{(b{+}b^{-1})^2(1{+}4t{+}t^2)+1{-}3t{-}t^2+2t(1{-}t)(\Delta_0^{\NS}{+}\Delta_{\infty}^{\NS})-2t(\Delta_t^{\NS}{+}\Delta_1^{\NS})}{(1-t)^3}t^{\frac12}\widetilde{\mathcal{F}_{\NS}}\\-\frac{t}{(1{-}t)^2}\,\frac{d^2}{d(\ln t)^2}\mathcal{F}_{\NS}
+\frac{(\Delta_0^{\NS}{+}\Delta_{\infty}^{\NS})(1{-}t)-(\Delta_t^{\NS}{+}\Delta_1^{\NS})(1{+}t)}{(1-t)^3}t\,\frac{d}{d\ln t}\mathcal{F}_{\NS}\\-\frac{(\Delta_t^{\NS}{-}\Delta_0^{\NS})(\Delta_1^{\NS}{-}\Delta_{\infty}^{\NS})-t(\Delta_t^{\NS}{+}\Delta_0^{\NS})(\Delta_1^{\NS}{+}\Delta_{\infty}^{\NS})}{(1-t)^3}t\mathcal{F}_{\NS}.
\end{multline}
Note that each relation can be decomposed in two independent ones.
Namely, as the conformal blocks behave around $t=0$ as follows
\begin{equation}
\mathcal{F}_c(\Delta|t)=t^{\Delta}\big(1+t\,\mathbb{C}[[t]]\big), \qquad  (\mathcal{F},\widetilde{\mathcal{F}})_{c^{\NS}}(\Delta^{\NS}|t)=t^{\Delta^{\NS}}\big(1+t^{\frac12}\,\mathbb{C}[[t^{\frac12}]]\big), 
\end{equation}
and for the prefactors of left and right hand sides there is relation
\begin{equation}\label{DNS=D+D}
\frac{\Delta^{\NS}(P+2nb)}{1-b^2}+\frac{\Delta^{\NS}(P+2nb^{-1})}{1-b^{-2}}=\Delta^{\NS}+2n^2    
\end{equation}
hence each such relation is a formal series in $t^{\frac12}$, up to common factor $t^{\Delta^{\NS}}$ . Thus we can split these relations in the two independent ones --- with integer and half-integer powers of $t$ respectively. Each sum of type \eqref{FhatHirota} therefore split in the two ones --- with integer and half-integer $n$ respectively. 

\subsection{Relations on conformal blocks as blowup relations}
\label{ssec:AGT}
\paragraph{AGT relation for Virasoro and Super Virasoro cases.}
According to the famous AGT correspondence~\cite{AGT09} the $4$-point $\mathsf{Vir}$ conformal block and the $N_f=4$ SUSY $\mathcal{N}=2$ $SU(2)$ partition functions are related by
\begin{equation}\label{AGT_block}
\mathcal{F}_c(\Delta_{0,t,1,\infty},\Delta|t)=(1{-}t)^{-\frac{2h}{\epsilon_1\epsilon_2}}\mathcal{Z}^{[4]}_{cl}(a;\epsilon_1,\epsilon_2|t)\mathcal{Z}^{[4]}_{inst}(a,m_{1,2,3,4};\epsilon_1,\epsilon_2|t)  
\end{equation}
where $4h=(m_1{+}m_3{+}\epsilon) (m_4{+}m_2{+}\epsilon)$ and identification dictionary is
\begin{equation}
\label{AGT_dictionary}   
c=1+6Q^2, \quad Q^2= \frac{\epsilon^2}{\epsilon_1\epsilon_2}, \qquad   \Delta=\frac{\epsilon
^2-4a^2}{4\epsilon_1\epsilon_2},
\end{equation}
\begin{equation}
\label{AGT_dictionary_masses}  
\Delta_0=\frac{\epsilon^2{-}(m_1{-}m_3)^2}{4\epsilon_1\epsilon_2}, \quad \Delta_t=\frac{\epsilon^2{-}(m_1{+}m_3)^2}{4\epsilon_1\epsilon_2}, \quad \Delta_1=\frac{\epsilon^2{-}(m_4{+}m_2)^2}{4\epsilon_1\epsilon_2}, \quad \Delta_{\infty}=\frac{\epsilon^2{-}(m_4{-}m_2)^2}{4\epsilon_1\epsilon_2}.
\end{equation}

Analogous AGT relation exists also for the $4$-point $\mathcal{N}=1$ Super Virasoro conformal block. Unlike the above  $\mathsf{Vir}$ case, where the conformal blocks correspond to the instanton partition functions on $\mathbb{C}^2$, Super Virasoro conformal blocks correspond to instanton partition functions on the minimal resolution of $\mathbb{C}^2/\mathbb{Z}_2$, which is called $X_2$. These partition functions are also calculated by the localization technique \cite{BBB11} and are given by a generalization of \eqref{Zinst}. In such generalization one needs to use specific selection rules for the contributing diagrams and for the contributing cells. To formulate these selection rules we assign a given  parity $\mathfrak{p}(i,j)=i{+}j\mod 2$ to each cell $(i,j)$ of a given diagram, i.e. colour a diagram as a chess-board, so cells of the diagram $Y$ split into two sets $Y_{\mathfrak{p}}$ with $\mathfrak{p}=0,1$. 
For given diagram $U$ we also define subset $Y(U)$ of cells $(i,j)\in Y$ for which
\begin{equation}
Y_i{-}j+\tilde{U}_j{-}i\mod 2 = 1.
\end{equation}
Then, instanton partition function $\mathcal{Z}^{X_2}_{inst}=\mathcal{Z}^{X_2,0}_{inst}+\mathcal{Z}^{X_2,1}_{inst}$ reads
\begin{equation}
\mathcal{Z}^{X_2,\mathfrak{p}}_{inst}(a,m_{1,2,3,4};\epsilon_1,\epsilon_2|t)=\sum_{Y^+,Y^-:\substack{|Y^+_0|{+}|Y^-_0|=N{+}\mathfrak{p}\\|Y^+_1|{+}|Y^-_1|=N}}\frac{\prod\limits_{f=1}^4 Z_{fund}^{X_2,\mathfrak{p}}(a,m_f|Y^+,Y^-)}{Z_{vec}^{X_2}(a|Y^+,Y^-)} t^{N+\frac{\mathfrak{p}}2},
\end{equation}
where \cite{BFMT02,FP02}
\begin{equation}
   Z_{fund}^{X_2,\mathfrak{p}}(a,m|Y^+,Y^-)=\prod_{\pm}\prod_{(i,j)\in Y^{\pm}_\mathfrak{p}} (m\pm a-\epsilon/2+\epsilon_1 i+\epsilon_2 j),
   \end{equation}
   \vspace{-0.7cm}
    \begin{multline}
  Z_{vec}^{X_2}(a|Y^+,Y^-)=\prod_{\pm}\left(\prod_{(i,j)\in Y^{\pm}(Y^\pm)}\Big(-\epsilon_1 (\tilde{Y}^{\pm}_j{-}i)+\epsilon_2 (Y^{\pm}_i{-}j{+}1)\Big)\prod_{(i,j)\in Y^{\pm}(Y^\pm)}\Big(\epsilon_1 (\tilde{Y}^{\pm}_j{-}i{+}1)-\epsilon_2 (Y^{\pm}_i{-}j)\Big)\right.  \\ \times  \left.\prod_{(i,j)\in Y^{\pm}(Y^\mp)}\Big(\pm 2a-\epsilon_1 (\tilde{Y}^{\mp}_j{-}i)+\epsilon_2 (Y^{\pm}_i{-}j{+}1)\Big)\prod_{(i,j)\in Y^{\mp}(Y^\pm)}\Big(\pm 2a+\epsilon_1 (\tilde{Y}^{\pm}_j{-}i{+}1)-\epsilon_2 (Y^{\mp}_i{-}j)\Big)\right).
\end{multline} 

According to \cite[(5.12)]{BBB11}, partition function $\mathcal{Z}_{inst}^{X_2,\mathfrak{p}}$ for $\mathfrak{p}=0$ equals to the integer power and for $\mathfrak{p}=1$ --- to the half-integer power parts of the series $\mathcal{F}_{c^{\NS}}$
\begin{equation}\label{AGT_block_X2}
t^{-\Delta^{\NS}}\mathcal{F}^\mathfrak{p}_{c^{\NS}}(\Delta^{\NS}_{0,t,1,\infty},\Delta^{\NS}|t)=(2\epsilon_1\epsilon_2)^\mathfrak{p}(1{-}t)^{-\frac{h}{\epsilon_1\epsilon_2}}\,\mathcal{Z}^{X_2,\mathfrak{p}}_{inst}(a,m_{1,2,3,4};\epsilon_1,\epsilon_2|t)    
\end{equation}
under dictionary
\begin{equation}
\label{AGTX2_dictionary} 
b=\frac{\epsilon_1}{\sqrt{\epsilon_1\epsilon_2}}, \quad b^{-1}=\frac{\epsilon_2}{\sqrt{\epsilon_1\epsilon_2}}, \qquad  P=\frac{a}{\sqrt{\epsilon_1\epsilon_2}}, 
\end{equation}
\begin{equation}
\label{AGTX2_dictionary_masses}  
P_0=\frac{m_1{-}m_3}{2\sqrt{\epsilon_1\epsilon_2}}, \quad P_t=\frac{m_1{+}m_3}{2\sqrt{\epsilon_1\epsilon_2}}, \quad P_1=\frac{m_4{+}m_2}{2\sqrt{\epsilon_1\epsilon_2}}, \quad 
P_{\infty}=\frac{m_4{-}m_2}{2\sqrt{\epsilon_1\epsilon_2}}.
\end{equation}
Notwithstanding that a precise gauge theoretical description of tilded conformal block $\widetilde{\mathcal{F}}_{c^{\NS}}$ is not presently known to us, we define corresponding partition function $\widetilde{\mathcal{Z}}^{X_2}_{inst}$ analogously to \eqref{AGT_block_X2} as
\begin{equation}
t^{-\Delta^{\NS}}\widetilde{\mathcal{F}}^\mathfrak{p}_{c^{\NS}}(\Delta^{\NS}_{0,t,1,\infty},\Delta^{\NS}|t)\equiv(2\epsilon_1\epsilon_2)^{1-\mathfrak{p}}(1{-}t)^{-\frac{h}{\epsilon_1\epsilon_2}}\widetilde{\mathcal{Z}}^{X_2,\mathfrak{p}}_{inst}(a,m_{1,2,3,4};\epsilon_1,\epsilon_2|t)    
\end{equation}

\paragraph{$\mathbb{C}^2/\mathbb{Z}_2$ blowup relations.}
Now we can write relations \eqref{Fhat0123}, \eqref{Fhat4} in terms of instanton partition functions. Using \eqref{blowup_coeff4} and \eqref{AGTX2_dictionary_masses} we identify
\begin{equation}\label{AGT_dictionary_l}
\mathfrak{l}_n^{\mathsf{CFT}}=(-2\epsilon_1\epsilon_2)^{\mathfrak{p}(2n)}\mathfrak{l}_n^{[4]}\, ,    
\end{equation}
while \eqref{FhatHirota} reads
\begin{equation}
\label{ZhatHirota}
\widehat{\mathcal{Z}}_k=\sum\limits_{2n\in\mathbb{Z}}
 D^k_{\epsilon_1,\epsilon_2[\ln t]}\Bigl(\mathcal{Z}^{[4]}(a{+}2n\epsilon_1;m_{1,2,3,4};2\epsilon_1,\epsilon_2{-}\epsilon_1|t), \mathcal{Z}^{[4]}(a{+}2n\epsilon_2;m_{1,2,3,4};\epsilon_1{-}\epsilon_2,2\epsilon_2|t) \Bigr).
 \end{equation}
Then, according to \eqref{AGT_block} and \eqref{AGT_block_X2} together with \eqref{AGT_dictionary_l} and \eqref{blowup_coeff_def} we obtain from \eqref{Fhat0123}, \eqref{Fhat4}  following relations 

\begin{equation}\label{Zhat02}
\mathcal{Z}^{X_2}_{tw}=\widehat{\mathcal{Z}_0}, \qquad \qquad \qquad \qquad
-\epsilon_1\epsilon_2\,t^{\frac12}\widetilde{\mathcal{Z}}^{X_2}_{tw}=(1{-}t)\widehat{\mathcal{Z}_2}-\frac{h t}{1{-}t}\widehat{\mathcal{Z}_0},
\end{equation}
\begin{equation}\label{Zhat13}
\widehat{\mathcal{Z}_1}=0, \qquad  \qquad \widehat{\mathcal{Z}_3}-\epsilon\frac{h t(1{+}t)}{(1{-}t)^3}\widehat{\mathcal{Z}_0}=-\epsilon\frac{1{+}t}{(1{-}t)^2}\left(\epsilon_1\epsilon_2t^{\frac12}\widetilde{\mathcal{Z}}^{X_2}_{tw}\right),  
\end{equation}
\begin{multline}\label{Zhat4}
\widehat{\mathcal{Z}_4}-\frac{6h t}{(1{-}t)^2}\widehat{\mathcal{Z}_2}+\frac{h t}{(1{-}t)^4}\big(3h t-(1{+}4t{+}t^2)(\epsilon^2{-}\epsilon_1\epsilon_2)\big)\widehat{\mathcal{Z}_0}=2\frac{1{+}t}{(1{-}t)^2}\left(\epsilon_1\epsilon_2\frac{d}{d\ln t}\right)\left(\epsilon_1\epsilon_2t^{\frac12}\widetilde{\mathcal{Z}}^{X_2}_{tw}\right)\\+\frac{t^2}{(1{-}t)^2}\left(\frac{6h}{t(1{-}t)}-\frac{\epsilon_1\epsilon_2{+}\epsilon^2}{t^2}-\frac{e_2^{[4]}{+}e_1^{[4]}\epsilon{+}6\epsilon^2{-}2\epsilon_1\epsilon_2}t-\frac{\frac14e_1^{[4]}(e_1^{[4]}{+}2\epsilon){+}6\epsilon^2{-}3\epsilon_1\epsilon_2}{1{-}t}\right) \left(\epsilon_1\epsilon_2t^{\frac12}\widetilde{\mathcal{Z}}^{X_2}_{tw}\right)\\-\frac{t}{(1{-}t)^2}\,\left(\epsilon_1\epsilon_2\frac{d}{d\ln t}\right)\mathcal{Z}^{X_2}_{tw}
-\frac{t^2}{4(1{-}t)^2}\left(\frac{8h(1{+}t)}{t(1{-}t)}-\frac{e_1^{[4]}(e_1^{[4]}{+}2\epsilon)}{1{-}t}-\frac{2(e_2^{[4]}{+}e_1^{[4]}\epsilon{+}\epsilon^2)}t\right)\left(\epsilon_1\epsilon_2\frac{d}{d\ln t}\right)\mathcal{Z}^{X_2}_{tw}\\-\frac{t^2}{4(1{-}t)^2}\left(\frac{4h t}{1{-}t}\left(\frac{3h{+}\epsilon_1\epsilon_2}{t(1{-}t)}-\frac{e_2^{[4]}{+}e_1^{[4]}\epsilon{+}\epsilon^2}{t}-\frac{e_1^{[4]}(e_1^{[4]}{+}2\epsilon)}{4(1{-}t)}\right)+\frac{e_4^{[4]}}t+\frac{e_1^{[4]}(e_3^{[4]}{+}e_2^{[4]}\epsilon{+}e_1^{[4]}\epsilon^2{+}\epsilon^3)}{2(1{-}t)}\right) \mathcal{Z}^{X_2}_{tw},
\end{multline}
where we introduced
\begin{equation}
\mathcal{Z}_{tw}^{X_2}=t^{\frac{\epsilon^2/4-a^2}{2\epsilon_1\epsilon_2}}\left(\mathcal{Z}^{X_2,0}_{inst}/C_0-\mathcal{Z}^{X_2,1}_{inst}/C_1\right), \qquad \widetilde{\mathcal{Z}}_{tw}^{X_2}=t^{\frac{\epsilon^2/4-a^2}{2\epsilon_1\epsilon_2}}\left(\widetilde{\mathcal{Z}}^{X_2,1}_{inst}/C_0-\widetilde{\mathcal{Z}}^{X_2,0}_{inst}/C_1\right).   
\end{equation}
Substituting $\mathcal{Z}_{tw}^{X_2}$, $\widetilde{\mathcal{Z}}_{tw}^{X_2}$ given by \eqref{Zhat02} into \eqref{Zhat13}, \eqref{Zhat4} we obtain 
\begin{equation}\label{blowup_gen_form}
\sum_{2n\in\mathbb{Z}}\mathfrak{D}_{\epsilon_1,\epsilon_2}^{[4],k}\Bigl(\mathcal{Z}^{[4]}(a+2n\epsilon_1;m_{1,2,3,4};2\epsilon_1,\epsilon_2{-}\epsilon_1|t), \mathcal{Z}^{[4]}(a+2n\epsilon_2;m_{1,2,3,4};\epsilon_1{-}\epsilon_2,2\epsilon_2|t) \Bigr)=0, \quad k=1,3,4     
\end{equation}
where $\mathfrak{D}$'s are given by \eqref{D13},\eqref{D4}. Each of these relations still can be split in two, namely \eqref{blowup_gen_form_split}, with respect to the power of $t$. From the gauge theory side this splitting is supported by an AGT-analog of~\eqref{DNS=D+D}
\begin{equation}
\mathcal{Z}_{cl}^{[4]}(a+2n\epsilon_1;2\epsilon_1,\epsilon_2{-}\epsilon_1|t) \mathcal{Z}_{cl}^{[4]}(a+2n\epsilon_2;\epsilon_1{-}\epsilon_2,2\epsilon_2|t)=t^{\frac{\epsilon^2/4-a^2}{2\epsilon_1\epsilon_2}} \cdot t^{2n^2}.
\end{equation}

\section{Painlev\'e equations coalescence}
\label{sec:Painleve}

Painlev\'e equations are nonautonomous Hamiltonian systems, which form a coalescence hierarchy with Painlev\'e VI equation as the initial, most general equation (see Fig. \ref{fig:Painleve_coalescence}). Here we present this well-known coalescence procedure in terms of Hamiltonians. We use these limits in the main text for the quantum Painlev\'e equations.   

\subsection{Painlev\'e VI}
\label{ssec:PVI}
\paragraph{Hamiltonian function and coordinate form.} The Painlev\'e VI Hamiltonian is given by 
\begin{multline}\label{Ham_PVI}
t(t{-}1)H_{\VI}(\theta_{0,t,1,\infty};p,q|t)=q(q{-}t)(q{-}1)p \left(p-\frac{2\theta_0}{q}-\frac{2\theta_1}{q{-}1}-\frac{2\theta_t{-}1}{q{-}t}\right)\\+(\theta_0{+}\theta_t{+}\theta_1{+}\theta_{\infty}) (\theta_0{+}\theta_t{+}\theta_1{-}\theta_{\infty}{-}1)q.   
\end{multline}
This Hamiltonian gives the Hamilton equations and then, after eliminating momentum $p$, Painlev\'e~VI equation, which is the following second order nonlinear ordinary differential equation
\begin{multline}
\label{PVI}
\frac{d^2q}{dt^2}=\frac12\left(\frac1{q}+\frac1{q{-}t}+\frac1{q{-}1}\right)\left(\frac{dq}{dt}\right)^2-\left(\frac1{t}+\frac1{t{-}1}+\frac1{q{-}t}\right)\frac{dq}{dt}+\\+\frac{2q(q{-}t)(q{-}1)}{t^2(t{-}1)^2}\left(\left(\theta_{\infty}{+}\frac12\right)^2-\theta_0^2\frac{t}{q^2}+\theta_1^2\frac{t{-}1}{(q{-}1)^2}-\left(\theta_t^2{-}\frac14\right)\frac{t(t{-}1)}{(q{-}t)^2}\right).
\end{multline}
This equation has a group of symmetries, namely the extended affine Weyl group of $D_4^{(1)}$ root system $S_4 \ltimes W\left(D_4^{(1)}\right)$, where $\mathrm{Aut}(D_4^{(1)})=S_4$. Its action on $p,q,t$ and the parameters can be found, for example, in \cite[(3.3)]{TOS05}. 

We now describe how to pass from this standard, coordinate form of Painlev\'e VI to the bilinear tau form via the Hamiltonian form.

\paragraph{Hamiltonian form.} One can write an equation on the nonautonomous Hamiltonian as a function of $t$. Namely, using the Hamilton equations of Hamiltonian \eqref{Ham_PVI} we obtain that function
\begin{equation}\label{zeta_VI}
\zeta(t)=t(t{-}1) H_{\VI}\left(\theta_0,\theta_t{+}\frac12,\theta_1,\theta_{\infty}{-}\frac12\Big|t\right)-(\theta_0{+}\theta_1)^2 t-(\theta_0{+}\theta_t)^2
\end{equation}
satisfies a nonlinear second-order differential equation
\begin{equation}\label{PVI_sigma}
(t(t{-}1)\ddot{\zeta})^2+4\dot{\zeta}(t\dot{\zeta}{-}\zeta) \left((t{-}1)\dot{\zeta}{-}\zeta{-}\frac{\mathrm{w}_2^{[4]}}2\right)+\left(\mathrm{w}_4^{[4]}{+}2\mathrm{e}_4^{[4]}{-}\frac{(\mathrm{w}_2^{[4]})^2}4\right)(t\dot{\zeta}{-}\zeta)-4\se_4^{[4]}\dot{\zeta}=\mathrm{w}_6^{[4]}{+}\mathrm{w}_2^{[4]}\se_4^{[4]},
\end{equation}
where dot means the differentiation by $t$ and we introduced new parameters defined in terms of mass variables
\begin{multline}\label{PVI_inv}
\mathrm{w}_2^{[4]}=\sm_1^2+\sm_2^2+\sm_3^2+\sm_4^2, \qquad 
\se_4^{[4]}=\sm_1\sm_2\sm_3\sm_4,\qquad
\mathrm{w}_4^{[4]}=\sm_1^2\sm_2^2+\sm_1^2\sm_3^2+\sm_1^2\sm_4^2+\sm_2^2\sm_3^2+\sm_2^2\sm_4^2+\sm_3^2\sm_4^2,\\
\mathrm{w}_6^{[4]}=\sm_1^2\sm_2^2\sm_3^2+\sm_1^2\sm_2^2\sm_4^2+\sm_1^2\sm_3^2\sm_4^2+\sm_2^2\sm_3^2\sm_4^2,
\end{multline}
\begin{equation}\label{masses_theta_4}
\textrm{where}\qquad \sm_1=\theta_t+\theta_0, \quad \sm_2=\theta_1-\theta_{\infty}, \quad \sm_3=\theta_t-\theta_0, \quad \sm_4=\theta_1+\theta_{\infty}.   
\end{equation}

Parameters $\mathrm{w}_2^{[4]},\se_4^{[4]},\mathrm{w}_4^{[4]},\mathrm{w}_6^{[4]}$ are invariant under permutations of the masses and an even number of the sign changes of the masses. This is the Weyl group of root system $D_4$ which corresponds to a finite part of the Painlev\'e VI symmetry root system $D_4^{(1)}$ mentioned above. 
These invariants also naturally parametrize coefficients \eqref{g2_4},\eqref{g3_4} of the Seiberg-Witten curve for the corresponding $\mathcal{N}=2$ four-dimensional SUSY $SU(2)$ gauge theory with $N_f=4$ fundamental hypermultiplets, which itself is level curve \eqref{zeta_SW_VI} of function $\zeta$.
Note that the masses in the main text differ by rescaling $m_f=2\epsilon_2 \sm_f|_ {f=1\ldots 4}$.

Equation \eqref{PVI_sigma} is called {\it Hamiltonian form} of Painlev\'e VI equation. The dynamics of coordinate $q$ and momentum $p$ can be restored back from function $\zeta$ by formulas
\begin{equation}
\frac{2t(t{-}1)}{q{-}t}=\frac{2\theta_{\infty}t(t{-}1)\ddot{\zeta}}{(\dot{\zeta}+(\theta_t{+}\theta_{\infty})^2) (\dot{\zeta}+(\theta_t{-}\theta_{\infty})^2)}+\sum_{\pm}\frac{\theta_t(\theta_{\infty}^2{-}\theta_1^2{-}\zeta)\pm \theta_{\infty}(\theta_t^2{-}\theta_0^2{-}\zeta)-(\theta_t{\pm}\theta_{\infty})^3t}{\theta_t(\dot{\zeta}+(\theta_t{\pm}\theta_{\infty})^2)},    
\end{equation}
\begin{equation}
2\theta_{\infty}p=\frac{(\theta_t{+}\theta_{\infty})^2}{q-t}+\frac{(\theta_1{+}\theta_{\infty})^2}{q-1}+\frac{2\theta_0\theta_{\infty}{-}\theta_t^2{-}\theta_1^2}{q}+\frac{t((t{-}1)\dot{\zeta}-\zeta)+\zeta q}{q(q-t)(q-1)}.
\end{equation}

\paragraph{Tau form.} Differentiating \eqref{PVI_sigma} by $t$, dividing the result by $\ddot{\zeta}$ and introducing the tau function by the formula
\begin{equation}
\zeta=t(t{-}1)\frac{d\ln \tau}{dt}-\frac{\mathrm{w}_2^{[4]}}4(1{+}t)    \, ,
\end{equation}
one obtains the bilinear {\it tau form} of Painlev\'e VI
\begin{multline}\label{PVI_tau}
D^4_{[\ln t]}(\tau,\tau)-2\frac{1{+}t}{1{-}t} \frac{d}{d\ln t}D^2_{[\ln t]}(\tau,\tau)+\frac{t}{(1{-}t)^2}\left(\frac{d}{d\ln t}\right)^2\tau^2+\left(1{-}\mathrm{w}_2^{[4]}{+}\frac{t}{(1{-}t)^2}\right)D^2_{[\ln t]}(\tau,\tau)-\\-\frac{t}{(1{-}t)^2}\left(\frac{(\mathrm{w}_2^{[4]})^2}4+\frac{\big(\mathrm{w}_4^{[4]}{+}2\mathrm{e}_4^{[4]}{-}\frac14(\mathrm{w}_2^{[4]})^2\big)t{-}4\mathrm{e}_4^{[4]}}{1-t}\right)\tau^2=0,
\end{multline}
where $D^n_{[x]}\equiv D^n_{1,-1[x]}$ is the standard Hirota derivative, given by \eqref{Hirota} with $\epsilon_1=-\epsilon_2=1$.
This is a differential equation of order $4$, which depends on three parameters $\mathrm{w}_2^{[4]},\se_4^{[4]},\mathrm{w}_4^{[4]}$.

The parameter $\mathrm{w}_6^{[4]}$ becomes an integration constant when we differentiated \eqref{PVI_sigma}, while the other new integration constant corresponds to the overall scaling of $\tau$ (see also \cite{HS23}). In this sense \eqref{PVI_tau} is a form of Painlev\'e VI equation. Besides the above Weyl group of $D_4$ symmetries it has also the Painlev\'e VI crossing symmetries, defined by Table~\ref{tab:PVI_tau_symm}.

\begin{table}[ht]
    \centering
    \begin{tabular}{|c||c|c|c|c|c|}
      \hline 
     Generator & $\mathrm{w}_2^{[4]}$ & $\mathrm{w}_4^{[4]}$ & $\mathrm{e}_4^{[4]}$ & $t$ & $\tau$ \\
     \hline
     $q_{0\infty}$ & $\mathrm{w}_2^{[4]}$ & $\frac38(\mathrm{w}_2^{[4]})^2{+}3\mathrm{e}_4^{[4]}{-}\frac12\mathrm{w}_4^{[4]}$ & $\frac12\mathrm{e}_4^{[4]}{+}\frac14\mathrm{w}_4^{[4]}{-}\frac1{16}(\mathrm{w}_2^{[4]})^2$  &   $\frac1{t}$ & $\tau_s$ \\
      \hline
    $q_{01}$ & $\mathrm{w}_2^{[4]}$ & $\frac38(\mathrm{w}_2^{[4]})^2{-}3\mathrm{e}_4^{[4]}{-}\frac12\mathrm{w}_4^{[4]}$  & $\frac12\mathrm{e}_4^{[4]}{-}\frac14\mathrm{w}_4^{[4]}{+}\frac1{16}(\mathrm{w}_2^{[4]})^2$ &  $1{-}t$ & $\left(\frac{t}{1{-}t}\right)^{\frac{\mathrm{w}_2^{[4]}}4}\tau$ \\
      \hline
    $q_{1\infty}$ & $\mathrm{w}_2^{[4]}$ & $\mathrm{w}_4^{[4]}$ & $-\mathrm{e}_4^{[4]}$ & $\frac{t}{t-1}$ & $(1{-}t)^{-\frac{\mathrm{w}_2^{[4]}}4}\tau$ \\
      \hline
    \end{tabular}
    \caption{Painlev\'e VI tau form symmetries}
    \label{tab:PVI_tau_symm}
\end{table}

\subsection{Coalescence to Painlev\'e V, III's}
\label{ssec:D-coal}

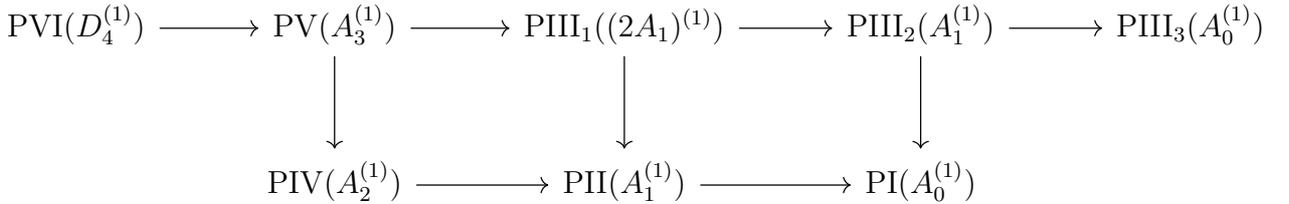
\begin{figure}[ht]
\begin{center}
\begin{tikzcd}[row sep=3em, column sep=3em]
\mathrm{PVI}(D_4^{(1)}) \arrow[r] & \mathrm{PV}(A_3^{(1)})  \arrow[r]  \arrow[d] & \mathrm{PIII}_1((2A_1)^{(1)}) \arrow[r]  \arrow[d] & \mathrm{PIII}_2(A_1^{(1)}) \arrow[r]  \arrow[d] & \mathrm{PIII}_3(A_0^{(1)}) \\
& \mathrm{PIV}(A_2^{(1)}) \arrow[r] & \mathrm{PII}(A_1^{(1)}) \arrow[r] & \mathrm{PI}(A_0^{(1)}) &
\end{tikzcd}
\end{center}

\caption{Painlev\'e coalescence diagram with symmetries}
	        
         \label{fig:Painleve_coalescence_sym}              
\end{figure}  

Here we follow the first line of the Painlev\'e coalescence diagram on Fig.~\ref{fig:Painleve_coalescence_sym}, where we denoted the symmetry type of each equation via the symmetry root lattices. We present the limits from the point of view of the Hamiltonians and extended canonical form $\Omega=dp\wedge dq-dH\wedge dt$.
We also present how such limits lead to the corresponding limits at the level of the Hamiltonian and the tau forms. Note that for the Painlev\'e equations except PVI we have scalings between classical Painlev\'e times and the gauge couplings $t$ of the main text, so for the Painlev\'e times we use notation $\mathrm{t}$, with above PVI time $t\equiv \mathrm{t}_{\VI}$.

\begin{itemize}
 
 \item {\bf PVI $\rightarrow $ PV.}
 Scaling $\sm_4\rightarrow\infty$ after substitution
 \begin{equation}
 q_{\VI}=\mathrm{t}_{\V} q_{\V}/\sm_4, \quad p_{\VI}=\sm_4 p_{\V}/\mathrm{t}_{\V}, \qquad  \mathrm{t}_{\VI}=\mathrm{t}_{\V}/\sm_4    
 \end{equation}
 one obtains that $\Omega_{\VI}\rightarrow\Omega_{\V}$ with
 \begin{equation}
 \mathrm{t} H_{\V}(\sm_{1,2,3};p,q|\mathrm{t})=p(p{+}\mathrm{t})q(q{-}1)-(2\sm_1q{+}\sm_3{-}\sm_1)p+(1{-}\sm_1{-}\sm_2)\mathrm{t} q.    
 \end{equation}
 The PV symmetry group is $\mathrm{Dih}_4 \ltimes W\left(A_3^{(1)}\right)$, where $\mathrm{Dih}_4=\mathrm{Aut}(A_3^{(1)})$, for details see, e.g. \cite[(4.3)]{TOS05}. It contains a time-rotation subgroup $\mathbb{Z}_2\subset\mathrm{Aut}(A_3^{(1)})$ whose generator ($\sigma$ in \cite[(4.3)]{TOS05}) acts as
 \begin{equation}
 q\mapsto1-q,\quad p\mapsto-p,\qquad \mathrm{t}\mapsto-\mathrm{t},\qquad \sm_3\mapsto-\sm_3.
 \end{equation}
 
 For the zeta function the corresponding limit is done via substitution
 \begin{equation}\label{zeta_V}
 \zeta_{\VI}(\mathrm{t}_{\VI})=-\frac{\sm_4\mathrm{t}_{\V}}4-\zeta_{\V}(\mathrm{t}_{\V}) \quad \textrm{for} \quad \zeta_{\V}(\mathrm{t})=\mathrm{t} H_{\V}(\sm_1,\sm_2{+}1,\sm_3|\mathrm{t})+\frac12(\sm_1{+}\sm_2{-}\sm_3)\mathrm{t}+\sm_1^2  
 \end{equation} 
  that gives the equation in the Hamiltonian form
 \begin{equation}
 \label{PV_sigma}
(\mathrm{t}\ddot{\zeta})^2-(\mathrm{t}\dot{\zeta}{-}\zeta) (\mathrm{t}\dot{\zeta}{-}\zeta{-}4\dot{\zeta}^2{+}\mathrm{w}_2^{[3]})+4\se_3^{[3]}\dot{\zeta}-\mathrm{w}_4^{[3]}=0
 \end{equation}
 with
 \begin{equation}\label{PV_inv}
 \mathrm{w}_2^{[3]}=\sm_1^2{+}\sm_2^2{+}\sm_3^2, \qquad  \se_3^{[3]}=\sm_1\sm_2\sm_3, \qquad \mathrm{w}_4^{[3]}=\sm_1^2\sm_2^2{+}\sm_1^2\sm_3^2{+}\sm_2^2\sm_3^2.   
 \end{equation}
Parameters $\mathrm{w}_2^{[3]}, \se_3^{[3]}, \mathrm{w}_4^{[3]}$  are invariant under permutations and an even number of the sign changes of the masses, these actions form Weyl group $W(A_3)=S_4$ corresponding to the above affine Weyl group. The above $\mathrm{t}\mapsto-\mathrm{t}$ generator preserves $\zeta_{\V}$ as well as $\mathrm{w}_{2,4}^{[3]}$ and changes the sign of $\se_3^{[3]}$.
 
For the tau function the corresponding limit is done via substitution
 \begin{equation}
  \tau_{\VI}(\mathrm{t}_{\VI})=\mathrm{t}_{\V}^{-\frac{\mathrm{w}_2^{[3]}{+}\sm_4^2}4}e^{-\frac18(2\sm_4{+}\mathrm{t}_{\sV})\mathrm{t}_{\sV}}\tau_{\V}(\mathrm{t}_{\V}) \quad \textrm{for} \quad \zeta_{\V}(\mathrm{t})=\frac{d\ln \tau_{\V}(\mathrm{t})}{d\ln \mathrm{t}} 
 \end{equation}
 that gives the equation in the tau form
 \begin{equation}
\label{PV_tau}
D^4_{[\ln \mathrm{t}]}(\tau,\tau)-2\frac{d}{d\ln \mathrm{t}}D^2_{[\ln \mathrm{t}]}(\tau,\tau)+\left(1{-}\mathrm{t}^2\right)D^2_{[\ln \mathrm{t}]}(\tau,\tau)+\mathrm{t}^2\frac{d}{d\ln \mathrm{t}}\tau^2+\left(4 \se_3^{[3]}{-}\mathrm{w}_2^{[3]}\mathrm{t}\right)\mathrm{t}\,\tau^2=0,
\end{equation}
so parameter $\mathrm{w}_4^{[3]}$ becomes an integration constant, analogously to the PVI case. 
This tau form is \eqref{quantum_PV} for $\epsilon=0$ under identification
\begin{equation}\label{PV_tau_cl_app}
\tau(-2\epsilon_2, 2\epsilon_2|t)=e^{\frac{\se_1^{[3]} \mathrm{t}_{\sV}}2}\tau_{\V}(\mathrm{t}_{\V}), \qquad t^{[3]}=2\epsilon_2 \mathrm{t}_{\V}, \qquad  e_k^{[3]}=(2\epsilon_2)^k\se_k^{[3]}|_{k=1,2,3}.  
\end{equation}
Therefore we verify that for the normalization in the main text for $\epsilon=0$ the limit is done without any rescaling of the tau function, so we keep such choice also for the coalescence limit at general $\epsilon$ (Sec. \ref{ssec:D-coal_q}). 
Our tau function definition is connected with the one of BLMST \cite[App. A.4]{BLMST16} by rescaling
\begin{equation}\label{PV_tau_lit}
\tau_{\V}=e^{-\theta_*\mathrm{t}_{\sV}/2}\tau^{\mathrm{BLMST}}_{\V}  
\end{equation}
under dictionary
\begin{equation}\label{PV_dct}
\sm_1=\theta_t+\theta_0, \qquad \sm_2=\theta_*, \qquad \sm_3=\theta_t-\theta_0.
\end{equation}

  \item {\bf PV $\rightarrow $ PIII$_1$.}
  Scaling $\sm_3\rightarrow\infty$ after substitution
 \begin{equation}
 q_{\V}=-\sm_3 q_{\III_1}/\mathrm{t}_{\III_1}, \quad p_{\V}=-\mathrm{t}_{\III_1} p_{\III_1}/\sm_3, \qquad  \mathrm{t}_{\V}=\mathrm{t}_{\III_1}/\sm_3    
 \end{equation}
 one obtains that $\Omega_{\V}\rightarrow\Omega_{\III_1}$ with
 \begin{equation}
 \mathrm{t} H_{\III_1}(\sm_{1,2};p,q|\mathrm{t})=p (p{-}1)q^2+((1{-}2\sm_1)q{+}\mathrm{t})p-(1{-}\sm_1{-}\sm_2)q.    
 \end{equation}
The PIII$_1$ symmetry group is $\mathrm{Dih}_4 \ltimes W\left((2A_1)^{(1)}\right)$, where $\mathrm{Dih}_4=\mathrm{Aut}((2A_1)^{(1)})$, for details see, e.g. \cite[(5.3)]{TOS05}.
It contains a time-rotation subgroup $\mathbb{Z}_2\subset\mathrm{Aut}((2A_1)^{(1)})$ whose generator ($\sigma_2$ in \cite[(5.3)]{TOS05}) acts as
 \begin{equation}
 q\mapsto q,\quad p\mapsto p-\frac{2\sm_1}q+\frac{\mathrm{t}}{q^2},\qquad \mathrm{t}\mapsto-\mathrm{t},\qquad \sm_1\mapsto-\sm_1.
 \end{equation}
 
 For the zeta function the corresponding limit is done via substitution
 \begin{equation}\label{zeta_III1}
 \zeta_{\V}(\mathrm{t}_{\V})=\zeta_{\III_1}(\mathrm{t}_{\III_1})     \quad \textrm{for} \quad \zeta_{\III_1}(\mathrm{t})=\mathrm{t} H_{\III_1}\left(\sm_1{+}\frac12,\sm_2{+}\frac12\Big|\mathrm{t}\right)-\mathrm{t}/2+\sm_1^2  
 \end{equation}
  that gives the equation in the Hamiltonian form
\begin{equation}
 \label{PIII1_sigma}
(\mathrm{t}\ddot{\zeta})^2+(4\dot{\zeta}^2{-}1)(\mathrm{t}\dot{\zeta}{-}\zeta) +4\sm_1\sm_2\dot{\zeta}-(\sm_1^2{+}\sm_2^2)=0
\end{equation}
with
\begin{equation}\label{PIII1_inv}
\se_2^{[2]}=\sm_1\sm_2, \qquad \mathrm{w}_2^{[2]}=\sm_1^2{+}\sm_2^2.
\end{equation}
Parameters $\se_2^{[2]},  \mathrm{w}_2^{[2]}$ are invariant under the transposition and an even number of the sign changes of the masses, these actions form Weyl group $W(2A_1)=\mathbb{Z}_2^2$ corresponding to the above affine Weyl group. The above $\mathrm{t}\mapsto-\mathrm{t}$ generator preserves $\zeta_{\III_1}$ as well as $\mathrm{w}_2^{[2]}$ and changes the sign of $\se_2^{[2]}$.
 
For the tau function the corresponding limit is done via substitution
 \begin{equation}
  \tau_{\V}(\mathrm{t}_{\V})=\tau_{\III_1}(\mathrm{t}_{\III_1}) \quad \textrm{for} \quad \zeta_{\III_1}(\mathrm{t})=\frac{d\ln \tau_{\III_1}(\mathrm{t})}{d\ln \mathrm{t}} 
 \end{equation}
  that gives the equation in the tau form
 \begin{equation}
\label{PIII1_tau}
D^4_{[\ln \mathrm{t}]}(\tau,\tau)-2\frac{d}{d\ln \mathrm{t}}D^2_{[\ln \mathrm{t}]}(\tau,\tau)+D^2_{[\ln \mathrm{t}]}(\tau,\tau)+\mathrm{t}(4\sm_1\sm_2{-}\mathrm{t})\tau^2=0,
\end{equation}
so parameter $\mathrm{w}_2^{[2]}$ becomes an integration constant, analogously to the previous cases. 
This tau form is
\eqref{quantum_PIII1} for $\epsilon=0$ under identification
\begin{equation}\label{PIII1_tau_cl_app}
\tau(-2\epsilon_2, 2\epsilon_2|t)=e^{\mathrm{t}_{\sIII_1}/2}\tau_{\III_1}(\mathrm{t}_{\III_1}), \qquad t^{[2]}=(2\epsilon_2)^2\mathrm{t}_{\III_1}, \qquad m_f=2\epsilon_2\sm_f|_{f=1,2}.
\end{equation}
Again we verify that for the normalization in the main text for $\epsilon=0$ the limit is done without any rescaling of the tau function, so we keep such choice also for the coalescence limit at general $\epsilon$ (Sec.~\ref{ssec:D-coal_q}). 
Our tau function definition coincides with the one of \cite[App. A.3]{BLMST16} 
under dictionary
\begin{equation}\label{PIII1_dct}
\sm_1=\theta_{\star}, \qquad \sm_2=\theta_*.
\end{equation}

   \item {\bf PIII$_1$ $\rightarrow $ PIII$_2$.}
   Scaling $\sm_2\rightarrow\infty$ after substitution
 \begin{equation}
 q_{\III_1}=q_{\III_2}/\sm_2, \quad p_{\III_1}=\sm_2 p_{\III_2}, \qquad  \mathrm{t}_{\III_1}=\mathrm{t}_{\III_2}/\sm_2    
 \end{equation}
 one obtains that $\Omega_{\III_1}\rightarrow\Omega_{\III_2}$ with
 \begin{equation}
 \mathrm{t} H_{\III_2}(\sm_1;p,q|\mathrm{t})=p^2q^2+(1{-}2\sm_1)pq+\mathrm{t}p+q.
 \end{equation}
The PIII$_2$ symmetry group is $\mathbb{Z}_2 \ltimes W\left(A_1^{(1)}\right)$, where  $\mathbb{Z}_2=\mathrm{Aut}(A_1^{(1)})$, for details see, e.g. \cite[(6.3)]{TOS05}. It contains a time-rotation subgroup $\mathbb{Z}_2\subset W\left(A_1^{(1)}\right)$ which generator ($s_0$ in \cite[(6.3)]{TOS05}\footnote{\color{red}It seems there are wrong signs in the formula for the $p$ transformation: the second and third signs should be opposite.}) acts as
 \begin{equation}
 q\mapsto q,\quad p\mapsto p-\frac{2\sm_1}q+\frac{\mathrm{t}}{q^2},\qquad \mathrm{t}\mapsto-\mathrm{t},\qquad \sm_1\mapsto-\sm_1.
 \end{equation}
 
For the zeta function the corresponding limit is done via substitution
 \begin{equation}\label{zeta_III2}
 \zeta_{\III_1}(\mathrm{t}_{\III_1})=\zeta_{\III_2}(\mathrm{t}_{\III_2})    \quad \textrm{for} \quad \zeta_{\III_2}(\mathrm{t})=\mathrm{t} H_{\III_2}\left(\sm_1{+}\frac12\Big|\mathrm{t}\right)+\sm_1^2  
 \end{equation}
  that gives the equation in the Hamiltonian form
\begin{equation}
 \label{PIII2_sigma}
(\mathrm{t}\ddot{\zeta})^2+4\dot{\zeta}^2(\mathrm{t}\dot{\zeta}{-}\zeta) +4\sm_1\dot{\zeta}-1=0.
 \end{equation}
The above $\mathrm{t}\mapsto-\mathrm{t}$ generator preserves $\zeta_{\III_2}$.

For the tau function the corresponding limit is done via substitution
 \begin{equation}
  \tau_{\III_1}(\mathrm{t}_{\III_1})=\tau_{\III_2}(\mathrm{t}_{\III_2}) \quad \textrm{for} \quad \zeta_{\III_2}(\mathrm{t})=\frac{d\ln \tau_{\III_2}(\mathrm{t})}{d\ln \mathrm{t}} 
 \end{equation}
  that gives the equation in the tau form
 \begin{equation}
\label{PIII2_tau}
D^4_{[\ln \mathrm{t}]}(\tau,\tau)-2\frac{d}{d\ln \mathrm{t}}D^2_{[\ln \mathrm{t}]}(\tau,\tau)+D^2_{[\ln \mathrm{t}]}(\tau,\tau)+4\sm_1\mathrm{t}\tau^2=0,
\end{equation}
so mass $\sm_1$ becomes an integration constant after rescaling $\mathrm{t}\mapsto \mathrm{t}/\sm_1$, analogously to the previous cases. 
This tau form is \eqref{quantum_PIII2} for $\epsilon=0$ under identification
\begin{equation}\label{PIII2_tau_cl_app}
\tau(-2\epsilon_2, 2\epsilon_2|t)=\tau_{\III_2}(\mathrm{t}_{\III_2}), \qquad t^{[1]}=(2\epsilon_2)^3\mathrm{t}_{\III_2}, \qquad m_1=2\epsilon_2\sm_1.    
\end{equation}
Again we verify that for the normalization in the main text for $\epsilon=0$ the limit is done without any rescaling of the tau function, so we keep such choice also for the coalescence limit at general $\epsilon$ (Sec.~\ref{ssec:D-coal_q}). Our tau function definition coincides with the one of \cite[App. A.2]{BLMST16} under dictionary
\begin{equation}\label{PIII2_dct}
\sm_1=\theta_*.    
\end{equation}

    \item {\bf PIII$_2$ $\rightarrow $ PIII$_3$.}
    Scaling $\sm_1\rightarrow\infty$ after substitution
 \begin{equation}
 q_{\III_2}=q_{\III_3}, \quad p_{\III_2}=p_{\III_3}+\frac{\sm_1}{q_{\III_3}}, \qquad  \mathrm{t}_{\III_2}=\mathrm{t}_{\III_3}/\sm_1    
 \end{equation}
 one obtains that $\Omega_{\III_2}\rightarrow\Omega_{\III_3}$ with
 \begin{equation}
 \mathrm{t} H_{\III_3}(p,q|\mathrm{t})=p^2q^2+pq+q+\frac{\mathrm{t}}{q}.
 \end{equation}
The PIII$_3$ symmetry group is $\mathbb{Z}_2$, for details see, e.g. \cite[(7.3)]{TOS05}\footnote{Note that our Hamiltonian system differs from one of \cite[Sec. 7]{TOS05} by rescaling $p=-2p^{\mathrm{TOS}},\, q=-q^{\mathrm{TOS}}/2,\, \mathrm{t}=t^{\mathrm{TOS}}/4$.}.
 
For the zeta function the corresponding limit is done via substitution
 \begin{equation}\label{zeta_III3}
 \zeta_{\III_2}(\mathrm{t}_{\III_2})=\zeta_{\III_3}(\mathrm{t}_{\III_3})    \quad \textrm{for} \quad \zeta_{\III_3}(\mathrm{t})=\mathrm{t} H_{\III_3}(\mathrm{t})-p q  
 \end{equation}
  that gives the equation in the Hamiltonian form
\begin{equation}
 \label{PIII3_sigma}
(\mathrm{t}\ddot{\zeta})^2+4\dot{\zeta}^2(\mathrm{t}\dot{\zeta}{-}\zeta) +4\dot{\zeta}=0.
 \end{equation}
 
For the tau function the corresponding limit is done via substitution
 \begin{equation}
  \tau_{\III_2}(\mathrm{t}_{\III_2})=\tau_{\III_3}(\mathrm{t}_{\III_3}) \quad \textrm{for} \quad \zeta_{\III_3}(\mathrm{t})=\frac{d\ln \tau_{\III_3}(\mathrm{t})}{d\ln \mathrm{t}} 
 \end{equation}
 that gives the equation in the tau form even without the limiting procedure
 \begin{equation}
\label{PIII3_tau}
D^4_{[\ln \mathrm{t}]}(\tau,\tau)-2\frac{d}{d\ln \mathrm{t}}D^2_{[\ln \mathrm{t}]}(\tau,\tau)+D^2_{[\ln \mathrm{t}]}(\tau,\tau)+4\mathrm{t}\tau^2=0.
\end{equation}
Actually Painlev\'e III$_2$ and Painlev\'e III$_3$ tau forms are the same, while the corresponding Hamiltonian forms differ by the constant $\sm_1^{-2}$ after the time rescaling of \eqref{PIII2_sigma}. 
The latter tau form \eqref{PIII3_tau} is \eqref{quantum_PIII3} for $\epsilon=0$ under identification
\begin{equation}\label{PIII3_tau_cl_app}
\tau(-2\epsilon_2, 2\epsilon_2|t)=\tau_{\III_3}(\mathrm{t}_{\III_3}), \qquad t^{[0]}=(2\epsilon_2)^4\mathrm{t}_{\III_3}.    
\end{equation}
Again we verify that for the normalization in the main text for $\epsilon=0$ the limit is done without any rescaling of the tau function, so we keep such choice also for the coalescence limit at general $\epsilon$ (Sec.~\ref{ssec:D-coal_q}). Our tau function definition coincides with the one of \cite[App. A.1]{BLMST16}.
 
\end{itemize}

\subsection{Coalescence to and among Painlev\'e IV, II, I}\label{ssec:E-coal}
\begin{itemize}
 
 \item {\bf PV $\rightarrow $ PIV.}
 Scaling
 $r\rightarrow0$ after substitution
 \begin{equation}
 \sm_1=-\frac1{2r^2}, \qquad \sm_2=\frac1{2r^2}+2(2\mathfrak{m}_1{+}\mathfrak{m}_2), \qquad \sm_3=\frac1{2r^2}+2(\mathfrak{m}_1{+}2\mathfrak{m}_2),
 \end{equation}
 \begin{equation}
 q_{\V}=1+r q_{\IV}, \quad p_{\V}=p_{\IV}/r, \qquad  \mathrm{t}_{\V}=-\frac1{r^2}-\frac{\mathrm{t}_{\IV}}r    
 \end{equation}
 one obtains that $\Omega_{\V}\rightarrow\Omega_{\IV}$ with
 \begin{equation}
 H_{\IV}(\mathfrak{m}_1,\mathfrak{m}_2;p,q|\mathrm{t})=pq(p{-}q{-}\mathrm{t})-2(\mathfrak{m}_1{+}2\mathfrak{m}_2)p-(1{-}2(2\mathfrak{m}_1{+}\mathfrak{m}_2))q. 
 \end{equation}
 Below we see that from the symmetry point of view it is useful to work with $3$ dependent mass parameters, such that $\mathfrak{m}_1{+}\mathfrak{m}_2{+}\mathfrak{m}_3=0$. 
 The PIV symmetry group is $\mathrm{Dic}_3 \ltimes W\left(A_2^{(1)}\right))$, 
 where $\mathrm{Dic}_3$ is a dicyclic extension of $S_3=\mathrm{Aut}(A_2^{(1)})$, for details see, e.g. \cite[(8.3)]{TOS05}\footnote{
 Note that our Hamiltonian system differs from one of \cite[Sec. 8]{TOS05} by rescaling $p=p^{\mathrm{TOS}}/\sqrt2,\, q=q^{\mathrm{TOS}}/\sqrt2,\, \mathrm{t}=t^{\mathrm{TOS}}\sqrt2$.}
 \footnote{With respect to the cited paper, we find that $\sigma_1$ and $\pi$ generate an extension of $S_3$ to $\mathrm{Dic}_3$ and that $\sigma_2$ should interchange $\alpha_0\leftrightarrow\alpha_1$.}. It contains a time-rotation subgroup $\mathbb{Z}_4\subset\mathrm{Dic}_3$ whose generator ($\pi^{-1}\sigma_1$ in \cite[(8.3)]{TOS05}) acts as
 \begin{equation}
 q\mapsto \ri(p{-}q{-}\mathrm{t}),\quad p\mapsto \ri p,\qquad \mathrm{t}\mapsto \ri \mathrm{t},\qquad \mathfrak{m}_1\mapsto-\mathfrak{m}_3, \quad \mathfrak{m}_2\mapsto-\mathfrak{m}_2, \quad \mathfrak{m}_3\mapsto-\mathfrak{m}_1.
 \end{equation}

 For the zeta function the corresponding limit is done via substitution
 \begin{equation}\label{zeta_IV}
 \zeta_{\V}(\mathrm{t}_{\V})=\frac1{2r^4}+\frac{\mathrm{t}_{\IV}}{4r^3}-\frac{3\mathfrak{m}_3}{r^2}+\frac{\zeta_{\IV}(\mathrm{t}_{\IV}){-}\mathfrak{m}_3\mathrm{t}_{\IV}}{r} \quad \textrm{for} \quad \zeta_{\IV}(\mathrm{t})= H_{\IV}\left(\mathfrak{m}_1{+}\frac13,\mathfrak{m}_2{-}\frac16\Big|\mathrm{t}\right)-2\mathfrak{m}_3\mathrm{t}  
 \end{equation} 
 that gives the equation in the Hamiltonian form
 \begin{equation}
 \label{PIV_sigma}
\ddot{\zeta}^2-(\mathrm{t}\dot{\zeta}{-}\zeta)^2 +4(\dot{\zeta}^3{+}4\mathfrak{e}_2\dot{\zeta}{+}8\mathfrak{e}_3)=0, 
 \end{equation}
with
 \begin{equation}\label{PIV_inv}
\mathfrak{e}_1=\mathfrak{m}_1{+}\mathfrak{m}_2{+}\mathfrak{m}_3=0, \quad
\mathfrak{e}_2=\mathfrak{m}_1\mathfrak{m}_2{+}\mathfrak{m}_1\mathfrak{m}_3{+}\mathfrak{m}_2\mathfrak{m}_3, \quad \mathfrak{e}_3=\mathfrak{m}_1\mathfrak{m}_2\mathfrak{m}_3.    
 \end{equation}
Parameters $\mathfrak{e}_{1,2,3}$ are invariant under permutations of masses, these permutations form Weyl group $W(A_2)=S_3$ corresponding to the above affine Weyl group. The above $\mathrm{t}\mapsto \ri \mathrm{t}$ generator maps $\zeta_{\IV}\mapsto-\ri\zeta_{\IV}$, keeps $\mathfrak{e}_1=0$, preserves $\mathfrak{e}_2$, and changes the sign of $\mathfrak{e}_3$.
 
For the tau function the corresponding limit is done via substitution
 \begin{equation}
  \tau_{\V}(\mathrm{t}_{\V})=e^{\frac{\mathrm{t}_{\sIV}}{2r^3}-\frac{\mathrm{t}_{\sIV}^2}{8r^2}+\frac{\mathrm{t}_{\sIV}^3/6{-}6\mathfrak{m}_3\mathrm{t}_{\sIV}}{2r}+\mathfrak{m}_3\mathrm{t}_{\sIV}^2{-}\mathrm{t}_{\sIV}^4/16}\tau_{\IV}(\mathrm{t}_{\IV}) \quad \textrm{for} \quad \zeta_{\IV}(\mathrm{t})=\frac{d\ln \tau_{\IV}(\mathrm{t})}{d\mathrm{t}} 
 \end{equation}
 that gives the equation in the tau form 
 \begin{equation}
\label{PIV_tau}
D^4_{[\mathrm{t}]}(\tau,\tau)-\mathrm{t}^2 D^2_{[\mathrm{t}]}(\tau,\tau)+\mathrm{t}\frac{d}{d\mathrm{t}}\tau^2+16\mathfrak{e}_2\tau^2=0,
\end{equation}
so parameter $\mathfrak{e}_3$ becomes an integration constant, analogously to the previous cases. 
This tau form is \eqref{quantum_PIV} at $\epsilon=0$ under identification
\begin{equation}\label{PIV_tau_cl_app}
\tau(-2\epsilon_2, 2\epsilon_2|t)=e^{\mathfrak{m}_3 \mathrm{t}_{\sIV}^2}\tau_{\IV}(\mathrm{t}_{\IV}), \qquad t^{[H_2]}=(2\epsilon_2)^{1/2}\mathrm{t}_{\IV}, \qquad \boldsymbol{m}_k=2\epsilon_2 \mathfrak{m}_k|_{k=1,2,3}.
\end{equation}
Then we see that for the normalization of the main text for $\epsilon=0$ the limit should be done after the following rescaling of the tau function in \eqref{D13irr}, \eqref{quantum_PV}
\begin{equation}\label{PV_IV_resc_cl}
\tau(-2\epsilon_2,2\epsilon_2|t)=t^{-\frac{e_2^{[3]}}{4\epsilon_2^2}}\tau_r(-2\epsilon_2,2\epsilon_2|t),
\end{equation}
while for general $\epsilon$ this rescaling should be deformed as in \eqref{PV_genlimresc}. 
Our tau function definition is connected with the one of BLMST \cite[Sec. 3.3]{BLMST16} by rescaling
\begin{equation}\label{PIV_tau_lit}
\tau_{\IV}=e^{-\frac{\theta_s{+}\theta_t}3 \mathrm{t}_{\sIV}^2}\tau_{\IV}^{\mathrm{BLMST}}
\end{equation}
under dictionary
\begin{equation}\label{PIV_dct}
\mathfrak{m}_1=\frac{\theta_t-2\theta_s}3, \qquad \mathfrak{m}_2=\frac{\theta_s-2\theta_t}3, \qquad \mathfrak{m}_3=\frac{\theta_s+\theta_t}3.
\end{equation}

 \item {\bf PIII$_1 \rightarrow $ PII.}
 Scaling
 $r\rightarrow0$ after substitution
 \begin{equation}
 \sm_1=\frac2{r^3}, \qquad \sm_2=\frac2{r^3}+4\mathfrak{m},
 \end{equation}
 \begin{equation}
 q_{\III_1}=\frac2{r^3}(1-r q_{\II}), \quad p_{\III_1}=1-p_{\II}r^2/2, \qquad  \mathrm{t}_{\III_1}=\frac4{r^6}+\frac{2\mathrm{t}_{\II}}{r^4}  
 \end{equation}
 one gets $\Omega_{\III_1}\rightarrow\Omega_{\II}$ with
 \begin{equation}
 H_{\II}(\mathfrak{m};p,q|\mathrm{t})=\frac12p^2-(q^2{+}\mathrm{t}/2)p-4\mathfrak{m}q. 
 \end{equation}
The PII symmetry group is $\mathbb{Z}_3\times(\mathbb{Z}_2 \ltimes W\left(A_1^{(1)}\right))$, $\mathbb{Z}_2=\mathrm{Aut}(A_1^{(1)})$, for details see, e.g. \cite[(9.3)]{TOS05}. In particular we find the extra symmetry time-rotation group $\mathbb{Z}_3$ whose generator acts as:
\begin{equation}
q\mapsto e^{-2\pi \ri/3}q, \quad p\mapsto e^{2\pi \ri/3}p, \qquad \mathrm{t}\mapsto e^{2\pi \ri/3}\mathrm{t}.
\end{equation}

For the zeta function the corresponding limit is done via substitution
 \begin{equation}\label{zeta_II}
 \zeta_{\III_1}(\mathrm{t}_{\III_1})=\frac6{r^6}+\frac{\mathrm{t}_{\II}}{r^4}+\frac{8\mathfrak{m}}{r^3}+\frac{2\zeta_{\II}(\mathrm{t}_{\II})}{r^2} \quad \textrm{for} \quad \zeta_{\II}(\mathrm{t})=H_{\II}(\mathrm{t})  
 \end{equation} 
  that gives the equation in the Hamiltonian form
 \begin{equation}
 \label{PII_sigma}
\ddot{\zeta}^2+2\dot{\zeta}(\mathrm{t}\dot{\zeta}{-}\zeta) +4\dot{\zeta}^3-4\mathfrak{m}^2=0.
\end{equation}
Parameter $\mathfrak{m}$ is invariant under the change of sign which forms Weyl group $W(A_1)=S_2$  corresponding to the above affine Weyl group. The above $\mathrm{t}\mapsto e^{2\pi\ri/3} \mathrm{t}$ generator maps $\zeta_{\II}\mapsto e^{-2\pi\ri/3}\zeta_{\II}$ and preserves $\mathfrak{m}$. 

For the tau function the corresponding limit is done via substitution
 \begin{equation}
  \tau_{\III_1}(\mathrm{t}_{\III_1})=e^{\frac{3\mathrm{t}_{\sII}}{r^4}-\frac{\mathrm{t}_{\sII}^2}{2r^2}+\frac{4\mathfrak{m}\mathrm{t}_{\sII}}{r}+\frac{\mathrm{t}_{\sII}^3}6}\tau_{\II}(\mathrm{t}_{\II}) \quad \textrm{for} \quad \zeta_{\II}(\mathrm{t})=\frac{d\ln \tau_{\II}(\mathrm{t})}{d\mathrm{t}} 
 \end{equation}
 that gives the equation in the tau form
 \begin{equation}
\label{PII_tau}
D^4_{[\mathrm{t}]}(\tau,\tau)+2\mathrm{t}D^2_{[\mathrm{t}]}(\tau,\tau)-\frac{d}{d\mathrm{t}}\tau^2=0,
\end{equation}
so mass $\mathfrak{m}$ becomes an integration constant, analogously to the previous cases. 
This tau form is \eqref{quantum_PII} for $\epsilon=0$ under identification
\begin{equation}\label{PII_tau_cl_app}
\tau(-2\epsilon_2, 2\epsilon_2|t)=\tau_{\II}(\mathrm{t}_{\II}), \qquad t^{[H_1]}=(2\epsilon_2)^{2/3} \mathrm{t}_{\II}, \qquad \boldsymbol{m}=2\epsilon_2\mathfrak{m}.
\end{equation}
Then we see that for the normalization of the main text for $\epsilon=0$ the limit should be done after the following rescaling of the tau function in \eqref{D13irr}, \eqref{quantum_PIII1}
\begin{equation}\label{PIII1_PII_resc_cl}
\tau(-2\epsilon_2,2\epsilon_2|t)=t^{\frac{m_1m_2}{4\epsilon_2^2}}e^{\frac{t}{4\epsilon_2^2}}\tau_r(-2\epsilon_2,2\epsilon_2|t),
\end{equation}
while for general $\epsilon$ this rescaling should be deformed as in \eqref{PIII1_genlimresc}. Our tau function definition coincides with the one of \cite[Sec. 3.2]{BLMST16} under dictionary
\begin{equation}\label{PII_dct}
\mathfrak{m}^2=\theta^2/16.
\end{equation}

 \item {\bf PIII$_2 \rightarrow $ PI.}
 Scaling
 $r\rightarrow0$ after substitution
 \begin{equation}
 \sm_1=-\frac3{r^5},
 \end{equation}
 \begin{equation}
 q_{\III_2}=\frac4{r^{10}}(1-r^2 q_{\I}), \quad p_{\III_2}=-\frac{r^5}4(1-r^2 q_{\I}+r^3 p_{\I}), \qquad  \mathrm{t}_{\III_2}=-\frac{16}{r^{15}}(1+r^4\mathrm{t}_{\I}/2)
 \end{equation}
 we obtain that $\Omega_{\III_2}\rightarrow\Omega_{\I}$ with
 \begin{equation}
 H_{\I}(p,q|\mathrm{t})=\frac{p^2}2-2q^3-\mathrm{t}q. 
 \end{equation}
The Painlev\'e I symmetry group is a time-rotation group $\mathbb{Z}_5$, whose generator acts as 
\begin{equation}
q\mapsto e^{-4\pi \ri/5}q, \quad p\mapsto e^{4\pi \ri/5}p, \qquad \mathrm{t}\mapsto e^{2\pi \ri/5}\mathrm{t}.
\end{equation}
 
For the zeta function the corresponding limit is done via substitution
 \begin{equation}\label{zeta_I}
 \zeta_{\III_2}(\mathrm{t}_{\III_2})=\frac{12}{r^{10}}+\frac{2\mathrm{t}_{\I}}{r^6}+\frac{2\zeta_{\I}(\mathrm{t}_{\I})}{r^4} \quad \textrm{for} \quad \zeta_{\I}(\mathrm{t})= H_{\I}(\mathrm{t})  
 \end{equation}
  that gives the equation in the Hamiltonian form
 \begin{equation}
 \label{PI_sigma}
\ddot{\zeta}^2+2(\mathrm{t}\dot{\zeta}{-}\zeta) +4\dot{\zeta}^3=0.
 \end{equation}
 
For the tau function the corresponding limit is done via substitution 
 \begin{equation}
  \tau_{\III_2}(\mathrm{t}_{\III_2})=e^{\frac{6\mathrm{t}_{\sI}}{r^6}-\frac{\mathrm{t}_{\sI}^2}{r^2}}\tau_{\I}(\mathrm{t}_{\I}) \quad \textrm{for} \quad \zeta_{\I}(\mathrm{t})=\frac{d\ln \tau_{\I}(\mathrm{t})}{d\mathrm{t}} 
 \end{equation}
 that gives the equation in the tau form
 \begin{equation}
\label{PI_tau}
D^4_{[\mathrm{t}]}(\tau,\tau)+2\mathrm{t}\tau^2=0.
\end{equation}
This tau form has a trivial symmetry given by multiplying $\tau$ on $e^{\alpha \mathrm{t}}$, which shifts $\zeta$ on $\alpha$. This gives an additional integration constant of the tau form. This tau form is \eqref{quantum_PI} for $\epsilon=0$ under identification
\begin{equation}\label{PI_tau_cl_app}
\tau(-2\epsilon_2, 2\epsilon_2|t)=\tau_{\I}(\mathrm{t}_{\I}), \qquad t^{[H_0]}=(2\epsilon_2)^{4/5}\mathrm{t}_{\I}.
\end{equation}
Then we see that for the normalization of the main text for $\epsilon=0$ the limit should be done after the following rescaling of the tau function
\begin{equation}\label{PIII2_PI_resc_cl}
\tau(-2\epsilon_2,2\epsilon_2|t)=t^{\frac{2m_1^2}{9\epsilon_2^2}}e^{\frac{3t}{16m_1\epsilon_2^2}}\tau_r(-2\epsilon_2,2\epsilon_2|t),
\end{equation}
while for general $\epsilon$ this rescaling should be deformed as \eqref{PIII2_genlimresc}. Our tau function definition coincides with the one of \cite[Sec. 3.1]{BLMST16}.

 \item {\bf PIV $\rightarrow$ PII.}
 Scaling
 $r\rightarrow0$ after substitution
 \begin{equation}
 \mathfrak{m}_1=-\frac1{24r^6}, \quad
 \mathfrak{m}_2=\frac1{12r^6}+2\mathfrak{m}, 
 \end{equation}
 \begin{equation}
 q_{\IV}=\frac1{2r^3}-q_{\II}/r, \quad p_{\IV}=(-p_{\II}{+}2q_{\II}^2{+}\mathrm{t}_{\II})r, \qquad  \mathrm{t}_{\IV}=-\frac1{r^3}{+}r\mathrm{t}_{\II}   
 \end{equation}
 one gets $\Omega_{\IV}\rightarrow\Omega_{\II}$.
 For the zeta and the tau function the corresponding limit is done via substitution
 \begin{equation}\label{zeta_tau_IV_II}
 \zeta_{\IV}(\mathrm{t}_{\IV})=-\frac1{12r^9}+\frac{\mathrm{t}_{\II}}{12r^5}-\frac{2\mathfrak{m}}{r^3}+\frac{\zeta_{\II}(\mathrm{t}_{\II})}{r}, \qquad \tau_{\IV}(\mathrm{t}_{\IV})=e^{-\frac{\mathrm{t}_{\sII}}{12r^8}+\frac{\mathrm{t}_{\sII}^2}{24r^4}-\frac{2\mathfrak{m}\mathrm{t}_{\sII}}{r^2}}\tau_{\II}(\mathrm{t}_{\II}).
 \end{equation} 
Then we see that for the normalization of the main text for $\epsilon=0$ the limit should be done after the following rescaling of the tau function
\begin{equation}\label{PIV_PII_resc_cl}
\tau(-2\epsilon_2,2\epsilon_2|t)=t^{-\frac{2(2\boldsymbol{m}_1{+}\boldsymbol{m}_2) (\boldsymbol{m}_1{+}2\boldsymbol{m}_2)}{\epsilon_2^2}}\tau_r(-2\epsilon_2,2\epsilon_2|t),
\end{equation}
while for general $\epsilon$ this rescaling should be deformed as \eqref{PIV_genlimresc}.

 \item {\bf PII$ \rightarrow $ PI.}
 Scaling
 $r\rightarrow0$ after substitution
 \begin{equation}
 \mathfrak{m}=\frac1{r^{15}},
 \end{equation}
 \begin{equation}
 q_{\II}=r^{-5}+rq_{\I}, \quad p_{\II}=-\frac2{r^{10}}+\frac{2q_{\I}}{r^4}+p_{\I}/r, \qquad  \mathrm{t}_{\II}=-\frac6{r^{10}}+r^2\mathrm{t}_{\I}  
 \end{equation}
 one gets $\Omega_{\II}\rightarrow\Omega_{\I}$. 
 For the zeta and the tau function the corresponding limit is done via substitution
 \begin{equation}\label{zeta_tau_II_I}
 \zeta_{\II}(\mathrm{t}_{\II})=-\frac6{r^{20}}+\frac{\mathrm{t}_{\I}}{r^8}+\frac{\zeta_{\I}(\mathrm{t}_{\I})}{r^2},  \qquad \tau_{\II}(\mathrm{t}_{\II})=e^{-\frac{6\mathrm{t}_{\sI}}{r^{18}}{+}\frac{\mathrm{t}_{\sI}^2}{2r^6}}\tau_{\I}(\mathrm{t}_{\I}).
 \end{equation} 
Then we see that for the normalization of the main text for $\epsilon=0$ the limit should be done after the following rescaling of the tau function
\begin{equation}\label{PII_PI_resc_cl}
\tau(-2\epsilon_2,2\epsilon_2|t)=t^{\frac{3\boldsymbol{m}^2}{\epsilon_2^2}}e^{-\frac{t^3}{108\epsilon_2^2}}\tau_r(-2\epsilon_2,2\epsilon_2|t),
\end{equation}
while for general $\epsilon$ this rescaling has to be deformed as \eqref{PII_genlimresc}.

\end{itemize}

\printbibliography[
heading=bibintoc,
title={References}
]

\end{document}